\documentclass[11pt]{article}
\usepackage{fullpage}
\usepackage{url}
\usepackage[colorlinks=true,linkcolor=Emerald,citecolor=RoyalBlue]{hyperref}
\usepackage{amsmath,nccmath, amsfonts,amsthm,multirow,mdframed,amssymb}

\usepackage{microtype}
\usepackage{paralist}
\usepackage{graphicx}
\usepackage{floatpag}
\usepackage{float}
\DeclareMathOperator{\supp}{supp}
\usepackage{bbold}
\usepackage{enumitem}
\usepackage{subcaption}
\usepackage[font=small, labelfont=bf]{caption}
\usepackage{soul}
\usepackage{comment}
\usepackage{calc}
\usepackage{upgreek}
\usepackage{mathtools}
\usepackage{color}
\usepackage[dvipsnames]{xcolor}
\usepackage{xspace}
\usepackage{tcolorbox}
\usepackage{cleveref}
\usepackage{stackengine}
\usepackage{tocloft}
\usepackage{physics}
\usepackage{tikz}
\usepackage{mathrsfs}

\usepackage{algorithm}
\usepackage{algorithmic}

\stackMath
\allowdisplaybreaks

\usepackage{xcolor}
\newtheorem{theorem}{Theorem}
\newtheorem*{theorem*}{Theorem}
\newtheorem{lemma}[theorem]{Lemma}
\newtheorem{corollary}[theorem]{Corollary}

\newtheorem{fact}{Fact}[section]

\crefname{fact}{Fact}{Facts}
\Crefname{fact}{Fact}{Facts}

\newcommand{\lr}[1]{\left(#1\right)}
\newcommand{\E}[1]{\mathbb{E}\left[#1\right]}
\newcommand{\EE}{\mathbb{E}}
\newcommand{\Var}[1]{\mathrm{Var}\left[#1\right]}
\newcommand{\C}{\mathbb{C}}

\newcommand{\R}{\mathbb{R}}

\newcommand{\cC}{\mathcal{C}}
\newcommand{\cD}{\mathcal{D}}

\newcommand{\cF}{\mathcal{F}}
\newcommand{\cG}{\mathcal{G}}

\newcommand{\cM}{\mathcal{M}}
\newcommand{\cN}{\mathcal{N}}

\newcommand{\cP}{\mathcal{P}}
\newcommand{\cQ}{\mathcal{Q}}
\newcommand{\cR}{\mathcal{R}}

\newcommand{\cT}{\mathcal{T}}
\newcommand{\cU}{\mathcal{U}}

\newcommand{\argmin}{\operatorname{argmin}}
\newcommand{\sgn}{\operatorname{sgn}}

\newcommand{\wh}[1]{\widehat{#1}}
\newcommand{\wt}[1]{\widetilde{#1}}
\newcommand{\ad}{\operatorname{ad}}

\newcommand{\He}{\mathrm{He}}

\newcommand{\proofstep}[1]{\par\medskip\noindent\textbf{#1.}\ }

\numberwithin{equation}{section}

\begin{document}
\title{Polynomial-time classical algorithms for mean-field models\\up to the glass transition}
\author{Alexander Schmidhuber$^*$ and Alexander Zlokapa\footnote{equal contribution} \\ \textit{Center for Theoretical Physics --- a Leinweber Institute, MIT}}
\date{}
\maketitle
\begin{abstract}
    \noindent
    The Sachdev-Ye-Kitaev model is a strongly interacting fermionic system that has been well-studied in condensed matter and high energy physics. It is highly quantum: Gaussian states are far from the thermal state (Hastings and O'Donnell, STOC'22) and representing the thermal state requires large polynomial-size quantum circuits (Anschuetz et al., QIP'25). Very recently, it was nonetheless proven that classical algorithms can estimate local thermal expectations at sufficiently high temperature in quasipolynomial time (Zlokapa, FOCS'26). We show that classical algorithms can in fact estimate local observables at all constant temperatures in polynomial time. Our techniques also extend straightforwardly to classical systems: we resolve an open question about computing thermal expectations of a classical spin glass up to its phase transition (Bencs et al., STOC'26). Our proof develops a fully rigorous quantum cavity method. Due to the success of the classical cavity method in optimization, sampling, inference and learning, we expect the quantum cavity method to find further applications of independent interest. As an example, we give a quantum algorithm that learns SYK Hamiltonians from the Gibbs state at any constant temperature with polynomial time and sample complexity.
\end{abstract}

\tableofcontents
\newpage

\section{Introduction}
We study algorithms for computing local expectations of \emph{mean-field models} in thermal equilibrium. Unlike geometrically local systems, every site in a mean-field model interacts with every other site. One of the most canonical examples is the classical Ising $p$-spin model,
\begin{align}\label{eq:ising}
    H_{\rm Ising} = \sqrt{\frac{p!}{2N^{p-1}}} \sum_{1 \leq i_1 < \cdots < i_p \leq N} J_{i_1\cdots i_p} Z_{i_1} \cdots Z_{i_p} \quad \text{for} \quad J_{i_1\cdots i_p} \overset{\rm iid}{\sim} \cN(0,1).
\end{align}
Here $Z_i$ is the Pauli $Z$ operator on spin $i$. When $p \geq 3$, this model is known to have a \emph{spin glass phase}~\cite{talagrand2000rigorous}. If the model is cooled from infinite temperature---for example, by weakly coupling it to a thermal bath---then at some temperature $1/\beta_d$, the cooling dynamics get stuck in a metastable state~\cite{ferrari2012two}. At this \emph{dynamical transition}, a large family of natural algorithms (e.g., low-degree or stable) also fail to sample from the Gibbs measure~\cite{alaoui2024near}.

Surprisingly, the work of \cite{bencs2025zeros} showed that quasipolynomial-time algorithms are in fact still able to estimate thermal expectations of local observables at some $\beta > \beta_d$. The algorithm of~\cite{bencs2025zeros}, contrary to most spin glass works, was based on Barvinok's interpolation method~\cite{barvinok2016combinatorics}. This technique typically succeeds at temperatures above a thermodynamic phase transition, when the free energy density becomes non-analytic in the $N\to\infty$ limit. For the Ising $p$-spin model, the phase transition occurs at a colder temperature $1/\beta_*$ below the dynamical transition $1/\beta_d$. It has remained open if thermal expectations can be efficiently estimated all the way to $\beta_*$.

Barvinok's algorithm also applies to quantum systems~\cite{harrow2020classical}. Recently, the techniques of~\cite{bencs2025zeros} were extended with a new cluster expansion for quantum systems to show quasipolynomial-time classical algorithms for the Sachdev-Ye-Kitaev (SYK) model~\cite{zlokapa2026rigorous}. The SYK model is arguably the most studied quantum mean-field system due to its relevance in condensed matter and holography~\cite{sachdev1993gapless,parcollet1999non,georges2000mean,georges2001quantum,kitaev2015SYK,maldacena2016remarks,maldacena2016bound,kitaev2018soft}. It is defined on Majorana fermions, which are Hermitian operators $\psi_i$ that satisfy the anticommutation relation $\{\psi_i,\psi_j\}=2\delta_{ij}I$. The standard SYK model is defined as
\begin{align}\label{eq:hsyk}
    H=\sqrt{\frac{3}{8N^3}}
    \sum_{1 \leq i_1 < i_2 < i_3 < i_4 \leq N}J_{i_1i_2i_3i_4} \psi_{i_1} \psi_{i_2} \psi_{i_3} \psi_{i_4}
\end{align}
for independent standard Gaussians $J_{i_1i_2i_3i_4}$. Unlike the Ising $p$-spin model, the SYK model has no thermodynamic phase transition at any temperature. Hence, an SYK instance weakly coupled to a quantum thermal bath is expected to cool quickly~\cite{almheiri2024universal,zhang2019evaporation,maldacena2021syk,schuster2025cooling}. Moreover, it has been suggested that the SYK model is classically hard due to various obstructions to classical algorithms~\cite{vidal2003efficient,troyer2005computational}, including a sign problem, entanglement, magic and the large quantum circuit complexity of its thermal state~\cite{liu2018quantum,pan2021yukawa,hastings2022optimizing,hastings2023field,anschuetz2025strongly,bettaque2026magic}.

However, the task of estimating local thermal expectations of the SYK model was recently understood to be \emph{classically easy}, albeit also in quasipolynomial time~\cite{zlokapa2026syk,zlokapa2026rigorous}. These works gave a non-rigorous prediction that quasipolynomial-time can be achieved at all constant temperatures, and rigorously proved the result for sufficiently large (constant) temperatures. Since both the Ising $p$-spin and SYK results are based on Barvinok's interpolation method, which has quasipolynomial overhead, it is a priori unclear if a polynomial-time algorithm exists. 
Moreover, the rigorous classical algorithm~\cite{zlokapa2026rigorous} is based on proving suitable zero-free regions of the partition function in the complex $\beta$-plane, but its techniques likely break down at constant $\beta$. Hence, it is desirable to find an approach that bypasses zero-freeness and gives a more natural algorithm using traditional methods from statistical physics.

In this work, we use the \emph{cavity method} to obtain a rigorous polynomial-time classical algorithm for estimating thermal expectations. This is done up to arbitrary inverse polynomial error, capturing instance-to-instance fluctuations due to the disorder. The cavity method was an early approach used to solve classical spin glasses~\cite{mezard1987spin} and (unlike other techniques) can be made fully rigorous in many settings~\cite{franz2003replica,panchenko2004bounds}. Unlike the prior works that used Barvinok's method to do expansions around infinite temperature, we use the cavity method to justify an expansion around the model's disorder-averaged, $N\to\infty$ solution at $\beta$ itself. For the SYK model, this algorithm efficiently reaches \emph{arbitrary} constant temperatures. We also apply the same algorithmic technique to the classical Ising $p$-spin model, where the algorithm succeeds all the way up to the thermodynamic (static 1RSB) transition $\beta_*$, resolving ``Open Problem 1'' and ``Open Problem 2'' of~\cite{bencs2025zeros}.  In the statement below, \emph{degree-$w$ thermal spin expectations} are the thermal expectation values of the operators $Z_A = \prod_{i\in A}Z_i$ with
$A\subseteq[N]=\{1,\ldots,N\}$ and $|A| = w$. 

\begin{theorem}[Ising $p$-spin thermal expectations]
\label{thm:pspin-intro}
Fix an integer $p\ge2$ and an inverse temperature $0\le\beta<\beta_*$,
where $\beta_*$ is the static glass transition of the zero-field Gaussian
Ising $p$-spin model in \cref{eq:ising}. 
 For every fixed observable degree $w$ and for any exponents $a,c>0$, there is a
deterministic classical algorithm which, with probability at least $1-N^{-a}$
over the Gaussian couplings, estimates all degree-at-most-$w$ thermal spin
expectations to simultaneous Euclidean error $N^{-c}$ and the logarithm of the
partition function to additive error $N^{-c}$. Its running time is $N^C$, where
$C$ is a constant that depends only on $p,\beta,w,a,c$. 
\end{theorem}
\Cref{sec:pspin-main} gives the algorithm and proof. An important special case is the Sherrington-Kirkpatrick model ($p=2$), for which the algorithm computes thermal expectations efficiently up to $\beta_*^{\mathrm{SK}} = 1$; its proof is different from the $p\geq 3$ case and is described in \Cref{sec:sk-main}. We describe the ideas for these proofs below when discussing our quantum generalization of the result.

While the predictions of spin glass theory have been rigorously confirmed for many classical mean-field models, the SYK model contains an additional barrier to obtaining formal results. This is most easily understood by considering the simplest thermodynamic quantity one can evaluate, the \emph{annealed partition function} $\EE \Tr e^{\beta H}$. Classically, knowledge of the eigenbasis allows one to evaluate $\EE \Tr e^{\beta H} = \sum_\sigma \prod_{i_1< \cdots < i_p} \EE \exp[\beta \sqrt{p!/2N^{p-1}} J_{i_1\cdots i_p} \sigma_{i_1}\cdots \sigma_{i_p}]$ directly using the Gaussian moment generating function. In comparison, the eigenbasis of the SYK model is unknown, preventing a direct sum over $\sigma\in\{\pm1\}^N$. Standard random matrix theory bounds are also unable to obtain tight bounds on this or other quantities~\cite{feng2019spectrum,gamarnik2026free,he2026spectral}. In the physics literature, the issue is circumvented via the non-rigorous \emph{path integral}~\cite{sachdev2023quantum}. Instead of summing over eigenstates, one integrates an overcomplete basis parameterized by paths $\gamma$ that replace the matrix $H$ with a (complex) scalar $H(\gamma)$. Schematically, one then evaluates
\begin{align}\label{eq:pathint}
    \EE \Tr e^{\beta H} = \int \EE \exp[\beta H(\gamma)]\,D\gamma
\end{align}
using an ansatz (called replica symmetry) that empirically holds in the $N\to\infty$ limit. For example, in a stoquastic Hamiltonian with a transverse field $h\sum_i X_i$, one could parameterize the above path integral by quantum Monte Carlo~\cite{kac1974stochastic,suzuki}: Trotterize $e^{\beta H}$ and insert the identity $\ketbra{0} + \ketbra{1}$ at each step, so $D\gamma$ is a probability measure that hops between computational basis states due to the transverse field. For non-stoquastic Hamiltonians, unfortunately, $D\gamma$ is not formally a measure and the integral is difficult to rigorously control (let alone other concerns about bringing the expectation inside the integral). Encouragingly, \cite{he2026spectral} recently gave a rigorous derivation of the SYK annealed free energy and obtained the same saddle point equations as the path integral solution. (In comparison, \cite{gamarnik2026free} rigorously derived the same free energy limit but with different techniques and in a different variational form.) In this work, our \emph{quantum cavity method} develops the required finite-$N$ control to progress from the asymptotic results of \cite{he2026spectral} to algorithmic applications.
\begin{figure}[htbp]
    \centering
    \includegraphics[width=0.9\textwidth]{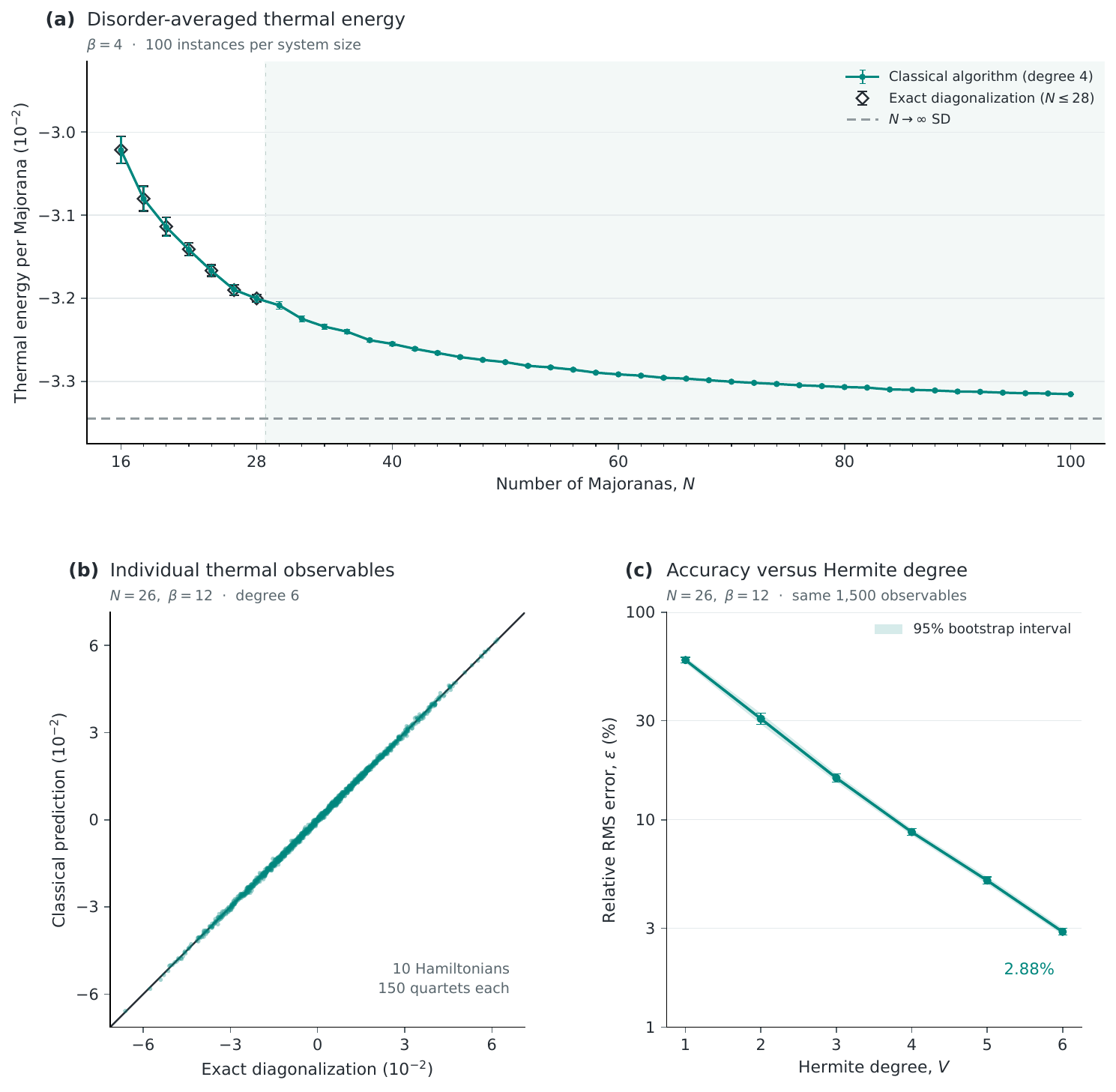}
    \caption{\textbf{Classical prediction of thermal observables in the $q=4$ SYK model.} \textbf{(a) Thermal energy at $\beta=4$.} We compare the thermal energy per Majorana predicted by our degree-4 classical algorithm (green, up to $N=100$) with exact diagonalization (black, up to $N=28$). Results are averaged over 100 independent disorder realizations at each even system size. Both methods use identical realizations at each overlapping size. Error bars show 95\% confidence intervals for the disorder mean due to the instance-to-instance fluctuations. Agreement between the means and the green and black error-bar widths indicates that the classical algorithm captures both the average thermal energy and the scale of disorder fluctuations. The dashed horizontal line shows the infinite-$N$ Schwinger–Dyson prediction.
\textbf{(b) Individual thermal observables.} Degree-6 classical predictions versus exact diagonalization at $N=26$ and $\beta=12$. Each point represents the thermal expectation of one of 150 fixed quartic observables in one of 10 independent disorder realizations, giving 1,500 points in total. The black line represents perfect agreement.
\textbf{(c) Accuracy versus Hermite degree. }Relative RMS error for degrees $V=1,\dots,6$, evaluated on the same 1,500 observables and normalized by the RMS of their ED expectations. Shading and error bars show 95\% bootstrap confidence intervals obtained from 10,000 resamples of whole disorder realizations, retaining all 150 observables within each realization. These intervals quantify sampling uncertainty. The numerical results is consistent with the analytically predicted exponential improvement in accuracy as Hermite degree $V$ is increased.}
    \label{fig:SYK}
\end{figure}

\begin{theorem}[SYK thermal expectations]
\label{thm:main}
Fix constant $\beta\geq0$ and a fixed
observable weight $w$. For $O=i^{w(w-1)/2}\psi_{i_1}\cdots\psi_{i_w}$, with $i_1<\cdots<i_w$, let $f_O(\beta, J) = \Tr(O e^{\beta H})/\Tr e^{\beta H}$ for the thermal state $\rho_\beta = e^{\beta H}/\Tr e^{\beta H}$ of the SYK Hamiltonian \eqref{eq:hsyk}. Fix exponents $a,c>0$. There is a
deterministic classical algorithm which, with probability at least
$1-N^{-a}$ over the Gaussian couplings, outputs
$\widehat f_O(\beta,J)$ satisfying
\begin{equation}
 |\widehat f_O(\beta,J)-f_O(\beta,J)|\leq N^{-c}
\end{equation}
in time at most $O(N^K)$, where $K$ is a constant that depends only on $\beta,w,a,c$.
\end{theorem}
We illustrate our classical algorithm in \cref{fig:SYK} for disorder-averaged quantities like the thermal energy as well as for instance to instance fluctuations of individual Majorana observables. 
Besides thermal expectations, we show that the quantum cavity method also gives a quantum algorithm to learn SYK Hamiltonians from the Gibbs state with polynomial sample and time complexity at any constant temperature.

\begin{theorem}[Learning the SYK Hamiltonian]
\label{thm:learn-main}
Fix a known constant $\beta > 0$, and set $a>0$,
$0<c<1$, and $0<\delta<1$.  There is an algorithm using $O_\beta\!\left(N^{5+2c}\log\frac{N}{\delta}\right)$ independent copies of $\rho_\beta$ and polynomial quantum and
classical time such that
\begin{align}
 \Pr_J\left[
   \Pr_{\rm alg}\left[
      \frac{\|\widehat J-J\|_2}{\sqrt{\binom N4}}
      \leq N^{-c}
      \,\middle|\,J
   \right]\geq1-\delta
 \right]
 \geq1-N^{-a}.
\end{align}
In particular, this suffices to efficiently learn the Hamiltonian to vanishing inverse polynomial error in the normalized Hilbert-Schmidt norm.
\end{theorem}

\subsection{Technical contributions}
Our algorithm is based on an expansion around the disorder-averaged solution to SYK in the $N\to\infty$ limit. This solution is described by the so-called ``saddle point equations'' that were first non-rigorously derived by~\cite{sachdev1993gapless} and only recently proven by~\cite{he2026spectral}. The main technical difficulty in proving the correctness (and polynomial runtime) of our algorithm is showing that typical SYK instances concentrate around the saddle point. (Note that they concentrate enough for the algorithm to work, but the algorithm will still recover instance-to-instance fluctuations.) We show concentration by introducing a quantum cavity method that can be rigorously controlled. Due to the success of the classical cavity method in optimization, sampling, inference and learning~\cite{mezard1987spin,charbonneau2023spin}---and indeed, given the applications in this work of the quantum cavity method for both quantum and classical algorithms---we expect the quantum cavity method to be a broadly useful primitive.

\paragraph{Quantum cavity method.} First developed in the beginning of spin glass theory \cite{mezard1987spin}, the cavity method is a method for computing the partition function $Z_N = \Tr e^{\beta H_N}$ on $N$ sites by evaluating the ratio of partition functions $Z_{N+1}/Z_N$ and iterating. To compute this ratio classically, one removes the couplings incident on $(N+1)$th spin, then attempts to find an ``effective field'' acting on the $(N+1)$th spin that reproduces its marginal distribution under the original Hamiltonian. One can write this effective field as a coupling between the $(N+1)$th spin and an auxiliary bath Hamiltonian composed of fresh spins that do not interact with the original Hamiltonian. The cavity method is a systematic way of deriving a self-consistent set of equations that determine the effective field, often via some iterative procedure such as message-passing (or belief propagation). This procedure converges under conditions such as strong spatial mixing (which conditions on an arbitrary boundary configuration)~\cite{gamarnik2009sequential}, allowing one to estimate local thermal expectations.

Several complications arise when extending the cavity method to the quantum setting. The most obvious is the appearance of a non-trivial self-overlap. Classically, an Ising spin configuration $\sigma \in \{\pm1\}^N$ satisfies $\Tr(\rho_\beta \sigma_i^2) = 1$, whereas in the quantum setting one can insert the spin at two different times: this leads to the imaginary-time-dependent function $G_i(\tau_1, \tau_2) = \langle \cT \,\psi_i(\tau_1) \psi_i(\tau_2) \rangle_\beta$ for $\psi_i(\tau) = e^{-\tau H} \psi_i e^{\tau H}$. This is not a problem when there is no sign problem, one can do the cavity method with the path integral; this was realized as early as~\cite{de1992cavity,biroli2000quantum}. Unfortunately, many interesting models (such as SYK) do have a sign problem. Operator-based approaches to a quantum cavity method~\cite{bapst2013quantum} such as quantum belief propagation~\cite{hastings2007quantum,leifer2008quantum} avoid this sign problem, but their efficient implementation requires additional structure (e.g., correlation decay established by Lieb-Robinson bounds or cluster expansions). These limitations often favor high-temperature or 1D systems, both of which are far from the all-temperature, mean-field setting of our work.

To illuminate the main ideas in our quantum cavity method, we consider the problem of proving the factorization
\begin{align}\label{eq:gg}
    \langle \cT \, \psi_i(\tau_1) \psi_i(\tau_2) \psi_j(\tau_3) \psi_j(\tau_4) \rangle_\beta \approx G_i(\tau_1, \tau_2) G_j(\tau_3, \tau_4) + O\lr{\frac{1}{N}}.
\end{align}
Using classical notation, this is like showing $\EE \prod_i G_i \approx \prod_i \EE\,G_i$, which is hard to do directly the quantum setting. We instead show something like $\EE_X[\prod_i G_i \mid X] \approx \EE_X \prod_i \E[G_i\mid X]$ for a random variable $X$. Our proof shows this factorization in three steps; note that these only describe steps of analysis, not our classical algorithm for estimating thermal expectations.
\begin{enumerate}
    \item We first make $\rho_\beta$ look block diagonal with a scalar in each block by rounding the spectrum. For an offset $u$ and some window size $\delta_E$, round all eigenvalues close to $u+j\delta_E$. If the rounded Gibbs state is $\rho(u)$, we sample $u$ randomly so that $\rho_\beta = \EE_u \rho(u)$.
    \item Measure on $\rho(u)$ estimates of at different times $\{t\}$ the two-point functions $e^{-itH}\psi_j e^{itH} \psi_j$ projected to be block diagonal in the above energy windows. After taking the expectation over $u$, these outcomes give estimates of $G_j(it)$. Recording the measurement outcomes in $x$, we prove that the post-measurement state $\rho_x$ satisfies $\Tr(\rho_x \prod_j G_j) \approx \prod_j \Tr(\rho_x G_j)$. The concentration occurs due to the spectral rounding: $\rho(u)$ looks like a classical object in the energy windows' block diagonalization, and the measurements commute with $\rho(u)$. Analytic continuation implies the factorization for $G_j(\tau)$.
    \item We cannot yet apply the cavity method since the post-measurement state $\rho_x$ is no longer a thermal state. If we act with a Majorana $\psi_i$ and it takes us from energy window $j$ to window $k$, the law will no longer satisfy detailed balance. We restore detailed balance with a projector and bound the error.
\end{enumerate}
We now have a state satisfying detailed balance and correlation decay. This allows us to follow a fairly standard cavity method analysis: we remove the $(N+1)$th fermion from the SYK Hamiltonian and couple it to an auxiliary bath, choosing the couplings to reproduce its $G_{N+1}(\tau)$ that we could isolate from other fermions via the factorization \eqref{eq:gg}. We refer to \Cref{sec:cavity} for a more detailed sketch of the above argument, and to \Cref{sec:concentration} for the main aspects of the proof. Equipped with a quantum cavity method, we can now discuss two of its applications.

\paragraph{Thermal expectations.} The same algorithmic principle applies to classical and quantum mean-field models. Once concentration around the saddle point is established (with the cavity method), we expand around the saddle point by inverting the Hessian. In the diagrammatic language of the SYK literature, this corresponds to resumming the ladder diagrams of~\cite{maldacena2016remarks}. The observation that this resummation leads to a good algorithm for thermal expectations was very recently made in \cite{swingle2026single}, which proposed and numerically verified a similar diagrammatic expansion to ours. However, controlling the error associated with the expansion requires first showing the concentration bound above, which appears to be difficult (even non-rigorously) without a quantum cavity method. In the classical case of the Ising $p$-spin model, a similar resummation is required for $p=2$ but not $p \geq 3$, which admits a simpler algorithm (due to having a trivial Hessian); however, both settings still require a cavity proof of concentration.

\paragraph{Learning.} Very recently, the work of~\cite{anshu2026learning} obtained a quasipolynomial-time algorithm that learns the coefficients of an SYK Hamiltonian given polynomially many copies of a high-temperature Gibbs state (as well as a polynomial sample complexity at all temperatures). Their algorithm is based on the high-temperature SYK cluster expansions of~\cite{zlokapa2026rigorous}. Here, we show that the same result follows at all temperatures and in polynomial time. Similarly to~\cite{anshu2026learning}, we measure Hamiltonian terms and account for the leading-order correction (in physics terms, the $1/N$ mean-field correction) to recover the couplings. Due to the quantum cavity method, our algorithm and its analysis (\Cref{sec:syk-learning}) are fairly straightforward. In diagrammatic language, we use the leading order and $1/N$ correction directly from the saddle point solution, then use the cavity method to bound the higher-order remainder.

\subsection{Discussion}
In this work, we focus on mean-field models because their dense random interactions are not captured by the locality assumptions used in many rigorous algorithms. Lieb-Robinson bounds underpin methods for estimating thermal expectations~\cite{hastings2007quantum,kuwahara2018polynomial} and Hamiltonian learning~\cite{anshu2020sample,bakshi2024learning,chen2025learning}. Cluster expansions likewise give thermal-expectation algorithms~\cite{kliesch2014locality,zlokapa2026rigorous} and learning algorithms~\cite{haah2022optimal,sanchez2025high,putterman2026quantum,anshu2026learning}. Recent work on SYK~\cite{zlokapa2026rigorous,anshu2026learning}, is based on a quasipolynomial-time algorithm and a high-temperature cluster expansion, but that cluster expansion appears to break at constant temperature. By constructing a formal quantum cavity method, this work narrows the gap between rigorous results and the predictions of mean-field physics, and provides the additional control required to produce classical and quantum algorithms with controlled performance guarantees.

Ironically, mean-field systems are often \emph{easier} than geometrically local system for non-rigorous methods. These include the replica trick~\cite{parisi1979infinite}, TAP equations (including its quantum version)~\cite{thouless1977solution,biroli2000quantum}, the Plefka and George-Yedidia expansions~\cite{plefka1982convergence,georges1991expand}. Indeed, the SYK model was originally solved using the replica trick~\cite{sachdev1993gapless}, which often only gives accurate predictions for mean-field models~\cite{mezard1987spin}. The model was later solved with diagrammatic approaches~\cite{maldacena2016remarks,kitaev2015SYK} which recently led to a diagrammatic algorithm for estimating single-instance thermal expectations to high accuracy~\cite{bettaque2026magic,swingle2026single}; we refer to references in the latter work for a more detailed review of the recent SYK literature.

In comparison to many of the above approaches, the (classical) cavity method can often be made fully rigorous and has thus seen many uses in the broader literature. It applies to random graphs~\cite{mezard2001bethe,gamarnik2006maximum}, lattices~\cite{gamarnik2009sequential}, and models without disorder~\cite{weitz2006counting,bandyopadhyay2008counting}. Algorithmically, the cavity method gives a derivation of belief propagation \cite{mezard2009information}, survey propagation for random $k$-SAT instances~\cite{braunstein2005survey}, and approximate message-passing algorithms introduced for compressed sensing~\cite{donoho2009message} and later used in high-dimensional inference~\cite{rangan2011generalized,javanmard2013state}. We expect that the quantum cavity method will provide analogous classical algorithms as well as new quantum algorithms. Beyond the applications of thermal expectations and Hamiltonian learning presented in this manuscript, we describe several interesting directions.
\begin{itemize}
    \item \emph{Planted inference.} A common task in learning theory seeks to distinguish between planted and null distributions. Approximate message-passing algorithms often achieve the algorithmic threshold predicted by low-degree lower bounds~\cite{zdeborova2016statistical,wein2025computational}. The analogous problem, defined on quantum states, likely requires similar tools.
    \item \emph{Gibbs sampling.} The cooling of a classical spin glass is governed by the Cugliandolo-Kurchan equations~\cite{cugliandolo1993analytical}, which can be rigorously derived by the cavity method~\cite{dandi2025sequential}. The quantum generalization of these open system dynamics is governed by analogous equations~\cite{cugliandolo1998quantum}, and thus we expect the thermalization dynamics of quantum systems such as SYK to be solvable via the quantum cavity method. Such a result would be especially interesting due to the relevance of strongly interacting fermionic systems such as the Fermi-Hubbard model.
    \item \emph{Real-time dynamics.} The recent work of~\cite{swingle2026single} (non-rigorously) showed that a similar algorithm to ours can also evaluate real-time dynamics with high accuracy. It is open if this technique can extend to times that grow with the system size (e.g., the scrambling time). We expect that formal control provided by the rigorous quantum cavity method will be especially useful for answering such questions. We also expect it to give rigorous classical algorithms for estimating out-of-time-order correlators of disordered systems, which are of experimental interest~\cite{google2025observation}.
\end{itemize}

The structure of the paper is as follows. After establishing notation and background for the SYK model in \Cref{sec:prelim}, we introduce the quantum cavity method (and briefly recall the classical cavity method) in \Cref{sec:cavity}. The algorithm for estimating SYK thermal expectations is presented in \Cref{sec:alg}. The application of the quantum cavity method required to prove the correctness of the SYK algorithm is given in \Cref{sec:concentration}. The algorithm for learning SYK Hamiltonians is in \Cref{sec:syk-learning}. Finally, we give the algorithms for estimating thermal expectations of the Ising $p$-spin model with $p \geq 3$ (\Cref{sec:pspin-main}) and $p=2$ (\Cref{sec:sk-main}).

\section{SYK Preliminaries}\label{sec:prelim}
We use $\Tr$ for the ordinary matrix trace and
$\tr=2^{-N/2}\Tr$ for the normalized trace. We write $\rho_\beta = e^{\beta H}/Z$ for $Z=\Tr e^{\beta H}$. Note that for SYK, one can recover the standard physics convention $e^{-\beta H}$ by substituting $J\to-J$, which leaves the Gaussian law unchanged. We denote canonically ordered Majorana strings with $\Gamma_A$, i.e.,
\begin{align}
    \{\psi_i,\psi_j\}=2\delta_{ij}I,\qquad
    \Gamma_A=i^{|A|(|A|-1)/2}\prod_{i\in A}^{\nearrow}\psi_i,
\end{align}
so the SYK model would be $H=\sqrt{\frac{3}{8N^3}}\sum_{A\in\binom{[N]}{4}}J_A\Gamma_A$.

We will use the probabilists' Hermite polynomials
\begin{align}
    e^{tx-t^2/2}
    =
    \sum_{m=0}^{\infty}\He_m(x)\frac{t^m}{m!},
    \qquad
    \EE_{x\sim\mathcal N(0,1)}
    [\He_m(x)\He_n(x)]
    =
    \delta_{mn}m!.
\end{align}
For $A_1,\ldots,A_v\in\binom{[N]}4$, write
\begin{align}\label{eq:wicknotation}
    \langle J_{A_1}\cdots J_{A_v}\rangle_J
    =
    \prod_{B\in\binom{[N]}4}
    \He_{\#\{r\in\{1,\ldots,v\}:A_r=B\}}(J_B).
\end{align}
For example, if $A\neq B$ then
$\langle J_AJ_B\rangle_J=J_AJ_B$, whereas
$\langle J_A^2\rangle_J=J_A^2-1$. We also use notation
\begin{align}
    \norm{X}_{2,\rho}^2 = \Tr(\rho X^\dagger X), \qquad \norm{X}_{\rm HS}^2 = \Tr(X^\dagger X).
\end{align}
We now provide a more extended description of the preliminaries for fermionic systems and SYK for readers less familiar with the physics literature.

\subsection{Free fermions}
Besides the Majorana operators $\psi_i$, we will also use Grassmann variables $\chi_i(t)$ labeled by a site $i$ and imaginary time $t$. These satisfy
\begin{align}
    \chi_i(u)\chi_j(v)
    =
    -\chi_j(v)\chi_i(u),
    \qquad
    \chi_i(u)^2=0.
\end{align}
By convention, these variables are also defined to be antiperiodic around the thermal circle: $\chi_i(t+1) = -\chi_i(t)$.
We denote the (normalized) Grassmann integration under the free fermion action by $\langle \cdot \rangle_0$. In the physics literature, this is defined non-rigorously via a path integral with the free fermion action:
\begin{align}
    \langle F(\chi) \rangle_0 = \frac{\int \cD\chi F(\chi) e^{-S_0[\chi]}}{\int \cD\chi  e^{-S_0[\chi]}}, \qquad S_0[\chi] = \frac{1}{4} \sum_i \int_0^1 dt\, \chi_i(t) \partial_t \chi_i(t).
\end{align}
Thus, for $u\neq v$, we have
\begin{align}
    \langle\chi_i(u)\chi_j(v)\rangle_0 = \delta_{ij}\sgn(u-v) = 2\delta_{ij}G_0(u,v) \quad \text{for} \text \quad G_0(u,v)=\frac12\operatorname{sgn}(u-v),
\end{align}
and set by symmetry $G_0(u,u)=0$. In the physics literature, $G_0$ is known as the free fermion propagator. Since fermions anticommute, $G_0(u,v)=-G_0(v,u)$. Since $\chi_i(t+1) = -\chi_i(t)$, we also extend the propagator to be antiperiodic, $G_0(u+1,v) = -G_0(u, v)$ and similarly for $v$. Wick's rule applied to free fermions gives
\begin{align}\label{eq:freewick}
    \left\langle
        \chi_{j_1}(t_1)\cdots\chi_{j_{2k}}(t_{2k})
    \right\rangle_0
    &=
    \sum_{\sigma\in\mathcal P_2([2k])}
    (-1)^{\operatorname{cr}(\sigma)}
    \prod_{\substack{\{r,s\}\in\sigma\\r<s}}
    2\delta_{j_rj_s}G_0(t_r,t_s).
\end{align}
Here $\mathcal P_2([2k])$ is the set of partitions of
$\{1,\ldots,2k\}$ into pairs, and $\operatorname{cr}(\sigma)$ is the number of crossings of pairs in $\sigma$. (For example, if $\{r,s\}$ and $\{u,v\}$ are in $\sigma$ with $r<s$ and $u<v$, they cross if $r<u<s<v$ or $u<r<v<s$.) The free fermion expectation vanishes for an odd number of variables.

To avoid the non-rigorous derivation of $\langle \cdot \rangle_0$, we define it directly through the Wick formula: set $\langle 1 \rangle_0 = 1$, $\langle \chi_{j_1}(t_1)\cdots\chi_{j_{2k+1}}(t_{2k+1})\rangle_0 = 0$, and \eqref{eq:freewick}. Observe that Majorana fermions also satisfy
\begin{align}
    \tr(\psi_{j_1}\cdots\psi_{j_{2k}}) = \sum_{\sigma\in\cP_2([2k])} (-1)^{\operatorname{cr}(\sigma)} \prod_{\{a,b\}\in\sigma} \delta_{j_aj_b}
\end{align}
For $A_1,\ldots,A_m\in\binom{[N]}4$ and $t_1>\cdots>t_m$,
the definition of $\langle\cdot\rangle_0$ therefore implies
\begin{align}
    \langle \chi_{A_1}(t_1) \cdots \chi_{A_m}(t_m) \rangle_0 = \tr(\psi_{A_1}\cdots\psi_{A_m}).
\end{align}
Here $\psi_A$ and $\chi_A$ denote the products, e.g., $\psi_A = \prod_{i \in A}^{\nearrow} \psi_i$. (Note that this omits the sign $i^{|A|(|A|-1)/2}$ to make the operator Hermitian, e.g., $\Gamma_A=-\psi_A$ for $|A|=4$.) Including all possible time orderings then gives
\begin{align}
    \int_{[0,1]^m} \langle \chi_{A_1}(t_1) \cdots \chi_{A_m}(t_m) \rangle_0\, d\vec t = \frac{1}{m!} \sum_{\pi\in S_m} \tr(\psi_{A_{\pi(1)}}\cdots\psi_{A_{\pi(m)}}),
\end{align}
where the sets where at least two vertex times coincide have measure zero. In particular, expanding the SYK partition function into a power series implies that for every choice of disorder and $N$,
\begin{align}\label{eq:syktofree}
    \tr e^{\beta H} = \left\langle \exp\left[
-\beta\sqrt{\frac{3}{8N^3}}
\sum_{A\in\binom{[N]}4}J_A
\int_0^1\chi_A(t)\,dt
\right] \right\rangle_0
\end{align}
since
\begin{align}
    \tr H^m = (-1)^m\sum_{A_1,\dots,A_m} \prod_{i=1}^m \sqrt{\frac{3}{8N^3}} J_{A_i} \int_{[0,1]^m} \langle \chi_{A_1}(t_1) \cdots \chi_{A_m}(t_m) \rangle_0\, d\vec t.
\end{align}
By analyzing the RHS of \eqref{eq:syktofree} rather than the original SYK Hamiltonian, we can evaluate the expectation over disorder rigorously. For $A \in \binom{[N]}{4}$, variables $\chi_A(t) = \prod_{i\in A}\chi_i(t)$ satisfy $\chi_A(t) \chi_B(s) = \chi_B(s) \chi_A(t)$. Hence, the quantities $\int_0^1 \chi_A(t) dt$ are commuting across different choices of $A$. By independence of $J_A$ and $\EE e^{xJ}=e^{x^2/2}$ applied to these commuting variables, we have
\begin{align}
\EE_J
\exp\left[
-\beta\sqrt{\frac{3}{8N^3}}
\sum_{A\in\binom{[N]}4}J_A
\int_0^1\chi_A(t)\,dt
\right] &= \exp\left[
\frac{3\beta^2}{16N^3}
\sum_{A\in\binom{[N]}4}\left(\int_0^1 \chi_A(t) \,dt\right)^2
\right] \\
&=
\exp\left[
\frac{3\beta^2}{16N^3}
\int_0^1\int_0^1
\sum_{A\in\binom{[N]}4}
\chi_A(u)\chi_A(v)\,du\,dv
\right].
\end{align}
To simplify this further, we introduce the quantity
\begin{align}\label{eq:quv}
    Q_{uv}
    =
    \frac{1}{2N}\sum_{i=1}^N\chi_i(u)\chi_i(v),
    \qquad
    Q_{uv}=-Q_{vu}.
\end{align}
Notice in particular that $\langle Q_{uv}\rangle_0=G_0(u,v)$. Moreover, the variables $Q_{uv}$ commute with one another because each is a polynomial of even degree. Using
\begin{align}
    (2NQ_{uv})^4 = \left(\sum_{i=1}^N \chi_i(u)\chi_i(v)\right)^4 = 4!\sum_{i_1 < i_2 < i_3 < i_4} \prod_{a=1}^4 \chi_{i_a}(u) \chi_{i_a}(v) = 4!\sum_{A\in\binom{[N]}4} \chi_A(u)\chi_A(v),
\end{align}
we find that
\begin{align}\label{eq:chitoq}
   \EE_J
\exp\left[
-\beta\sqrt{\frac{3}{8N^3}}
\sum_{A\in\binom{[N]}4}J_A
\int_0^1\chi_A(t)\,dt
\right] = \exp\left[
\frac{N\beta^2}{8}
\int_0^1\int_0^1Q_{uv}^4\,du\,dv
\right].
\end{align}
For any polynomial $P(Q)$ in finitely many of the variables $Q_{uv}$,
such as $P(Q)=Q_{uv}Q_{st}$, define the following ratio of free fermion expectations:
\begin{align}\label{eq:nexp}
    \langle P(Q)\rangle_N
    =
    \frac{
        \left\langle
        P(Q)\exp\left[
        \frac{N\beta^2}{8}
        \int_0^1\!\!\int_0^1Q_{uv}^4\,du\,dv
        \right]
        \right\rangle_0
    }{
        \left\langle
        \exp\left[
        \frac{N\beta^2}{8}
        \int_0^1\!\!\int_0^1Q_{uv}^4\,du\,dv
        \right]
        \right\rangle_0
    }.
\end{align}
The same quotient defines $\langle F(\chi)\rangle_N$ for every fixed
Grassmann polynomial $F$; it is not restricted to polynomials in $Q$.
For $m = O(1)$ distinct times $1>t_1>\cdots>t_m>0$, combining \eqref{eq:chitoq} with the derivation of \eqref{eq:syktofree} implies that the SYK Hamiltonian satisfies
\begin{align}\label{eq:chin}
    \frac{
        \EE_J\Tr\left[
        e^{(1-t_1)\beta H}
        \psi_{j_1}e^{(t_1-t_2)\beta H}
        \psi_{j_2}\cdots
        e^{(t_{m-1}-t_m)\beta H}
        \psi_{j_m}e^{t_m\beta H}
        \right]
    }{
        \EE_J\Tr e^{\beta H}
    } = \left\langle
        \chi_{j_1}(t_1)\cdots\chi_{j_m}(t_m)
    \right\rangle_N.
\end{align}
For another ordering of distinct times, first reorder the variables
and include the sign obtained by permuting fermions. The case where times coincide is discussed when we do this derivation more carefully in \Cref{lem:contact-dictionary}.

\subsection{SYK saddle point equations}
Our algorithm will consist of an expansion around the solution $G_*,\Sigma_*$ to the SYK saddle point equations. Here, $G_*$ plays a similar role to $G_0$ above, but it will only depend on the difference between times: we will sometimes write $G_*(u,v)=G_*(u-v)$. These are defined on $[0,1]$ by
\begin{align}\label{eq:sd}
    \Sigma_*(t)=\beta^2G_*(t)^3,
    \quad
    \wh G_*(n)
    =
    \lr{-i\omega_n-\wh\Sigma_*(n)}^{-1},
    \quad
    \omega_n=(2n+1)\pi.
\end{align}
Here $\wh f(n)=\int_0^1e^{i\omega_nt}f(t)\,dt$, and $G_*$ is normalized
as the two-point function of $\psi_i/\sqrt{2}$, so its one-sided endpoint
values are $G_*(0^+)=G_*(1^-)=1/2$. In complex-time expressions, $G_*(z)$ denotes the continuation of this positive-time branch, with value $1/2$ at $z=0$. This is distinct from $G_*(u,u)=0$.

In the physics literature, there are two equivalent non-rigorous ways to obtain these saddle point equations: one can insert the ansatz of replica symmetry and treat $G$ as an overlap matrix (with imaginary time dependence), or one can sum over the so-called melonic diagrams. Since we will be expanding around $G_*$, we need to first compute the saddle point solution. To do this in polynomial time, we introduce the convex optimization problem
\begin{align}\label{eq:convexsd}
    (u_0,\ldots,u_{M-1})
    =
    \mathop{\operatorname{argmin}}_{u_0,\ldots,u_{M-1}>0}
    \left\{
    \sum_{n=0}^{M-1}(u_n-\log u_n-1)
    +
    \frac{\beta^2}{8}
    \int_0^1
    \left[
        \frac12+
        2\sum_{n=0}^{M-1}
        \frac{u_n-1}{\omega_n}\sin(\omega_nt)
    \right]^4dt
    \right\},
\end{align}
and define the truncation
\begin{align}\label{eq:gtrunc}
    \widetilde G_*(t)
    =
    \frac12+
    2\sum_{n=0}^{M-1}
    \frac{u_n-1}{\omega_n}\sin(\omega_nt)
\end{align}
for some cutoff $M$ that will be chosen later.
The variables $u_n$ approximate the ratios
$\wh G_*(n)/(i/\omega_n)$. Bounding the approximation error due to this cutoff is fairly straightforward, so we postpone it to \Cref{app:saddle_trunc}.

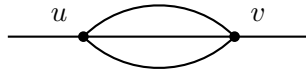
\begin{figure}[H]
    \centering
    \begin{tikzpicture}[baseline=-0.5ex, line width=0.8pt]
        \coordinate (u) at (0,0);
        \coordinate (v) at (2,0);

        \draw (-1,0) -- (u);
        \draw (v) -- (3,0);

        \draw (u) to[out=45,in=135] (v);
        \draw (u) -- (v);
        \draw (u) to[out=-45,in=-135] (v);

        \fill (u) circle (2pt);
        \fill (v) circle (2pt);

        \node[above left=2pt] at (u) {$u$};
        \node[above right=2pt] at (v) {$v$};
    \end{tikzpicture}
    \caption{A melonic diagram with interaction vertices at times $u$ and $v$.}
    \label{fig:melon}
\end{figure}

The following argument is non-rigorous, but it will provide intuition for the saddle point equations, as well as for the correct Hessian that we will eventually analyze rigorously. First, we rewrite the second saddle point equation $\wh G_*(n) = \lr{-i\omega_n-\wh\Sigma_*(n)}^{-1}$ as $G_*^{-1} = G_0^{-1} - \Sigma_*$. (In the physics literature, $G_0^{-1}$ is more commonly written as $\partial_\tau$.) A Neumann expansion would then suggest that
\begin{align}\label{eq:gneu}
    G_* = G_0 + G_0 \Sigma_* G_0 + G_0 \Sigma_* G_0 \Sigma_* G_0 + \cdots,
\end{align}
where multiplication suppresses integration over the intermediate times. Suppose we start by trying to set $G_*$ to the free fermion propagator. Then the first saddle point equation reads $\Sigma(u,v) = \beta^2 G_0(u,v)^3$, which we represent by \Cref{fig:melon}. If we apply this to \eqref{eq:gneu}, we obtain the first correction to $G_0$ by integrating $G_0\Sigma G_0$:
\begin{align}
    G(\tau_1,\tau_2) = G_0(\tau_1,\tau_2) + \beta^2 \int du\,dv\, G_0(\tau_1, u)G_0(u,v)^3 G_0(v,\tau_2).
\end{align}
The cubic ``self-energy'' is shown in \Cref{fig:melon}, where each line corresponds to a copy of $G_0$. To obtain the ``dressed'' propagator $G_*$, one repeats the above process recursively, replacing each line with another melon.

\subsection{SYK ladder kernel}

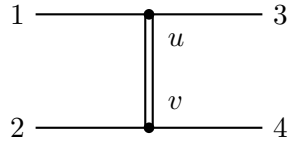
\begin{figure}[H]
    \centering
    \begin{tikzpicture}[line width=0.8pt]
        \coordinate (u) at (0,0);
        \coordinate (v) at (0,-1.5);

        \draw[double, double distance=2pt] (u) -- (v);

        \draw (-1.5,0) node[left] {$1$}
            -- (1.5,0) node[right] {$3$};
        \draw (-1.5,-1.5) node[left] {$2$}
            -- (1.5,-1.5) node[right] {$4$};

        \fill (u) circle (2pt);
        \fill (v) circle (2pt);

        \node[below right=3pt] at (u) {$u$};
        \node[above right=3pt] at (v) {$v$};
    \end{tikzpicture}
    \caption{A one-rung ladder diagram with internal vertices at $u$ and $v$.}
    \label{fig:ladder}
\end{figure}

We recall the informal physics derivation of the ladder kernel to motivate the appropriate definition for a Hessian. Consider perturbing the saddle point equation \eqref{eq:sd} by $\delta G_*$. Denoting the propagator between sites 3 and 4 by $G_*(3,4)$, a perturbation $\delta G_*(3,4)$ leads to a perturbation $\delta \Sigma_*(3,4) = 3\beta^2 G_*(3,4)^2 \delta G_*(3,4)$. Plugging this into \eqref{eq:gneu} and resumming to obtain dressed propagators, we also find that $\delta G_* = G_* \delta \Sigma_* G_*$ to first order. Enforcing stationarity, we have
\begin{align}
    (G \delta \Sigma G)(1,2) &=  \int d\tau_3\,d\tau_4\, G(1,3)  \big[3\beta^2G(3,4)^2 \delta G(3,4)\big] G(4,2) \\
    &= -3\beta^2 \int d\tau_3\,d\tau_4\, G(1,3) G(2,4) G(3,4)^2 \delta G(3,4).
\end{align}
Writing this as $(K \delta G)(1,2)$, we arrive at the definition of the \emph{ladder kernel}:
\begin{align}
    K(1,2;3,4) = -3\beta^2 G(1,3) G(2,4) G(3,4)^2.
\end{align}
Pictorially, $K$ acts on two parallel lines $G(1,3)$ and $G(2,4)$ by adding a ``rung'' consisting of two propagators $G(3,4)^2$ between the rungs (\Cref{fig:ladder}).
It is straightforward to show that repeatedly acting with $K$ does not change the order of the contribution. A ladder diagram with $r > 1$ rungs has labels $i \to k_1 \to \cdots \to k_{r-1} \to j$ on both the top and the bottom. At rung $s$, there are labels $a_s, b_s$. Schematically, a ladder with $r$ rungs thus contributes
\begin{align}
    \sum_{k_1,\dots,k_{r-1}} \sum_{a_1,b_1,\dots,a_r,b_r} \prod_{s=1}^r \frac{3}{8N^3} \E{J^2_{k_{s-1} k_s a_s b_s}}.
\end{align}
Each contracted pair contributes the variance of order $N^{-3}$, giving a factor of $N^{-3r}$; there are $N^2$ choices for $a_s, b_s$, giving a factor of $N^{2r}$; finally, there are $N^{r-1}$ choices for $k_1,\dots,k_{r-1}$. Each ladder diagram thus contributes $1/N$, independent of $r$. The Neumann series $\sum_r K^r = (I-K)^{-1}$ suggests that in order to expand around the saddle point solution, we must invert the operator defined as $\mathfrak L_\beta = I-K$, i.e.,
\begin{align}\label{eq:lbeta}
    (\mathfrak L_\beta X)(u,v)
    =
    X(u,v)
    -
    3\beta^2
    \int_0^1\!\!\int_0^1
    G_*(u,s)G_*(s,t)^2X(s,t)G_*(t,v)\,ds\,dt.
\end{align}

\section{Quantum cavity method}\label{sec:cavity}
The main technical step of the proof is to show concentration of the empirical two-point field around $G_*$ to justify our expansion around the saddle point solution. Formally, the concentration bound that we need is as follows.
\begin{theorem}[Concentration around saddle solution]\label{thm:mbound}
    Let $B$ be a Hermitian linear combination of finitely many operators of the form
    \begin{align}
        B(t) = \frac1{2N}\sum_{i=1}^N
        e^{-it\beta H}\psi_i e^{it\beta H}\psi_i-G_*(it)I
    \end{align}
    at fixed real times, together with their adjoints (e.g., $B = B(t) + B(t)^\dagger$). There exist $\delta > 0$ and an integer $r \geq 1$ such that, for every fixed integer $m\geq1$,
    \begin{align}
        \frac{\EE_J\Tr(e^{\beta H}B^{2m})}{\EE_J\Tr e^{\beta H}} \leq C_mN^{- \delta m/(4r+4)}.
        \label{eq:even-physical-moments}
    \end{align}
\end{theorem}
We prove this in \Cref{sec:concentration} using the cavity method. Roughly, the cavity method proceeds by iteratively reducing from $N+1$ to $N$ sites: each fermion (or spin, classically) is removed from interacting with the rest of the Hamiltonian and is instead coupled to a free bath. This will ultimately lead to a decomposition of correlators into two-point functions. Before describing the cavity method in more detail, we describe basic properties of these two-point functions. For diagonalization $-\beta H \ket{a} = E_a \ket{a}$, the averaged two-point function is
\begin{align}\label{eq:gdiag}
    G(t) = \frac{1}{2N} \sum_i \Tr(\rho_\beta e^{-t\beta H} \psi_i e^{t\beta H} \psi_i) = \frac{1}{2NZ} \sum_i \sum_{a,b} e^{-E_a} e^{-t(E_b-E_a)} |\bra{a}\psi_i\ket{b}|^2 = \frac{1}{2}\int_\R e^{-t\omega} \nu(d\omega)
\end{align}
for probability distribution $\nu = \frac{1}{NZ} \sum_i \sum_{a,b} e^{-E_a} |\bra{a}\psi_i\ket{b}|^2 \delta_{E_b-E_a}$. In particular, the detailed balance (KMS) condition $\nu(-d\omega) = e^{-\omega}\nu(d\omega)$ for $\omega > 0$ implies that the distribution of absolute frequency $\mu(S) = \nu(\{\omega:|\omega|\in S\})$ for $S\subseteq [0,\infty)$ allows us to rewrite the two-point function as
\begin{align}\label{eq:gmu}
 G(t)=\frac{1}{2}\int_0^\infty
       \frac{\cosh(x(1/2-t))}{\cosh(x/2)}\,\mu(dx),\qquad 0\leq t\leq1.
\end{align}
Moreover, $G$ satisfies $G(t) = G(1-t)$. We will sometimes refer to this according to the measure in \eqref{eq:gmu}, e.g., $G_\mu$. To see how this relates to free fermions, recall the two-point function of a free fermion Hamiltonian $\frac i2\sum_j x_j \eta_j \zeta_j$ with coefficients $x_j$ and Majorana fermions $\{\eta_j, \eta_k\} = \{\zeta_j, \zeta_k\} = 2\delta_{jk}$ and $\{\eta_j, \zeta_k\}=0$:
\begin{align}
    \frac{\Tr(e^{-(1-w) \frac{i}{2}\sum_j x_j \eta_j \zeta_j}\eta_\alpha e^{-w\frac{i}{2}\sum_j x_j \eta_j \zeta_j} \eta_{\alpha'})}{\Tr e^{-\frac{i}{2}\sum_j x_j \eta_j \zeta_j}} = \delta_{\alpha\alpha'} \frac{\cosh(x_\alpha(1/2-w))}{\cosh(x_\alpha/2)}.
\end{align}
In the cavity computation, we imagine adding a fermion $\psi_{N+1}$ to the SYK model with fresh disorder. We want to show that the contribution of this cavity fermion (e.g., to the partition function or correlators) is well-approximated by coupling $\psi_{N+1}$ that does not interact with the other SYK fermions but is instead coupled to a free fermion bath consisting of $M_{\rm bath}$ modes of energy $x_j$ and couplings $v_j$ to $\psi_{N+1}$. (Note that this $M_{\rm bath}$ is unrelated to the Fourier cutoff $M$ in \eqref{eq:gtrunc}.) Since $\psi_{N+1}$ is only part of a free fermion Hamiltonian, its contribution to the partition function can be solved exactly. In particular, its two-point function can be written as
\begin{align}
    \mathcal D(\Sigma)(u)
    =\frac{1}{2}
    \frac{\Tr\lr{e^{-(1-u)\frac{i}{2}\sum_j
    (x_j\eta_j\zeta_j+v_j\psi_{N+1}\eta_j)}
    \psi_{N+1} e^{-u\frac{i}{2}\sum_j
    (x_j\eta_j\zeta_j+v_j\psi_{N+1}\eta_j)}\psi_{N+1}}}
    {\Tr\exp[-\frac{i}{2}\sum_j
    (x_j\eta_j\zeta_j+v_j\psi_{N+1}\eta_j)]}
\end{align}
for
\begin{align}\label{eq:sigmaguess}
    \Sigma(u)=\frac{1}{2}\sum_{j=1}^{M_{\rm bath}} v_j^2
    \frac{\cosh(x_j(1/2-u))}{\cosh(x_j/2)}, \qquad \widehat{\mathcal D(\Sigma)}(n)
    =\int_0^1e^{i\omega_nu}\mathcal D(\Sigma)(u)\,du
      =\lr{-i\omega_n-\widehat\Sigma(n)}^{-1}.
\end{align}
Comparing to \eqref{eq:gmu}, this corresponds to setting $\Sigma = G_\sigma$ for the measure $\sigma = \sum_j v_j^2 \delta_{x_j}$. The purpose of the cavity computation is to show that we can choose $x_j$ and $v_j$ such that this $\Sigma = G_\sigma$ approximately satisfies the SYK saddle point equations \eqref{eq:sd}. Once we obtain $G,\Sigma$ that approximate the $N\to\infty$ solution for $G_*, \Sigma_*$, we can expand around $G_*,\Sigma_*$ using the ladder kernel.

To highlight the technical difficulty in making a rigorous quantum cavity method, we first describe the \emph{classical} cavity method in more detail. We will later use a variant of this classical cavity method to efficiently predict thermal expectation values of the SK and $p\geq3$ Ising spin model in \cref{sec:pspin-main,sec:sk-main}. Consider the classical Ising $p$-spin model $H_N$ on $N$ spins \eqref{eq:ising}. We denote a spin configuration by $\sigma \in \{\pm1\}^N$ and write $\sigma_A = \prod_{i\in A} \sigma_i$ for the product of spins, so the Hamiltonian is
\begin{align}
    H_N(\sigma) = \sqrt{\frac{p!}{2N^{p-1}}} \sum_{A\in\binom{[N]}{p}} J_A \sigma_A.
\end{align}
The cavity method adds a new spin $\sigma_{N+1}\in\{\pm 1\}$ on site $N+1$ and couples it with fresh disorder, so
\begin{align}
    H_{N+1}(\sigma, \sigma_{N+1}) &= \sqrt{\frac{p!}{2(N+1)^{p-1}}} \sum_{A\in\binom{[N]}{p}} J_A \sigma_A + \sqrt{\frac{p!}{2(N+1)^{p-1}}} \sum_{B\in\binom{[N]}{p-1}} J_B \sigma_{N+1}\sigma_B \\
    &= \alpha_N H_N(\sigma) + \sigma_{N+1} g_N(\sigma)
\end{align}
for
\begin{align}
    \alpha_N = \lr{\frac{N}{N+1}}^{(p-1)/2}, \qquad g_N(\sigma) = \sqrt{\frac{p!}{2(N+1)^{p-1}}} \sum_{B\in\binom{[N]}{p-1}} J_B \sigma_B
\end{align}
The idea of the cavity method is to extend the old partition function $Z_N(\beta) = \sum_\sigma e^{\beta H(\sigma)}$ to include the new spin by computing only the marginal on $\sigma_{N+1}$:
\begin{align}
    Z_{N+1}(\beta) = Z_N(\alpha_N \beta) \times \frac{\sum_{\sigma_{N+1} = \pm 1} \sum_\sigma e^{\beta \sigma_{N+1} g_N(\sigma) + \beta \alpha_N H_N(\sigma)} }{Z_N(\alpha_N \beta)} = Z_N(\alpha_N \beta) \langle 2\cosh(\beta g_N)\rangle_{\rm cavity}
\end{align}
for $\langle f \rangle_{\rm cavity} = \frac{\sum_\sigma f(\sigma) e^{\beta \alpha_N H_N(\sigma)}}{Z_N(\alpha_N \beta)}$. This doesn't appear to buy us anything unless we can somehow compute the cavity expectation easily. The insight is that if there is some version of correlation decay, e.g., for local observables $F_j$ one has
\begin{align}
    \langle F_1 F_2 F_3 F_4 \rangle_\beta \approx \langle F_1 \rangle_\beta \langle F_2 \rangle_\beta \langle F_3 \rangle_\beta \langle F_4 \rangle_\beta,
\end{align}
then the cavity spin only feels an ``effective'' Hamiltonian that can be easily solved, allowing $\langle f \rangle_{\rm cavity}$ to be evaluated. Schematically, its interactions with the other spins can be replaced by cutting it out from $H_{N+1}$ (creating a cavity) and coupling it to new independent spins $\eta_j$. By iterating this procedure from $Z_{N+1}$ to $Z_N$, the entire partition function (or local expectations) can be computed.

In our setting, the observables $F_j$ are constructed from the $B(t_j)$ operators above (e.g., $F_j = \Re B(t_j)$). However, showing $\Tr(\rho_\beta F_1 F_2 F_3 F_4) \approx \Tr(\rho_\beta F_1) \Tr(\rho_\beta F_2) \Tr(\rho_\beta F_3) \Tr(\rho_\beta F_4)$ is difficult to do directly for the SYK model. Non-rigorously, it can be done to show factorization of correlators into two-point functions~\cite{maldacena2016remarks}. Rigorously, it was also done recently for the annealed free energy of the SYK model in the $N\to\infty$ limit: \cite{he2026spectral} originally overcame this issue by using a de Finetti theorem for operator algebras. In our setting, we need bounds whose decay in $N$ improves with the moment order, so that the expansion can achieve arbitrary inverse polynomial error. Consequently, although our bath construction shares some similarities to the cavity method of~\cite{he2026spectral}, the spectral rounding, measurement and restoration of detailed balance below are distinct to our construction, and appear to be necessary to achieve sufficient concentration for the algorithm to work.

We modify the classical cavity procedure to show concentration after (classically speaking) conditioning. For classical random variables with $|F_j|\leq1$, conditioning on information $X$ gives the telescoping identity
\begin{align}
    \prod_{j=1}^q F_j - \prod_{j=1}^q \E{F_j\mid X} = \sum_{j=1}^q \lr{\prod_{k < j} F_k} \lr{F_j - \E{F_j \mid X}} \lr{\prod_{k > j} \E{F_k \mid X}}
\end{align}
so, conditioned on $X$, the expectation factorizes if there is concentration:
\begin{align}\label{eq:clvar}
    \abs{\E{\prod_{j=1}^q F_j \mid X} - \prod_{j=1}^q \E{F_j \mid X}} \leq \sum_{j=1}^q \E{\abs{F_j-\E{F_j\mid X}} \mid X} \leq \sum_{j=1}^q \sqrt{\Var{F_j \mid X}}.
\end{align}
Taking the expectation over $X$, we find that
\begin{align}
    \E{\prod_{j=1}^q F_j} \approx \EE_X\left[\prod_{j=1}^q \E{F_j \mid X}\right].
\end{align}
In the quantum setting, conditioning on an energy eigenstate need not
make the different $F_j$ simultaneously concentrated (and we can't control the eigenstates directly anyway). We instead use
weak measurements that commute with a rounded Gibbs state, while
controlling their disturbance of the state. This yields the conditional
concentration needed for the cavity method.
\begin{enumerate}
    \item \emph{Spectral rounding.} We round the spectrum of the original Hamiltonian based on a random offset $u$. In the $j$th window of size $\delta_E$, all energies between $u + j\delta_E$ and $u + (j+1)\delta_E$ are rounded to $u+j\delta_E$. We can then find a distribution over $u$ such that $\rho_\beta = \EE_u \rho(u)$ for density matrices $\rho(u)$. Let $P_j(u)$ be the projector onto the $j$th window, and let $\Pi_u(X) = \sum_j P_j(u)XP_j(u)$ remove off-diagonal elements between different energy windows.
    \item \emph{Measurement.} Since $\rho(u)$ is scalar within each subspace $P_j(u)$, it commutes with any measurement that acts separately on separate blocks: $[\rho(u), \Pi_u B] = 0$ for any $B$. In particular, such measurements look classical, allowing us to show a concentration inequality like \eqref{eq:clvar}. We measure operators that estimate quantities like $\Pi_u B$, but also two-point functions $\frac{1}{N} \sum_i \Pi_u \Re \psi_i(t) \psi_i(0)$ for $\psi_i(t) = e^{-it\beta H} \psi_i e^{it\beta H}$ at a grid of times $t_1,\dots,t_R$. After measuring $\rho(u)$ and getting outcomes $x_1,\dots,x_R$, denote the post-measurement state by $\rho_x$. We show a concentration inequality for weight $w(x) = |\Tr(\rho_x \Pi_u B)|^{2m}$:
    \begin{align}
        \E{\abs{\Tr(\rho_x F_1 \cdots F_q) - \prod_{j=1}^q \Tr(\rho_x F_j)}w(x)} \leq \sqrt{\EE w(x)} \sum_{j=1}^q \sqrt{\E{w(x)\mathrm{Var}_{\rho_x}(F_j)}}.
    \end{align}
    \item \emph{Detailed balance.} On the state $\rho_x$, consider transitions from energy window $j$ to energy window $k$ caused by acting with some Majorana fermion $\psi_i$. Since $\rho_x$ is not a thermal state, the distribution $\nu_x(\omega)$ of energy changes $\omega = E_k-E_j$ may give a transition law that does not satisfy detailed balance, i.e., $\nu_x(-\omega) \neq e^{-\omega} \nu_x(+\omega)$. This is problematic for the cavity method: if we set $G(\tau) = \frac{1}{2}\int_\R e^{-\tau \omega} \nu_x(d\omega)$, then the KMS condition would be violated: $G(\tau) \neq G(1-\tau)$. This prevents us from instantiating a thermal bath to replace a cavity fermion. Hence, we introduce a projector $\cP$ that enforces detailed balance while preserving $\nu_x(\omega) + \nu_x(-\omega)$, and we bound the effect of the projector.
    \item \emph{Cavity method.} Finally, we can apply the cavity argument described above, albeit using $\cP \nu_x$ rather than the transitions given by the true thermal state's $\nu$. More explicitly, we apply the cavity method for the function
    \begin{align}
        G(t) = \frac12\int_{\mathbb R}e^{-z\omega}(\mathcal P\nu_x)(d\omega)
    \end{align}
    rather than in the eigenbasis \eqref{eq:gdiag}.
\end{enumerate}
Once we know that $G$ concentrates around $G_*$, we can perform the actual expansion by inverting $\mathfrak L_\beta$ constructed from the ladder kernel. (In physics language, this corresponds to resumming the ladder diagrams.) We now show how to do this in the following algorithm. 

\section{Estimating thermal expectations of the SYK model}\label{sec:alg}

The algorithm estimates the coefficients (using shorthand $s_0=1-\sum_{r=1}^vs_r$)
\begin{align}\label{eq:foa}
    F_{O;A_1,\ldots,A_v}
    &=
    \sum_{\pi\in S_v}
    \int_{\substack{s_1,\ldots,s_v\geq0\\
                    s_1+\cdots+s_v\leq1}}
    \frac{
        \EE_J\Tr\left[
        e^{s_0\beta H}
        \Gamma_{A_{\pi(1)}}e^{s_1\beta H}
        \cdots
        \Gamma_{A_{\pi(v)}}e^{s_v\beta H}O
        \right]
    }{
        \EE_J\Tr e^{\beta H}
    }
    \,ds_1\cdots ds_v.
\end{align}
Then for Hermite expansion truncated at order $V$, define
\begin{align}\label{eq:no}
    \cN_O(J)
    &=
    \sum_{v=0}^V
    \frac{1}{v!}
    \lr{\beta\sqrt{\frac{3}{8N^3}}}^{v}
    \sum_{A_1,\ldots,A_v\in\binom{[N]}4}
    \langle J_{A_1}\cdots J_{A_v}\rangle_J
    F_{O;A_1,\ldots,A_v}.
\end{align}
We output
\begin{align}
    \widehat f_O(\beta,J)
    =
    \begin{cases}
        \displaystyle
        \max\left\{-1,\min\left\{1,
        \frac{\cN_O(J)}{\cN_I(J)}
        \right\}\right\},
        & \cN_I(J)\geq\frac14,
        \\
        0,
        & \cN_I(J)<\frac14.
    \end{cases}
\end{align}
If we can exactly estimate coefficients $F_{O;A_1,\dots,A_v}$, this succeeds due to the following result.
\begin{lemma}[Hermite truncation]
\label{lem:hermite-tail}
At truncation order $V$, the following bounds hold:
\begin{align}
    \EE_J\left|
    \frac{\Tr(Oe^{\beta H})}
         {\EE_J\Tr e^{\beta H}}
    -\cN_O(J)
    \right|^2
    \leq
    C_{\beta,V,w}N^{-V-1-w/2}, \quad \EE_J\left|
    \frac{\Tr e^{\beta H}}
         {\EE_J\Tr e^{\beta H}}
    -\cN_I(J)
    \right|^2
    \leq
    C_{\beta,V}N^{-V-1}.
    \label{eq:analysis-hermite-tail}
\end{align}
\end{lemma}
The proof of \Cref{lem:hermite-tail} is straightforward and provided in \Cref{app:hermite}.
The main difficulty lies in estimating $F_{O;A_1,\dots,A_v}$. Equivalently, we have to estimate the ratios in \eqref{eq:foa} of Majoranas inserted at various terms. By \eqref{eq:chin}, it suffices to estimate quantities of the form $\left\langle\chi_{j_1}(t_1)\cdots\chi_{j_{2k}}(t_{2k})\right\rangle_N$. For the matrix
\begin{align}\label{eq:gram}
    (\mathcal G_{k,N})_{\pi\sigma}
    =
    N^{\#\{\text{connected components of }\pi\cup\sigma\}-k},
    \qquad
    \pi,\sigma\in\mathcal P_2([2k]),
\end{align}
these correlators satisfy (\Cref{lem:orthogonal-reconstruction})
\begin{align}\label{eq:chiprod}
    &\left\langle
        \chi_{j_1}(t_1)\cdots\chi_{j_{2k}}(t_{2k})
    \right\rangle_N =
    2^k
    \sum_{\pi,\sigma\in\mathcal P_2([2k])}
    \left(
        \prod_{\{r,s\}\in\pi}\delta_{j_rj_s}
    \right)
    (\mathcal G_{k,N}^{-1})_{\pi\sigma}
    (-1)^{\operatorname{cr}(\sigma)}
    \left\langle
        \prod_{\substack{\{r,s\}\in\sigma\\r<s}}
        Q_{t_rt_s}
    \right\rangle_N
\end{align}
To evaluate the expectation of $Q_{t_rt_s}$, we expand around the saddle point solution:
\begin{align}
    T_{uv}
    =
    Q_{uv}-G_*(u,v),
    \qquad
    \mu_p((u_1,v_1),\ldots,(u_p,v_p))
    =
    \left\langle
        \prod_{j=1}^pT_{u_jv_j}
    \right\rangle_N,
    \qquad
    \mu_0=1.
\end{align}
If we can compute $\mu_p$, then we can evaluate
\begin{align}\label{eq:qprod}
    \left\langle
        \prod_{j=1}^kQ_{u_jv_j}
    \right\rangle_N
    =
    \sum_{S\subseteq[k]}
    \left(
        \prod_{j\notin S}G_*(u_j,v_j)
    \right)
    \mu_{|S|}\big((u_j,v_j)_{j\in S}\big).
\end{align}
We estimate $\mu_p$ using the ladder kernel. For $p\geq1$, hold $(u_2,v_2),\ldots,(u_p,v_p)$ fixed and apply
$\mathfrak L_\beta$ to the dependence on the first pair $(u,v)$. Writing this action and all the moments explicitly gives
\begin{align}\label{eq:explicit-collective-hierarchy}
    &\mu_p\big((u,v),(u_j,v_j)_{j=2}^p\big)
    -
    3\beta^2
    \int_0^1\!\!\int_0^1
    G_*(u,s)G_*(s,t)^2
    \mu_p\big((s,t),(u_j,v_j)_{j=2}^p\big)
    G_*(t,v)\,ds\,dt
    \nonumber\\
    &\quad=
    \operatorname{Alt}_{u,v}
    \Bigg\{
    \beta^2\int_0^1\!\!\int_0^1
    G_*(u,s)
    \Big[
        3G_*(s,t)G_*(t,v)
        \mu_{p+1}\big((s,t),(s,t),(u_j,v_j)_{j=2}^p\big)
    \nonumber\\
    &\hspace{30mm}
        +
        G_*(t,v)
        \mu_{p+2}\big((s,t),(s,t),(s,t),(u_j,v_j)_{j=2}^p\big)
    \nonumber\\
    &\hspace{30mm}
        +
        3G_*(s,t)^2
        \mu_{p+1}\big((s,t),(t,v),(u_j,v_j)_{j=2}^p\big)
    \nonumber\\
    &\hspace{30mm}
        +
        3G_*(s,t)
        \mu_{p+2}\big((s,t),(s,t),(t,v),(u_j,v_j)_{j=2}^p\big)
    \nonumber\\
    &\hspace{30mm}
        +
        \mu_{p+3}\big((s,t),(s,t),(s,t),(t,v),
                     (u_j,v_j)_{j=2}^p\big)
    \Big]\,ds\,dt
    \nonumber\\
    &\hspace{12mm}
    +
    \frac1N\sum_{j=2}^p
    \Big[
        \lr{
            G_*(u,u_j)G_*(v_j,v)
            -
            G_*(u,v_j)G_*(u_j,v)
        }
        \mu_{p-2}\big((u_r,v_r)_{r\in\{2,\ldots,p\}\setminus\{j\}}\big)
    \nonumber\\
    &\hspace{30mm}
        +
        G_*(u,u_j)
        \mu_{p-1}\big((v_j,v),
                     (u_r,v_r)_{r\in\{2,\ldots,p\}\setminus\{j\}}\big)
    \nonumber\\
    &\hspace{30mm}
        -
        G_*(u,v_j)
        \mu_{p-1}\big((u_j,v),
                     (u_r,v_r)_{r\in\{2,\ldots,p\}\setminus\{j\}}\big)
    \Big]
    \Bigg\}.
\end{align}
Here $(u_j,v_j)_{j\in S}$ denotes the list of pairs whose indices
belong to $S$, in increasing index order, and
\begin{align}
    \operatorname{Alt}_{u,v}F(u,v)
    =
    \frac{F(u,v)-F(v,u)}2,
\end{align}
with all other arguments held fixed.

Equation \eqref{eq:explicit-collective-hierarchy} is proved by Grassmann
integration by parts in \Cref{lem:exact-hierarchy}. We apply $\mathfrak L_\beta^{-1}$ to its first pair of variables, which requires the following control over the inverse (proven in \Cref{app:ladder}).
\begin{lemma}[Ladder kernel inversion]
\label{lem:full-bilocal-inversion}
For every constant $\beta\geq0$, the operator $\mathfrak L_\beta$
is invertible and satisfies
\begin{align}
    \|\mathfrak L_\beta^{-1}X\|_\infty
    \leq C_\beta\|X\|_\infty.
    \label{eq:analysis-invertibility}
\end{align}
\end{lemma}
Observe that in the $O(1)$ terms, we have five terms involving $\mu_{p+1},\mu_{p+2}$ and $\mu_{p+3}$; in the $O(1/N)$ correction, we have $\mu_{p-1}$ and $\mu_{p-2}$. Hence, the inversion yields an equation of the form
\begin{align}
    \mu_p = \cU_{p,p+1} \mu_{p+1} + \cU_{p,p+2} \mu_{p+2} + \cU_{p,p+3} \mu_{p+3} + \frac{1}{N}\lr{\cC_{p,p-1} \mu_{p-1} + \cC_{p,p-2} \mu_{p-2}},
\end{align}
where for example,
\begin{align}
    &(\cU_{p,p+3}\nu)(u,v;(u_j,v_j)_{j=2}^p)\nonumber\\
    &\quad=\Big[\mathfrak L_\beta^{-1}\operatorname{Alt}_{a,b}
      \beta^2\int_0^1\!\int_0^1G_*(a,s)
      \nu\bigl((s,t),(s,t),(s,t),(t,b),(u_j,v_j)_{j=2}^p\bigr)
      \,ds\,dt\Big](u,v).
\end{align}
Note that when evaluating these terms, $G_*$ is approximated by solving the convex optimization problem \eqref{eq:convexsd} defined on $M$ Fourier modes. We evaluate this estimate solving $\argmin_{u_j > 0} \cF_M$ (where $\cF_M$ is the objective in \eqref{eq:convexsd}) until $\norm{\nabla \cF_M}_2 \leq M^{-2}$. Similarly, $\mathfrak L_\beta^{-1}$ is inverted in discretized form, where the imaginary time interval $[0,1]$ is divided into $M^4$ steps. The numerical stability of these approximations is shown in \Cref{lem:linv}.

We solve the above recursion with $\mu_0=1$. Writing the remainder in square brackets, this is
\begin{align}
    \mu_p = \sum_{\substack{d=1,2,3\\ p+d < K}} \cU_{p,p+d} \mu_{p+d} + \frac{1}{N}\lr{\cC_{p,p-1} \mu_{p-1} + \cC_{p,p-2} \mu_{p-2}} + \left[\sum_{\substack{d=1,2,3\\ p+d \geq K}} \cU_{p,p+d} \mu_{p+d}\right].
\end{align}
For each fixed $p$, concentration supplies an inverse power of $N$ proportional to $p$. This is stated as follows as a corollary of \Cref{thm:mbound} (and proven in \Cref{app:mubound}).
\begin{corollary}[Concentration of $\mu_p$]
\label{lem:collective-concentration}
For every constant $\beta\geq0$, there is $\theta>0$ such
that for every constant $p\geq1$,
\begin{align}
    \|\mu_p\|_\infty
    \leq C_{\beta,p}N^{-\theta p}.
    \label{eq:analysis-concentration}
\end{align}
\end{corollary}
Due to this concentration, we can truncate the recursion at some order $K$. Since all $\cU_{p,q}$ are bounded operators at fixed $K$, the square brackets have size at most $O(N^{-\theta K})$. To solve these recursive equations, we evaluate
\begin{align}
    \lr{I-A_0-\frac{1}{N}A_1}\boldsymbol \mu = e_0 + O\lr{N^{-\theta K}}
\end{align}
for $\boldsymbol \mu = (\mu_0,\dots,\mu_{K-1})$, $e_0=(1,0,\dots,0)$ and matrices of operators
\begin{align}
    A_0=
\begin{pmatrix}
0&0&0&0&0&\cdots\\
0&0&\mathcal U_{1,2}&\mathcal U_{1,3}&\mathcal U_{1,4}&\cdots\\
0&0&0&\mathcal U_{2,3}&\mathcal U_{2,4}&\cdots\\
0&0&0&0&\mathcal U_{3,4}&\cdots\\
\vdots&&&&\ddots&\ddots
\end{pmatrix}, \quad A_1=
\begin{pmatrix}
0&0&0&0&0&\cdots\\
0&0&0&0&0&\cdots\\
\mathcal C_{2,0}&\mathcal C_{2,1}&0&0&0&\cdots\\
0&\mathcal C_{3,1}&\mathcal C_{3,2}&0&0&\cdots\\
0&0&\mathcal C_{4,2}&\mathcal C_{4,3}&0&\cdots\\
\vdots&&\ddots&\ddots&\ddots
\end{pmatrix}.
\end{align}
Each component of $A_0\boldsymbol\mu$ depends only on moments of
strictly higher order. Consequently $K$ successive applications
give zero, and $(I-A_0)^{-1}=\sum_{r=0}^{K-1}A_0^r$.
We estimate $\mu_p$ by iterating
\begin{align}
    \widetilde{\boldsymbol\mu}^{(0)}
    &=
    e_0,
    \\
    \widetilde{\boldsymbol\mu}^{(j+1)}
    &=
    e_0+
    \frac1N
    \left(\sum_{r=0}^{K-1}A_0^r\right)
    A_1\widetilde{\boldsymbol\mu}^{(j)}.
\end{align}
The error associated with truncating at order $K$ is bounded by the following lemma.
\begin{lemma}[Recursion truncation]
\label{lem:hierarchy-error}
For fixed $K,j$ and sufficiently large $N$, the following bounds hold:
\begin{align}
    \|\boldsymbol\mu-
    \widetilde{\boldsymbol\mu}^{(j)}\|_\infty
    \leq
    C_{\beta,K}N^{-\theta K}
    +C_{\beta,K}
      \left(\frac{C_{\beta,K}}{N}\right)^{j+1}
    +C_{\beta,K,j}M^{-2}.
    \label{eq:analysis-moment-error}
\end{align}
\end{lemma}
The proof (given in \Cref{app:recurse}) is fairly short and elementary. We can now prove the correctness and bound the cost of the algorithm.

\subsection{Correctness and cost}

\begin{theorem}[Correctness and polynomial cost]
\label{thm:algorithm-correctness-cost}
Fix constant $\beta\geq0$, fixed $a,c>0$, and a fixed even
weight $w$.  Choose
\begin{align}
    V
    &=\lceil a+2c+4\rceil,
    \\
    K
    &\geq
    \max\left\{
    2V+\frac w2+3,
    \left\lceil
    \frac{6V+a+2c+w+20}{\theta}
    \right\rceil
    \right\},
    \\
    j+1
    &\geq 6V+a+2c+w+20,
    \\
    M
    &=
    \left\lceil N^{(6V+a+2c+w+20)/2}\right\rceil,
    \label{eq:final-parameter-choices}
\end{align}
where $\theta$ is supplied by
\Cref{lem:collective-concentration}.
Then, for all sufficiently large $N$,
\begin{align}
    \Pr_J\left[
    |\widehat f_O(\beta,J)-f_O(\beta,J)|>N^{-c}
    \right]
    \leq N^{-a}.
    \label{eq:final-correctness}
\end{align}
The number of operations is at most
$N^{C_{\beta,a,c,w}}$.
\end{theorem}

\begin{proof}
For $X\in\{I,O\}$, write $\widehat F_{X;A_1,\ldots,A_v}$ and
$\widehat{\cN}_X$ for the computed versions of \eqref{eq:foa} and
\eqref{eq:no}.

\proofstep{Accuracy}
We condition on $\Tr e^{\beta H}/\EE_J\Tr e^{\beta H}\geq1/2$, which occurs with high probability (\Cref{lem:partition-concentration}, which follows directly from~\cite{anschuetz2025strongly}). We also condition on
\begin{align}
    \max_{X\in\{I,O\}}
    \left|\widehat{\cN}_X(J)
       -\frac{\Tr(Xe^{\beta H})}{\EE_J\Tr e^{\beta H}}\right|
    &\leq\frac1{32}N^{-c},
    \label{eq:ncass}
\end{align}
whose probability of failure we will show shortly.
Then $\widehat{\cN}_I(J)\geq1/2-N^{-c}/32\geq1/4$.
Since $|f_O(\beta,J)|\leq1$ and the algorithm clips $\wh f_O$ to $[-1,1]$, the triangle inequality gives
\begin{align}
    |\widehat f_O(\beta,J)-f_O(\beta,J)|
    &\leq\frac{|\widehat{\cN}_O(J)
                     -f_O(\beta,J)\widehat{\cN}_I(J)|}
                    {\widehat{\cN}_I(J)} \leq4\sum_{X\in\{I,O\}}
       \left|\widehat{\cN}_X(J)
         -\frac{\Tr(Xe^{\beta H})}{\EE_J\Tr e^{\beta H}}\right|
      \leq\frac14N^{-c}.
    \label{eq:ratio-error}
\end{align}
We will shortly prove the coefficient bound
\begin{align}
    \max_{\substack{X\in\{I,O\},\ 0\leq v\leq V\\
                    A_1,\ldots,A_v\in\binom{[N]}4}}
    |\widehat F_{X;A_1,\ldots,A_v}-F_{X;A_1,\ldots,A_v}|
    &\leq C_{\beta,K,j,V,w}N^{-(6V+a+2c+w+20)}.
    \label{eq:final-coefficient-accuracy}
\end{align}
By \eqref{eq:no},
$\widehat{\cN}_X-\cN_X$ has Hermite degree at most $V$ and is orthogonal
to the omitted tail. Hermite orthogonality, Cauchy--Schwarz over
permutations of each tuple, and \eqref{eq:analysis-hermite-tail} give
\begin{align}
    &\EE_J\left|\widehat{\cN}_X(J)
       -\frac{\Tr(Xe^{\beta H})}{\EE_J\Tr e^{\beta H}}\right|^2
       =\EE_J|\widehat{\cN}_X(J)-\cN_X(J)|^2
       +\EE_J\left|\cN_X(J)
         -\frac{\Tr(Xe^{\beta H})}{\EE_J\Tr e^{\beta H}}\right|^2
       \nonumber\\
    &\leq\sum_{v=0}^V\frac1{v!}
       \left(\frac{3\beta^2}{8N^3}\right)^v
       \sum_{A_1,\ldots,A_v\in\binom{[N]}4}
       |\widehat F_{X;A_1,\ldots,A_v}-F_{X;A_1,\ldots,A_v}|^2
       +C_{\beta,V,w}N^{-V-1}\nonumber\\
    &\leq C_{\beta,K,j,V,w}
       \left[N^{V-2(6V+a+2c+w+20)}+N^{-V-1}\right]
       \leq C_{\beta,K,j,V,w}N^{-V-1}.
    \label{eq:computed-hermite-error}
\end{align}
By Chebyshev's inequality, \eqref{eq:computed-hermite-error}, and
$V\geq a+2c+4$, the probability that \eqref{eq:ncass} fails is bounded by
\begin{align}
    &\Pr_J\left[
      \max_{X\in\{I,O\}}
      \left|\widehat{\cN}_X(J)
        -\frac{\Tr(Xe^{\beta H})}{\EE_J\Tr e^{\beta H}}\right|
                    >\frac1{32}N^{-c}\right]\nonumber\\
    &\quad\leq32^2N^{2c}\sum_{X\in\{I,O\}}
       \EE_J\left|\widehat{\cN}_X(J)
         -\frac{\Tr(Xe^{\beta H})}{\EE_J\Tr e^{\beta H}}\right|^2
       \leq C_{\beta,K,j,V,w}N^{-a-5}.
    \label{eq:ncass-probability}
\end{align}
Combining \eqref{eq:ratio-error}, \eqref{eq:ncass-probability}, and
our conditioning $\Tr e^{\beta H}/\EE_J\Tr e^{\beta H}\geq1/2$ (\Cref{lem:partition-concentration}) yields
\begin{align}
    \Pr_J[|\widehat f_O(\beta,J)-f_O(\beta,J)|>N^{-c}]
    &\leq C_\beta e^{-N/C_\beta}
           +C_{\beta,K,j,V,w}N^{-a-5}\leq N^{-a}.
\end{align}
It remains to prove \eqref{eq:ncass}, \eqref{eq:final-coefficient-accuracy} and bound the cost.

\proofstep{Coefficient error}
By \eqref{eq:chin} and its definition \eqref{eq:foa}, a coefficient
$F_{O;A_1,\ldots,A_v}$ uses $4v+w$ Grassmann variables. (Formally, we deal with the possibility of coincident times in \Cref{lem:contact-dictionary}.) Thus
\eqref{eq:chiprod} and \eqref{eq:qprod} require moments only through
\begin{align}
    k&=2v+\frac w2\leq2V+\frac w2<K.
\end{align}
For $X=I$, replace $w$ by zero.
By \eqref{eq:gram},
\begin{align}
    \|\mathcal G_{k,N}-I\|_{\infty\to\infty}
    &\leq\frac{(2k-1)!!-1}{N}\leq\frac12,\nonumber\\
    \|\mathcal G_{k,N}^{-1}\|_{\infty\to\infty}
    &\leq\frac1{1-\|\mathcal G_{k,N}-I\|_{\infty\to\infty}}\leq2.
    \label{eq:grambound}
\end{align}
By
\eqref{eq:grambound}, its coefficients satisfy
\begin{align}
    2^k\sum_{\pi,\sigma\in\mathcal P_2([2k])}
       \left(\prod_{\{r,s\}\in\pi}\delta_{j_rj_s}\right)
       |(\mathcal G_{k,N}^{-1})_{\pi\sigma}|
    &\leq2^k(2k-1)!!\|\mathcal G_{k,N}^{-1}\|_{\infty\to\infty} \leq2^{k+1}(2k-1)!!.
    \label{eq:reconstruction-row-sum}
\end{align}
We have to control the numerical approximation due to having an estimate $\wt G_*$ of the true saddle $G_*$ obtained by the convex optimization \eqref{eq:convexsd}. This is done in
\eqref{eq:saddle-pointwise-bounds}, \eqref{eq:analysis-saddle-error},
and the cavity concentration bound \eqref{eq:analysis-concentration}, which guarantee bounds
\begin{align}
    \|G_*\|_\infty&\leq\frac12,
    &\|\widetilde G_*\|_\infty&\leq1,
    &\max_{0\leq p<K}\|\mu_p\|_\infty&\leq C_{\beta,K}.
\end{align}
Telescoping the products in \eqref{eq:qprod} gives
\begin{align}
    &\left\|\sum_{S\subseteq[k]}
       \left(\prod_{r\notin S}\widetilde G_*(u_r,v_r)\right)
       \widetilde\mu_{|S|}^{(j)}\big((u_r,v_r)_{r\in S}\big)
       -\left\langle\prod_{r=1}^kQ_{u_rv_r}\right\rangle_N
       \right\|_\infty\nonumber\\
    &\quad\leq\sum_{S\subseteq[k]}
       \left[\|\widetilde\mu_{|S|}^{(j)}-\mu_{|S|}\|_\infty
        +(k-|S|)\|\widetilde G_*-G_*\|_\infty
                         \|\mu_{|S|}\|_\infty\right]\nonumber\\
    &\quad\leq2^k\|\widetilde{\boldsymbol\mu}^{(j)}
                              -\boldsymbol\mu\|_\infty
       +C_{\beta,K}k2^k\|\widetilde G_*-G_*\|_\infty.
    \label{eq:Q-error-from-moments}
\end{align}
We bound the error from the recursion truncation
(\eqref{eq:A0A1-numerical-error} from the proof of \Cref{lem:hierarchy-error})
\begin{align}
    \left\|\widetilde{\boldsymbol\mu}^{(j)}
      -\sum_{q=0}^j
        \left[\frac1N\left(\sum_{r=0}^{K-1}A_0^r\right)A_1\right]^q
             e_0\right\|_\infty
    &\leq C_{\beta,K,j}M^{-2}.
    \label{eq:coefficient-quadrature-regularity}
\end{align}
Consequently, \eqref{eq:reconstruction-row-sum},
\eqref{eq:Q-error-from-moments}, and
\eqref{eq:coefficient-quadrature-regularity} imply
\begin{align}
    |\widehat F_{X;A_1,\ldots,A_v}-F_{X;A_1,\ldots,A_v}|
    &\leq C_{\beta,K,j,V,w}
       \left[\|\widetilde{\boldsymbol\mu}^{(j)}-\boldsymbol\mu\|_\infty
             +\|\widetilde G_*-G_*\|_\infty+M^{-2}\right]\nonumber\\
    &\leq C_{\beta,K,j,V,w}
       \left[N^{-\theta K}
           +\left(\frac{C_{\beta,K}}N\right)^{j+1}+M^{-2}\right]
       \nonumber\\
    &\leq C_{\beta,K,j,V,w}N^{-(6V+a+2c+w+20)}.\nonumber
\end{align}
The second inequality uses the recursion truncation error \eqref{eq:analysis-moment-error} and
the bound \eqref{eq:analysis-saddle-error} on the numerical stability of estimating $G_*$ and inverting $\mathfrak L_\beta$; the last uses
\eqref{eq:final-parameter-choices}. This proves
\eqref{eq:final-coefficient-accuracy}.

\proofstep{Cost}
Take the least admissible $K,j$ in \eqref{eq:final-parameter-choices}.
The saddle computation, inverse, and $j$ iterations cost
$M^{C_{\beta,K,j}}$ by \Cref{lem:saddle,lem:linv} and the proof of
\Cref{lem:hierarchy-error}. In particular, the moment grids have at most
\begin{align}
    C_K\sum_{p=0}^{K-1}(M^4+1)^{2p}&\leq C_KM^{8K}
\end{align}
entries. The matrices $\mathcal G_{k,N}$ have dimension at most
$(4V+w-1)!!$. Each integral in \eqref{eq:foa} has at most $V$ variables, so
\eqref{eq:ordered-piece-quadrature} gives cost $M^{C_{\beta,K,j,V,w}}$
per coefficient. There are at most
\begin{align}
    2\sum_{v=0}^V\binom N4^v&\leq2(V+1)N^{4V}
\end{align}
coefficients, and each Wick polynomial takes $O_V(1)$ operations to
evaluate. With $M$ as in \eqref{eq:final-parameter-choices}, the total
number of operations is at most
\begin{align}
    C_{\beta,K,j,V,w}N^{4V}M^{C_{\beta,K,j,V,w}}
    &\leq N^{C_{\beta,a,c,w}}.
\end{align}
\end{proof}

\section{Concentration at the saddle point}
\label{sec:concentration}
We describe in more technical detail the proof strategy outlined in \Cref{sec:cavity}. This section contains proofs of the more conceptually interesting steps, while the more tedious steps are left for the appendices.
We will use the notation in the statement of \Cref{thm:mbound}, as well as the shorthand
\begin{align}
 \Phi_N(X)=\frac{\EE_J\Tr(e^{\beta H}X_J)}
                    {\EE_J\Tr e^{\beta H}},
 \qquad M_m=\Phi_N(B^{2m})
 \label{eq:concentration-annealed-state}
\end{align}
with $M_0=1$. Since rescaling $B$ only changes constants in \Cref{thm:mbound}, we assume $\norm{B} \leq 1$; hence, $0 \leq M_m \leq M_{m-1} \leq 1$.

\subsection{Spectral rounding}
We first record the notation and properties of the spectral rounding technique described in \Cref{sec:cavity}. Fix $0<\delta_E\leq1$ and offset $u \in \R$. For $0\leq u<\delta_E$, let $P_j(u)=\mathbf1_{[u+j\delta_E,u+(j+1)\delta_E)}(-\beta H)$ project onto the $j$th energy block, and let $\Pi_uX=\sum_jP_j(u)XP_j(u)$ project away support between different blocks. Define the rounded Hamiltonian $\wt H(u)$ and its Gibbs state by
\begin{align*}
 -\beta \wt H(u)=\sum_j(u+j\delta_E)P_j(u),
 \qquad \rho(u)=\frac{e^{\beta\widetilde H(u)}}{\Tr e^{\beta\widetilde H(u)}}.
\end{align*}

\begin{lemma}[Exact averaging of rounded Gibbs states]
\label{lem:exact-rounded-gibbs}
The rounded Hamiltonian satisfies $\|\beta H-\beta\widetilde H(u)\|\leq\delta_E$ and $\int_0^{\delta_E}e^{\beta\widetilde H(u)}\,du=(e^{\delta_E}-1)e^{\beta H}$. If $d\gamma(J)$ is the original Gaussian law, define the joint law of $J,u$ by
\begin{align}
 d\mathbb P(J,u)
 &=\frac{\Tr e^{\beta\widetilde H_J(u)}}
 {(e^{\delta_E}-1)\EE_J\Tr e^{\beta H_J}}\,d\gamma(J)\,du.
 \label{eq:joint-rounded-law}
\end{align}
For every shift-independent $X_J$,
\begin{align}
 \EE_{J,u}\Tr(\rho(u)X_J)=\Phi_N(X).
 \label{eq:gibbs-as-rounded-average}
\end{align}
\end{lemma}
\begin{proof}
For an eigenvalue $E$ of $-\beta H$, its distance above the left
endpoint of its bin ranges uniformly over $[0,\delta_E)$ as the
shift varies.  Thus the corresponding eigenvalue of the integral is
\begin{align*}
 e^{-E}\int_0^{\delta_E}e^s\,ds=(e^{\delta_E}-1)e^{-E},
\end{align*}
proving the first claim.
Taking a trace and averaging over $J$ verifies that
\eqref{eq:joint-rounded-law} has mass one; inserting $X_J$ proves
\eqref{eq:gibbs-as-rounded-average}.
\end{proof}

Let $c>0$ be the exponent in \eqref{eq:source-curvature} of
\Cref{lem:bbounds}. Choose
\begin{align}
 \delta_E=N^{-\varepsilon},\qquad 0<2\varepsilon<c,\qquad
 c_0=\frac{c-2\varepsilon}{4}>0.
 \label{eq:energy-width-choice}
\end{align}
Then \Cref{lem:source-power-pinching} shows
that for every fixed odd $m$, one can approximate moments $M_m$ by projecting each power of $B$ with $\Pi_u$, i.e.,
\begin{align}
 0\leq M_m-\EE_{J,u}\Tr\left(\rho(u)(\Pi_uB)^{2m}\right)
 &\leq C_m\left(N^{-c_0}M_{m-1}+N^{-mc_0}\right),\label{eq:mm1}\\
 \EE_{J,u}\|B^m-(\Pi_uB)^m\|_{2,\rho(u)}^2
 &\leq C_m\left(N^{-c_0}M_{m-1}+N^{-mc_0}\right). \label{eq:mm2}
\end{align}
Crucially, the RHS of these bounds will decay with $m$. In the following subsection, we introduce the measurement procedure and show that similar inequalities hold after measurement as well.

\subsection{Measurements and the posterior mean}
For a real grid $t_1,\ldots,t_{L/2}$, including the times in $B$, let
$C_1,\ldots,C_L$ be the Hermitian real and imaginary parts of
\begin{align}
 C(t)&=\frac1N\sum_{i=1}^N
       e^{-it\beta H}\psi_i e^{it\beta H}\psi_i.
 \label{eq:measured-probes}
\end{align}
The $L/2$ times in this grid include the times in $B$ and the times in an equal discretization of $[-T_N, T_N]$, such that
\begin{align}\label{eq:stnl}
    T_N = \tau_m \sqrt{\log(N+2)}, \qquad L \leq (\log(N+2))^{C_{m,L_0}}
\end{align}
for $\tau_m > 0$ and $L_0 > 0$ that we will choose later.

We measure $\Pi_uB,\Pi_uC_1,\ldots,\Pi_uC_L$ in this order or in reverse order, each with probability $1/2$. For a Hermitian matrix $F$, we define the Kraus operator $K_x$ with width $\sigma = N^{-c_1/16}$, and the operator $O_{\mathbf x}$, as
\begin{align}
    K_x(F)&=(2\pi\sigma^2)^{-1/4} \exp[-(x-F)^2/(4\sigma^2)], \quad O_{\mathbf x}=K_{x_L}(\Pi_uC_L)\cdots K_{x_1}(\Pi_uC_1)K_{x_0}(\Pi_uB).
\end{align}
We choose $c_1$ to be no larger than the exponent in \Cref{lem:measurement-physical-input} and than $\min(c_0,\epsilon)$
Here, $O_{\mathbf x}$ corresponds to measurement in the forward order, and $O_{\mathbf x}^\dagger$ describes the reverse order. Since every factor in $O_{\mathbf x}$ commutes with $\rho(u)$, the post-measurement state is
\begin{align}
    \rho_x=\frac{\rho(u)(O_{\mathbf x}O_{\mathbf x}^\dagger
     +O_{\mathbf x}^\dagger O_{\mathbf x})}
    {2p_u(\mathbf x)}, \qquad p_u(\mathbf x)=\Tr(\rho(u)O_{\mathbf x}^\dagger O_{\mathbf x})=\Tr(\rho(u)O_{\mathbf x}O_{\mathbf x}^\dagger).
\end{align}
Finally, we define the quantity
\begin{align}\label{eq:def-posterior-source}
    a&=\Tr(\rho_xB)=\Tr(\rho_x\Pi_uB)
\end{align}
that will be repeatedly used.

\begin{lemma}[The posterior state]\label{lem:common-gaussian-instrument}
The density $p_u$ is the outcome density for the equal mixture of the forwards and reverse orders. After forgetting the order, the conditional post-measurement state is $\rho_x$, and
\begin{align}
 \int p_u(\mathbf x)\rho_x\,d\mathbf x=\rho(u),\qquad
 \EE|a|^{2j}\leq M_j\quad(j\geq1),
 \label{eq:posterior-jensen}
\end{align}
where the expectation $\EE$ is over 
\begin{align}
    d\mathbb P_m(J,u,\mathbf x) = p_u(\mathbf x)\,d\mathbf x\,d\mathbb P(J,u). \label{eq:joint-measurement-law}
\end{align}
\end{lemma}
\begin{proof}
All Kraus factors commute with $\rho(u)$ and preserve each energy
block.  They are positive and invertible, so $p_u(\mathbf x)>0$.
Completeness gives $\int K_x(F)^2\,dx =I$ and $\int O_{\mathbf x}^\dagger O_{\mathbf x}\,d\mathbf x = \int O_{\mathbf x}O_{\mathbf x}^\dagger\,d\mathbf x=I$, so
\begin{align}
 \int p_u(\mathbf x)\rho_x\,d\mathbf x
 &=\frac12\rho(u)\int
    (O_{\mathbf x}O_{\mathbf x}^\dagger+
       O_{\mathbf x}^\dagger O_{\mathbf x})\,d\mathbf x=\rho(u).
\end{align}
Jensen applied to the spectral measure of $B$ followed by \eqref{eq:gibbs-as-rounded-average} gives
\begin{align}
 \EE|a|^{2j}
 &\leq\EE\Tr(\rho_xB^{2j})
  =\EE_{J,u}\Tr(\rho(u)B^{2j})=M_j.
\end{align}
\end{proof}
The point of \eqref{eq:posterior-jensen} is
that the state $\rho(u)$ is exactly preserved after all measurement outcomes are forgotten. This yields concentration inequalities (\Cref{lem:sequential-measurement}) similar to \eqref{eq:mm1} and \eqref{eq:mm2}, such as
\begin{align}
    0\leq M_m-\EE|a|^{2m}
 &\leq C_m(N^{-b}M_{m-1}+N^{-mb}).\label{eq:main-posterior-moment-comparison}
\end{align}
and
\begin{align}
    \frac1{2N}\EE_{J,u}\int|a|^{2m}\sum_{i=1}^N
  \left(\|[\psi_i,O_{\mathbf x}]\|_{{\rm h},u}^2
       +\|[\psi_i,O_{\mathbf x}^\dagger]\|_{{\rm h},u}^2\right)
           \,d\mathbf x \leq C_m(N^{-b}M_{m-1}+N^{-mb})\label{eq:main-posterior-harmonic-bound},
\end{align}
where we used notation
\begin{align}
 \rho(u)&=\sum_jw_jP_j(u),\quad
 w_j=\frac{e^{-(u+j\delta_E)}}{\Tr e^{\beta\widetilde H(u)}},\quad  \|X\|_{{\rm h},u}^2
 =\sum_{j,k}\frac{w_jw_k}{w_j+w_k}
                \|P_k(u)XP_j(u)\|_{\rm HS}^2.
 \label{eq:main-harmonic-gibbs-norm}
\end{align}
\subsection{Conditional detailed balance}
We consider the transition law $\nu_x$ associated with the following procedure:
choose $i$ uniformly, measure the initial energy block $j$, apply the
unitary $\psi_i$, measure the final block $k$, and report
$E_k(u)-E_j(u)$, where $E_j(u)=u+j\delta_E$ is the energy of the $j$th energy block.

\begin{lemma}[The conditional transition law]
\label{lem:posterior-transition-law}
Define
\begin{align}
 \nu_x&=\frac1N\sum_{i=1}^N\sum_{j,k\in\mathbb Z}
   \Tr\left(\rho_xP_j(u)\psi_iP_k(u)\psi_iP_j(u)\right)
                    \delta_{E_k(u)-E_j(u)},
 \label{eq:raw-posterior-transition-law}\\
 \int f\,d\mathcal P\nu_x
 &=\int_{\mathbb R}
   \frac{f(|\omega|)+e^{-|\omega|}f(-|\omega|)}{1+e^{-|\omega|}}
                         \nu_x(d\omega).
 \label{eq:mass-preserving-db-projection}
\end{align}
Both measures are probabilities. The second preserves the distribution
of $|\omega|$ and satisfies detailed balance. In particular,
\begin{align}
 G(z)&=\frac12\int_{\mathbb R}e^{-z\omega}(\mathcal P\nu_x)(d\omega)
   =\int_{\mathbb R}
       \frac{\cosh(|\omega|(1/2-z))}{2\cosh(|\omega|/2)}\nu_x(d\omega),
 \label{eq:def-posterior-kernel}
\end{align}
satisfies $G(z)=G(1-z)$, and $D=\|\nu_x-\mathcal P\nu_x\|_{\rm TV}\leq2$.
Here the total variation norm is the full variation norm.
\end{lemma}
\begin{proof}
Each weight in \eqref{eq:raw-posterior-transition-law} is nonnegative,
since
\begin{align}
 P_j\psi_iP_k\psi_iP_j&=(P_k\psi_iP_j)^\dagger(P_k\psi_iP_j),\\
 \frac1N\sum_{i,j,k}\Tr(\rho_xP_j\psi_iP_k\psi_iP_j)
 &=\frac1N\sum_{i,j}\Tr(\rho_xP_j\psi_i^2P_j)=1.
\end{align}
The two coefficients in \eqref{eq:mass-preserving-db-projection}
are positive and sum to one. At frequency zero they combine to one.
At positive frequency their ratio gives
$(\mathcal P\nu_x)(-d\omega)=e^{-\omega}(\mathcal P\nu_x)(d\omega)$.
Substituting $f(\omega)=e^{-z\omega}$ proves the kernel formula and
its thermal reflection symmetry.
\end{proof}

Although the rounded Gibbs state satisfies detailed balance, conditioning can spoil it. We bound this change after weighting outcomes by $|a|^{2m}$. This leads to our first main bound, which will directly be used to show \Cref{thm:main}.

\begin{lemma}[Weighted detailed balance]\label{lem:conditional-db}
For the distribution \eqref{eq:joint-measurement-law} and every fixed
odd $m\geq1$, there is $b>0$, independent of $m$, such that, for each
integer $q\geq1$,
\begin{align}
 \EE\left[|a|^{2m}
   \|\nu_x^{\otimes q}-(\mathcal P\nu_x)^{\otimes q}\|_{\rm TV}\right]
 &\leq q\EE[|a|^{2m}D] \leq qC_m\sqrt{M_m\left(N^{-b}M_{m-1}+N^{-mb}\right)}.
 \label{eq:conditional-tensor-db}
\end{align}
\end{lemma}
\begin{proof}
Fix $(J,u,\mathbf x)$. Use the weights $w_j$ in
\eqref{eq:main-harmonic-gibbs-norm} and suppress $u$ in the projectors.
Let $q_{ijk}$ be the probability of choosing site $i$ and observing blocks $j,k$, so $\nu_x=\sum_{i,j,k}q_{ijk}\delta_{(k-j)\delta_E}$ for $q_{ijk} = \frac1N\Tr(\rho_xP_j\psi_iP_k\psi_iP_j)$. We can rewrite $q_{ijk}$ as
\begin{align}
 q_{ijk}
 &=\frac{w_j}{2Np_u(\mathbf x)}\left(
     \|P_k\psi_iP_jO_{\mathbf x}P_j\|_{\rm HS}^2
    +\|P_k\psi_iP_jO_{\mathbf x}^\dagger P_j\|_{\rm HS}^2\right)\label{eq:db-two-orientation-amplitudes}
\end{align}
and similarly define  $q_{ikj}=\frac{w_k}{2Np_u(\mathbf x)}\left(\|P_kO_{\mathbf x}P_k\psi_iP_j\|_{\rm HS}^2+\|P_kO_{\mathbf x}^\dagger P_k\psi_iP_j\|_{\rm HS}^2\right)$.
For $j<k$, the projection $\mathcal P$ sends $q_{ijk}$ to $\frac{q_{ijk}+q_{ikj}}{w_j+w_k}w_j$ (and similarly for $j \leftrightarrow k$). Using $w_k/w_j = e^{-(k-j)\delta_E}$ and the triangle inequality, we find that
\begin{align}
 D&\leq\sum_{i=1}^N\sum_{j<k}
           \frac{2|w_kq_{ijk}-w_jq_{ikj}|}{w_j+w_k}.
 \label{eq:db-redistribution}
\end{align}
For either $V=O_{\mathbf x}$ or $V=O_{\mathbf x}^\dagger$,
block diagonality implies $P_k\psi_iP_jVP_j-P_kVP_k\psi_iP_j=P_k[\psi_i,V]P_j$ and thus, by Cauchy--Schwarz,
\begin{align}
 \left|\|P_k\psi_iP_jVP_j\|_{\rm HS}^2 -\|P_kVP_k\psi_iP_j\|_{\rm HS}^2\right|
 &\leq\|P_k[\psi_i,V]P_j\|_{\rm HS}\left(\|P_k\psi_iP_jVP_j\|_{\rm HS} +\|P_kVP_k\psi_iP_j\|_{\rm HS}\right).
 \label{eq:db-amplitude-difference}
\end{align}
Using the definition of $p_u(\mathbf x)$, we bound
\begin{align}
 \sum_{i,j<k}\frac{w_jw_k}{w_j+w_k}
                   \|P_k\psi_iP_jVP_j\|_{\rm HS}^2
 \leq\sum_{i,j,k}w_j\|P_k\psi_iP_jVP_j\|_{\rm HS}^2=N\Tr(\rho(u)V^\dagger V)=Np_u(\mathbf x).
 \label{eq:db-amplitude-sum}
\end{align}
Substitute \eqref{eq:db-two-orientation-amplitudes} into
\eqref{eq:db-redistribution}, apply \eqref{eq:db-amplitude-difference},
and use Cauchy--Schwarz over $i,j,k$ and the two choices of $V$:
\begin{align}
 p_u(\mathbf x)D^2
 &\leq\frac8N\sum_{i=1}^N\sum_{j<k}\frac{w_jw_k}{w_j+w_k}
    \left(\|P_k[\psi_i,O_{\mathbf x}]P_j\|_{\rm HS}^2
        +\|P_k[\psi_i,O_{\mathbf x}^\dagger]P_j\|_{\rm HS}^2\right)
 \nonumber\\
 &\leq\frac8N\sum_{i=1}^N
    \left(\|[\psi_i,O_{\mathbf x}]\|_{{\rm h},u}^2
        +\|[\psi_i,O_{\mathbf x}^\dagger]\|_{{\rm h},u}^2\right).
 \label{eq:pointwise-db-kraus-bound}
\end{align}
The factor $8$ follows from $(x+y)^2\leq2(x^2+y^2)$ and \eqref{eq:db-amplitude-sum} (also with $j \leftrightarrow k$). A second Cauchy--Schwarz inequality, now under
\eqref{eq:joint-measurement-law}, gives
\begin{align}
 \bigl(\EE[|a|^{2m}D]\bigr)^2 \leq\EE|a|^{2m}\,\EE[|a|^{2m}D^2] \leq M_m\EE_{J,u}\int |a|^{2m}p_u(\mathbf x)D^2\,d\mathbf x.
\end{align}
Applying \eqref{eq:pointwise-db-kraus-bound} and \eqref{eq:main-posterior-harmonic-bound} gives the second inequality of \eqref{eq:conditional-tensor-db}, i.e.,
\begin{align}
 \bigl(\EE[|a|^{2m}D]\bigr)^2 \leq C_mM_m(N^{-b}M_{m-1}+N^{-mb}).
 \label{eq:balance-defect-by-kraus-commutator}
\end{align}
To show the first inequality, we rewrite $\nu^{\otimes q}-(\nu')^{\otimes q}=\sum_{j=1}^q\nu^{\otimes(j-1)}\otimes(\nu-\nu')\otimes(\nu')^{\otimes(q-j)}$ and use $\|\nu^{\otimes(j-1)}\otimes(\nu-\nu') \otimes(\nu')^{\otimes(q-j)}\|_{\rm TV} =\|\nu-\nu'\|_{\rm TV}$ to obtain $ \|\nu_x^{\otimes q}-(\mathcal P\nu_x)^{\otimes q}\|_{\rm TV} \leq q\|\nu_x-\mathcal P\nu_x\|_{\rm TV}=qD$. This implies the first inequality of \eqref{eq:conditional-tensor-db}.
\end{proof}

To show $M_m$ is small it suffices to show that $\EE |a|^{2m}$ is small by \eqref{eq:main-posterior-moment-comparison}. Later on, the cavity method will later allow us to use detailed balance to control $\EE |a|^{2m} \norm{G-G_*}_2^2$. We need more fine-grained control over the error caused by violating detailed balance. In particular, we will now show that if $\norm{G-G_*}_2^2$ is small then $a$ is also small, up the detailed balance approximation error $D$.

Before showing this fact, we introduce a useful lemma proven in \Cref{app:recovery}.
\begin{lemma}[From imaginary time to real time]
\label{lem:heat-to-real-recovery}
For every fixed $\beta \geq 0$ and $T_0<\infty$, there are $C<\infty$ and $\zeta>0$
such that every normalized positive spectral kernel $G$ satisfies
\begin{align}
 \sup_{|t|\leq T_0}|G(it)-G_*(it)|
 \leq C\|G-G_*\|_{L^2(0,1)}^\zeta.
 \label{eq:complex-time-recovery}
\end{align}
The constants depend only on $\beta$ and $T_0$.
\end{lemma}
This lemma gives us control over $G$ applied at imaginary times (corresponding to two-point functions in real time). Naively, while large frequencies are exponentially suppressed in Euclidean time evolution, they can contribute large oscillations in real time evolution. The key to proving \Cref{lem:heat-to-real-recovery} is a tail bound \eqref{eq:saddle-exponential-moment} on $G_*$: for $\beta>0$, the probability measure $\mu_{*,\beta}$ representing $G_*$ satisfies (\Cref{lem:positive-saddle})
\begin{align}
 \int e^{x/(2\beta)}\mu_{*,\beta}(dx)&\leq1+\sqrt2.
\end{align}
We can now show the bound we ultimately need to prove \Cref{thm:mbound}.

\begin{lemma}[Recovery of the posterior mean]
\label{lem:raw-source-recovery}
There is a fixed integer $r\geq1$, independent of $m$, such that the
posterior mean satisfies
\begin{align}
 |a|^{2r}\leq C\|G-G_*\|_2^2+CD+C\delta_E^{2r}.
 \label{eq:raw-source-recovery-power}
\end{align}
\end{lemma}
\begin{proof}
Write the Hermitian linear combination in \Cref{thm:mbound} as
\begin{align}
 B&=\sum_{\ell=1}^s
       \bigl(c_\ell B(t_\ell)+\overline{c_\ell}B(t_\ell)^\dagger\bigr)
\end{align}
We bound
\begin{align}
    |a| &\leq 2\sum_{\ell=1}^s |c_\ell| \lr{|G(it_\ell) - G_*(it_\ell)| + \left|\Tr(\rho_xB(t))-(G(it)-G_*(it))\right|}\\
    &\leq C\Big(\sup_{|t|\leq \max_\ell|t_\ell|}|G(it)-G_*(it)| + \|e^{-it\beta H}\psi_i e^{it\beta H}
       -e^{-it\beta\widetilde H(u)}\psi_ie^{it\beta\widetilde H(u)}\| \nonumber\\
       &\qquad + \left|\frac12\int e^{-it\omega}\nu_x(d\omega)-G(it)\right|\Big),
\end{align}
where in the second line we used the fact that $\rho_x$ preserves every energy block and $-\beta\wt H(u) P_j(u) = E_j(u) P_j(u)$, so expansion in the projectors gives, for every real $t$,
\begin{align}
 \frac1{2N}\sum_i\Tr\left(
 \rho_xe^{-it\beta\widetilde H(u)}\psi_i
                  e^{it\beta\widetilde H(u)}\psi_i\right)
 &=\frac1{2N}\sum_{i,j,k}
       e^{-it(E_k(u)-E_j(u))}
       \Tr\bigl(\rho_xP_j(u)\psi_iP_k(u)\psi_iP_j(u)\bigr)
       \\
 &=\frac12\int e^{-it\omega}\nu_x(d\omega)
\end{align}
and thus the centering terms $G_*(it)$ cancel, so
\begin{align}
 &\Tr(\rho_xB(t))-(G(it)-G_*(it))\nonumber\\*
 &\quad=\frac1{2N}\sum_i\Tr\left[\rho_x
 \left(e^{-it\beta H}\psi_i e^{it\beta H}
       -e^{-it\beta\widetilde H(u)}\psi_i
          e^{it\beta\widetilde H(u)}\right)\psi_i\right]
       +\frac12\int e^{-it\omega}(\nu_x-\mathcal P\nu_x)(d\omega).
 \label{eq:raw-source-rounded-decomposition}
\end{align}
\Cref{lem:exact-rounded-gibbs} and Duhamel's formula imply
\begin{align}
 \|e^{-it\beta H}\psi_i e^{it\beta H}
       -e^{-it\beta\widetilde H(u)}\psi_i
                           e^{it\beta\widetilde H(u)}\|
 &\leq2\|e^{-it\beta H}-e^{-it\beta\widetilde H(u)}\| \leq 2|t|\|\beta H-\beta\widetilde H(u)\|
 \leq2|t|\delta_E.
 \label{eq:rounded-real-time-error}
\end{align}
Since $\left|\frac12\int e^{-it\omega}\nu_x(d\omega)-G(it)\right|=\frac12\left|\int e^{-it\omega}(\nu_x-\mathcal P\nu_x)(d\omega)\right|\leq\frac D2$, we have
\begin{align}
 |a| \leq C\sup_{|t|\leq \max_\ell|t_\ell|}|G(it)-G_*(it)|+CD+C\delta_E \leq C\|G-G_*\|_2^\zeta+CD+C\delta_E,
\end{align}
where the last inequality is \Cref{lem:heat-to-real-recovery}.  Choose
$r=\max\{1,\lceil1/\zeta\rceil\}$.  Since
$\|G-G_*\|_2\leq1$ and $0\leq D\leq2$,
\begin{align}
 |a|^{2r}
 &\leq C\|G-G_*\|_2^{2r\zeta}+CD^{2r}+C\delta_E^{2r} \leq C\|G-G_*\|_2^2+C2^{2r-1}D+C\delta_E^{2r}.
\end{align}
Absorbing $2^{2r-1}$ into $C$ proves the claim.
\end{proof}

\subsection{Adding a noninteracting cavity fermion}
Before applying the cavity argument, we show that imaginary-time correlators between disjoint Majorana strings can be rewritten in terms of the $G$ functions of \eqref{eq:def-posterior-kernel} (i.e., defined from $\cP\nu_x$) up to small error. The main fact we need relates $G$ to two-point functions. Assume $|s_j|,|t_j|,|s_j-t_j|\leq T_N = \tau_m \sqrt{\log(N+2)}$, where $\tau_m$ is sufficiently small to satisfy
\begin{align}\label{eq:taum}
    (L+1)^4(1+T_N)^{C_m}e^{C_m(1+T_N)^2} \leq C_mN^{c_1/32}.
\end{align}
\Cref{lem:posterior-product} gives, for fixed or logarithmically growing $r$ and some $b > 0$ independent of $m$
\begin{align}
 &\left|\EE_{J,u}\Tr\left[\rho(u)(\Pi_uB)^{2m} \prod_{j=1}^r \lr{\frac1N\sum_i
     (e^{-is_j\beta H}\psi_i e^{is_j\beta H})
     (e^{-it_j\beta H}\psi_i e^{it_j\beta H})}\right]\right.\nonumber\\*
 &\qquad\left.-\EE\left[|a|^{2m}\prod_{j=1}^r2G(i(s_j-t_j))\right]\right|
 \nonumber\\*
 &\qquad\leq C_m(1+r)^{C_m}
                  (N^{-b}M_{m-1}+N^{-mb})+C_mr^2N^{-L_0}.
 \label{eq:main-continuous-conditional-product}
\end{align}
We will use this real-time bound and continue it to imaginary time. To do this, we use the following fact.
\begin{fact}[Three-lines theorem]\label{fact:three-lines}
Let $F$ be bounded and holomorphic on $0<\Re z<1$, and continuous on
its closure. If $|F(j+iv)|\leq A_j$ for $j=0,1$ and every $v\in\mathbb R$,
then
\begin{align}
 |F(t+iv)|\leq A_0^{1-t}A_1^t
 \qquad(0\leq t\leq1,\ v\in\mathbb R).
 \label{eq:three-lines-fact}
\end{align}
\end{fact}

For later use in the cavity method, we consider correlators that act on an enlarged system: we add fermion $\psi_{N+1}$ decoupled from the rest of the system (so $[H,\psi_{N+1}] = 0$). In the next subsection, we will use the below representation in $G$ functions to show that interactions between the cavity fermion and the SYK Hamiltonian can effectively be replaced by interactions between the cavity fermion and a free fermion bath, without changing the values of any local correlators. Crucially, the analysis is restricted to coupling $\psi_{N+1}$ with sites $T_e \in \binom{[N]}{3}$ that are mutually disjoint. Since every $T_e$ appears twice, each site will appear exactly twice; this allows us to cleanly decompose everything into two-point functions. This restriction is a good approximation to choosing $T_e$ uniformly at random, and we will bound the error associated with the approximation.

\begin{lemma}[Replacing two-point functions with $G$]
\label{lem:cavity-coefficient-comparison}
There is $a_0>0$, independent of the fixed odd integer $m\geq1$, with
the following property. Fix parameters \eqref{eq:stnl} with $L_0>ma_0+10$ and $T_N$ satisfying \eqref{eq:taum}, and let $P_N=\lceil C_{m,L_0}\log(N+2)\rceil$. For $A\subseteq[N+1]$ with $|A|\leq4$, $d\geq0$, and $\pi\in\mathcal P_2([2d])$, set $\pi(j)$ to be the unique $e \in \pi$ such that $j \in e$.
Let
\begin{align}
 \mathcal R_{A,\pi}
 =\left\{(T_e)_{e\in\pi}\in\prod_{e\in\pi}\binom{[N]}3:
 \begin{array}{ll}
 T_e\cap T_f=\varnothing& (e\ne f),\\
 T_e\cap(A\cap[N])=\varnothing& (e\in\pi)
 \end{array}\right\}.
 \label{eq:cavity-retained-assignments}
\end{align}
For an arbitrary assignment $\mathbf T$,
$1>t_1>\cdots>t_{2d}>0$, and $0<t<1$, set $\Gamma_E(s)=e^{-s\beta H}\Gamma_Ee^{s\beta H}$ and
\begin{align}
 c_{A,\pi}(\mathbf T;\mathbf t,t)
 &=\frac{\EE_J\tr\left[B^{2m}e^{\beta H}\mathcal T\left(
 \Gamma_A(t)\Gamma_A(0)
 \prod_{j=1}^{2d}\Gamma_{T_{\pi(j)}\cup\{N+1\}}(t_j)
 \right)\right]}{\EE_J\tr e^{\beta H}},
 \label{eq:literal-cavity-coefficient}\\
 \ell_{A,\pi}(\mathbf t,t)
 &=\epsilon_{A,\pi}\EE\left[|a|^{2m}(2G(t))^{|A\cap[N]|}
 \prod_{\substack{\{j,k\}\in\pi\\j<k}}(2G(t_j-t_k))^3\right],
 \label{eq:ellapi}\\
 \epsilon_{A,\pi}
 &=(-1)^{\operatorname{cr}(\pi)+
 \mathbf1_{N+1\in A}\#\{\{j,k\}\in\pi:t_j>t>t_k\}}.
 \label{eq:scalar-cavity-coefficient}
\end{align}
For $d\leq P_N$ and
$\mathbf T\in\mathcal R_{A,\pi}$,
\begin{align}
 |c_{A,\pi}(\mathbf T;\mathbf t,t)-\ell_{A,\pi}(\mathbf t,t)|
 &\leq C_me^{C_md}(1+d)^{C_m}
 \left[N^{a_0/4}(N^{-a_0}M_{m-1}+N^{-ma_0})+N^{-L_0}\right].
 \label{eq:cavity-positive-heat-comparison}
\end{align}
The constants may depend on the fixed $L_0$.
\end{lemma}
\begin{proof}
If $\mathbf T,\mathbf T'\in\mathcal R_{A,\pi}$, a permutation of $[N]$
fixing $A\cap[N]$ maps each $T_e$ to $T'_e$. Due to the invariance of the disorder
and the empirical source $B$ under this relabeling, we have
\begin{align}
 c_{A,\pi}(\mathbf T;\mathbf t,t)
 =c_{A,\pi}(\mathbf T';\mathbf t,t).
 \label{eq:cavity-assignment-invariance}
\end{align}
Hence, we will suppress writing the $\mathbf T$ dependence in $c_{A,\pi}$.

If the ordered times are $\tau_1>\cdots>\tau_s=0$, with
$s=2d+2$, and $X_j$ is the operator applied at time $\tau_j$, the numerator is of the form
\begin{align}
 \EE_J\tr\left(B^{2m}e^{(1-\tau_1)\beta H}X_1
        e^{(\tau_1-\tau_2)\beta H}X_2\cdots
        e^{(\tau_{s-1}-\tau_s)\beta H}X_s\right),
\end{align}
where the operators $X_j$ are static (i.e., do not involve real or imaginary time evolution under the Hamiltonian). We reparameterize these times with the differences
\begin{align}
 h_1=1-\tau_1,\qquad h_j=\tau_{j-1}-\tau_j\ (2\leq j\leq s),
 \qquad \sum_{j=1}^s h_j=1,
 \label{eq:cavity-insertion-intervals}
\end{align}
so
\begin{align}
 c_{A,\pi}(\mathbf t,t)
 =\frac{\EE_J\tr\left(B^{2m}e^{h_1\beta H}X_1
 e^{h_2\beta H}X_2\cdots e^{h_s\beta H}X_s\right)}
 {\EE_J\tr e^{\beta H}}.
 \label{eq:cavity-ordered-word}
\end{align}
For this ordering, we will consider the more general version with complex times:
\begin{align}
 c_{A,\pi}(\mathbf z)
 =\frac{\EE_J\tr\left(B^{2m}\prod_{j=1}^s e^{z_j\beta H}X_j\right)}
 {\EE_J\tr e^{\beta H}},\qquad
 \Re z_j\geq0,\quad\sum_{j=1}^s z_j=1.
 \label{eq:cavity-complex-coefficient}
\end{align}
If the two external insertions have positions $p_A<q_A$, and the
interaction at $t_j$ has position $p(j)$, the corresponding complex version of $\ell_{A,\pi}$ is
\begin{align}
 \ell_{A,\pi}(\mathbf z)
 =\epsilon_{A,\pi}\EE\Bigg[|a|^{2m}
 \left(2G\left(\sum_{r=p_A+1}^{q_A}z_r\right)\right)^{|A\cap[N]|}
 \prod_{\substack{\{j,k\}\in\pi\\j<k}}
 \left(2G\left(\sum_{r=p(j)+1}^{p(k)}z_r\right)\right)^3\Bigg].
 \label{eq:cavity-scalar-complex-coefficient}
\end{align}
We first analyze these complex versions of $c_{A,\pi}$ and $\ell_{A,\pi}$ for $\mathbf z$ satisfying, for some $b \in [s]$ and real $v_j$,
\begin{align}\label{eq:vt}
    z_j = \mathbf 1_{j=b} + iv_j, \quad \sum_j v_j = 0, \quad T = \sum_j |v_j|.
\end{align}
Hence, the $b$th interval carries all the imaginary-time evolution. For this choice of $\mathbf z$, cyclicity implies $c_{A,\pi}(\mathbf z) = \Phi_N(U_b B^{2m} V_b)$ for unitaries
\begin{align}\label{eq:cavity-boundary-unitaries}
    U_b = \prod_{j=b}^s e^{iv_j \beta H}X_j, \qquad V_b = \prod_{j=1}^{b-1} e^{iv_j\beta H} X_j.
\end{align}
To use \eqref{eq:main-continuous-conditional-product} to relate our correlators to $G$ functions, we first need to go from SYK thermal states to expectations on $\rho(u)$ via spectral rounding. Using cyclicity and \Cref{lem:exact-rounded-gibbs} for the exact equalities and \Cref{lem:cavity-source-rounding} for the approximations, we have
\begin{align}
    \Phi_N(U_b B^{2m} V_b) \approx \Phi_N(B^{2m} U_b V_b) = \EE_{J,u} \Tr[\rho(u) B^{2m} U_b V_b] \approx \EE_{J,u} \Tr[\rho(u) (\Pi_u B)^{2m} U_b V_b].
\end{align}
The proof of \Cref{lem:cavity-source-rounding} uses \eqref{eq:mm1}--\eqref{eq:mm2} and a bound \eqref{eq:source-power-commutator} on commutators.

\proofstep{Two-point functions}
To obtain two-point functions that can turn into $G$ functions, we rewrite $U_b V_b$. For $(k_1, \dots, k_s) = (b,\dots,s,1,\dots,b-1)$ and $\alpha_\ell = \sum_{h=1}^\ell v_{k_h}$, we have
\begin{align}\label{eq:uv}
    U_b V_b = \prod_{\ell=1}^s e^{i\alpha_\ell \beta H} X_{k_\ell} e^{-i\alpha_\ell \beta H}, \quad |\alpha_\ell| \leq T, \quad \sum_{\ell=1}^{s-1} |\alpha_{\ell+1}-\alpha_\ell| \leq T.
\end{align}
The factors $X_{k_\ell}$ are now reordered. For distinct $e \neq f$, these static factors satisfy
\begin{align}
 \Gamma_{T_e\cup\{N+1\}}\Gamma_{T_f\cup\{N+1\}}
 =-\Gamma_{T_f\cup\{N+1\}}\Gamma_{T_e\cup\{N+1\}},\quad
 \Gamma_A\Gamma_{T_e\cup\{N+1\}} =(-1)^{\mathbf1_{N+1\in A}}
           \Gamma_{T_e\cup\{N+1\}}\Gamma_A.
\end{align}
Counting signs from crossings, we observe that
\begin{align}
 X_1\cdots X_s=\epsilon_{A,\pi}I,
 \qquad X_b\cdots X_sX_1\cdots X_{b-1}=\epsilon_{A,\pi}I.
 \label{eq:cavity-clifford-signs}
\end{align}
The sign $\epsilon_{A,\pi}$ is acquired from reordering $X_1,\dots,X_s$ into $X_{k_1},\dots,X_{k_s}$. However, $U_bV_b$ also has time evolution. This is controlled more carefully in \Cref{lem:cavity-majorana-pairing} to show that, up to a $T$-dependent approximation error, the real-time evolved operators $\psi_i(t) = e^{-it\beta H}\psi_i e^{it\beta H}$ satisfy
\begin{align}
    \EE_{J,u} \Tr[\rho(u) (\Pi_u B)^{2m} U_b V_b] \approx \epsilon_{A,\pi} \EE_{J,u} \Tr[\rho(u) (\Pi_u B)^{2m} \prod_{\substack{i \in A \cap [N]\\ i \in \cup_{e\in\pi} T_e}} \psi_i(-\alpha_{p_i}) \psi_i(-\alpha_{q_i})],
\end{align}
where each $i$ corresponds to exactly two of the static operators $X_{k_1},\dots,X_{k_s}$, and we denote their positions as $p_i, q_i \in [s]$ ordered as $p_i < q_i$. The product over $i$ is ordered according to increasing $q_i$ and takes at most
\begin{align}
    r = |A\cap[N]| + 3d \leq 4 + 3P_N = O_{m,L_0}(\log(N+2))
\end{align}
values. This is close to sampling with replacement: by permutation invariance, the labels can be averaged uniformly without replacement, with collisions introducing $O(r^2 M_m/N)$ error. In \Cref{lem:cavity-site-average} we thus show that
\begin{align}
    \EE_{J,u} \Tr[\rho(u) (\Pi_u B)^{2m} U_b V_b] \approx \epsilon_{A,\pi} \EE_{J,u} \Tr[\rho(u) (\Pi_u B)^{2m} \prod_{\substack{i \in A \cap [N]\\ i \in \cup_{e\in\pi} T_e}} \lr{\frac{1}{N}\sum_{j=1}^N \psi_j(-\alpha_{p_i}) \psi_j(-\alpha_{q_i})}].
\end{align}
Finally, if $T$ is sufficiently small, we can relate these two-point functions to $G$: by \eqref{eq:uv} when $T\leq T_N$, we can apply \eqref{eq:main-continuous-conditional-product} (shown in \Cref{lem:posterior-product}). Finally, we sum all the errors from the approximations above and use the identity
\begin{align}
 G\left(\sum_{j=p+1}^qz_j\right)
 &=\begin{cases}
 G\left(i\sum_{j=p+1}^qv_j\right),&b\notin\{p+1,\ldots,q\},\\
 G\left(-i\sum_{j=p+1}^qv_j\right),&b\in\{p+1,\ldots,q\}
 \end{cases}
 \label{eq:cavity-pair-orientation}
\end{align}
from detailed balance to obtain, for some $a_0>0$ independent of $d$ and $m$, that
\begin{align}
    \abs{c_{A,\pi}(\mathbf z) - \ell_{A,\pi}(\mathbf z)} &\leq C_m[(1+d)(1+T)\log(N+2)]^{C_m}e^{C_m(1+T)^2}
       (N^{-a_0}M_{m-1}+N^{-ma_0})\nonumber\\
       &\quad +C_m(1+d)^2N^{-L_0}\label{eq:cavity-vertex-comparison}
\end{align}
when $d \leq P_N$ and $T \leq T_N$.

\proofstep{Analytic continuation} So far, we have only analyzed that $c_{A,\pi}(\mathbf z) \approx \ell_{A,\pi}(\mathbf z)$ for our particular choice of $z_j = 1\{j=b\}+iv_j$. We now extend this to a target $\mathbf h$ that satisfies $h_j>0$ and $\sum_jh_j=1$. Define for some $\lambda > 0$ that we will choose shortly
\begin{align}
 f(\mathbf z)&=(c_{A,\pi}(\mathbf z)-\ell_{A,\pi}(\mathbf z))
       \exp\left[-\lambda\sum_{j=1}^s\frac{z_j(1-z_j)}{h_j}\right].
 \label{eq:cavity-gaussian-damping}
\end{align}
For our choice of $\mathbf z$, it is easy to check that this satisfies
\begin{align}
 |f(\mathbf z)|\leq|c_{A,\pi}(\mathbf z)-\ell_{A,\pi}(\mathbf z)| e^{-\lambda\|\mathbf v\|_1^2}
\end{align}
using $\Re \sum_j z_j(1-z_j)/h_j \geq \norm{\mathbf v}_1^2$.
Choose $L_0>ma_0+10$, then decrease $\tau_m$ so that
\begin{align}
 [(1+T_N)\log(N+2)]^{C_m}e^{C_m(1+T_N)^2} \leq C_mN^{a_0/4}.
\end{align}
Then choose $\lambda$ such that $\lambda\tau_m^2>L_0+2$.
Using \eqref{eq:cavity-vertex-comparison} for $\norm{\mathbf v}_1\leq T_N$ (and otherwise using that $|c_{A,\pi}(\mathbf z)|, |\ell_{A,\pi}(\mathbf z)|\leq1$), we find for our choice $\mathbf z = e_b + i\mathbf v$ (where $e_b=(0,\dots,0,1,0,\dots0)$ has 1 at $j=b$) that
\begin{align}
 \sup_{\substack{\mathbf v\in\mathbb R^s\\\sum_jv_j=0}}
        |f(e_b+i\mathbf v)|
 &\leq C_m(1+d)^{C_m}
   \left[N^{a_0/4}(N^{-a_0}M_{m-1}+N^{-ma_0})+N^{-L_0}\right].
 \label{eq:cavity-damped-boundary}
\end{align}
For two real vectors $\mathbf x,\mathbf y$ in the simplex, apply \Cref{fact:three-lines} to $z \mapsto f((1-z)\mathbf x + z\mathbf y + i\mathbf v)$. Taking the supremum over $\mathbf v$ gives
\begin{align}
 \sup_{\sum_jv_j=0}|f((1-t)\mathbf x+t\mathbf y+i\mathbf v)|
 &\leq\left(\sup_{\sum_jv_j=0}|f(\mathbf x+i\mathbf v)|\right)^{1-t}
       \left(\sup_{\sum_jv_j=0}|f(\mathbf y+i\mathbf v)|\right)^t,
       \quad0\leq t\leq1
\end{align}
and
\begin{align}
 |f(\mathbf h)|
 &\leq\prod_{b=1}^s
       \left(\sup_{\sum_jv_j=0}|f(e_b+i\mathbf v)|\right)^{h_b}
 \leq\max_b\sup_{\sum_jv_j=0}|f(e_b+i\mathbf v)|.
\end{align}
Finally, using $\exp\left[\lambda\sum_j\frac{h_j(1-h_j)}{h_j}\right]=e^{\lambda(s-1)}\leq e^{\lambda(2d+1)}$, we find $|c_{A,\pi}(\mathbf h)-\ell_{A,\pi}(\mathbf h)| \leq e^{\lambda(s-1)}|f(\mathbf h)|$ and thus
\begin{align}
 |c_{A,\pi}(\mathbf h)-\ell_{A,\pi}(\mathbf h)|
 &\leq C_me^{C_md}(1+d)^{C_m}
   \left[N^{a_0/4}(N^{-a_0}M_{m-1}+N^{-ma_0})+N^{-L_0}\right]
\end{align}
for all $\mathbf h$ with $h_j > 0$ and $\sum_j h_j = 1$.
\end{proof}

\subsection{Adding the cavity interaction}
Couple the cavity Majorana $\psi_{N+1}$ to the original $N$ sites using independent disorder $J_T^+\stackrel{\rm iid}{\sim}\mathcal N(0,1)$:
\begin{align}
 H^+&=H+\sqrt{\frac3{8N^3}}
        \sum_{T\in\binom{[N]}3}J_T^+\Gamma_{T\cup\{N+1\}}.
 \label{eq:cavity-hamiltonian-split}
\end{align}
As discussed in \Cref{sec:cavity}, the cavity method replaces the interacting bath $H$ of $\psi_{N+1}$ with a quadratic bath using fermions $\eta_j, \zeta_j$, i.e.,
\begin{align}\label{eq:main-quadratic-hamiltonian}
    H_{\rm quad}=H_{\rm bath}+\frac i2\sum_jv_j\psi_{N+1}\eta_j, \qquad H_{\rm bath} =\frac i2\sum_jx_j\eta_j\zeta_j.
\end{align}
For measure $\mu$ on $[0,\infty)$, define the measure $\nu_\mu$ on $\R$ by
\begin{align}
 \int f\,d\nu_\mu
 =\int_0^\infty\frac{f(x)+e^{-x}f(-x)}{1+e^{-x}}\,\mu(dx).
\end{align}
It has the same mass as $\mu$ and by construction satisfies detailed balance. Recall from \eqref{eq:gmu} that we then have
\begin{align}
 G_\mu(t)=\frac{1}{2}\int e^{-t\omega} \nu_\mu(d\omega) = \int_0^\infty
       \frac{\cosh(x(1/2-t))}{2\cosh(x/2)}\,\mu(dx),\qquad 0\leq t\leq1,
\end{align}
As suggested by \eqref{eq:sigmaguess}, we now choose the couplings $x_j, v_j$ so that $\Sigma = G_\sigma = \beta^2 G^3$, where we set measure $\sigma=\sum_jv_j^2\delta_{x_j}$ and defined $G$ from \eqref{eq:def-posterior-kernel} (i.e., defined with respect to $\cP \nu_x$). In the notation of the following lemma, the SYK saddle point equations \eqref{eq:sd} would be written as $\Sigma = \beta^2 G^3$ (which we satisfy by construction) and $G = \cD(\Sigma)$; we show that $\cD(\Sigma)$ can be represented as a two-point function in the free fermion bath.
\begin{lemma}[Free fermion two-point function]
\label{lem:quadratic-dyson-representation}
For every posterior kernel $G$ in \eqref{eq:def-posterior-kernel},
there are $x_j\geq0$ and $v_j\geq0$ such that the measure $\sigma=\sum_jv_j^2\delta_{x_j}$ normalized as $\sigma([0,\infty))=\frac{\beta^2}{4}$ satisfies $G_\sigma(t)=\beta^2G(t)^3$. Set $\Sigma=G_\sigma$ and define
$\widehat{\mathcal D(\Sigma)}(n)
=(-i\omega_n-\widehat\Sigma(n))^{-1}$. Then the quadratic Hamiltonian of \eqref{eq:main-quadratic-hamiltonian} satisfies
\begin{align}
 \mathcal D(\Sigma)(t)
 &=\frac{\Tr(e^{-(1-t)H_{\rm quad}}\psi_{N+1}
                         e^{-tH_{\rm quad}}\psi_{N+1})}
          {2\Tr e^{-H_{\rm quad}}},\qquad 0\leq t\leq1.
 \label{eq:xsd}
\end{align}
\end{lemma}
\begin{proof}
The projected law $\mathcal P\nu_x$ in
\eqref{eq:def-posterior-kernel} is a finitely supported probability
satisfying detailed balance. Consequently, for every bounded Borel $f$,
\begin{align}
 \int f(-\omega)\,(\mathcal P\nu_x)^{*3}(d\omega) &=\int_{\mathbb R^3}f(-\omega_1-\omega_2-\omega_3)
                  \prod_{j=1}^3(\mathcal P\nu_x)(d\omega_j)\nonumber\\*
 &=\int_{\mathbb R^3}f(\omega_1+\omega_2+\omega_3)
       e^{-(\omega_1+\omega_2+\omega_3)}
                  \prod_{j=1}^3(\mathcal P\nu_x)(d\omega_j)\nonumber\\*
 &=\int f(\omega)e^{-\omega}
                         (\mathcal P\nu_x)^{*3}(d\omega).
 \label{eq:convolution-detailed-balance}
\end{align}
Define $\mu(S)=(\mathcal P\nu_x)^{*3}
(\{\omega:|\omega|\in S\})$ for Borel $S\subseteq[0,\infty)$ and observe that $\nu_\mu = (\mathcal P\nu_x)^{*3}$ and
\begin{align}
 G(t)^3
 &=\frac18\int e^{-t\omega}(\mathcal P\nu_x)^{*3}(d\omega)
 =\frac14G_\mu(t).
 \label{eq:cubic-positive-source}
\end{align}
Thus $\sigma=(\beta^2/4)\mu$ as claimed and, since it has finite support, it can be written as $\sigma=\sum_jv_j^2\delta_{x_j}$ by choosing $v_j^2 = (\beta^2/4)\mu(\{x_j\})$.

To show \eqref{eq:xsd}, define a matrix $K$ indexed
by $(\psi_{N+1},\eta_1,\zeta_1,\ldots)$, with upper entries
$K_{0,2j-1}=v_j$ and $K_{2j-1,2j}=x_j$, all other upper entries zero.
Let $P_y$ project onto the sum of its skew blocks of frequency $y\geq0$. A single block gives
\begin{align}
 \frac{\Tr(e^{-(1-t)\frac{iy}{2}\eta\zeta}\eta
                     e^{-t\frac{iy}{2}\eta\zeta}\eta)}
      {2\Tr e^{-\frac{iy}{2}\eta\zeta}}
 &=\frac{\cosh(y(1/2-t))}{2\cosh(y/2)},\nonumber\\*
 \frac{\Tr(e^{-(1-t)H_{\rm quad}}\psi_{N+1}
                     e^{-tH_{\rm quad}}\psi_{N+1})}
      {2\Tr e^{-H_{\rm quad}}}
 &=\sum_y\|P_ye_0\|^2
                   \frac{\cosh(y(1/2-t))}{2\cosh(y/2)}.
 \label{eq:quadratic-block-traces}
\end{align}
Here $e_0$ is the unit vector of the cavity coordinate and
$\sum_y\|P_ye_0\|^2=1$. Taking the Fourier coefficient of the last
line and using the Schur complement \eqref{eq:schur-fact} gives
\begin{align}
 &\int_0^1e^{i\omega_nt}
   \frac{\Tr(e^{-(1-t)H_{\rm quad}}\psi_{N+1}
                     e^{-tH_{\rm quad}}\psi_{N+1})}
      {2\Tr e^{-H_{\rm quad}}}\,dt =i\sum_y\|P_ye_0\|^2
                  \frac{\omega_n}{\omega_n^2+y^2}
       =i[(\omega_nI-K)^{-1}]_{00}\nonumber\\*
 &\quad=\frac{i}{\omega_n\left(1+\displaystyle\sum_j
                           \frac{v_j^2}{\omega_n^2+x_j^2}\right)}
       =\bigl(-i\omega_n-\widehat\Sigma(n)\bigr)^{-1}.
 \label{eq:main-quadratic-resolvent}
\end{align}
The last equality uses
$\widehat\Sigma(n)=i\omega_n\sum_jv_j^2/(\omega_n^2+x_j^2)$.
Equality of Fourier coefficients proves \eqref{eq:xsd}.
\end{proof}

We introduce function $W$ acting on $\Sigma=G_\sigma$ to keep track of the change in the partition function when the couplings $v_j$ are turned on. We show in \Cref{lem:quadratic-cut} that the following definitions are equivalent:
\begin{align}
 W(\Sigma)&=\frac{\Tr e^{-H_{\rm quad}}}{\Tr e^{-H_{\rm bath}}}\\
 &= \prod_{n\geq0} \left(1+\int\frac{\sigma(dx)}{\omega_n^2+x^2}\right)\\
 &=\sum_{d\geq0}2^{-d}\sum_{\pi\in\mathcal P_2([2d])}
   (-1)^{\operatorname{cr}(\pi)}
   \int_{1>t_1>\cdots>t_{2d}>0}
       \prod_{\substack{\{a,b\}\in\pi\\a<b}}
               \Sigma(t_a-t_b)\,d\mathbf t.
 \label{eq:main-quadratic-vacuum-series}
\end{align}
We also show in \Cref{lem:quadratic-cut} that
\begin{align}
 2W(\Sigma)\mathcal D(\Sigma)(u)
 &=\sum_{d\geq0}2^{-d}\sum_{\pi\in\mathcal P_2([2d])}
   \int_{1>t_1>\cdots>t_{2d}>0}
     \hspace{-5em}  (-1)^{\operatorname{cr}(\pi)+\#\{\{a,b\}\in\pi:a<b,\ t_b<u<t_a\}}
       \prod_{\substack{\{a,b\}\in\pi\\a<b}}
               \Sigma(t_a-t_b)\,d\mathbf t.
 \label{eq:main-quadratic-two-point-series}
\end{align}
Finally, define $B^+$ by the same Hermitian linear combination as $B$, replacing
each $B(t)$ by
\begin{align}
 \frac1{2N}\sum_{i=1}^{N+1}
       e^{-it\beta H^+}\psi_i e^{it\beta H^+}\psi_i
                   -G_*(it)I-\frac1{2N}I.
 \label{eq:cavity-full-source}
\end{align}
Note that $B^+$ treats all $N+1$ interacting sites symmetrically, unlike $B$. We now do a cavity computation on quantities of the form
\begin{align}
    \frac{\EE_{J,J^+}\tr\left[
     (B^+)^{2m}e^{(1-t)\beta H^+}\Gamma_Ae^{t\beta H^+}\Gamma_A\right]}
     {\EE_J\tr e^{\beta H}}
\end{align}
for different choices of $A$ (including and excluding the cavity fermion).

\begin{lemma}[Concentration at the saddle point]
\label{lem:weighted-cavity-concentration}
For each fixed odd $m\geq1$, use the joint law in
\Cref{lem:cavity-coefficient-comparison}. For its posterior mean
$a=\Tr(\rho_xB)$ and kernel $G$ from \eqref{eq:def-posterior-kernel},
there is $b>0$, independent of $m$, such that
\begin{align}
 \EE\left[|a|^{2m}\|G-G_*\|_{L^2(0,1)}^2\right]
 &\leq C_m\left(N^{-b}M_{m-1}+N^{-mb}\right).
 \label{eq:weighted-cavity-residual}
\end{align}
\end{lemma}
\begin{proof}
For $F=B$ or $B^+$, set
\begin{align}
 U_A(F;t)
 &=\frac{\EE_{J,J^+}\tr\left[
 F^{2m}e^{(1-t)\beta H^+}\Gamma_Ae^{t\beta H^+}\Gamma_A\right]}
 {\EE_J\tr e^{\beta H}},\qquad 0\leq t\leq1.
 \label{eq:cavity-four-correlators}
\end{align}
We first apply \Cref{lem:cavity-coefficient-comparison} when just using $B$ instead of $B^+$. Wick's rule on $J^+$ and a Duhamel expansion gives
\begin{align}
 U_A(B;t)
 &=\sum_{d\geq0}\left(\frac{3\beta^2}{8N^3}\right)^d
 \sum_{\pi\in\mathcal P_2([2d])}
 \sum_{\mathbf T\in\prod_{e\in\pi}\binom{[N]}3}
 \int_{1>t_1>\cdots>t_{2d}>0}
 c_{A,\pi}(\mathbf T;\mathbf t,t)\,d\mathbf t.
 \label{eq:exact-star-gaussian-expansion}
\end{align}
To use the truncated degree $d \leq P_N$ and the restricted set of $T_e$ considered in \eqref{eq:cavity-retained-assignments} of \Cref{lem:cavity-coefficient-comparison}, we bound the error from removing terms with $d > P_N$ or $\mathbf T \notin \cR_{A,\pi}$ in \Cref{lem:cavity-omitted}. For the remaining terms, \Cref{lem:cavity-coefficient-comparison} successfully replaces $c_{A,\pi}$ by $\ell_{A,\pi}$. This gives (\Cref{lem:cavity-coefficient-sum})
\begin{align}
 U_A(B;t)
 &=\sum_{d\geq0}\left(\frac{\beta^2}{16}\right)^d
 \sum_{\pi\in\mathcal P_2([2d])}
 \int_{1>t_1>\cdots>t_{2d}>0}
 \ell_{A,\pi}(\mathbf t,t)\,d\mathbf t
 +O_m(N^{-b}M_{m-1}+N^{-mb}).
 \label{eq:main-cavity-scalar-series}
\end{align}
For each measurement outcome, we use $\Sigma = \beta^2 G^3$ so $(\beta^2/16)^d \prod (2G(t_j-t_k))^3$ in \eqref{eq:ellapi} becomes $2^{-d} \prod \Sigma(t_j-t_k)$. Interchanging the expectation, integrals and sum using $|a| \leq 1, |2G| \leq 1$ and \eqref{eq:quadratic-order-majorant}, we have
\begin{align}
 &\sum_{d\geq0}\left(\frac{\beta^2}{16}\right)^d
   \sum_{\pi\in\mathcal P_2([2d])}
   \int_{1>t_1>\cdots>t_{2d}>0}
             \ell_{A,\pi}(\mathbf t,t)\,d\mathbf t\nonumber\\*
 &\quad=\EE\Bigg[|a|^{2m}(2G(t))^{|A\cap[N]|}
       \sum_{d\geq0}2^{-d}\sum_{\pi\in\mathcal P_2([2d])}
       \int_{1>t_1>\cdots>t_{2d}>0}
          \epsilon_{A,\pi}
          \prod_{\substack{\{j,k\}\in\pi\\j<k}}
                   \Sigma(t_j-t_k)\,d\mathbf t\Bigg].
 \label{eq:cavity-quadratic-series-matching}
\end{align}
By \eqref{eq:scalar-cavity-coefficient}, the inner series is
$W(\Sigma)$ when $N+1\notin A$, by
\eqref{eq:main-quadratic-vacuum-series}; it is
$2W(\Sigma)\mathcal D(\Sigma)(t)$ when $N+1\in A$, by
\eqref{eq:main-quadratic-two-point-series}. Therefore
\begin{align}
 U_A(B;t)
 &=\EE\left[|a|^{2m}(2G(t))^{|A\cap[N]|}W(\beta^2G^3)
 \bigl(2\mathcal D(\beta^2G^3)(t)\bigr)^{\mathbf1_{N+1\in A}}\right] +O_m(N^{-b}M_{m-1}+N^{-mb}).
 \label{eq:cavity-general-comparison}
\end{align}
\Cref{lem:cavity-source-change} replaces $B$ by $B^+$ with the same
form of error. Different choices of $A$ give
\begin{align}
 U_{\{N+1\}}(B^+;t)
 &=2\EE\left[|a|^{2m}W(\beta^2G^3)\mathcal D(\beta^2G^3)(t)\right]
 +O_m(N^{-b}M_{m-1}+N^{-mb}),\\
 U_{\{1\}}(B^+;t)
 &=2\EE\left[|a|^{2m}W(\beta^2G^3)G(t)\right]
 +O_m(N^{-b}M_{m-1}+N^{-mb}),\\
 U_{\{N+1,2,3,4\}}(B^+;t)
 &=16\EE\left[|a|^{2m}W(\beta^2G^3)
 \mathcal D(\beta^2G^3)(t)G(t)^3\right]
 +O_m(N^{-b}M_{m-1}+N^{-mb}),\\
 U_{\{1,2,3,4\}}(B^+;t)
 &=16\EE\left[|a|^{2m}W(\beta^2G^3)G(t)^4\right]
 +O_m(N^{-b}M_{m-1}+N^{-mb}).
 \label{eq:cavity-four-role-formulas}
\end{align}
Permutation invariance of $H^+$ and $B^+$ also gives the identities
\begin{align}
 U_{\{N+1\}}(B^+;t)=U_{\{1\}}(B^+;t),\qquad
 U_{\{N+1,2,3,4\}}(B^+;t)=U_{\{1,2,3,4\}}(B^+;t).
 \label{eq:exact-cavity-permutation-identities}
\end{align}
Subtract the corresponding comparisons and cancel the factors $2,16$:
\begin{align}
 \left|\EE\left[|a|^{2m}W(\beta^2G^3)
 (\mathcal D(\beta^2G^3)(t)-G(t))\right]\right|
 &\leq C_m(N^{-b}M_{m-1}+N^{-mb}),
 \label{eq:cavity-two-residuals-1}\\
 \left|\EE\left[|a|^{2m}W(\beta^2G^3)
 (\mathcal D(\beta^2G^3)(t)-G(t))G(t)^3\right]\right|
 &\leq C_m(N^{-b}M_{m-1}+N^{-mb}).
 \label{eq:cavity-two-residuals-2}
\end{align}
Using that for each outcome \eqref{eq:scalar-coercivity} holds (shown in \Cref{lem:positive-saddle}), i.e.,
\begin{align}
 \frac14e^{-\beta/2}\|G-G_*\|_2^2
 &\leq-\int_0^1(\mathcal D(\beta^2G^3)-G)(G^3-G_*^3)\,dt,
\end{align}
and that $W(\beta^2 G^3) \geq 1$ from \eqref{eq:quadratic-order-majorant}, we find
\begin{align}
 \frac14e^{-\beta/2}\EE\left[|a|^{2m}\|G-G_*\|_2^2\right]
 &\leq-\EE\left[|a|^{2m}W(\beta^2G^3)
 \int_0^1(\mathcal D(\beta^2G^3)-G)(G^3-G_*^3)\,dt\right]\nonumber\\
 &=-\int_0^1\EE\left[|a|^{2m}W(\beta^2G^3)
 (\mathcal D(\beta^2G^3)(t)-G(t))G(t)^3\right]dt\nonumber\\*
 &\quad+\int_0^1G_*(t)^3\EE\left[|a|^{2m}W(\beta^2G^3)
 (\mathcal D(\beta^2G^3)(t)-G(t))\right]dt\nonumber\\
 &\leq C_m(N^{-b}M_{m-1}+N^{-mb}).
\end{align}
The last inequality uses \eqref{eq:cavity-two-residuals-1}, \eqref{eq:cavity-two-residuals-2} and
$0\leq G_*\leq1/2$.
\end{proof}

\subsection{Completing the proof}
We can now show the main moment bound.

\begin{proof}[Proof of \Cref{thm:mbound}]
So far, we have shown that there are $\delta>0$ and an integer $r\geq1$, independent of the moment index, such that for each fixed odd $m\geq1$,
\begin{align}
 0\leq M_m-\EE|a|^{2m}
 &\leq C_m(N^{-\delta}M_{m-1}+N^{-m\delta}),
 \label{eq:finite-source-moment-comparison}\\
 \EE[|a|^{2m}\|G-G_*\|_2^2]
 &\leq C_m(N^{-\delta}M_{m-1}+N^{-m\delta}),
 \label{eq:finite-source-dyson}\\
 \EE[|a|^{2m}D]
 &\leq C_m\sqrt{M_m(N^{-\delta}M_{m-1}+N^{-m\delta})},
 \label{eq:finite-source-balance}\\
 |a|^{2r}&\leq C\|G-G_*\|_2^2+CD+CN^{-r\delta}.
 \label{eq:finite-source-recovery}
\end{align}
The law and constants may depend on the fixed $m$, but $\delta,r$ do not.
The first three inequalities are \eqref{eq:main-posterior-moment-comparison}, \Cref{lem:weighted-cavity-concentration} and \Cref{lem:conditional-db}.
The last is \Cref{lem:raw-source-recovery} with
$\delta_E=N^{-\varepsilon}$. Take $\delta$ no larger than their three
common comparison exponents and $2\varepsilon$.

Fix an odd $m\geq1$ and write $X=M_m^{1/m}$. Scalar moment monotonicity
in the positive state $\Phi_N$ gives
\begin{align}
 M_{m-1}&\leq X^{m-1},&0\leq X&\leq1.
\end{align}
If $X\leq K_mN^{-\delta}$, the assertion already follows. Otherwise
choose $K_m$ large enough that
\eqref{eq:finite-source-moment-comparison} implies
\begin{align}
 N^{-\delta}M_{m-1}+N^{-m\delta}
 &\leq2N^{-\delta}X^{m-1},
 &\EE|a|^{2m}&\geq\tfrac12 X^m.
\end{align}
Multiply \eqref{eq:finite-source-recovery} by $|a|^{2m}$ and use the
other two estimates at this same index $m$. Jensen's inequality yields
\begin{align}
 2^{-(m+r)/m}X^{m+r}
 &\leq(\EE|a|^{2m})^{(m+r)/m}
      \leq\EE|a|^{2m+2r}\nonumber\\
 &\leq C_m\left(N^{-\delta}X^{m-1}
          +N^{-\delta/2}X^{m-1/2}
          +N^{-r\delta}X^m\right)\nonumber\\
 &\leq C_mN^{-\delta/2}X^{m-1/2},\\
 M_m^{1/m}=X&\leq C_mN^{-\delta/(2r+1)}.
 \label{eq:moment-absorption}
\end{align}
In the penultimate line use $N^{-\delta}\leq X\leq1$ and $r\geq1$.
For an arbitrary $j\geq1$, choose the next odd $m\geq j$ and obtain
\begin{align}
 M_j&\leq M_m^{j/m}\leq C_jN^{-\delta j/(2r+1)}
                  \leq C_jN^{-\delta j/(4r+4)}
\end{align}
as claimed.
\end{proof}

\section{Learning the SYK Hamiltonian}
\label{sec:syk-learning}

As an application of the expansion used by our algorithm (\Cref{sec:alg}), we show how to recover the SYK Hamiltonian couplings from copies of the Gibbs state with polynomial sample and time complexity.

\begin{theorem}[Learning from an SYK Gibbs state]
\label{thm:syk-learning}
Fix $0<\beta<\infty$. For all sufficiently large even $N$ and
$\varepsilon=2^{-b}$, $\delta=2^{-r}$, with integers $b,r \geq 1$,
there is an algorithm using
\begin{align}
    S&\leq C_\beta N^5\varepsilon^{-2}\log\frac{N}{\delta}
    \label{eq:syk-learning-samples}
\end{align}
independent copies of $\rho_\beta$ and polynomially many elementary
quantum gates and classical bit operations in
$N,\varepsilon^{-1},\log(\delta^{-1})$. Its output
$\widehat J\in\mathbb R^{\binom{[N]}4}$ satisfies, for every fixed $p\geq2$,
\begin{align}
    \Pr_J\left[
      \Pr_{\rm alg}\left[
        \frac{\|\widehat J-J\|_2}{\sqrt{\binom N4}}>\varepsilon
        \,\middle|\,J\right]>\delta
    \right]
    &\leq C_{\beta,p}(N\varepsilon)^{-p}+2e^{-N^2},
    \label{eq:syk-learning-quantitative-probability}
\end{align}
where $\Pr_{\rm alg}$ includes measurement outcomes and classical randomness. In particular, for every fixed $a>0$ and $0<c<1$, the algorithm uses
\begin{align}
    S&\leq C_\beta N^{5+2c}\log\frac{N}{\delta}
    \label{eq:syk-learning-power-samples}
\end{align}
samples and succeeds with probability at least $1-N^{-a}$ over $J$, and with probability at least $1-\delta$ over classical randomness and measurement outcomes, outputs $\wh J$ such that
\begin{align}
    \frac{\|\widehat J-J\|_2}{\sqrt{\binom N4}}&\leq N^{-c}.
    \label{eq:syk-learning-power-error}
\end{align}
The quantum computer only measures expectations of pairs of Majorana on the Gibbs state.
\end{theorem}
\begin{corollary}[Normalized Hilbert--Schmidt error]
\label{cor:syk-learning-hamiltonian-error}
Let $\widehat H=\sqrt{\frac{3}{8N^3}}\sum_A\widehat J_A\Gamma_A$. On the success event in \eqref{eq:syk-learning-power-error},
\begin{align}
    \sqrt{\tr[(\widehat H-H)^2]}
    &=\sqrt{\frac{3}{8N^3}}\|\widehat J-J\|_2
      \leq\frac18N^{1/2-c}.
    \label{eq:syk-learning-hamiltonian-error}
\end{align}
In particular, $1/2<c<1$ gives vanishing inverse-polynomial error in this norm at every fixed $\beta > 0$.
\end{corollary}
\begin{proof}
Since $\tr(\Gamma_A\Gamma_B)=\delta_{AB}$, \eqref{eq:syk-learning-power-error} gives $\tr[(\widehat H-H)^2]=\frac3{8N^3}\sum_A(\widehat J_A-J_A)^2\leq\frac3{8N^3}\binom N4 N^{-2c}\leq\frac1{64}N^{1-2c}$.
\end{proof}

\subsection{Preliminaries}
The learning algorithm proceeds by expanding around the coefficients
\begin{align}
    a_\beta&=16\int_0^1G_*(t)^4\,dt,\qquad
    b_\beta=128\beta\int_0^1\int_0^u
       G_*(u)^2G_*(v)^2G_*(u-v)^2\,dv\,du.
    \label{eq:syk-learning-response-coefficients}
\end{align}
In our algorithm's Hermite expansion, the first order will look like $\alpha_\beta J_A$; the second order will look like $b_\beta \cQ_A(J)$ for the quantity
\begin{align}
    \mathcal Q_A(x)
    &=\frac12\sqrt{\frac{3}{8N^3}}
       \sum_{\substack{B,C\in\binom{[N]}4\\
                       B\mathbin\triangle C=A}}
       \tr(\Gamma_A\Gamma_B\Gamma_C)x_Bx_C,
    \label{eq:syk-learning-quadratic}
\end{align}
where the sum is over ordered pairs. Since $B\triangle C = A$, an unordered pair $\{B,C\}$ is specified by a partition of $A$ into two
pairs and a pair in $[N]\setminus A$. Since there are $3 \binom{N-4}{2}$ such choices, Wick's theorem gives
\begin{align}
    \EE_J\mathcal Q_A(J)&=0,
    \qquad \EE_J|\mathcal Q_A(J)|^2
    =\frac{9}{8N^3}\binom{N-4}{2}\leq\frac{C}{N}.
    \label{eq:syk-learning-quadratic-second-moment}
\end{align}
It will also be useful to have a matrix representation of $\cQ$. Extend $x\in\mathbb R^{\binom{[N]}4}$ to a completely antisymmetric
tensor $x_{ijkl}$, zero when indices repeat, and define
\begin{align}
    \mathsf T(x)_{ij,kl}&=x_{ijkl}\qquad(i<j,\ k<l).
    \label{eq:syk-learning-pair-matrix}
\end{align}
This is a real symmetric matrix indexed by pairs. Its adjoint, from
symmetric matrices with the Hilbert--Schmidt inner product to coefficient
vectors with the Euclidean inner product, is
\begin{align}
    (\mathsf T^*W)_{ijkl}
    &=2(W_{ij,kl}-W_{ik,jl}+W_{il,jk})
      \qquad(i<j<k<l).
    \label{eq:syk-learning-pair-operators}
\end{align}

\begin{lemma}[Matrix form of $\cQ$]
\label{lem:syk-learning-pair-identities}
The map $\mathsf T$ satisfies $\|\mathsf T(x)\|_{\rm HS}=\sqrt6\,\|x\|_2$, $\|\mathsf T^*\|_{{\rm HS}\to2}=\sqrt6$ and
\begin{align}
    \mathcal Q(x)
    &=\frac12\sqrt{\frac{3}{8N^3}}
                     \mathsf T^*(\mathsf T(x)^2).
    \label{eq:syk-learning-quadratic-matrix-identity}
\end{align}
\end{lemma}
\begin{proof}
Each $x$ occurs in six signed matrix entries, so $\|\mathsf T(x)\|_{\rm HS}^2=6\sum_{i<j<k<l}x_{ijkl}^2$. Writing out
\begin{align}
    \langle\mathsf T(x),W\rangle_{\rm HS}
    &=2\sum_{i<j<k<l}x_{ijkl}
             (W_{ij,kl}-W_{ik,jl}+W_{il,jk})
      =\langle x,\mathsf T^*W\rangle,
\end{align}
we find that $\mathsf T^*\mathsf T=6I$ and thus $\|\mathsf T^*\|_{{\rm HS}\to2}=\|\mathsf T\|_{2\to{\rm HS}}=\sqrt6$.
For six distinct indices, we use $(-\psi_i\psi_j\psi_r\psi_s)(-\psi_k\psi_l\psi_r\psi_s) =-\psi_i\psi_j\psi_k\psi_l$ to find that
the three partitions of $\{i,j,k,l\}$ into pairs yield
\begin{align}
    \mathcal Q_{\{i,j,k,l\}}(x)
    &=\sqrt{\frac{3}{8N^3}}\sum_{r<s}
       (x_{ijrs}x_{klrs}-x_{ikrs}x_{jlrs}+x_{ilrs}x_{jkrs})\nonumber\\
    &=\sqrt{\frac{3}{8N^3}}
       \bigl[(\mathsf T(x)^2)_{ij,kl}
             -(\mathsf T(x)^2)_{ik,jl}
             +(\mathsf T(x)^2)_{il,jk}\bigr]\nonumber\\
    &=\frac12\sqrt{\frac{3}{8N^3}}
                          (\mathsf T^*(\mathsf T(x)^2))_{ijkl}.
\end{align}
\end{proof}

\begin{lemma}[Norm of the Gaussian matrix]
\label{lem:syk-learning-gaussian-matrix}
The Gaussian couplings satisfy
\begin{align}
    \Pr_J\bigl[\|\mathsf T(J)\|>16N\bigr]&\leq2e^{-N^2}.
    \label{eq:syk-learning-gaussian-matrix-tail}
\end{align}
\end{lemma}
\begin{proof}
For real unit vectors $u,v$ indexed by pairs,
$u^{\mathsf T}\mathsf T(J)v$ is centered Gaussian. Each coupling has
six signed matrix entries. Thus
\begin{align}
    \EE_J|u^{\mathsf T}\mathsf T(J)v|^2
    &\leq6\sum_{\substack{P,Q\in\binom{[N]}2\\P\cap Q=\varnothing}}
                  u_P^2v_Q^2
      \leq6\|u\|_2^2\|v\|_2^2=6,\qquad 
    \Pr_J\bigl[|u^{\mathsf T}\mathsf T(J)v|>t\bigr]
    \leq2e^{-t^2/12}.
\end{align}
Let $\mathcal V$ be a maximal $1/4$-separated subset of the unit sphere
in $\mathbb R^{\binom N2}$. Its radius-$1/8$ balls are disjoint and lie
in the radius-$9/8$ ball. We bound the norm $\|\mathsf T(J)\|\leq\max_{u,v\in\mathcal V}|u^{\mathsf T}\mathsf T(J)v|+\tfrac12\|\mathsf T(J)\|$ then use $|\mathcal V|\leq9^{\binom N2}$ and $\sup_{\|u\|_2=1}\min_{v\in\mathcal V}\|u-v\|_2\leq\tfrac14$ to obtain
\begin{align}
    \Pr_J\bigl[\|\mathsf T(J)\|>16N\bigr]
    &\leq2\exp\left(2\binom N2\log9-\frac{(8N)^2}{12}\right)
      \leq2e^{-N^2}.
\end{align}
\end{proof}

We will use the following Gaussian moment bound, writing $\|X\|_{L^p(J)}=(\EE_J|X|^p)^{1/p}$ for scalar random variables.

\begin{fact}[Moments of a fixed Hermite degree, Theorem 5.10 of~\cite{janson1997gaussian}]
\label{fact:syk-learning-gaussian-moments}
For a real polynomial $P_d$ in Hermite degree $d$ of finitely many
independent standard Gaussian variables, and every fixed $p\geq2$,
\begin{align}
    \|P_d\|_{L^p}&\leq C_{d,p}\|P_d\|_{L^2}.
    \label{eq:syk-learning-chaos-moments}
\end{align}
The constant is independent of the number of Gaussian variables.
\end{fact}

\subsection{Hermite truncation}
We bound the error associated with truncating the Hermite expansion of \Cref{sec:alg} at the first two orders, and we show that this truncation is efficiently computable (using the same method as our main algorithm). The proof is largely the same as in \Cref{sec:alg}.
\begin{lemma}[Truncating the Hermite expansion]
\label{lem:syk-learning-response}
For every fixed $p\geq2$,
\begin{align}
    \max_{A\in\binom{[N]}4}
    \left\|\frac1\beta\sqrt{\frac{8N^3}{3}}\Tr(\rho_\beta \Gamma_A)
               -a_\beta J_A-b_\beta\mathcal Q_A(J)\right\|_{L^p(J)}
    &\leq\frac{C_{\beta,p}}{N}.
    \label{eq:syk-learning-response}
\end{align}
Consequently, the vector $f(J) = \lr{\Tr(\rho_\beta \Gamma_A)}_{A\in\binom{[N]}{4}}$ satisfies
\begin{align}
    \left\|
      \frac{\|\frac1\beta\sqrt{\frac{8N^3}{3}}f(J)
                    -a_\beta J-b_\beta\mathcal Q(J)\|_2}
           {\sqrt{\binom N4}}
    \right\|_{L^p(J)}
    &\leq\frac{C_{\beta,p}}{N},\qquad
    \left\|\frac{\|\mathcal Q(J)\|_2}{\sqrt{\binom N4}}\right\|_{L^p(J)}
    \leq\frac{C_p}{\sqrt N}.
    \label{eq:syk-learning-vector-moments}
\end{align}
Moreover, the coefficients $a_\beta,b_\beta$ can be computed to absolute error $\tau$ in time polynomial in $\tau^{-1}$ at fixed $\beta$.
\end{lemma}
\begin{proof}
\proofstep{First two Hermite orders} As in \Cref{sec:alg}, we denote the Hermite expansions, truncated at order $V$, by
\begin{align}
    g_O(J)&=\frac{\Tr(Oe^{\beta H})}{\EE_J Z}, \qquad g_O=\sum_{d\geq0}g_{O,d},
    \qquad \mathcal N_O=\sum_{d=0}^Vg_{O,d}.
    \label{eq:syk-learning-annealed-normalization}
\end{align}
where $g_{O,d}$ is the $d$th Hermite component. From the Hermite expansions of \Cref{app:hermite} we have that (see \eqref{eq:quartet-parity-condition} and
\eqref{eq:hermite-coefficient-identification})
\begin{align}
    g_{\Gamma_A,0}&=0,\qquad
    g_{\Gamma_A,1}
    =\beta\sqrt{\frac{3}{8N^3}}F_{\Gamma_A;A}J_A,\qquad
    g_{\Gamma_A,2}
    =\frac{3\beta^2}{16N^3}
      \sum_{\substack{B,C\in\binom{[N]}4\\B\mathbin\triangle C=A}}
          F_{\Gamma_A;B,C}J_BJ_C
    \label{eq:syk-learning-two-hermite-orders}
\end{align}
for coefficients $F$ defined in our main algorithm as (reproducing \eqref{eq:foa})
\begin{align}\label{eq:foa2}
    F_{O;A_1,\ldots,A_v}
    &=
    \sum_{\pi\in S_v}
    \int_{\substack{s_1,\ldots,s_v\geq0\\
                    s_1+\cdots+s_v\leq1}}
    \frac{
        \EE_J\Tr\left[
        e^{s_0\beta H}
        \Gamma_{A_{\pi(1)}}e^{s_1\beta H}
        \cdots
        \Gamma_{A_{\pi(v)}}e^{s_v\beta H}O
        \right]
    }{
        \EE_J\Tr e^{\beta H}
    }
    \,ds_1\cdots ds_v.
\end{align}
To bound our estimates of the coefficients $F$ using coefficients $a_\beta, b_\beta$, we first show that
\begin{align}
    &\left\langle\chi_{j_1}(t_1)\cdots
                         \chi_{j_{2k}}(t_{2k})\right\rangle_N =\sum_{\pi\in\mathcal P_2([2k])}
       (-1)^{\operatorname{cr}(\pi)}
       \prod_{\substack{\{r,s\}\in\pi\\r<s}}
                  2\delta_{j_rj_s}G_*(t_r,t_s)
       +O_{\beta,k}(N^{-1}).
    \label{eq:syk-learning-fixed-wick}
\end{align}
This estimate can be shown from the exact expansion of $\left\langle\chi_{j_1}(t_1)\cdots\chi_{j_{2k}}(t_{2k})\right\rangle_N$ given by \eqref{eq:chiprod}, reproduced here:
\begin{align}\label{eq:chiprod2}
    &\left\langle
        \chi_{j_1}(t_1)\cdots\chi_{j_{2k}}(t_{2k})
    \right\rangle_N =
    2^k
    \sum_{\pi,\sigma\in\mathcal P_2([2k])}
    \left(
        \prod_{\{r,s\}\in\pi}\delta_{j_rj_s}
    \right)
    (\mathcal G_{k,N}^{-1})_{\pi\sigma}
    (-1)^{\operatorname{cr}(\sigma)}
    \left\langle
        \prod_{\substack{\{r,s\}\in\sigma\\r<s}}
        Q_{t_rt_s}
    \right\rangle_N
\end{align}
and proven in \Cref{lem:orthogonal-reconstruction}.
For fixed $k$, choose $K>k$ with $\theta K\geq2$. By \Cref{lem:collective-concentration}, we have that (see \eqref{eq:hierarchy-fixed-point})
\begin{align}
    \|\boldsymbol\mu-e_0\|_\infty
    &\leq C_{\beta,K}N^{-\theta K}+C_{\beta,K}N^{-1},\qquad
    \max_{1\leq r\leq k}\|\mu_r\|_\infty
    \leq C_{\beta,k}N^{-1}
\end{align}
and hence \eqref{eq:qprod} implies that the two-point function $Q_{uv}$ \eqref{eq:quv} satisfies
\begin{align}\label{eq:quvgstar}
    \left\langle\prod_{j=1}^kQ_{u_jv_j}\right\rangle_N
    &=\sum_{S\subseteq[k]}
       \left(\prod_{j\notin S}G_*(u_j,v_j)\right)
       \mu_{|S|}\bigl((u_j,v_j)_{j\in S}\bigr) =\prod_{j=1}^kG_*(u_j,v_j)+O_{\beta,k}(N^{-1}).
\end{align}
Moreover, by \eqref{eq:grambound} the matrix $\cG$ \eqref{eq:gram} satisfies
\begin{align}
    \|\mathcal G_{k,N}-I\|_{\infty\to\infty}
    &\leq\frac{(2k-1)!!-1}{N},\qquad
    \|\mathcal G_{k,N}^{-1}-I\|_{\infty\to\infty}
    \leq\sum_{r\geq1}
       \|\mathcal G_{k,N}-I\|_{\infty\to\infty}^r
      \leq\frac{C_k}{N}
    \label{eq:syk-learning-pairing-inverse}
\end{align}
for sufficiently large $N$. Substituting \eqref{eq:quvgstar} and \eqref{eq:syk-learning-pairing-inverse} into \eqref{eq:chiprod2} gives \eqref{eq:syk-learning-fixed-wick}.
Consequently, \eqref{eq:foa2} gives
\begin{align}
    F_{\Gamma_A;A}
    &=\int_0^1[2G_*(t)]^4\,dt+O_\beta(N^{-1})
      =a_\beta+O_\beta(N^{-1}).
    \label{eq:syk-learning-linear-coefficient}
\end{align}
and (setting $u=s_1+s_2$, $v=s_2$)
\begin{align}
    F_{\Gamma_A;B,C}
    &=128\tr(\Gamma_A\Gamma_B\Gamma_C)
      \int_0^1\int_0^uG_*(u)^2G_*(v)^2G_*(u-v)^2\,dv\,du
      +O_\beta(N^{-1})\nonumber\\
    &=\frac{b_\beta}{\beta}\tr(\Gamma_A\Gamma_B\Gamma_C)
      +O_\beta(N^{-1}).
    \label{eq:syk-learning-quadratic-coefficient}
\end{align}
Hence, Hermite orthogonality implies (using \eqref{eq:syk-learning-quadratic-second-moment})
\begin{align}
    \left\|\frac1\beta\sqrt{\frac{8N^3}{3}}g_{\Gamma_A,1}
                   -a_\beta J_A\right\|_{L^2(J)}
    &\leq C_\beta N^{-1},\nonumber\\
    \left\|\frac1\beta\sqrt{\frac{8N^3}{3}}g_{\Gamma_A,2}
                   -b_\beta\mathcal Q_A(J)\right\|_{L^2(J)}^2
    &\leq \frac{C_\beta}{N^2}\EE\abs{\cQ_A(J)}^2 \leq C_\beta\frac{3\beta^2}{8N^5}
                     3\binom{N-4}{2}
      \leq C_\beta N^{-3}.
    \label{eq:syk-learning-coefficient-errors}
\end{align}
By \Cref{fact:syk-learning-gaussian-moments}, for every fixed $p\geq2$,
\begin{align}
    \left\|\frac1\beta\sqrt{\frac{8N^3}{3}}
                   (g_{\Gamma_A,1}+g_{\Gamma_A,2})
                   -a_\beta J_A-b_\beta\mathcal Q_A(J)\right\|_{L^p(J)}
    &\leq C_{\beta,p}N^{-1},\label{eq:syk-learning-low-order-lp-1}\\
    \|a_\beta J_A+b_\beta\mathcal Q_A(J)\|_{L^p(J)}
    \leq a_\beta\|J_A\|_{L^p(J)}
        +b_\beta C_{2,p}\|\mathcal Q_A(J)\|_{L^2(J)}
      &\leq C_{\beta,p}.
    \label{eq:syk-learning-low-order-lp}
\end{align}
This bounds how close $a_\beta J_A + b_\beta \cQ_A(J)$ is to the first two orders of the Hermite expansion; we next bound away the error from this Hermite truncation.

\proofstep{Higher Hermite orders}
For fixed $d\geq1$ and $V$, \Cref{lem:hermite-tail} and
\eqref{eq:low-degree-hermite-bound} imply
\begin{align}
    \|g_{\Gamma_A,d}\|_{L^2(J)}&\leq C_{\beta,d}N^{-(d+2)/2},
    &\|g_{\Gamma_A}-\mathcal N_{\Gamma_A}\|_{L^2(J)}
    &\leq C_{\beta,V}N^{-(V+3)/2},\nonumber\\
    \|g_{I,d}\|_{L^2(J)}&\leq C_{\beta,d}N^{-d/2},
    &\|g_I-\mathcal N_I\|_{L^2(J)}
    &\leq C_{\beta,V}N^{-(V+1)/2}.
    \label{eq:syk-learning-hermite-bounds}
\end{align}
Also $g_{I,0}=1$ and $g_{I,1}=0$. For every real $q$,
\eqref{eq:hermite-logZ} gives
\begin{align}
    \EE_J g_I^q
    &=e^{q(\EE_J\log Z-\log\EE_J Z)}
      \EE_J e^{q(\log Z-\EE_J\log Z)}
      \leq e^{C_\beta(q^2+|q|)/N}.
    \label{eq:syk-learning-denominator-moments}
\end{align}
Thus, for every fixed $q\geq2$,
\begin{align}
    |g_{\Gamma_A}|&\leq g_I,
    &\|g_I\|_{L^q(J)}+\|g_I^{-1}\|_{L^q(J)}&\leq C_{\beta,q}.
    \label{eq:syk-learning-crude-moments}
\end{align}
Fix $p\geq2$ and an integer $V\geq6p$. For $O=I$ or $O=\Gamma_A$,
\begin{align}
    \|g_O-\mathcal N_O\|_{L^{2p}(J)}
    &\leq\|g_I\|_{L^{2p}(J)}
       +\sum_{d=0}^V C_{d,2p}\|g_{O,d}\|_{L^2(J)}
      \leq C_{\beta,p,V},\nonumber\\
    \|g_O-\mathcal N_O\|_{L^p(J)}
    &\leq\|g_O-\mathcal N_O\|_{L^2(J)}^{1/(p-1)}
           \|g_O-\mathcal N_O\|_{L^{2p}(J)}^{(p-2)/(p-1)},\nonumber\\
    \|g_{\Gamma_A}-\mathcal N_{\Gamma_A}\|_{L^p(J)}
    &\leq C_{\beta,p,V}N^{-(V+3)/(2(p-1))}
      \leq C_{\beta,p}N^{-5/2},\nonumber\\
    \|g_I-\mathcal N_I\|_{L^p(J)}
    &\leq C_{\beta,p,V}N^{-(V+1)/(2(p-1))}
      \leq C_{\beta,p}N^{-1}.
\end{align}
The triangle inequality gives
\begin{align}
    \|g_{\Gamma_A}-g_{\Gamma_A,1}-g_{\Gamma_A,2}\|_{L^p(J)}
    &\leq\|g_{\Gamma_A}-\mathcal N_{\Gamma_A}\|_{L^p(J)}
       +\sum_{d=3}^V C_{d,p}\|g_{\Gamma_A,d}\|_{L^2(J)} \leq C_{\beta,p}N^{-5/2},
    \label{eq:syk-learning-numerator-lp-tail}\\
    \|g_I-1\|_{L^p(J)}
    &\leq\|g_I-\mathcal N_I\|_{L^p(J)}
       +\sum_{d=2}^V C_{d,p}\|g_{I,d}\|_{L^2(J)}
      \leq C_{\beta,p}N^{-1}
    \label{eq:syk-learning-denominator-lp-tail}
\end{align}
and in particular by \eqref{eq:syk-learning-low-order-lp-1},
\begin{align}
    \left\|\frac1\beta\sqrt{\frac{8N^3}{3}}g_{\Gamma_A}
                  -a_\beta J_A-b_\beta\mathcal Q_A(J)\right\|_{L^p(J)}
    &\leq C_{\beta,p}N^{-1}.
    \label{eq:syk-learning-annealed-response}
\end{align}
To complete the proof, rewrite
\begin{align}
    \frac1\beta\sqrt{\frac{8N^3}{3}}\Tr(\rho_\beta\Gamma_A)-a_\beta J_A-b_\beta\mathcal Q_A
    &=\frac{\frac1\beta\sqrt{\frac{8N^3}{3}}g_{\Gamma_A}
                   -a_\beta J_A-b_\beta\mathcal Q_A}{g_I}+\frac{(a_\beta J_A+b_\beta\mathcal Q_A)(1-g_I)}{g_I}
\end{align}
and apply H\"older's inequality to obtain by \eqref{eq:syk-learning-low-order-lp} \eqref{eq:syk-learning-crude-moments}, \eqref{eq:syk-learning-denominator-lp-tail} and \eqref{eq:syk-learning-annealed-response} that
\begin{align}
    \left\|\frac1\beta\sqrt{\frac{8N^3}{3}}\Tr(\rho_\beta \Gamma_A)
                   -a_\beta J_A-b_\beta\mathcal Q_A\right\|_{L^p(J)}&\leq
      \left\|\frac1\beta\sqrt{\frac{8N^3}{3}}g_{\Gamma_A}
                   -a_\beta J_A-b_\beta\mathcal Q_A\right\|_{L^{2p}(J)}
      \|g_I^{-1}\|_{L^{2p}(J)}\nonumber\\
    &\quad+
      \|a_\beta J_A+b_\beta\mathcal Q_A\|_{L^{3p}(J)}
      \|g_I-1\|_{L^{3p}(J)}\|g_I^{-1}\|_{L^{3p}(J)}\nonumber\\
    &\leq C_{\beta,p}N^{-1}.
    \label{eq:syk-learning-response-remainder}
\end{align}
For any vector of random variables $X_A$, Minkowski's inequality gives
\begin{align}
    \left\|\frac{\|X\|_2}{\sqrt{\binom N4}}\right\|_{L^p(J)}^2
    &=\left\|\binom N4^{-1}\sum_A|X_A|^2\right\|_{L^{p/2}(J)}
      \leq\binom N4^{-1}\sum_A\|X_A\|_{L^p(J)}^2.
    \label{eq:syk-learning-vector-minkowski}
\end{align}
Apply this to the vector in \eqref{eq:syk-learning-response} and to
$\mathcal Q(J)$, using \eqref{eq:syk-learning-quadratic-second-moment}
and \Cref{fact:syk-learning-gaussian-moments}, to obtain
\eqref{eq:syk-learning-vector-moments}.

\proofstep{Computing the coefficients}
Finally, we show that $a_\beta, b_\beta$ are efficiently computable. By \eqref{eq:positive-spectral-kernel} and \Cref{lem:positive-saddle}, the saddle point satisfies
\begin{align}
    |G_*'(t)|
    &\leq\frac12\int_0^\infty x\,\mu_{*,\beta}(dx)
      \leq\frac12
         \left(\int_0^\infty x^2\,\mu_{*,\beta}(dx)\right)^{1/2}
      =\frac\beta4,\nonumber\\
    |G_*(t)-G_*(s)|&\leq\frac\beta4|t-s|
      \qquad(s,t\in[0,1]).
\end{align}
Products are controlled by
\begin{align}
    \prod_{j=1}^r x_j-\prod_{j=1}^r y_j
    &=\sum_{j=1}^r
      \left(\prod_{i<j}x_i\right)(x_j-y_j)
      \left(\prod_{i>j}y_i\right),\nonumber\\
    \left|\prod_{j=1}^r x_j-\prod_{j=1}^r y_j\right|
    &\leq r\max_j|x_j-y_j|
      \qquad(|x_j|,|y_j|\leq1).
\end{align}
Thus the integrands in \eqref{eq:syk-learning-response-coefficients}
are bounded and Lipschitz, with constants depending only on $\beta$.
Compute $\widetilde G_*$ by \Cref{lem:saddle,lem:ladder-certificate}.
Clip its values to $[0,1/2]$, which cannot increase the error.
For any cell $Q$ of mesh width $h$ and any point $t_Q\in Q$,
\begin{align}
    \left|\int_Q F(t)\,dt-|Q|\widetilde F(t_Q)\right|
    &\leq |Q|\sup_{t\in Q}|F(t)-F(t_Q)|
          +|Q|\|F-\widetilde F\|_\infty \leq C_\beta|Q|\bigl(h+\|G_*-\widetilde G_*\|_\infty\bigr).
\end{align}
Here $F,\widetilde F$ are either displayed integrand, before and after
replacing $G_*$ by $\widetilde G_*$. Summing over the interval cells or
the cells intersected with $0\leq v\leq u\leq1$ gives total error
$C_\beta(h+\|G_*-\widetilde G_*\|_\infty)$. Choose both errors below
a sufficiently small fixed multiple of $\tau$.
There are polynomially many cells in $\tau^{-1}$. For each fixed $p$, the finitely many excluded even sizes are absorbed
by increasing $C_{\beta,p}$: $|\Tr(\rho_\beta\Gamma_A)|\leq1$, and $J_A,\mathcal Q_A(J)$
have finite Gaussian moments. The bounds therefore hold for all even
$N\geq8$.
\end{proof}

\subsection{Learning from quartic expectations}

By \Cref{lem:syk-learning-response}, the quartic expectations are close to $a_\beta J+b_\beta\cQ(J)$. We now show how to recover $J$ from this approximation when given estimates $\widehat f_A\in[-1,1]$ of $f_A=\Tr(\rho_\beta\Gamma_A)$. We will compute
\begin{align}
    Y_A&=\frac1\beta\sqrt{\frac{8N^3}{3}}\widehat f_A,\qquad
    W=\mathop{\operatorname{argmin}}_{X=X^{\mathsf T},\,\|X\|\leq16N}
                  \|X-\mathsf T(Y)\|_{\rm HS}^2.
    \label{eq:syk-learning-reconstruction}
\end{align}
and estimate the couplings by approximating
\begin{align}
    \frac{Y}{a_\beta}
       -\frac{b_\beta}{2a_\beta^3}\sqrt{\frac{3}{8N^3}}
                                      \mathsf T^*(W^2).
    \label{eq:syk-learning-quadratic-correction}
\end{align}
We bound the error of this estimate.

\begin{lemma}[Recovering the couplings]
\label{lem:syk-learning-reconstruction}
For $Y,W$ in \eqref{eq:syk-learning-reconstruction}, all sufficiently large $N$, and $J$ satisfying $\|\mathsf T(J)\|\leq16N$,
\begin{align}
    \left\|\frac{Y}{a_\beta}
       -\frac{b_\beta}{2a_\beta^3}\sqrt{\frac{3}{8N^3}}
                          \mathsf T^*(W^2)-J\right\|_2
    &\leq\frac2{a_\beta}\|Y-a_\beta J-b_\beta\cQ(J)\|_2
           +\frac{C_\beta}{\sqrt N}\|\cQ(J)\|_2.
    \label{eq:syk-learning-deterministic-error}
\end{align}
For every $0<\varepsilon<1$, a vector approximating \eqref{eq:syk-learning-quadratic-correction} to normalized $L_2$ error $\varepsilon/8$ can be computed in time polynomial in $N,\varepsilon^{-1}$.
\end{lemma}
\begin{proof}
To use the truncation bound in \Cref{lem:syk-learning-response}, we write the error in terms of $Y-a_\beta J-b_\beta\cQ(J)$. By \eqref{eq:syk-learning-quadratic-matrix-identity},
\begin{align}
    \frac{Y}{a_\beta}
       -\frac{b_\beta}{2a_\beta^3}\sqrt{\frac{3}{8N^3}}
                           \mathsf T^*(W^2)-J
    &=\frac{Y-a_\beta J-b_\beta\cQ(J)}{a_\beta} -\frac{b_\beta}{2a_\beta^3}\sqrt{\frac{3}{8N^3}}
               \mathsf T^*(W^2-a_\beta^2\mathsf T(J)^2).
    \label{eq:syk-learning-exact-error-identity}
\end{align}
It remains to bound the last term. For two inputs $Y,Y'$ and their minimizers $W,W'$, we use $\langle\mathsf T(Y)-W,W'-W\rangle_{\rm HS} \leq 0$ and $\langle\mathsf T(Y')-W',W-W'\rangle_{\rm HS}\leq0$ to obtain
\begin{align}
    \|W-W'\|_{\rm HS}^2
    &\leq\langle W-W',\mathsf T(Y)-\mathsf T(Y')\rangle_{\rm HS} \leq\|W-W'\|_{\rm HS}\,\|\mathsf T(Y)-\mathsf T(Y')\|_{\rm HS}.
    \label{eq:syk-learning-projection-nonexpansive}
\end{align}
Since $a_\beta\leq1$ and $\|\mathsf T(J)\|\leq16N$, the input $Y'=a_\beta J$ has minimizer $W'=a_\beta\mathsf T(J)$. Thus, by \Cref{lem:syk-learning-pair-identities},
\begin{align}
    \|W-a_\beta\mathsf T(J)\|_{\rm HS}
    &\leq\|\mathsf T(Y)-a_\beta\mathsf T(J)\|_{\rm HS}
      =\sqrt6\,\|Y-a_\beta J\|_2.
    \label{eq:syk-learning-projection-error}
\end{align}
Using $W^2-a_\beta^2\mathsf T(J)^2=W(W-a_\beta\mathsf T(J))+(W-a_\beta\mathsf T(J))a_\beta\mathsf T(J)$ and $\|\mathsf T^*\|_{{\rm HS}\to2}=\sqrt6$, we obtain
\begin{align}
    \|\mathsf T^*(W^2-a_\beta^2\mathsf T(J)^2)\|_2
    &\leq\sqrt6\,(\|W\|+a_\beta\|\mathsf T(J)\|)
                        \|W-a_\beta\mathsf T(J)\|_{\rm HS} \leq192N\|Y-a_\beta J\|_2.
    \label{eq:syk-learning-square-difference}
\end{align}
Substituting \eqref{eq:syk-learning-square-difference} into \eqref{eq:syk-learning-exact-error-identity}, then applying the triangle inequality to $Y-a_\beta J$, gives
\begin{align}
    &\left\|\frac{Y}{a_\beta}
       -\frac{b_\beta}{2a_\beta^3}\sqrt{\frac{3}{8N^3}}
                              \mathsf T^*(W^2)-J\right\|_2\nonumber\\
    &\quad\leq\frac1{a_\beta}\|Y-a_\beta J-b_\beta\cQ(J)\|_2
       +\frac{96b_\beta}{a_\beta^3}\sqrt{\frac{3}{8N}}
                              \|Y-a_\beta J\|_2\nonumber\\
    &\quad\leq\left(\frac1{a_\beta}
           +\frac{96b_\beta}{a_\beta^3}\sqrt{\frac{3}{8N}}\right)
                              \|Y-a_\beta J-b_\beta\cQ(J)\|_2
       +\frac{96b_\beta^2}{a_\beta^3}\sqrt{\frac{3}{8N}}\|\cQ(J)\|_2.
\end{align}
For sufficiently large $N$, $96b_\beta a_\beta^{-2}\sqrt{3/(8N)}\leq1$, proving \eqref{eq:syk-learning-deterministic-error}. By \Cref{lem:syk-learning-response}, the coefficients $a_\beta, b_\beta$ are efficient to compute.
\end{proof}

\subsection{Measuring all quartic expectations}

We next show how to measure the estimates $\wh f_A$ of $\Tr(\rho_\beta \Gamma_A)$ by measuring disjoint Majorana pairs.

\begin{lemma}[Estimating quartic expectations]
\label{lem:syk-learning-measurements}
For any state $\rho$ on $N$ Majoranas, with even $N\geq4$, and $0<\eta\leq1$, $0<\delta<1$, there is an algorithm using
\begin{align}
    S&=\left\lceil\frac{4(N-1)(N-3)}{3\eta^2}
                     \log\frac{2\binom N4}{\delta}\right\rceil
    \label{eq:syk-learning-matching-samples}
\end{align}
independent copies to compute estimates $\widehat f_A\in[-1,1]$ satisfying
\begin{align}
    \Pr\left[\max_{A\in\binom{[N]}4}
           |\widehat f_A-\Tr(\rho\Gamma_A)|>\eta\right]&\leq\delta.
    \label{eq:syk-learning-simultaneous-estimation}
\end{align}
Each copy uses a joint measurement of $N/2$ commuting Majorana pairs, implemented by $O(N^2)$ elementary Clifford gates and one reusable ancillary qubit in the Jordan--Wigner representation.
\end{lemma}
\begin{proof}
Choose an independent uniformly random perfect matching $\mathcal M$ of $[N]$ for each copy, and jointly measure $\Gamma_{\{i,j\}}=i\psi_i\psi_j$ on its pairs $\{i,j\}$ with $i<j$. These operators commute because the pairs are disjoint. For the outcomes $z_P\in\{-1,1\}$ and disjoint pairs $P,Q$ with $P\cup Q=A$, we use the following estimator
\begin{align}
    X_A&=\begin{cases}
       \displaystyle\frac{(N-1)(N-3)}3
          \tr(\Gamma_A\Gamma_P\Gamma_Q)z_Pz_Q,
          &A=P\cup Q,\quad P,Q\in\mathcal M,\\
       0,&\text{otherwise}.
       \end{cases}
\end{align}
and output, after $S$ independent samples $X_A^{(1)}, \dots, X_A^{(S)}$,
\begin{align}
    \widehat f_A =\max\left\{-1,\min\left\{1,\frac1S\sum_{s=1}^S X_A^{(s)}\right\}\right\}
\end{align}
The prefactor is chosen since in the matching $\cM$, we have $\Pr[P,Q\in\mathcal M]=[(N-1)(N-3)]^{-1}$ and there are three partitions of $A$ into disjoint pairs. Hence, $\EE X_A = \Tr(\rho \Gamma_A)$. Using
\begin{align}
    \EE X_A^2=\left(\frac{(N-1)(N-3)}3\right)^2  \frac3{(N-1)(N-3)}=\frac{(N-1)(N-3)}3
\end{align}
and $|X_A-\EE X_A|\leq\frac{2(N-1)(N-3)}3$, we can apply Bernstein's inequality to the independent copies $X_A^{(1)},\ldots,X_A^{(S)}$ to obtain
\begin{align}
    \Pr\left[\left|\frac1S\sum_{s=1}^S X_A^{(s)}-\Tr(\rho\Gamma_A)\right|>\eta\right]
    &\leq2\exp\left[-\frac{3S\eta^2}{(N-1)(N-3)(2+4\eta/3)}\right]\nonumber\\
    &\leq2\exp\left[-\frac{3S\eta^2}{4(N-1)(N-3)}\right].
    \label{eq:syk-learning-estimator-tail}
\end{align}
Thus the estimates and their joint failure probability satisfy
\begin{align}
    \Pr\left[\max_A|\widehat f_A-\Tr(\rho\Gamma_A)|>\eta\right]
    &\leq2\binom N4\exp\left[-\frac{3S\eta^2}{4(N-1)(N-3)}\right]\leq\delta,
\end{align}
where the last inequality uses \eqref{eq:syk-learning-matching-samples}.

In the Jordan--Wigner representation, each Majorana pair is a Pauli string. Its nondestructive measurement uses local Clifford basis changes, CNOTs accumulating parity in an ancillary qubit, measurement and reset of the ancilla, and inverse basis changes. Each measurement uses $O(N)$ gates, so measuring $N/2$ commuting pairs costs $O(N^2)$ gates per copy and gives their joint measurement.
\end{proof}

\subsection{Correctness and cost of learning}

We can now state the learning algorithm and prove \Cref{thm:syk-learning}.

\begin{proof}[Proof of \Cref{thm:syk-learning}]
Using \Cref{lem:syk-learning-measurements}, measure copies of the Gibbs state to estimate $\wh f_A$ for each $A \in \binom{[N]}{4}$ with failure probability $\delta/2$ and tolerance $\eta = (\beta e^{-\beta}/24)\sqrt{3/8N^3}\epsilon$. Since $ \tfrac12e^{-\beta/4}\leq G_*(t)\leq\tfrac12$ by \Cref{lem:positive-saddle}, the coefficients $a_\beta, b_\beta$ satisfy $e^{-\beta}\leq a_\beta\leq1$ and $0\leq b_\beta\leq\beta$. Hence, our choice of $\eta$ implies that the true expectations $f_A=\Tr(\rho_\beta \Gamma_A)$ are estimated with error
\begin{align}
    \frac2{\beta a_\beta}\sqrt{\frac{8N^3}{3}}
                \frac{\|\widehat f-f(J)\|_2}{\sqrt{\binom N4}} \leq \frac2{\beta a_\beta}\sqrt{\frac{8N^3}{3}}\max_A \abs{\wh f_A - f_A(J)} \leq \frac2{\beta a_\beta}\sqrt{\frac{8N^3}{3}}\eta \leq\frac{e^{-\beta}}{12a_\beta}\varepsilon\leq\frac\varepsilon{12}
    \label{eq:syk-learning-measurement-error-budget}
\end{align}
conditioned on measurement success. Apply \Cref{lem:syk-learning-reconstruction} to compute $Y,W$ in \eqref{eq:syk-learning-reconstruction}.
Since
\begin{align}
    Y-a_\beta J-b_\beta\cQ(J)
    &=\frac1\beta\sqrt{\frac{8N^3}{3}}f(J)-a_\beta J-b_\beta\cQ(J)
       +\frac1\beta\sqrt{\frac{8N^3}{3}}(\widehat f-f(J)),
\end{align}
\eqref{eq:syk-learning-deterministic-error} and \eqref{eq:syk-learning-measurement-error-budget} imply that
\begin{align}
    \frac2{a_\beta}
       \frac{\|\frac1\beta\sqrt{\frac{8N^3}{3}}f(J)
                          -a_\beta J-b_\beta\cQ(J)\|_2}{\sqrt{\binom N4}}
       +\frac{C_\beta}{\sqrt N}\frac{\|\cQ(J)\|_2}{\sqrt{\binom N4}}
    &\leq\frac\varepsilon2
    \label{eq:syk-learning-good-disorder}
\end{align}
are sufficient for
\begin{align}
    \frac{\|\widehat J-J\|_2}{\sqrt{\binom N4}}
    &\leq\frac\varepsilon2+\frac\varepsilon{12}+\frac\varepsilon8<\varepsilon.
\end{align}
Here $C_\beta$ is the constant from \eqref{eq:syk-learning-deterministic-error}. To bound the probability that \eqref{eq:syk-learning-good-disorder} fails, apply \Cref{lem:syk-learning-gaussian-matrix} and Markov's inequality, then use the two moment bounds in \eqref{eq:syk-learning-vector-moments}:
\begin{align}
    &\Pr_J\left[\Pr_{\rm alg}\left[
        \frac{\|\widehat J-J\|_2}{\sqrt{\binom N4}}>\varepsilon
                   \,\middle|\,J\right]>\delta\right]\nonumber\\
    &\quad\leq2e^{-N^2}
      +\left(\frac2\varepsilon\right)^p\EE_J\Bigg[
         \frac2{a_\beta}
         \frac{\|\frac1\beta\sqrt{\frac{8N^3}{3}}f(J)
                            -a_\beta J-b_\beta\cQ(J)\|_2}{\sqrt{\binom N4}}
         +\frac{C_\beta}{\sqrt N}\frac{\|\cQ(J)\|_2}{\sqrt{\binom N4}}
                          \Bigg]^p\nonumber\\
    &\quad\leq2e^{-N^2}+C_{\beta,p}(N\varepsilon)^{-p}.
    \label{eq:syk-learning-good-disorder-probability}
\end{align}
This proves \eqref{eq:syk-learning-quantitative-probability}.

\proofstep{Sample and time complexity} Substituting our choice for $\eta$ into \eqref{eq:syk-learning-matching-samples} gives
\begin{align}
    S&\leq C\frac{(N-1)(N-3)}3
            \frac{8N^3e^{2\beta}}{3\beta^2\varepsilon^2}
                            \log\frac{4\binom N4}{\delta}+1
      \leq C_\beta N^5\varepsilon^{-2}\log\frac N\delta
\end{align}
copies of the Gibbs state. Together with the polynomial measurement cost of \Cref{lem:syk-learning-measurements} and the cost of \Cref{lem:syk-learning-reconstruction}, this gives polynomial complexity in $N,\varepsilon^{-1},\log(\delta^{-1})$.

Finally, fix $a>0$ and $0<c<1$. For any fixed $p\geq2$ with $p(1-c)>a+1$, 
\begin{align}
    C_{\beta,p}(N\varepsilon)^{-p}+2e^{-N^2}
    &\leq C_{\beta,p}N^{-p(1-c)}+2e^{-N^2}\leq N^{-a},\nonumber\\
    C_\beta N^5\varepsilon^{-2}\log\frac N\delta
    &\leq C_\beta N^{5+2c}\log\frac N\delta
\end{align}
for sufficiently large $N$, proving \eqref{eq:syk-learning-power-samples}--\eqref{eq:syk-learning-power-error}.
\end{proof}

\section{Estimating thermal expectations of Ising spin glasses}
\label{sec:pspin-main}

We now describe the classical algorithm for Ising models with interactions of
order at least three. These algorithms are, in a certain sense, simplified versions of the quantum cavity method. The algorithm computes a finite Hermite expansion in the
couplings. Concentration lets us truncate the same recursion as in the SYK algorithm. This concentration is proven with the cavity method; the computations are easier because the cavity method is classical, and because some of the cavity computations have been done before in the literature (e.g., concentration of the overlap matrix in the replica formalism~\cite{CHL21}). However, unlike the $\mathfrak L_\beta$ obtained that corresponds to the resummed ladder diagrams in the SYK model, there is no similar operator that acts on a moment without changing its $1/N$ order. Hence, we can simplify the algorithm to avoid the $\mathfrak L_\beta^{-1}$ step; indeed, the Hessian in the $p \geq 3$ model is simply the identity. Such a step will reappear in the SK model algorithm in \Cref{sec:sk-main}.

\subsection{Models and main result}
\label{sec:pspin-model}

Let $\mathcal P\subseteq\{3,\ldots,P\}$ be a fixed nonempty set of interaction
orders, and let $\lambda_p>0$ for $p\in\mathcal P$ be fixed coefficients. Write
$\sigma_A=\prod_{i\in A}\sigma_i$ for $\sigma\in\{-1,1\}^N$, with
$\sigma_\varnothing=1$. We use the Hamiltonian
\begin{equation}
 H_N(\sigma)=\sum_{p\in\mathcal P}a_{p,N}
       \sum_{E\in\binom{[N]}p}J_{p,E}\sigma_E,
 \qquad
 a_{p,N}=\lambda_p\sqrt{\frac{p!}{2N^{p-1}}},
 \qquad J_{p,E}\overset{\mathrm{iid}}{\sim}\mathcal N(0,1).
 \label{eq:pspin-model}
\end{equation}
The pure $p$-spin model is the case $\mathcal P=\{p\}$ and $\lambda_p=1$.
For convenience, index the couplings by
$\mathcal I_N=\{(p,E):p\in\mathcal P,\ E\in\binom{[N]}p\}$ and write
$a_e=a_{p,N}$ and $\sigma_e=\sigma_E$ when $e=(p,E)$. Thus
$H_N=\sum_{e\in\mathcal I_N}a_eJ_e\sigma_e$. Define
\begin{equation}
 Z_N(\beta,J)=\sum_{\sigma\in\{-1,1\}^N}e^{\beta H_N(\sigma)},
 \qquad
 \langle O\rangle_{\beta,J}
   =\frac{\sum_\sigma O(\sigma)e^{\beta H_N(\sigma)}}{Z_N(\beta,J)}.
 \label{eq:pspin-gibbs}
\end{equation}
Put
\begin{equation}
 \xi(q)=\frac12\sum_{p\in\mathcal P}\lambda_p^2q^p,
 \qquad p_* =\min\mathcal P,
 \qquad \eta=\frac{p_*-2}{2}>0,
 \qquad \kappa=\frac{p_*-1}{2}\ge1.
 \label{eq:pspin-exponents}
\end{equation}
Let $\beta_*$ be the static transition along this mixture: it is the endpoint
of the interval on which the limiting pressure equals
$\log2+\beta^2\xi(1)/2$. For a pure model, this value is
$\log2+\beta^2/4$. The overlap theorems cited below identify this interval as the
zero-overlap replica-symmetric phase. In particular, $\beta_*$ is the static
transition, not a dynamical threshold or a second-moment threshold.

\begin{theorem}[Ising thermal expectations and partition function]
\label{thm:pspin-main}
Fix the mixture in \eqref{eq:pspin-model}, an inverse temperature
$0\le\beta<\beta_*$, an observable degree $\ell\ge0$, and exponents $a,c>0$.
There is a deterministic classical algorithm that outputs
$\widehat m_A(\beta,J)$ for all $|A|\le\ell$ and $\widehat L(\beta,J)$ such that,
for all sufficiently large $N$,
\begin{equation}
 \Pr_J\!\left[
  \sum_{|A|\le\ell}
   |\widehat m_A-\langle\sigma_A\rangle_{\beta,J}|^2>N^{-2c}
  \ \text{or}\ 
  |\widehat L-\log Z_N(\beta,J)|>N^{-c}
 \right]\le N^{-a}.
 \label{eq:pspin-guarantee}
\end{equation}
Its number of operations is at most $N^C$, where $C$ depends only on the
fixed parameters.
\end{theorem}

The same event controls any observable
$O_J=\sum_{|A|\le\ell}u_A(J)\sigma_A$ with $\sum_A|u_A(J)|^2\le1$:
Cauchy--Schwarz gives error at most $N^{-c}$ in
$\sum_Au_A(J)\widehat m_A$. The coefficients may be chosen after seeing $J$.
Moreover, exponentiating $\widehat L$ approximates $Z_N$ within a factor
$e^{\pm N^{-c}}$. This suffices to resolve instance-to-instance fluctuations.

\subsection{Concentration of replica observables}
\label{sec:pspin-concentration}
We start by using the known facts from the rigorous replica solution of the Ising $p$-spin model. However, since our algorithm will expand to resolve fluctuations beyond the standard replica concentration statement, we will still require a cavity argument later to control the remainder.

We use independent Gibbs samples $\sigma^1,\ldots,\sigma^n$ conditional on the
same disorder, called replicas. For a function of finitely many replicas, set
\begin{equation}
 \nu_\beta(f)=\mathbb E_J\langle f\rangle_{\beta,J},
 \qquad
 R_{ab}=\frac1N\sum_{i=1}^N\sigma_i^a\sigma_i^b.
 \label{eq:pspin-replicas}
\end{equation}
Replica notation here denotes finite products of probability measures. In
particular, $\nu_\beta$ averages the normalized Gibbs expectation; it is not a
ratio of disorder-averaged numerators and denominators.

The statistical input is concentration of the overlap throughout the static
replica-symmetric phase. For every fixed $b<\beta_*$ and integer $r\ge1$,
\begin{equation}
 \sup_{0\le\beta\le b}\nu_\beta(|R_{12}|^{2r})
       \le C_{b,r}N^{-r}.
 \label{eq:pspin-overlap}
\end{equation}
For the pure ordered model, this is Theorem~3.1 of
\cite{CHL21}; Theorems~3--4 and Remark~3 of \cite{Che17} give the mixed-model
extension with minimum interaction order at least three.
Section~\ref{sec:pspin-normalization} records their normalizations and transfers
the estimate to \eqref{eq:pspin-model}. Section~\ref{sec:pspin-overlap-upgrade}
also gives the cavity argument that upgrades macroscopic overlap concentration
to \eqref{eq:pspin-overlap}.

The covariance of the Hamiltonian is
\begin{equation}
 K(\sigma,\tau)=\mathbb E_J H_N(\sigma)H_N(\tau)
     =\sum_ea_e^2\sigma_e\tau_e,
 \qquad K_{ab}=K(\sigma^a,\sigma^b).
 \label{eq:pspin-covariance}
\end{equation}
For different replicas, \eqref{eq:pspin-overlap} implies
\begin{equation}
 \sup_{0\le\beta\le b}\nu_\beta(|K_{12}|^r)
       \le C_{b,r}N^{-r\eta}
 \qquad (r\ge1\text{ fixed}).
 \label{eq:pspin-covariance-moments}
\end{equation}
The proof uses Newton's identities to retain repeated-label corrections in the
distinct-site covariance; it is given in Section~\ref{sec:pspin-covariance-proof}.
The exponent $\eta>0$ is what will make a fixed Hermite cutoff sufficient.

To describe the cavity recursion, we will refer to a \emph{replica monomial} as
\begin{equation}
 f=\prod_{j=1}^{s}\sigma_{i_j}^{B_j},
 \qquad \sigma_i^B=\prod_{a\in B}\sigma_i^a,
 \qquad i_1,\ldots,i_s\text{ distinct},\quad
 \varnothing\ne B_j\subseteq[n],
 \label{eq:pspin-monomial}
\end{equation}
whose sites $i_1,\ldots,i_s$ are its \emph{active sites}. We use notation
\begin{align}
    \supp(f) = \{i_1,\dots,i_s\}, \qquad s(f) = s, \qquad n(f) = n.
\end{align}
We will store both the monomial and its replica count $n$, including replica labels that do not occur in the product. Whenever spin factors are multiplied, even spin powers cancel $(\sigma_i^a)^2=1$ and thus put the result back in the form of \eqref{eq:pspin-monomial}; a site is removed from the support if none its spin is squared to one. Each replica monomial satisfies the following bound (proved in \Cref{sec:pspin-support-proof}).

\begin{lemma}[Concentration by active-site order]
\label{lem:pspin-support}
For fixed $s,n$ and $b<\beta_*$, every monomial in
\eqref{eq:pspin-monomial} satisfies
\begin{equation}
 \sup_{0\le\beta\le b}|\nu_\beta(f)|\le C_{s,n,b}N^{-s/2}.
 \label{eq:pspin-support-bound}
\end{equation}
If any $|B_j|$ is odd, then $\nu_\beta(f)=0$ exactly.
\end{lemma}

The odd case follows by changing the sign of a spin and its incident
Gaussian couplings, causing $\nu_\beta(f)$ to vanish. For even $B$, moments of the multi-overlap
$N^{-1}\sum_i\sigma_i^B$ are bounded by the corresponding moments of
$R_{12}$. Permutation symmetry and inclusion--exclusion for distinct site labels
then give \eqref{eq:pspin-support-bound}.

\subsection{The cavity recursion at the target temperature}
\label{sec:pspin-cavity}

Choose an active site $i$ in a nonconstant monomial $f$. Split the Hamiltonian
into the interactions avoiding $i$ and those incident on $i$:
\begin{equation}
 H_N=H^{\setminus i}+H^{(i)},
 \qquad
 H_t=H^{\setminus i}+\sqrt t\,H^{(i)},
 \qquad 0\le t\le1.
 \label{eq:pspin-local-path}
\end{equation}
Let $\nu_t$ denote the disorder-averaged normalized replica expectation for
$H_t$ at inverse temperature $\beta$. We keep the ambient normalization $N$
fixed. At $t=0$, only the selected spin is decoupled; the bath remains at
$\beta$. Since $i$ is active, its independent uniform replica spins imply
$\nu_0(f)=0$.

The covariance of the deleted interactions is
\begin{equation}
 \Delta_i^{ab}
   =\sum_{p\in\mathcal P}a_{p,N}^2
       \sum_{\substack{E\in\binom{[N]}p\\i\in E}}
          \sigma_E^a\sigma_E^b,
 \qquad
 \Lambda_{i,N}=\sum_{p\in\mathcal P}a_{p,N}^2
                   \binom{N-1}{p-1}
       \le\frac12\sum_{p\in\mathcal P}p\lambda_p^2.
 \label{eq:pspin-local-covariance}
\end{equation}
In particular, $\Delta_i^{aa}=\Lambda_{i,N}$ is a scalar. For a function of
$n$ replicas, define
\begin{equation}
 \mathscr D_i f=f\left[
   \sum_{a<b\le n}\Delta_i^{ab}
   -n\sum_{a\le n}\Delta_i^{a,n+1}
   +\frac{n(n+1)}2\Delta_i^{n+1,n+2}
 \right].
 \label{eq:pspin-local-operator}
\end{equation}
The output has $n+2$ declared replicas. Iterating this rule always uses two
new replica labels.

Gaussian integration by parts gives
\begin{equation}
 \frac{d}{dt}\nu_t(f)=\beta^2\nu_t(\mathscr D_i f),
 \qquad
 \frac{d^k}{dt^k}\nu_t(f)=\beta^{2k}\nu_t(\mathscr D_i^kf).
 \label{eq:pspin-local-derivative}
\end{equation}
The diagonal covariance cancels against differentiation of the Gibbs
normalizer. Section~\ref{sec:pspin-gaussian} proves this identity, including its
endpoint derivatives. Expanding from $t=1$ to the known value at $t=0$ gives
\begin{equation}
 \nu_\beta(f)=
   \sum_{k=1}^{L-1}\frac{(-1)^{k+1}\beta^{2k}}{k!}
                       \nu_\beta(\mathscr D_i^kf)+r_L(f),
 \qquad L\ge2.
 \label{eq:pspin-cavity-row}
\end{equation}
We call \eqref{eq:pspin-cavity-row} the cavity equation for $f$. Every nontrivial expectation on the right-hand side is in the full model at the target
$\beta$. We prove the following bound on the remainder $r_L$ in \Cref{sec:pspin-local-proof}.

\begin{lemma}[Cavity remainder]
\label{lem:pspin-local-remainder}
For every fixed $n,L$ and $b<\beta_*$, the remainder in
\eqref{eq:pspin-cavity-row} satisfies
\begin{equation}
 |r_L(f)|\le C_{n,L,b}N^{-L\kappa}
 \qquad (0\le\beta\le b),
 \label{eq:pspin-local-error}
\end{equation}
uniformly over the physical labels in $f$.
\end{lemma}
To evaluate the cavity equation, expand each $\mathscr D_i^kf$ by choosing
one replica-pair term from \eqref{eq:pspin-local-operator} and one incident
interaction from \eqref{eq:pspin-local-covariance} at each application.
Let $\mathcal T(f)$ index these choices, retaining their order. For $1 \leq k < L$, we denote an element of $\cT(f)$ by $\tau=\bigl(k;(E_j,a_j,b_j)_{j=1}^k\bigr)$. For $i \in E_j$ where $|E_j| \in \cP$ (e.g., $|E_j| = p$ for the pure $p$-spin model), we expand the cavity equation as
\begin{align}
 \sum_{k=1}^{L-1}\frac{(-1)^{k+1}\beta^{2k}}{k!}\mathscr D_i^kf
 &=\sum_{\tau\in\mathcal T(f)}w_\tau f_\tau \quad \text{for} \quad
 f_\tau=f\prod_{j=1}^k\sigma_{E_j}^{a_j}\sigma_{E_j}^{b_j}.
 \label{eq:pspin-expanded-cavity-equation}
\end{align}
At step $j$, the pair $(a_j,b_j)$ ranges over the three sums in
\eqref{eq:pspin-local-operator} with $n+2(j-1)$ in place of $n$.
The coefficient $w_\tau$ is the product of the Taylor coefficient, the
selected replica coefficients, and $\prod_{j=1}^k a_{|E_j|,N}^2$.
The product $f_\tau$ is reduced using $\sigma_i^2=1$ and
stored with $n(f_\tau)=n(f)+2k$.

Call a summand \emph{collision-free} if
\begin{align}
 (E_j\setminus\{i\})\cap\operatorname{supp}(f)&=\varnothing
       &&(1\le j\le k),\nonumber\\
 (E_j\setminus\{i\})\cap(E_h\setminus\{i\})&=\varnothing
       &&(1\le j<h\le k).
 \label{eq:pspin-collision-free-conditions}
\end{align}
Otherwise call it a \emph{collision term}. Write $\chi_\tau=0$ in the
first case and $\chi_\tau=1$ in the second. In a collision-free
summand, each new site occurs in exactly one pair of distinct replicas
and therefore remains active. Every old active site other than $i$
remains unchanged, so
\begin{align}
 s(f_\tau)&\ge s(f)-1+\sum_{j=1}^{k}(|E_j|-1)
           \ge s(f)+1
           \qquad(\chi_\tau=0).
 \label{eq:pspin-support-increase}
\end{align}
The selected site may disappear, but each interaction introduces at
least two new sites. For fixed $s,n,L$,
\begin{align}
 \sum_{\tau\in\mathcal T(f)}|w_\tau|&\le C_{s,n,L,b},
 &\sum_{\tau:\chi_\tau=1}|w_\tau|&\le\frac{C_{s,n,L,b}}{N}.
 \label{eq:pspin-row-mass}
\end{align}
Indeed, the total incident interaction weight is bounded by
\eqref{eq:pspin-local-covariance}, while a contact with an old active
site or an intersection between two new sets costs a factor $1/N$.
\Cref{sec:pspin-recursion-proof} gives the counting argument.

Fix an active-site cutoff $S$, a collision cutoff $R$, and a local
expansion order $L$. Define $\operatorname{Eval}(f,r)$, where $r$ counts
the collision terms selected to reach the current call, as follows.
\begin{enumerate}
 \item Cancel pairs of identical spin factors using
 $\sigma_i^2 = 1$. If $f=1$, return one. If some
 active site $i_j$ has $|B_j|$ odd, return zero.
 \item If $s(f)\ge S$ or $r\ge R$, return zero.
 \item Otherwise choose the least active site $i$, generate the
 summands indexed by $\mathcal T(f)$ in
 \eqref{eq:pspin-expanded-cavity-equation}, and return
 \begin{align}
   \operatorname{Eval}(f,r)&=
       \sum_{\tau\in\mathcal T(f)}w_\tau
          \operatorname{Eval}(f_\tau,r+\chi_\tau).
   \label{eq:pspin-evaluator}
 \end{align}
\end{enumerate}
Write $\widehat\nu_\beta(f)=\operatorname{Eval}(f,0)$ and extend this
notation linearly to finite sums of monomials.

These calls form a rooted tree. Its root is the initial call, and a
summand $\tau$ creates an edge, with weight $w_\tau$, to the call
$\operatorname{Eval}(f_\tau,r+\chi_\tau)$. Calls that return at either
of the first two steps are leaves. A path from the root to a leaf
selects one summand at each successive expansion; its weight is the
product of the selected coefficients. Distinct summands remain distinct
edges, even if their calls have identical inputs.

\begin{lemma}[Recursion truncation]
\label{lem:pspin-recursion}
For fixed $S\ge2$, $R\ge1$, $L\ge2$, and initial replica count at most $n_0$,
every path from the root to a leaf has at most $T$ expansions, and every
call has at most $n_{\max}$ declared replicas, where
\begin{align}
 T&=(R+1)(S+1),
 &n_{\max}&=n_0+2(L-1)T.
 \label{eq:pspin-depth}
\end{align}
For every initial monomial with $s(f)<S$ and $n(f)\le n_0$,
\begin{align}
 |\widehat\nu_\beta(f)-\nu_\beta(f)|
 &\le C_{S,R,L,n_0,b}
       \left(N^{-S/2}+N^{-R}+N^{-L\kappa}\right).
 \label{eq:pspin-evaluator-error}
\end{align}
The number of arithmetic operations is polynomial in $N$. Its output is a
polynomial in $\beta$ of degree at most $2(L-1)T$.
\end{lemma}

Along consecutive edges with $\chi_\tau=0$, the number of active sites
increases by at least one until it reaches $S$. A path has at most $R$
edges with $\chi_\tau=1$, and each edge adds at most $2(L-1)$ declared
replicas. This gives \eqref{eq:pspin-depth}. For the error, repeatedly
substitute the exact identity \eqref{eq:pspin-cavity-row} along the same
tree, retaining the true expectations at its leaves. Leaves stopped at
$s(f)\ge S$ have expectation $O(N^{-S/2})$ by
\Cref{lem:pspin-support}. The sum of absolute path weights for leaves
stopped at $r=R$ is $O(N^{-R})$ by \eqref{eq:pspin-row-mass}.
Each expansion also contributes a remainder $O(N^{-L\kappa})$.
The coefficient bounds already sum over all interaction labels, so
there is no additional factor from the number of label assignments.
\Cref{sec:pspin-recursion-proof} proves the resulting error bound.

The distinction from SYK is that terms that do not increase the number
of active sites have total coefficient weight $O(N^{-1})$:
\begin{align}
 \sum_{\substack{\tau\in\mathcal T(f)\\s(f_\tau)\le s(f)}}|w_\tau|
 &\le\sum_{\tau:\chi_\tau=1}|w_\tau|
 \le\frac{C_{s,n,L,b}}{N}.
 \label{eq:pspin-same-support-weight}
\end{align}
The collision-free terms need not have total weight below one, but
they strictly increase $s(f)$ and hence admit only finitely many
successive substitutions before the cutoff. In the SYK hierarchy,
$\mathfrak L_\beta$ acts on a moment without increasing its order;
its inverse is needed before the remaining order-one terms increase
the moment order, as in \Cref{lem:exact-hierarchy}. Here no such
same-order operator remains to be inverted.

Increasing the declared replica count from $n$ to $n+2k$ implements
differentiation of the normalized Gibbs expectation. It does not itself
supply a power of $1/N$: some of those replicas may be unused in a
summand. Termination instead follows from active-site growth and the
collision cutoff. For two-spin interactions, a collision-free term can
replace the selected site by just one new site, preserving $s(f)$
despite the increase in replica count. This is why the SK construction
in \Cref{sec:sk-main} requires a separate resummation step.

\subsection{Reconstructing the thermal expectations}
\label{sec:pspin-reconstruction}

For a repeated list of interaction labels $\mathbf e=(e_1,\ldots,e_d)$,
define the function on $d+1$ replicas
\begin{equation}
 \mathcal F_{A;\mathbf e}
   =\sigma_A^1\prod_{j=1}^{d}
       \left(\sum_{u=1}^{j}\sigma_{e_j}^u
                         -j\sigma_{e_j}^{j+1}\right).
 \label{eq:pspin-insertion}
\end{equation}
At $d=0$, this is $\sigma_A^1$. Differentiating the normalized Gibbs
expectation gives
\begin{equation}
 \partial_{e_1}\cdots\partial_{e_d}F_A(J;\beta)
    =\beta^d\left(\prod_{j=1}^{d}a_{e_j}\right)
                      \langle\mathcal F_{A;\mathbf e}\rangle_{\beta,J},
 \qquad F_A(J;\beta)=\langle\sigma_A\rangle_{\beta,J}.
 \label{eq:pspin-disorder-derivative}
\end{equation}
There are finitely many monomials in \eqref{eq:pspin-insertion} at fixed $d$,
and their total absolute coefficient is at most $2^dd!$.

We use notation
\begin{equation}
 h_\alpha(J)=\frac{\prod_e\operatorname{He}_{\alpha_e}(J_e)}{\sqrt{\alpha!}},
 \qquad \alpha!=\prod_e\alpha_e!,\qquad |\alpha|=\sum_e\alpha_e.
 \label{eq:pspin-hermites}
\end{equation}
Gaussian integration by parts and \eqref{eq:pspin-disorder-derivative} identify
the exact coefficients:
\begin{equation}
 c_{A,\alpha}(\beta)
  =\mathbb E_J[F_A(J;\beta)h_\alpha(J)]
  =\frac{\beta^d\prod_ea_e^{\alpha_e}}{\sqrt{\alpha!}}
                     \nu_\beta(\mathcal F_{A;\mathbf e}),
 \qquad d=|\alpha|.
 \label{eq:pspin-exact-coefficient}
\end{equation}
Choose a fixed ordering of the repeated list representing $\alpha$ when
computing an approximation. The exact value is independent of that ordering.
The cavity algorithm estimates the last expectation, yielding
\begin{equation}
 \widehat c_{A,\alpha}(\beta)
  =\frac{\beta^d\prod_ea_e^{\alpha_e}}{\sqrt{\alpha!}}
                     \widehat\nu_\beta(\mathcal F_{A;\mathbf e}),
 \qquad
 P_A(J;\beta)=\sum_{|\alpha|\le D}\widehat c_{A,\alpha}(\beta)h_\alpha(J).
 \label{eq:pspin-reconstruction}
\end{equation}
The initial monomials have at most $\ell+PD$ active sites and $D+1$
replicas. Both are fixed once the Hermite cutoff $D$ is fixed.

\begin{lemma}[Hermite truncation and coefficient error]
\label{lem:pspin-hermite}
For every fixed $D$,
\begin{equation}
 \sup_{0\le\beta\le b}\sum_{|\alpha|>D}|c_{A,\alpha}(\beta)|^2
       \le C_{D,b}N^{-(D+1)\eta}.
 \label{eq:pspin-hermite-tail}
\end{equation}
This is proven in Section~\ref{sec:pspin-hermite-proof}. If the cavity solver estimates every monomial in
\eqref{eq:pspin-insertion} to error $O(N^{-Q})$, uniformly for $d\le D$, then
\begin{align}
 |\widehat c_{A,\alpha}-c_{A,\alpha}|
   &\le C_{D,Q,b}\frac{\prod_e|a_e|^{\alpha_e}}{\sqrt{\alpha!}}N^{-Q},
       && |\alpha|\le D, \label{eq:pspin-coefficient-error}\\
 \sup_{0\le\beta\le b}\mathbb E_J|P_A-F_A|^2
   &\le C_{D,Q,b}\left(N^{-(D+1)\eta}+N^{D-2Q}\right).
       \label{eq:pspin-total-l2}
\end{align}
The constants are uniform over $A$ of fixed bounded size.
\end{lemma}

To see the small parameter in the tail bound, square
\eqref{eq:pspin-disorder-derivative} using two disjoint groups of replicas and
sum over the derivative labels. Every summed label gives a covariance $K_{ab}$
between the two groups, so $a\ne b$. Equation~\eqref{eq:pspin-covariance-moments}
then gives
\begin{equation}
 \mathbb E_J\|D^dF_A\|_{\mathrm{HS}}^2\le C_{d,b}N^{-d\eta}.
 \label{eq:pspin-gaussian-sobolev}
\end{equation}
Here the Hilbert--Schmidt norm sums over ordered derivative indices. Bessel's
inequality (\Cref{fact:bessel}) applied to these derivatives bounds the entire Hermite tail. The
coefficient errors are summed using
\begin{equation}
 \sum_{|\alpha|=d}\frac{\prod_ea_e^{2\alpha_e}}{\alpha!}
       =\frac{(\sum_ea_e^2)^d}{d!}\le C_DN^d.
 \label{eq:pspin-multinomial}
\end{equation}

\subsection{Correctness, counting, and cost}
\label{sec:pspin-correctness}

We now choose all cutoffs. Let $\ell_0=\max(\ell,P)$ and set
\begin{equation}
 \Gamma=a+2c+\ell_0+P+20,
 \qquad
 D=\left\lceil\frac{\Gamma+2}{\eta}\right\rceil,
 \qquad
 Q=\left\lceil\frac{\Gamma+D+2}{2}\right\rceil.
 \label{eq:pspin-error-choices}
\end{equation}
For the cavity solver, take
\begin{equation}
 S=\max(\ell_0+PD+1,2Q+2),\qquad R=Q+1,\qquad
 L=\max\left(2,\left\lceil\frac{Q+1}{\kappa}\right\rceil\right).
 \label{eq:pspin-cavity-choices}
\end{equation}
All these quantities are independent of $N$. Lemmas~\ref{lem:pspin-recursion}
and~\ref{lem:pspin-hermite} imply
\begin{equation}
 \sup_{0\le u\le b}\sup_{|A|\le\ell_0}
     \mathbb E_J|P_A(J;u)-\langle\sigma_A\rangle_{u,J}|^2
        \le C N^{-\Gamma-2}.
 \label{eq:pspin-final-ms}
\end{equation}
In particular, the same polynomials give sufficiently accurate predictions for
all interaction monomials.

\begin{proof}[Proof of Theorem~\ref{thm:pspin-main}]
Choose $b$ with $\beta\le b<\beta_*$. Summing
\eqref{eq:pspin-final-ms} over $O(N^\ell)$ targets and applying Markov's
inequality bounds the observable failure probability at threshold
$\tfrac12N^{-c}$ by
\begin{equation}
 C N^{\ell+2c-\Gamma-2}.
 \label{eq:pspin-observable-markov}
\end{equation}
This is smaller than any required constant multiple of $N^{-a}$ for all
sufficiently large $N$.

For the partition function, use the identity
\begin{equation}
 \log Z_N(\beta,J)=N\log2+\int_0^\beta\langle H_N\rangle_{u,J}\,du
 \label{eq:pspin-thermodynamic-integration}
\end{equation}
and define
\begin{equation}
 \widetilde L(\beta,J)=N\log2+
             \sum_{e\in\mathcal I_N}a_eJ_e\int_0^\beta P_e(J;u)\,du.
 \label{eq:pspin-counting-output}
\end{equation}
Here $P_e=P_E$ for $e=(p,E)$. Each cavity row is a polynomial in $u$, so every
integral is evaluated term by term. We use the un-clipped polynomials in this
formula.

Let $\mathcal G=\{\sum_ea_e^2J_e^2\le C_0N\}$ for a sufficiently large fixed
$C_0$. For each interaction order, the sum of the squared couplings is
chi-squared. Its moment generating function gives
$\Pr(\mathcal G^c)\le C e^{-c_0N^{p_*}}$; see
Section~\ref{sec:pspin-bits}. On $\mathcal G$, Cauchy--Schwarz over the
interaction labels and over $u$ gives
\begin{align}
 |\widetilde L-\log Z_N|^2
 &\le C_0N\beta\int_0^\beta
       \sum_e|P_e(J;u)-\langle\sigma_e\rangle_{u,J}|^2\,du,
       \label{eq:pspin-counting-cs}\\
 \mathbb E_J\!\left[\mathbf1_{\mathcal G}
                         |\widetilde L-\log Z_N|^2\right]
 &\le C N^{P+1-\Gamma-2}.
       \label{eq:pspin-counting-ms}
\end{align}
Markov's inequality at threshold $\tfrac12N^{-c}$ bounds this failure
probability by $C N^{P+1+2c-\Gamma-2}$, in addition to
$\Pr(\mathcal G^c)$. Uniform expected error along the interval is enough; we
have not assumed a single good event for every real temperature.

There are $O(N^{PD})$ Hermite indices. A cavity row chooses at most $L-1$
incident interactions, and a path has at most $T=(R+1)(S+1)$ rows. A
conservative arithmetic bound for all degree-at-most-$\ell_0$ targets and
counting is
\begin{equation}
 N^{\ell_0+PD+(P-1)(L-1)T+O(1)}.
 \label{eq:pspin-arithmetic-cost}
\end{equation}
The replica choices only affect its constant, since
$n_{\max}\le D+1+2(L-1)T$. All coefficients are generated by the explicit
recursion; no table of thermal expectations is an input.

Finally, Section~\ref{sec:pspin-bits} shows that $O(\log N)$ fractional bits
suffice for inputs and arithmetic on a Gaussian box. Choose the fixed
precision exponent so the numerical error in the observable vector and in
\eqref{eq:pspin-counting-output} is at most $\tfrac14N^{-c}$. Set
$\widehat m_A$ to the computed $P_A$ clipped to $[-1,1]$, and let
$\widehat L$ be the computed counting expression. Clipping cannot increase
error to a spin expectation. The box failure and the two ideal failure
probabilities have total at most $N^{-a}$ for sufficiently large $N$, proving
\eqref{eq:pspin-guarantee}. At $\beta=0$, return the exact uniform-spin
expectations and $N\log2$.
\end{proof}

\section{Estimating thermal expectations of the SK model}
\label{sec:sk-main}

In the Sherrington--Kirkpatrick (SK) model, each pair of spins
interacts through a coupling $J_{ij}/\sqrt N$, where $J_{ji}=J_{ij}$
and the $J_{ij}$ for $i<j$ are independent standard Gaussians.
Although each coupling is small, successive interactions need not
produce additional powers of $1/N$ after summing over intermediate
spins. For example, the expansion of
$\langle\sigma_i\sigma_j\rangle_{\beta,J}$ about $\beta=0$ contains
the terms
\begin{align}
 \frac{\beta}{\sqrt N}J_{ij},
 \qquad
 \frac{\beta^2}{N}\sum_{k\notin\{i,j\}}J_{ik}J_{kj}.
\end{align}
The first describes the direct interaction between $i$ and $j$;
the second describes interactions mediated by another spin $k$.
Their second moments are
\begin{align}
 \mathbb E_J\left|\frac{\beta}{\sqrt N}J_{ij}\right|^2
 &=\frac{\beta^2}{N},&
 \mathbb E_J\left|
   \frac{\beta^2}{N}\sum_{k\notin\{i,j\}}J_{ik}J_{kj}
 \right|^2
 &=\frac{\beta^4(N-2)}{N^2}.
\end{align}
Thus both have root-mean-square size of order $N^{-1/2}$ at fixed
$\beta>0$. More generally, a sequence of $m$ interactions between
$i$ and $j$, summed over distinct intermediate spins, has
asymptotically $N^{m-1}$ summands. Independence and centering make
different summands orthogonal, so its second moment is asymptotic
to
\begin{align}
 N^{m-1}\left(\frac{\beta^2}{N}\right)^m
 =\frac{\beta^{2m}}{N}.
\end{align}
The length of these sequences does not change the power of $1/N$, analogous to the ladder diagrams in the SYK model.

Unlike the $p\geq 3$ and SYK algorithms, the fixed-degree Hermite truncation also fails: $\eta=(p_*-2)/2$ in \Cref{lem:pspin-hermite} is zero at $p_*=2$. We instead have to truncate at degree $O(\log N)$, which requires slightly more delicate control of the concentration statements compared to the $p\geq 3$ and SYK proofs. After these modifications, we obtain an algorithm that succeeds at every fixed $\beta<\beta_*=1$.

\begin{theorem}[SK thermal expectations and partition function]
\label{thm:sk-main}
Fix an inverse temperature $0\le\beta<1$, an observable degree
$\ell\ge0$, and exponents $a,c>0$. There is a deterministic classical
algorithm that outputs $\widehat m_A(\beta,J)$ for all $|A|\le\ell$ and
$\widehat L(\beta,J)$ such that, for all sufficiently large $N$,
\begin{equation}
 \Pr_J\!\left[
  \sum_{|A|\le\ell}|\widehat m_A-\langle\sigma_A\rangle_{\beta,J}|^2
       >N^{-2c}
  \ \text{or}\ 
  |\widehat L-\log Z_N(\beta,J)|>N^{-c}
 \right]\le N^{-a}.
 \label{eq:sk-guarantee}
\end{equation}
The number of bit operations is at most $N^C$, where $C$ depends only on
the fixed parameters.
\end{theorem}

\begin{figure}[t]
    \centering
    \includegraphics[width=\linewidth]{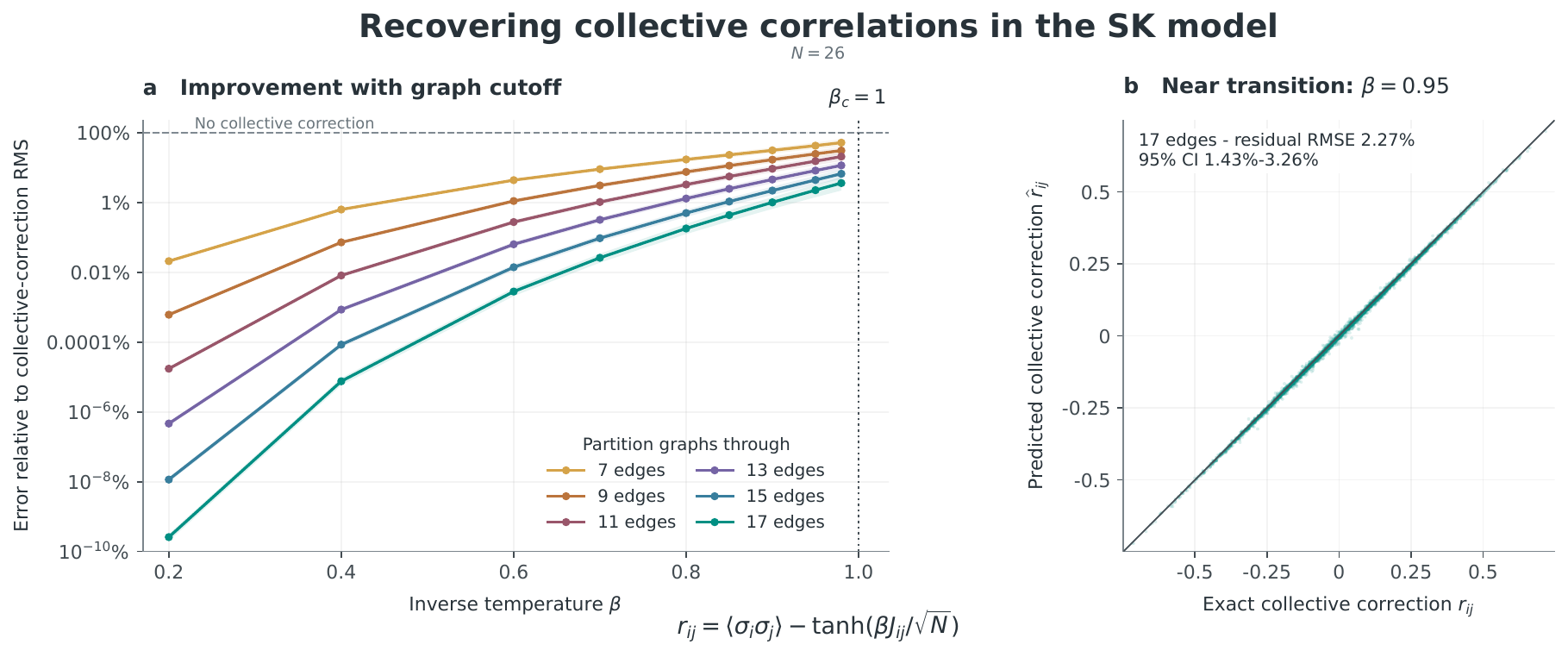}
    \caption{\textbf{Recovering collective correlations in the Sherrington--Kirkpatrick model.} Results for $N=26$ spins and 32 independent Gaussian Hamiltonians, using all 325 spin pairs per realization. We subtract the independent-edge contribution to obtain $r_{ij}=\langle\sigma_i\sigma_j\rangle-\tanh(\beta J_{ij}/\sqrt{N})$.
    \textbf{(a)} RMS prediction error, normalized by the RMS exact collective correction, for partition-graph cutoffs of 7, 9, 11, 13, 15 and 17 edges. The dashed horizontal line corresponds to predicting zero collective correction; the dotted vertical line marks the thermodynamic transition at $\beta_c=1$.
    \textbf{(b)} Predictions at the 17-edge cutoff versus exact collective corrections at $\beta=0.95$. The diagonal indicates perfect agreement. The relative RMS error is $2.27\%$ (95\% confidence interval 1.43--3.26\%). Shaded bands and the reported interval use 4{,}000 bootstrap resamples of whole Hamiltonians. }
    \label{fig:SK_model}
\end{figure}

The performance of the classical algorithm for the SK model is visualized in \cref{fig:SK_model}. Here, we compare to the exact collective correction for a pair of spins, which is $r_{ij}=\langle\sigma_i\sigma_j\rangle_\beta-\tanh(\beta J_{ij}/\sqrt N)$, where the thermal expectation is taken in the full SK Gibbs distribution for a fixed realization of the couplings. The subtracted term is the correlation that the pair would have if its direct coupling were the only interaction present. Thus, $r_{ij}$ measures how the surrounding interaction network modifies the pair correlation through indirect interactions and feedback. In the figure, the exact thermal expectations are obtained by summing the Boltzmann weights over all spin configurations, up to floating-point precision, providing a reference against which to compare the classical algorithm’s predictions.

We now describe the algorithm and its proof in some detail.

\subsection{An exact one-site cavity identity}
\label{sec:sk-cavity}

Let $E_N=\binom{[N]}2$, $M=|E_N|$, and set
\begin{equation}
 x_e=\frac{\beta J_e}{\sqrt N},\qquad
 t_e=\tanh x_e,\qquad q_N=\mathbb E_J t_e^2\le\frac{\beta^2}{N}.
 \label{eq:sk-edge-weights}
\end{equation}
The variables $t_e$ are independent and centered. We identify a simple graph
$G$ with an edge subset of $E_N$. Write $V(G)$ for its nonisolated vertices,
$\partial G$ for its odd-degree vertices, and
$w(G)=\prod_{e\in G}t_e$. The empty graph has weight one and no active
vertices. For $A\subseteq U\subseteq[N]$, define
\begin{equation}
 S_A(U)=\sum_{\substack{G\subseteq\binom U2\\\partial G=A}}w(G),
 \qquad S_A=S_A([N]),\qquad D=S_\varnothing.
 \label{eq:sk-graph-sums}
\end{equation}
We set $S_A(U)=0$ if $A\nsubseteq U$. The base value is
$S_\varnothing(\varnothing)=1$.

The identity $e^{x\sigma_i\sigma_j}
=\cosh(x)(1+\tanh(x)\sigma_i\sigma_j)$ and the uniform-spin parity rule give
\begin{equation}
 Z_N=2^N\left(\prod_{e\in E_N}\cosh x_e\right)D,
 \qquad \langle\sigma_A\rangle_{\beta,J}=\frac{S_A}{D}.
 \label{eq:sk-spin-graph}
\end{equation}
Indeed, on expanding the product, a graph contributes to the numerator
exactly when every spin in $\sigma_A\prod_{e\in G}\sigma_e$ has even
multiplicity. This is the condition $\partial G=A$. In particular,
$D>0$, $|S_A|\le D$, and $S_A=0$ when $|A|$ is odd.

\begin{lemma}[Vertex deletion]
\label{lem:sk-deletion}
For every $A\subseteq U\subseteq[N]$ and $i\in U$,
\begin{equation}
 S_A(U)=
 \sum_{\substack{B\subseteq U\setminus\{i\}\\
          |B|\equiv\mathbf1_{\{i\in A\}}\ ({\rm mod}\ 2)}}
     \left(\prod_{j\in B}t_{ij}\right)
     S_{(A\setminus\{i\})\triangle B}(U\setminus\{i\}).
 \label{eq:sk-deletion}
\end{equation}
Every edge weight retains the ambient normalization $N^{-1/2}$.
\end{lemma}

\begin{proof}
Given a graph with boundary $A$, let $B$ be the neighbor set of $i$.
The degree parity at $i$ gives the constraint in the sum. Deleting the
incident edges removes $i$ and toggles the parity at precisely the vertices
in $B$, so the remaining boundary is
$(A\setminus\{i\})\triangle B$. Conversely, adjoining the edges
$\{\{i,j\}:j\in B\}$ to a graph with that boundary reconstructs the
original graph uniquely. The weights factor as displayed.
\end{proof}

Equation~\eqref{eq:sk-deletion} is an exact cavity identity at the chosen
$\beta$. Its naive recursion is exponential. We will instead compute a
controlled truncation of its full graph expansion. 

For deterministic real coefficients $b_G$, independence and centering give
\begin{equation}
 \mathbb E_J[w(G)w(G')]=\mathbf1_{\{G=G'\}}q_N^{|G|},
 \qquad
 \mathbb E_J\left|\sum_Gb_Gw(G)\right|^2
       =\sum_Gb_G^2q_N^{|G|}.
 \label{eq:sk-orthogonality}
\end{equation}
If $G\ne G'$, an edge in their symmetric difference appears to the first
power and its mean is zero. This orthogonality retains the cancellations
between signed graph weights.

\subsection{Summing chains and cycles before truncation}
\label{sec:sk-truncation}
We first quantify the length dependence discussed above. Write
$(n)_a=n(n-1)\cdots(n-a+1)$, with $(n)_0=1$. For fixed distinct
$i,j$ and fixed $m$, \eqref{eq:sk-orthogonality} gives
\begin{align}
 \mathbb E_J\left|
   \sum_{\substack{G\text{ a simple path from }i\text{ to }j\\|G|=m}}
          w(G)\right|^2
 &= (N-2)_{m-1}q_N^m
    \sim \frac{\beta^{2m}}{N}
       \qquad(m\ge1),\nonumber\\
 \mathbb E_J\left|
   \sum_{\substack{G\text{ a simple cycle}\\|G|=m}}w(G)\right|^2
 &=\frac{(N)_m}{2m}q_N^m
    \sim\frac{\beta^{2m}}{2m}
       \qquad(m\ge3).
 \label{eq:sk-chain-cycle-moments}
\end{align}
The factor $2m$ removes the choices of starting vertex and orientation
of a cycle. Both asymptotics are for $N\to\infty$ at fixed $m$ and
$\beta>0$, using $Nq_N\to\beta^2$. Thus truncating at a fixed edge
count omits part of the leading-order contribution; we will truncate at degree $O(\log N)$.

Fix a boundary $A$ of size $r=2h$. If $G$ has $m$ edges and $v$ active
vertices outside $A$, its degrees have the form
\begin{equation}
 d_i=2+2s_i\quad(i\in V(G)\setminus A),\qquad
 d_i=1+2s_i\quad(i\in A),\qquad s_i\ge0.
 \label{eq:sk-degrees}
\end{equation}
The excess above these minimum degrees is
\begin{equation}
 \delta(G):=m-v-h=m-|V(G)|+h=\sum_{i\in V(G)}s_i\ge0.
 \label{eq:sk-excess}
\end{equation}
The vertices in $A$ are the \emph{marked vertices}. At a fixed graph
shape, the $v$ unmarked vertex labels contribute a factor of order
$N^v$ to its embedding count, while its expected squared weight contributes
a factor of order $N^{-m}$. The resulting power is
\begin{align*}
 N^{v-m}&=N^{-h-\delta(G)}.
\end{align*}
Subdividing a chain adds one edge and one vertex, and therefore leaves
$\delta(G)$ unchanged. Adding a disjoint simple cycle also leaves it
unchanged. Graphs with $\delta(G)=0$ have degree one at each marked
vertex and degree two at every other active vertex. Their connected
components are therefore paths pairing the marked vertices and simple
cycles, with all components vertex-disjoint. For $A=\varnothing$,
this also includes the empty graph.

For integers $k,B\ge1$, let
\begin{equation}
 \mathcal G_A^{k,B}
   =\{G\subseteq E_N:\partial G=A,\ |G|\le k,\ \delta(G)<B\},
 \qquad
 T_A^{k,B}=\sum_{G\in\mathcal G_A^{k,B}}w(G).
 \label{eq:sk-retained}
\end{equation}
We retain logarithmically many edges but only a fixed amount of excess.

\begin{lemma}[Graph truncation]
\label{lem:sk-truncation}
Fix $0<\beta<\beta'<1$ and put $\rho=(\beta/\beta')^2$. For every fixed
even $r=2h$, uniformly over $|A|=r$,
\begin{align}
 \mathbb E_J\left|\sum_{\substack{\partial G=A\\|G|>k}}w(G)\right|^2
     &\le C_{\beta'}\rho^k,
       \label{eq:sk-edge-tail}\\
 \mathbb E_J\left|
  \sum_{\substack{\partial G=A,\ |G|\le k\\\delta(G)\ge B}}w(G)
                \right|^2
     &\le C_{\beta,r}N^{-h}
          \left(\frac{C_rk^2}{N}\right)^B e^{C_rk^2/N}.
       \label{eq:sk-excess-tail}
\end{align}
The two omitted sums are orthogonal, so the sum of these bounds controls
$\mathbb E_J|S_A-T_A^{k,B}|^2$.
\end{lemma}

Section~\ref{sec:sk-graph-bounds} proves the lemma. Its role is
analogous to \Cref{lem:hermite-tail} in the SYK proof: orthogonality
bounds the squared error from discarding terms of the disorder
expansion. The excess cutoff retains a fixed number of orders in $1/N$, analogous to the finite-size truncation in \Cref{lem:hierarchy-error} after the SYK
ladder response has been summed. A sufficiently large fixed $B$
makes this second error inverse polynomial as well.

Enumerating every assignment of physical vertices along the retained
chains and cycles would still be too expensive since $k=\Theta(\log N)$. The following color-subset recurrences perform the sum efficiently.

\subsection{Computing the retained graph sum}
\label{sec:sk-evaluation}

A coloring $\chi:[N]\to[s]$ makes a graph \emph{colorful} if its active
vertices have distinct colors. For a graph with $s$ active vertices this
means that every color is used exactly once. Define
\begin{equation}
 C_{A,s}(\chi;t)=
    \sum_{\substack{G\in\mathcal G_A^{k,B},\ |V(G)|=s\\
                    \chi|_{V(G)}\ \mathrm{injective}}}w(G).
 \label{eq:sk-colorful-sum}
\end{equation}
By \eqref{eq:sk-excess}, only
$s\le s_{\max}:=\min(N,k+\lfloor\ell/2\rfloor)$ is needed for the
boundaries $|A|\le\ell$.

To evaluate \eqref{eq:sk-colorful-sum}, retain the marked vertices and all
unmarked vertices of degree at least four. Suppress each maximal chain
whose internal vertices have degree two, and put aside components that
are simple cycles. If there are $b$ unmarked branching vertices, the
resulting core has $r+b$ vertices and
\begin{equation}
 b\le\delta(G)<B,\qquad
 q=h+b+\delta(G)\le h+2\delta(G)
 \label{eq:sk-core-size}
\end{equation}
edges. Loops and parallel edges in the core represent different chains
in the original simple graph. There are only finitely many core types at
fixed $B$ and $\ell$.

For a fixed core embedding, let $\mathcal R$ be its set of colors. A path
from a core vertex $u$ with subsequent color set
$S\subseteq[s]\setminus\mathcal R$ is computed by
\begin{equation}
 F_u(\varnothing,x)=\mathbf1_{\{x=u\}},\qquad
 F_u(S,x)=
 \begin{cases}
  \displaystyle\sum_{y=1}^N F_u(S\setminus\{\chi(x)\},y)t_{yx},
                                      &\chi(x)\in S,\\
  0,                                  &\chi(x)\notin S,
 \end{cases}
 \label{eq:sk-path-recursion}
\end{equation}
for $S\ne\varnothing$, with $t_{xx}=0$. The terminal vertex is included
in $S$, whereas $u$ is not. Distinct colors enforce distinct vertices.
For a fixed $u$, all values cost $O(2^sN^2)$ arithmetic operations.
Disjoint-subset convolution joins these paths at the bounded core and
adds the free cycles. We prove in Section~\ref{sec:sk-core-proof} the following result.

\begin{lemma}[Exact colorful evaluation]
\label{lem:sk-colorful}
For a fixed coloring and signed edge weights, the sum
$C_{A,s}(\chi;t)$ can be evaluated exactly in
\begin{equation}
 2^{O_{B,\ell}(s)}N^{O(B+\ell+1)}
 \label{eq:sk-colorful-cost}
\end{equation}
arithmetic operations. Every retained colorful simple graph has
coefficient one.
\end{lemma}

We remove the coloring restriction by a deterministic average. For
$s\ge1$, a uniform random coloring makes a fixed $s$-set colorful with
probability $p_s=s!/s^s$. For a list
$\mathcal C=(\chi_1,\ldots,\chi_m)$, put
\begin{equation}
 d_{\mathcal C}(V)=\frac1{mp_s}\sum_{j=1}^m
               \mathbf1_{\{\chi_j|_V\ \mathrm{injective}\}}-1,
 \qquad
 Q_s(\mathcal C)=\binom Ns^{-1}\sum_{|V|=s}d_{\mathcal C}(V)^2.
 \label{eq:sk-color-discrepancy}
\end{equation}

\begin{lemma}[Deterministic color families]
\label{lem:sk-color-family}
For integers $1\le s\le N$ and $m\ge1$, a deterministic algorithm
constructs a list $\mathcal C$ of $m$ colorings such that
\begin{equation}
 Q_s(\mathcal C)\le\frac{p_s^{-1}-1}{m}.
 \label{eq:sk-color-family-bound}
\end{equation}
It uses at most $m^2N^2\operatorname{poly}(s)2^s$ arithmetic operations
on rationals of polynomial bit length in
$s\log(N+2)+s\log(s+1)+\log(m+1)$.
\end{lemma}

The proof in Section~\ref{sec:sk-color-proof} chooses each vertex color
by conditional expectation. The conditional objective is computable by
an elementary symmetric polynomial and an $s\times s$ permanent; the
latter is evaluated by a $2^s$ subset formula. Since $p_s^{-1}\le e^s$,
this construction is polynomial when $s=O(\log N)$ and $m=N^{O(1)}$.

Choose one such list $\mathcal C_s$ for each $1\le s\le s_{\max}$,
independently of the disorder, and use it for every target boundary. The
ideal real-weight estimator is
\begin{equation}
 \widehat T_A(t)=\mathbf1_{\{A=\varnothing\}}
   +\sum_{s=1}^{s_{\max}}\frac1{m_sp_s}
                      \sum_{\chi\in\mathcal C_s}C_{A,s}(\chi;t).
 \label{eq:sk-estimator}
\end{equation}
Each retained nonempty graph is multiplied by
$1+d_{\mathcal C_{|V(G)|}}(V(G))$. Let
\begin{equation}
 \mathcal A_\ell=\{A\subseteq[N]:|A|\le\ell,\ |A|\ \mathrm{even}\},
 \qquad R_\ell=|\mathcal A_\ell|\le C_\ell N^\ell.
 \label{eq:sk-target-family}
\end{equation}
The denominator needs the extra weight $R_\ell$ because it is used in
every ratio. Orthogonality and permutation invariance give
\begin{align}
 \mathbb E_J\left[\sum_{A\in\mathcal A_\ell}
             |\widehat T_A-T_A^{k,B}|^2
       +R_\ell|\widehat T_\varnothing-T_\varnothing^{k,B}|^2\right]
 &\le 2C_\beta R_\ell\max_{1\le s\le s_{\max}}Q_s(\mathcal C_s).
 \label{eq:sk-color-error}
\end{align}
Section~\ref{sec:sk-color-error-proof} proves this estimate.

\subsection{Correctness, counting, and cost}
\label{sec:sk-correctness}

We now specify the cutoffs. Choose an integer $T>a+2c+\ell+12$ and
rationals $\beta<\bar\beta<\beta'<1$. Set $\bar\rho=(\bar\beta/\beta')^2$
and take
\begin{equation}
 k=\min\{j\ge1:\bar\rho^j\le N^{-T-2}\},\qquad
 B=T+5,\qquad
 m_s=\left\lceil p_s^{-1}N^{T+2}\right\rceil.
 \label{eq:sk-cutoffs}
\end{equation}
Here $k=O_\beta(\log(N+2))$, and
$Q_s(\mathcal C_s)\le N^{-T-2}$. Define the ideal weighted squared error
\begin{equation}
 \mathcal E=
    \sum_{A\in\mathcal A_\ell}|\widehat T_A-S_A|^2
            +R_\ell|\widehat T_\varnothing-D|^2.
 \label{eq:sk-ideal-error}
\end{equation}
The coloring error is supported on retained graph monomials, and the
omission error is supported on their complement. They are orthogonal by
\eqref{eq:sk-orthogonality}. Lemma~\ref{lem:sk-truncation},
\eqref{eq:sk-color-error}, and
$(C\log^2N/N)^B=O(N^{-B+1})$ for fixed $B$ imply
\begin{equation}
 \mathbb E_J\mathcal E\le C_{\beta,\ell,T}N^{\ell-T-2}.
 \label{eq:sk-total-mean-square}
\end{equation}

Let $\widetilde T_A$ denote the implemented estimator after approximating
the edge weights and evaluating the graph recurrences. Choose the
precision so that, on the input box,
\begin{equation}
 \sum_{A\in\mathcal A_\ell}|\widetilde T_A-\widehat T_A|^2
       +R_\ell|\widetilde T_\varnothing-\widehat T_\varnothing|^2
       \le\frac{N^{-2c-2}}{1024}.
 \label{eq:sk-arithmetic-budget}
\end{equation}
On the event
\begin{equation}
 D\ge N^{-1},\qquad
 \mathcal E\le\frac{N^{-2c-2}}{1024},
 \qquad \text{the input lies in the prescribed box},
 \label{eq:sk-success-event}
\end{equation}
the combined weighted squared error, denoted by $\widetilde{\mathcal E}$,
is at most $N^{-2c-2}/256$. In particular,
\begin{equation}
 \left|\frac{\widetilde T_\varnothing}{D}-1\right|
        \le\frac{N^{-c}}{16},\qquad
 \widetilde T_\varnothing\ge\frac1{2N}.
 \label{eq:sk-relative-denominator}
\end{equation}
Since $|S_A|\le D$, direct subtraction of the two ratios gives
\begin{align}
 \sum_{A\in\mathcal A_\ell}
  \left|\frac{\widetilde T_A}{\widetilde T_\varnothing}
                   -\frac{S_A}{D}\right|^2
 &\le\frac8{D^2}\left(
       \sum_{A\in\mathcal A_\ell}|\widetilde T_A-S_A|^2
                 +R_\ell|\widetilde T_\varnothing-D|^2\right)
        \le 8N^2\widetilde{\mathcal E}
       \le \frac1{32}N^{-2c}.
 \label{eq:sk-stable-ratios}
\end{align}
For example, each scalar difference is at most
$2D^{-1}(|\widetilde T_A-S_A|
              +|\widetilde T_\varnothing-D|)$, which proves the first
inequality after squaring and summing.

The algorithm outputs $\widehat m_\varnothing=1$ and
$\widehat m_A=0$ for odd $|A|$. For nonempty even $A$, it divides by
$\widetilde T_\varnothing$ only when
$\widetilde T_\varnothing\ge1/(2N)$, and clips the ratio to $[-1,1]$.
If the input overflows or the denominator fails this test, it returns
zero for the remaining outputs. On \eqref{eq:sk-success-event}, clipping
cannot increase the error.

The same denominator supplies the counting output. On passing the safe
threshold, evaluate
\begin{equation}
 L_{\mathrm{calc}}
   =N\log2+\sum_{e\in E_N}\log\cosh x_e
                                  +\log\widetilde T_\varnothing.
 \label{eq:sk-log-output}
\end{equation}
The change in the last term relative to $\log D$ is at most $N^{-c}/8$
by \eqref{eq:sk-relative-denominator} and
$|\log(1+u)|\le2|u|$ for $|u|\le1/2$. Compute
\eqref{eq:sk-log-output} from the rounded inputs with total additional
error at most $N^{-c}/4$, including evaluation of $N\log2$.
The resulting $\widehat L$ has the required error. The explicit factor
$2^N$ in \eqref{eq:sk-spin-graph} is not absorbed into the graph denominator.

Markov's inequality and \eqref{eq:sk-total-mean-square} bound failure of
the ideal error threshold by
$1024C_{\beta,\ell,T}N^{\ell-T+2c}$. The remaining failure event in \eqref{eq:sk-success-event} is when the denominator $D$ in \eqref{eq:sk-spin-graph} is too small.

\begin{lemma}[Lower tail of the denominator]
\label{lem:sk-denominator}
For every fixed $0<\beta<1$ and $s>0$, there are positive constants
$C_{\beta,s},c_{\beta,s}$ such that, for all sufficiently large $N$,
\begin{equation}
 \Pr_J(D<N^{-s})
       \le C_{\beta,s}\exp[-c_{\beta,s}(\log N)^2].
 \label{eq:sk-denominator-tail}
\end{equation}
\end{lemma}

The proof in Section~\ref{sec:sk-denominator-proof} uses a second-moment
bound, an averaged gradient bound, and convexity of
$\log(Z_N/\mathbb E_JZ_N)$. The sum of all the failure events
is at most $N^{-a}$ for sufficiently large $N$. Equations
\eqref{eq:sk-stable-ratios}--\eqref{eq:sk-log-output} prove
\eqref{eq:sk-guarantee} on the same event.

Finally, $s\le s_{\max}=O(\log N)$ and $p_s^{-1}\le e^s$ make every
$m_s$ polynomial in $N$. Lemmas~\ref{lem:sk-colorful}
and~\ref{lem:sk-color-family}, summed over polynomially many boundaries
and support sizes, give polynomial arithmetic cost.
Section~\ref{sec:sk-bits} proves polynomial bit cost and completes
Theorem~\ref{thm:sk-main}.

\section*{Acknowledgments}
We would like to thank Brian Swingle for discussions and sharing an early draft of~\cite{swingle2026single}, and Brice Huang for related discussions about the Ising $p$-spin model and quantum spin glasses. 

We acknowledge substantial interactions with large language models in developing the results in this paper. The core proof idea was discovered by 40 instances of GPT 6 Astra Pro working collaboratively over the span of several days. We have also used large language models to assist with writing and proofreading. The authors verified the arguments, rewrote and simplified the proofs, and wrote the manuscript. We take full responsibility for its content.

AS is supported by a Google PhD Fellowship and the Simons Foundation. AZ is supported by a Hertz Fellowship and the Simons Foundation (MP-SIP-00001553, AWH).

\bibliography{References}
\bibliographystyle{alpha}

\clearpage
\appendix

\section{Fourier truncation error of the saddle solution}\label{app:saddle_trunc}

\begin{fact}[Parseval's identity]\label{fact:parseval}
For $f,g\in L^2(0,1)$ and
$\widehat f(n)=\int_0^1e^{i\omega_nt}f(t)\,dt$, $n\in\mathbb Z$,
\begin{align}
 \int_0^1\overline{f(t)}g(t)\,dt
 &=\sum_{n\in\mathbb Z}\overline{\widehat f(n)}\widehat g(n),
 &\|f\|_2^2&=\sum_{n\in\mathbb Z}|\widehat f(n)|^2.
 \label{eq:parseval-fact}
\end{align}
In particular, if $f(1-t)=f(t)$, then
\begin{align}
 f(t)&=2\sum_{n\geq0}\left(\int_0^1f(s)\sin(\omega_ns)\,ds\right)
                    \sin(\omega_nt)\quad\text{in }L^2,\nonumber\\
 \|f\|_2^2&=2\sum_{n\geq0}
       \left|\int_0^1f(t)\sin(\omega_nt)\,dt\right|^2.
 \label{eq:sine-parseval}
\end{align}
\end{fact}
\begin{fact}[Bessel's inequality]\label{fact:bessel}
For an orthonormal family $(e_n)$ in a Hilbert space,
\begin{align}
 \sum_n|\langle e_n,f\rangle|^2&\leq\|f\|^2,
 &\sum_{n\geq0}\left|\int_0^1f(t)\sin(\omega_nt)\,dt\right|^2
 &\leq\tfrac12\|f\|_2^2.
 \label{eq:bessel-fact}
\end{align}
The second assertion uses $e_n(t)=\sqrt2\sin(\omega_nt)$.
\end{fact}

For $M\geq1$, let $\mathcal F_M$ denote the objective in
\eqref{eq:convexsd}. Numerical coordinates are restricted to the explicit box
\begin{align}
 \mathcal B_{\beta,M}
 :=\left[
 \frac12\min\left\{e^{-1-\beta^2/128},
                  \frac1{1+\beta^2/(4\pi^2)}\right\},
 4(1+\beta^2/128)\right]^M.
 \label{eq:explicit-saddle-box}
\end{align}
For any $u\in\mathcal B_{\beta,M}$, define $\widetilde G_*$ by
\eqref{eq:convexsd}'s trigonometric expression, with these coordinates.
Its \emph{saddle residual} is
\begin{align}
 \|\nabla\mathcal F_M(u)\|_2
 =\left(\sum_{n=0}^{M-1}\left|
 1-u_n^{-1}+\frac{\beta^2}{\omega_n}
      \int_0^1\widetilde G_*(t)^3\sin(\omega_nt)\,dt
                   \right|^2\right)^{1/2}.
 \label{eq:definition-saddle-residual}
\end{align}

\begin{lemma}[Saddle approximation]\label{lem:saddle}
For every $u\in\mathcal B_{\beta,M}$,
\begin{align}
 \|\widetilde G_*-G_*\|_\infty
 &\leq C_\beta M^{-2}+C_\beta\|\nabla\mathcal F_M(u)\|_2.
 \label{eq:analysis-saddle-error}
\end{align}
Coordinates with residual at most $M^{-2}$ can be computed in time
polynomial in $M$.
\end{lemma}
\begin{proof}
Let $u^{(M)}$ minimize $\mathcal F_M$ over positive coordinates. Since
$u-\log u-1\geq\max\{-\log u-1,u/2-1\}$,
\begin{align}
 \mathcal F_M(u^{(M)})&\leq\mathcal F_M(\mathbf1)=\beta^2/128,
 &e^{-1-\beta^2/128}\leq u_n^{(M)}&\leq2(1+\beta^2/128).
 \label{eq:saddle-coordinate-box}
\end{align}
For the exact Fourier ratios $u_n^*=\widehat G_*(n)/(i/\omega_n)$,
\eqref{eq:positive-kernel-fourier} and \eqref{eq:saddle-pointwise-bounds} give
\begin{align}
 \frac1{1+\beta^2/(4\omega_n^2)}\leq u_n^*&\leq1,
 &|u_n^*-1|&\leq\frac{\beta^2}{4\omega_n^2}.
 \label{eq:analysis-coordinate-decay}
\end{align}
Thus $u^{(M)}$ and $(u_n^*)_{n<M}$ lie strictly inside
\eqref{eq:explicit-saddle-box}. For $v\in\mathbb R^M$,
\begin{align}
 v^\mathsf T\nabla^2\mathcal F_M(u)v
 &=\sum_{n<M}\frac{v_n^2}{u_n^2}
   +6\beta^2\int_0^1\widetilde G_*(t)^2
      \left(\sum_{n<M}\frac{v_n}{\omega_n}\sin(\omega_nt)\right)^2dt,\nonumber\\
 c_\beta I&\preceq\nabla^2\mathcal F_M(u)
       \preceq C_\beta(1+\log(M+1))^2I
       \qquad(u\in\mathcal B_{\beta,M}).
 \label{eq:saddle-hessian-bounds}
\end{align}
Here $\|\widetilde G_*\|_\infty\leq C_\beta(1+\log(M+1))$ and
$\sum_{n\geq0}\omega_n^{-2}=1/8$.

Set $G_*^{(M)}(t)=\frac12+2\sum_{n<M}(u_n^*-1)\sin(\omega_nt)/\omega_n$.
By \Cref{fact:parseval,fact:bessel} and \eqref{eq:sd},
\begin{align}
 G_*^{(M)}(t)-G_*(t)
 &=-2\sum_{n\geq M}\frac{u_n^*-1}{\omega_n}\sin(\omega_nt),\nonumber\\
 \|G_*^{(M)}-G_*\|_\infty
 &\leq\frac{\beta^2}{2}\sum_{n\geq M}\omega_n^{-3}
 \leq C_\beta M^{-2},\nonumber\\
 \|G_*^{(M)}-G_*\|_2^2
 &=2\sum_{n\geq M}\frac{|u_n^*-1|^2}{\omega_n^2}
 \leq\frac{\beta^4}{8}\sum_{n\geq M}\omega_n^{-6}
 \leq C_\beta M^{-5},
 \label{eq:saddle-tail-two-norms}\\
 \partial_n\mathcal F_M(u^*)
 &=\frac{\beta^2}{\omega_n}\int_0^1
       \bigl((G_*^{(M)})^3-G_*^3\bigr)\sin(\omega_nt)\,dt,\nonumber\\
 \|\nabla\mathcal F_M(u^*)\|_2^2
 &\leq\frac{\beta^4}{2\pi^2}
              \|(G_*^{(M)})^3-G_*^3\|_2^2
 \leq C_\beta M^{-5}.
 \label{eq:finite-saddle-gradient}
\end{align}
The last inequality uses the uniform bound on $G_*^{(M)}$ supplied by
its uniform tail estimate. Integrating the Hessian on the segment
from $u^*$ to $u$ gives
\begin{align}
 c_\beta\|u-u^*\|_2^2
 &\leq\langle u-u^*,\nabla\mathcal F_M(u)-\nabla\mathcal F_M(u^*)\rangle,\nonumber\\
 \|u-u^*\|_2
 &\leq c_\beta^{-1}
       \bigl(\|\nabla\mathcal F_M(u)\|_2+
                         \|\nabla\mathcal F_M(u^*)\|_2\bigr),\nonumber\\
 \|\widetilde G_*-G_*^{(M)}\|_\infty
 &\leq2\left(\sum_{n<M}\omega_n^{-2}\right)^{1/2}\|u-u^*\|_2.
 \label{eq:coordinate-to-kernel}
\end{align}
This proves \eqref{eq:analysis-saddle-error}.

For the numerical claim, take
$L_M=C_\beta(1+\log(M+1))^2$ as an upper Hessian bound and use
projected gradient descent. Theorem~3.10 of~\cite{bubeck2015convex} gives
\begin{align}
 u^{(0)}&=\mathbf1,\nonumber\\
 u^{(k+1)}&=\operatorname{proj}_{\mathcal B_{\beta,M}}
       \bigl(u^{(k)}-L_M^{-1}\nabla\mathcal F_M(u^{(k)})\bigr),\nonumber\\
 \|u^{(k)}-u^{(M)}\|_2^2
 &\leq e^{-kc_\beta/L_M}\|\mathbf1-u^{(M)}\|_2^2
 \leq C_\beta M e^{-kc_\beta/L_M},\nonumber\\
 \|\nabla\mathcal F_M(u^{(k)})\|_2
 &\leq L_M\|u^{(k)}-u^{(M)}\|_2\leq\tfrac12M^{-2}
 \quad\text{for }k\geq C_\beta(1+\log(M+1))^3.
\end{align}
The quartic trigonometric polynomial and its derivatives are finite sums
of elementary integrals. Computing each update to error
$M^{-O(1)}$ makes the accumulated coordinate error at most
$M^{-2}/(2L_M)$: projection and the exact gradient update are
nonexpansive on the box. The computed residual is then at most $M^{-2}$.
The number of terms, iterations, and arithmetic bits is polynomial in $M$.
\end{proof}

For $d$ time variables, a \emph{cell} is a set
\begin{align}
 \{\mathbf t\in[0,1]^d:0<t_{\pi(1)}<\cdots<t_{\pi(d)}<1\},
 \qquad\pi\in S_d.
 \label{eq:time-cell-definition}
\end{align}
A \emph{face} is a nonempty set obtained by replacing any of these strict
inequalities by equalities, leaving the others strict. Equivalently,
partition the indices into ordered nonempty sets $I_1,\ldots,I_q$ and impose
\begin{align}
 t_i&=\tau_a\quad(i\in I_a),
 &0\leq\tau_1&<\cdots<\tau_q\leq1,
 \label{eq:time-face-definition}
\end{align}
with each endpoint either fixed to $0$ or $1$, respectively, or required
to lie strictly inside $(0,1)$. On a face we store each prescribed ordered
limit separately, as well as the alternating value of
\Cref{lem:contact-dictionary}. These values need not equal one another.
We write $F$ for a cell or a face together with its specified prescription,
and $X_F$ for the resulting function of its free coordinates.

\begin{lemma}[Stable numerical inversion]\label{lem:linv}
Choose $u\in\mathcal B_{\beta,M}$ with
$\|\nabla\mathcal F_M(u)\|_2\leq M^{-2}$. Extend its approximation by
\begin{align}
 \widetilde G_*(u,v)&=G_0(u,v)+
       2\sum_{n<M}\frac{u_n-1}{\omega_n}\sin(\omega_n(u-v)),
       \qquad 0\leq u,v\leq1,
 \label{eq:exact-jump-saddle-approximation}
\end{align}
using $G_0(u,u)=0$ and the one-sided values of $G_0$ when an ordered
limit is specified. Thus the discontinuous part is not discretized.
Let $\widetilde K$ be the kernel of $I-\widetilde{\mathfrak L}_\beta$,
obtained by replacing $G_*$ with $\widetilde G_*$ in \eqref{eq:lbeta}.
Approximate the bounded kernel $R$ of
$K\mathfrak L_\beta^{-1}K$, $K=I-\mathfrak L_\beta$, using the certified
finite inverse of \Cref{lem:ladder-certificate} at mesh width at most
$M^{-4}$. Define
\begin{align}
 (\widetilde{\mathfrak L_\beta^{-1}}X)(u,v)
 &=X(u,v)+\int_{[0,1]^2}
     \bigl(\widetilde K(u,v;s,t)+\widetilde R(u,v;s,t)\bigr)X(s,t)\,ds\,dt.
 \label{eq:explicit-numerical-inverse}
\end{align}
The term $X(u,v)$ is evaluated exactly; prescribed boundary values are
limits, or finite alternating averages of limits, of the entire expression.
With exact integration, for every bounded input with these prescriptions,
\begin{align}
 \|\widetilde{\mathfrak L_\beta^{-1}}X-\mathfrak L_\beta^{-1}X\|_\infty
 &\leq C_\beta M^{-2}\|X\|_\infty,
 \label{eq:numerical-inverse-error}\\
 \|\widetilde{\mathfrak L_\beta^{-1}}X\|_\infty
 &\leq C_\beta\|X\|_\infty.
 \label{eq:numerical-inverse-bound}
\end{align}
For quadrature, suppose each $X_F$ extends continuously to the closure of
its own domain and
\begin{align}
 \|X\|_{1/2,\mathrm{pw}}
 :=\max_F\left(\|X_F\|_\infty+
   \sup_{x\ne y\in F}\frac{|X_F(x)-X_F(y)|}{\|x-y\|_\infty^{1/2}}\right)
 <\infty.
 \label{eq:piecewise-holder-norm}
\end{align}
The supremum is zero when $F$ has no free coordinates. A mesh of width $h$
gives a reconstructed output satisfying
\begin{align}
 \| (\mathfrak L_\beta^{-1}X)_{\mathrm{grid}}
                    -\mathfrak L_\beta^{-1}X\|_\infty
 &\leq C_\beta(M^{-2}+h^{1/2})\|X\|_{1/2,\mathrm{pw}}.
 \label{eq:numerical-inverse-quadrature-error}
\end{align}
Taking $h=M^{-4}$ has polynomial cost in $M$ for a fixed number of
variables, given polynomial-cost evaluation of the input functions.
Every fixed hierarchy iterate has a uniformly bounded norm
\eqref{eq:piecewise-holder-norm}.
\end{lemma}
\begin{proof}
Telescoping the four factors in the ladder kernel and applying
\Cref{lem:saddle} gives
\begin{align}
 \|\widetilde{\mathfrak L}_\beta-\mathfrak L_\beta\|_{\infty\to\infty}
 +\|\widetilde{\mathfrak L}_\beta-\mathfrak L_\beta\|_{2\to2}
 &\leq C_\beta M^{-2}.
 \label{eq:saddle-to-operator-error}
\end{align}
For sufficiently large $M$, in either norm,
\begin{align}
 \|\widetilde{\mathfrak L}_\beta^{-1}\|
 &\leq2\|\mathfrak L_\beta^{-1}\|,\nonumber\\
 \widetilde{\mathfrak L}_\beta^{-1}-\mathfrak L_\beta^{-1}
 &=\widetilde{\mathfrak L}_\beta^{-1}
       (\mathfrak L_\beta-\widetilde{\mathfrak L}_\beta)
                  \mathfrak L_\beta^{-1}.
 \label{eq:inverse-resolvent-identity}
\end{align}
The row, inverse, and column approximations in
\Cref{lem:ladder-certificate}, at mesh width $M^{-4}$, imply
\begin{align}
 \|\widetilde R-R\|_\infty&\leq C_\beta M^{-2},\nonumber\\
 \|\widetilde{\mathfrak L_\beta^{-1}}X-\mathfrak L_\beta^{-1}X\|_\infty
 &\leq\bigl(\|\widetilde K-K\|_\infty+
                  \|\widetilde R-R\|_\infty\bigr)\|X\|_\infty.
\end{align}
This proves \eqref{eq:numerical-inverse-error} and
\eqref{eq:numerical-inverse-bound}.

For fixed external variables $u,v$, intersect the mesh boxes in $(s,t)$
with all relevant inequalities in \eqref{eq:time-cell-definition}.
In particular, subdivide at
\begin{align}
 s=u,\qquad t=v,\qquad s=t,
 \label{eq:quadrature-jump-locations}
\end{align}
and at $s=t_a,t=t_a$ for every passive time variable $t_a$ occurring
in $X$. Thus neither $G_*(u,s)$ nor $G_*(t,v)$ changes sign inside
any resulting piece. Use the exact area $|Q|$ and a point $y_Q\in Q$
for each nonempty piece. The same construction applies on each face,
using its free coordinates. For any such integrand $f$,
\begin{align}
 \left|\int_Q f(y)\,dy-|Q|f(y_Q)\right|
 &\leq |Q|\sup_{y\in Q}|f(y)-f(y_Q)|
 \leq |Q|h^{1/2}\|f\|_{1/2,\mathrm{pw}},\nonumber\\
 \left|\int f(y)\,dy-\sum_Q|Q|f(y_Q)\right|
 &\leq h^{1/2}\|f\|_{1/2,\mathrm{pw}}\sum_Q|Q|
 \leq h^{1/2}\|f\|_{1/2,\mathrm{pw}},\nonumber\\
 \|X-X_{\mathrm{grid}}\|_\infty
 &\leq h^{1/2}\|X\|_{1/2,\mathrm{pw}}.
 \label{eq:ordered-piece-quadrature}
\end{align}
The exact inverse has the form $I+K+R$. Its kernel factors, and the
finite approximations after replacing them by their exact counterparts,
satisfy the row and column estimates in
\eqref{eq:inverse-kernel-holder}. Their positive quadrature weights give
\begin{align}
 \left\|\sum_Q|Q|
   (\widetilde K+\widetilde R)(\,\cdot\,;y_Q)X(y_Q)\right\|_\infty
 &\leq C_\beta\|X\|_\infty,\nonumber\\
 \| (\mathfrak L_\beta^{-1}X)_{\mathrm{grid}}
                    -\mathfrak L_\beta^{-1}X\|_\infty
 &\leq C_\beta M^{-2}\|X\|_\infty
       +C_\beta h^{1/2}\|X\|_{1/2,\mathrm{pw}}.
\end{align}
Here the first term also bounds the kernel replacement at the quadrature
points; the second includes interpolation of the external coordinates.
The estimate is applied to each prescribed limit before taking any finite
alternating average. Finally, \eqref{eq:hierarchy-holder-stability},
starting from $e_0$, bounds every exact intermediate expression in a
fixed number of hierarchy operations. There are polynomially many pieces
in $h^{-1}$ at fixed dimension.
\end{proof}

\section{Hermite truncation error}\label{app:hermite}
\begin{lemma}[Partition-function concentration]\label{lem:partition-concentration}
For fixed $\beta$, with $Z(J)=\Tr e^{\beta H}$,
\begin{align}
 \EE e^{t(\log Z-\EE\log Z)}&\leq e^{C_\beta t^2/N},
 &0\leq\log\EE Z-\EE\log Z&\leq C_\beta/N,\\
 \EE Z^2/(\EE Z)^2&\leq e^{C_\beta/N},
 &\Pr[Z<\tfrac12\EE Z]&\leq C_\beta e^{-N/C_\beta}.
\end{align}
\end{lemma}
\begin{proof}
For a parity-preserving density matrix $\rho$, the Majorana relations give
\begin{align}
 (\psi_i\otimes\psi_i)^2&=I,\nonumber\\
 (\psi_i\otimes\psi_i)(\psi_j\otimes\psi_j)
 &=(\psi_i\psi_j)\otimes(\psi_i\psi_j)
  =(\psi_j\psi_i)\otimes(\psi_j\psi_i),\nonumber\\
 &\left[\prod_i(I+ix\psi_i\otimes\psi_i)\right]
 \left[\prod_i(I+ix\psi_i\otimes\psi_i)\right]^\dagger
 =(1+x^2)^NI,\nonumber\\
 \sum_{A\ {\rm even}}x^{|A|}|\Tr(\rho\Gamma_A)|^2
 &=\Tr\left[(\rho\otimes\rho)\prod_i(I+ix\psi_i\otimes\psi_i)\right]
 \leq(1+x^2)^{N/2}\quad(x\geq0).
 \label{eq:unconjugated-frame}
\end{align}
Theorem~II.1 of~\cite{anschuetz2025strongly} gives the corresponding
$O(N^2)$ bound for the sum over quartic expectations. The elementary
calculation above also supplies a sufficient explicit constant:
\begin{align}
 \sum_{|A|=4}|\Tr(\rho\Gamma_A)|^2
 &\leq\frac{N^2}{16}(1+4/N)^{N/2}\leq\frac{e^2N^2}{16},\nonumber\\
 \partial_{J_A}Z
 &=\beta\sqrt{\frac3{8N^3}}\int_0^1
   \Tr(e^{(1-s)\beta H}\Gamma_Ae^{s\beta H})\,ds
  =\beta\sqrt{\frac3{8N^3}}\Tr(\Gamma_Ae^{\beta H}),\nonumber\\
 \|\nabla\log Z\|_2^2
 &=\frac{3\beta^2}{8N^3}
       \sum_{|A|=4}|\Tr(\rho_\beta\Gamma_A)|^2
 \leq\frac{3e^2\beta^2}{128N}.
\end{align}
Fact~C.1 and Lemma~C.1 of~\cite{anschuetz2025strongly} give Gaussian
concentration and its moment-generating-function form. Consequently,
for a constant $C_\beta$ and every real $t$,
\begin{align}
 \EE e^{t(\log Z-\EE\log Z)}&\leq e^{C_\beta t^2/N},\nonumber\\
 0\leq\log\EE Z-\EE\log Z
 &=\log\EE e^{\log Z-\EE\log Z}\leq C_\beta/N.
 \label{eq:hermite-logZ}
\end{align}
The second assertion is also the $q=4$ specialization of their
Theorem~I.2, with the Hamiltonian normalization adjusted. The remaining
claims follow directly:
\begin{align}
 \frac{\EE Z^2}{(\EE Z)^2}
 &=\EE e^{2(\log Z-\EE\log Z)}
       e^{-2(\log\EE Z-\EE\log Z)}\leq e^{4C_\beta/N},\nonumber\\
 \Pr[Z<\tfrac12\EE Z]
 &\leq\Pr[\log Z-\EE\log Z<-(\log2-C_\beta/N)]\nonumber\\
 &\leq\inf_{t>0}
       \exp\left[-t(\log2-C_\beta/N)+C_\beta t^2/N\right]\nonumber\\
 &=\exp\left[-\frac{N(\log2-C_\beta/N)^2}{4C_\beta}\right]
 \leq C_\beta e^{-N/C_\beta}.
 \label{eq:Z-second-moment-and-lower-tail}
\end{align}
For the equality take $N$ large enough that $\log2>C_\beta/N$;
increase the constants for smaller $N$. At $\beta=0$ all claims are immediate.
\end{proof}

\begin{proof}[Proof of \Cref{lem:hermite-tail}]
Let $g_{O,d}$ be the degree-$d$ Hermite component of
$g_O(J)=\Tr(Oe^{\beta H})/\EE Z$, where $Z=\Tr e^{\beta H}$.
For a multi-index $\alpha=(\alpha_A)_{A\in\binom{[N]}4}$, write
$\alpha!=\prod_A\alpha_A!$ and
$\He_\alpha(J)=\prod_A\He_{\alpha_A}(J_A)$. Gaussian integration by
parts and differentiation of the exponential give
\begin{align}
 \EE[g_O\He_\alpha]&=\EE[\partial^\alpha g_O],\nonumber\\
 \EE[\partial_{A_1}\cdots\partial_{A_d}g_O]
 &=\left(\beta\sqrt{\frac3{8N^3}}\right)^d F_{O;A_1,\ldots,A_d},\nonumber\\
 \sum_{d=0}^Vg_{O,d}(J)
 &=\sum_{|\alpha|\leq V}\frac{\EE[\partial^\alpha g_O]}{\alpha!}
                                  \He_\alpha(J)\nonumber\\
 &=\sum_{d=0}^V\frac1{d!}
       \left(\beta\sqrt{\frac3{8N^3}}\right)^d
       \sum_{A_1,\ldots,A_d}
           \langle J_{A_1}\cdots J_{A_d}\rangle_J
                     F_{O;A_1,\ldots,A_d}
 =\mathcal N_O(J).
 \label{eq:hermite-coefficient-identification}
\end{align}
In the last equality a multiset with multiplicities $\alpha$ occurs
$d!/\alpha!$ times. Gaussian integrability of the finite-dimensional
matrix exponential justifies the differentiations.

\proofstep{A bound at every Hermite degree}
For $u>0$, use the conjugate tensor product and set
\begin{align}
 W=\sum_{|A|=4}\Gamma_A\otimes\overline{\Gamma_A}, \quad S=\sum_i\psi_i\otimes\overline{\psi_i},\quad F_O(u)=\Tr\left[(O\otimes\overline O)
 e^{\beta H\otimes I+I\otimes\overline{\beta H}+uW}\right].
\end{align}
For $\Delta_r=\{\mathbf s\geq0:\sum_{j=1}^rs_j\leq1\}$ and
$s_0=1-\sum_{j=1}^rs_j$, Duhamel expansion gives
\begin{align}
 [u^r]F_O(u)
 &=\sum_{A_1,\ldots,A_r}\int_{\Delta_r}
 \left|\Tr\left(Oe^{s_0\beta H}\Gamma_{A_1}e^{s_1\beta H}
          \cdots\Gamma_{A_r}e^{s_r\beta H}\right)\right|^2d\mathbf s\geq0,
 \label{eq:doubled-duhamel-positive}\\
 \partial_{A_1}\cdots\partial_{A_r}g_O
 &=\frac{(\beta\sqrt{3/(8N^3)})^r}{\EE Z}
 \sum_{\pi\in S_r}\int_{\Delta_r}
 \Tr\left(Oe^{s_0\beta H}\Gamma_{A_{\pi(1)}}e^{s_1\beta H}
            \cdots\Gamma_{A_{\pi(r)}}e^{s_r\beta H}\right)d\mathbf s,
 \label{eq:single-copy-derivative}\\
 \|D^rg_O\|_{\rm HS}^2
 &\leq\frac{r!}{(\EE Z)^2}\left(\frac{3\beta^2}{8N^3}\right)^r
 [u^r]F_O(u)
 \leq\frac{r!}{(\EE Z)^2}\left(\frac{3\beta^2}{8N^3u}\right)^rF_O(u).
 \label{eq:derivative-to-doubled-trace}
\end{align}
Here $r!\operatorname{vol}(\Delta_r)=1$, so Cauchy--Schwarz on the
permuted integrals proves the last line. With $\rho=e^{\beta H}/Z$, the contraction bound, Golden--Thompson,
and \eqref{eq:unconjugated-frame} give
\begin{align}
 0\leq F_O(u)
 &\leq\|O\otimes\overline O\|\,
       \Tr e^{\beta H\otimes I+I\otimes\overline{\beta H}+uW}\nonumber\\
 &\leq\Tr\left(e^{\beta H\otimes I+I\otimes\overline{\beta H}}e^{uW}\right)
 =Z^2\Tr[(\rho\otimes\overline\rho)e^{uW}],
 \label{eq:doubled-trace-domination}\\
 e^{tS}&=\prod_{i=1}^N
       (\cosh t\,I+\sinh t\,\psi_i\otimes\overline{\psi_i}),\nonumber\\
 \Tr[(\rho\otimes\overline\rho)e^{tS}]
 &=(\cosh t)^N\sum_{A\ {\rm even}}(\tanh t)^{|A|}
                                      |\Tr(\rho\Gamma_A)|^2\nonumber\\
 &\leq(\cosh t)^N(1+\tanh^2t)^{N/2}
  =(\cosh2t)^{N/2}\leq e^{Nt^2}.
 \label{eq:S-subgaussian}
\end{align}

The frame estimate uses $\rho\otimes\rho$; the displayed absolute
squares use $\rho\otimes\overline\rho$. Since the summands of $S$ are
commuting involutions, expansion of $S^4$ and $|S|\leq N$ imply
\begin{align}
 W&=\frac{S^4-(6N-8)S^2+3N^2-6N}{24},\\
 W/N^3&\leq S^2/(24N)+1/(8N),
 \label{eq:W-polynomial-S}\\
 \Tr[(\rho\otimes\overline\rho)e^{W/N^3}]
 &\leq e^{1/(8N)}\EE_{g\sim\mathcal N(0,1)}
       \Tr[(\rho\otimes\overline\rho)e^{gS/\sqrt{12N}}] \leq e^{1/(8N)}\EE_g e^{g^2/12}\leq C.
 \label{eq:W-exponential-bound}
\end{align}
Set $u=N^{-3}$ in \eqref{eq:derivative-to-doubled-trace} and use
\Cref{lem:partition-concentration}. Hermite orthogonality then gives
\begin{align}
 \EE\|D^rg_O\|_{\rm HS}^2&\leq C_\beta r!C_\beta^r,
 \label{eq:all-hermite-derivatives}\\
 \sum_{d\geq0}(1+t)^d\|g_{O,d}\|_2^2
 &=\sum_{r\geq0}\frac{t^r}{r!}\EE\|D^rg_O\|_{\rm HS}^2
 \leq C_\beta\qquad(0<t=t_\beta<1/C_\beta).
 \label{eq:hermite-exponential-tail}
\end{align}
The derivative identity follows first for finite Hermite sums, then by
closure of the Gaussian Sobolev derivative; nonnegative summands justify
the interchange.

\proofstep{The additional powers of $N$}
Changing the sign of $\psi_i$ preserves the trace and the Gaussian law,
so a nonzero coefficient must satisfy
\begin{align}
 \sum_{q=1}^d\mathbf1_{i\in A_q}
 &\equiv\mathbf1_{i\in\operatorname{supp}(O)}\pmod2
 \quad(i\in[N]).
 \label{eq:quartet-parity-condition}
\end{align}
The admissible ordered tuples are precisely
\begin{align}
 \mathcal A_{d,O}:=\left\{(A_1,\ldots,A_d)\in\binom{[N]}4^d:
 A_1\mathbin\triangle\cdots\mathbin\triangle A_d=\supp(O)\right\}.
\end{align}
This set is empty if $4d<w$. Otherwise write each quartet increasingly,
regarding the tuple as $4d$ positions. Select one position for each of the
$w$ prescribed indices, and pair the remaining positions carrying each
site. Every admissible tuple has at least one such description, so
\begin{align}
 |\mathcal A_{d,O}|
 &\leq\underbrace{(4d)^w}_{\text{selected positions}}
       \underbrace{(4d-w-1)!!}_{\text{pairings of the other positions}}
       \underbrace{N^{(4d-w)/2}}_{\text{one label per pair}}\nonumber\\
 &\leq C_w(Cd)^{2d+w}N^{2d-w/2},
 \label{eq:ordered-quartet-count}\\
 \left|\Tr\left(Oe^{s_0\beta H}\Gamma_{A_1}e^{s_1\beta H}
           \cdots\Gamma_{A_d}e^{s_d\beta H}\right)\right|
 &\leq\|O\|\prod_{r=1}^d\|\Gamma_{A_r}\|
       \prod_{r=0}^d\|e^{s_r\beta H}\|_{1/s_r}
 =Z\qquad(\mathbf s\in\Delta_d),\nonumber\\
 |\EE\partial_{A_1}\cdots\partial_{A_d}g_O|
 &\leq\left(\beta\sqrt{\frac3{8N^3}}\right)^d
       \sum_{\pi\in S_d}\int_{\Delta_d}\frac{\EE Z}{\EE Z}\,d\mathbf s
 =\left(\beta\sqrt{\frac3{8N^3}}\right)^d,
 \label{eq:single-hermite-coefficient-bound}\\
 \|g_{O,d}\|_2^2
 &=\frac1{d!}\sum_{(A_1,\ldots,A_d)\in\mathcal A_{d,O}}
           |\EE\partial_{A_1}\cdots\partial_{A_d}g_O|^2\nonumber\\
 &\leq\frac{|\mathcal A_{d,O}|}{d!}
             \left(\frac{3\beta^2}{8N^3}\right)^d
 \leq C_wd^wN^{-w/2}(C\beta^2d/N)^d.
 \label{eq:low-degree-hermite-bound}
\end{align}
Here a zero heat length uses the operator norm, and $(-1)!!=1$.
No condition $A_r\ne A_s$ is imposed in $\mathcal A_{d,O}$; labels assigned
to different pairs may also coincide. Thus the enumeration includes
every tuple satisfying \eqref{eq:quartet-parity-condition}.

Choose $C\beta^2\eta_\beta\leq1/2$.
Then
\begin{align}
 \sum_{V<d\leq\eta_\beta N}\|g_{O,d}\|_2^2
 &\leq C_{\beta,V,w}N^{-V-1-w/2}
       \sum_{d>V}d^{w+V+1}2^{-(d-V-1)}
 \leq C_{\beta,V,w}N^{-V-1-w/2},\\
 \sum_{d>\eta_\beta N}\|g_{O,d}\|_2^2
 &\leq C_\beta(1+t_\beta)^{-\eta_\beta N}
 \leq C_\beta e^{-N/C_\beta}.
\end{align}
Their sum proves \eqref{eq:analysis-hermite-tail}, including $O=I$.
At $\beta=0$ the Hermite tail vanishes.
\end{proof}

\section{Recursion truncation error}\label{app:recurse}

\begin{proof}[Proof of \Cref{lem:hierarchy-error}]
The omitted moments have orders $K,K+1,K+2$. By
\Cref{lem:collective-concentration},
\begin{align}
 (I-A_0-N^{-1}A_1)\boldsymbol\mu&=e_0+R_K,
 &\|R_K\|_\infty&\leq C_{\beta,K}N^{-\theta K}.
 \label{eq:finite-hierarchy-remainder}
\end{align}
Since $A_0^K=0$ and $A_0e_0=0$, multiplication by
$\sum_{r<K}A_0^r$ gives
\begin{align}
 \boldsymbol\mu
 &=e_0+\frac1N\left(\sum_{r<K}A_0^r\right)A_1\boldsymbol\mu
       +O_{\beta,K}(N^{-\theta K}),
 \label{eq:hierarchy-fixed-point}\\
 \left\|\left(\sum_{r<K}A_0^r\right)A_1\right\|&\leq C_{\beta,K},\\
 \left\|\boldsymbol\mu-
   \sum_{q=0}^j\left[\frac1N\left(\sum_{r<K}A_0^r\right)A_1\right]^qe_0
          \right\|_\infty
 &\leq C_{\beta,K}N^{-\theta K}
       +2(C_{\beta,K}/N)^{j+1}\qquad(N\geq2C_{\beta,K}).
\end{align}
For computation, \Cref{lem:ladder-certificate} bounds the piecewise
$1/2$-H\"older norm of each exact finite iterate and each intermediate
expression by $C_{\beta,K,j}$, independently of $N$. The numerical
blocks are stable in the supremum norm and consistent on these inputs:
\begin{align}
 \|\widetilde A_kX\|_\infty&\leq C_{\beta,K}\|X\|_\infty\quad(k=0,1),\\
 \|(\widetilde A_0-A_0)X\|_\infty
   +\|(\widetilde A_1-A_1)X\|_\infty
 &\leq C_{\beta,K}M^{-2}\|X\|_{1/2,\mathrm{pw}}.
 \label{eq:A0A1-numerical-error}
\end{align}
Each input and output is evaluated separately on the domains
\eqref{eq:time-cell-definition}--\eqref{eq:time-face-definition}.
For any fixed product of blocks $T_1,\ldots,T_l\in\{A_0,A_1\}$,
\begin{align}
 \widetilde T_l\cdots\widetilde T_1e_0-T_l\cdots T_1e_0
 &=\sum_{r=1}^l\widetilde T_l\cdots\widetilde T_{r+1}
             (\widetilde T_r-T_r)T_{r-1}\cdots T_1e_0,\nonumber\\
 \|\widetilde T_l\cdots\widetilde T_1e_0-T_l\cdots T_1e_0\|_\infty
 &\leq C_{\beta,K}M^{-2}\sum_{r=1}^l
       C_{\beta,K}^{l-r}\|T_{r-1}\cdots T_1e_0\|_{1/2,\mathrm{pw}}\nonumber\\
 &\leq C_{\beta,K,l}M^{-2},\nonumber\\
 \left\|\widetilde{\boldsymbol\mu}^{(j)}-
   \sum_{q=0}^j\left[\frac1N\left(\sum_{r<K}A_0^r\right)A_1\right]^qe_0
          \right\|_\infty
 &\leq C_{\beta,K,j}M^{-2}.
\end{align}
The last line expands the two finite operator polynomials and applies
the preceding bound to each product; all its lengths are at most $Kj$.
Only the exact vectors $T_{r-1}\cdots T_1e_0$ require a H\"older bound.

The numerical blocks preserve the zero pattern, hence
$\widetilde A_0^K=0$. Choose arithmetic precision so its accumulated
error also satisfies the last bound. Combining the two errors proves
\eqref{eq:analysis-moment-error}. At fixed $K,j$ the number of variables
is fixed, giving cost $M^{C_{K,j}}$.
\end{proof}

\section{SYK saddle point equations}

For a finite positive Borel measure $\mu$ on $[0,\infty)$, write
\begin{align}
 G_\mu(t)=\int_0^\infty
       \frac{\cosh(x(1/2-t))}{2\cosh(x/2)}\,\mu(dx),\qquad 0\leq t\leq1.
 \label{eq:positive-spectral-kernel}
\end{align}
A normalized positive spectral kernel is one for which $\mu$ is a
probability measure.  The extension outside this interval is
antisymmetric and antiperiodic, as in the preliminaries. Below, we use the fact that integration gives
\begin{align}
 \widehat G_\mu(n)&=i\omega_n\int_0^\infty
                 \frac{\mu(dx)}{\omega_n^2+x^2}.
 \label{eq:positive-kernel-fourier}
\end{align}
\begin{fact}[Schur complement]\label{fact:schur}
For a block matrix $\begin{pmatrix}A&B\\C&D\end{pmatrix}$ with $D$
and $A-BD^{-1}C$ invertible,
\begin{align}
 \begin{pmatrix}A&B\\C&D\end{pmatrix}
 &=\begin{pmatrix}I&BD^{-1}\\0&I\end{pmatrix}
   \begin{pmatrix}A-BD^{-1}C&0\\C&D\end{pmatrix},\nonumber\\
 \left[\begin{pmatrix}A&B\\C&D\end{pmatrix}^{-1}\right]_{11}
 &=(A-BD^{-1}C)^{-1},\nonumber\\
 \det\begin{pmatrix}A&B\\C&D\end{pmatrix}
 &=\det D\,\det(A-BD^{-1}C).
 \label{eq:schur-fact}
\end{align}
\end{fact}

\begin{lemma}[The positive scalar saddle]\label{lem:positive-saddle}
For every $\beta\geq0$, the equations \eqref{eq:sd} have a unique
normalized positive spectral solution $G_*=G_{\mu_{*,\beta}}$.
For $\beta>0$,
\begin{align}
 \int x^2\mu_{*,\beta}(dx)&=\beta^2/4,
 &\tfrac12e^{-\beta/4}\leq G_*(t)&\leq\tfrac12,
 \label{eq:saddle-pointwise-bounds}\\
 \int e^{x/(2\beta)}\mu_{*,\beta}(dx)&\leq1+\sqrt2.
 \label{eq:saddle-exponential-moment}
\end{align}
At $\beta=0$, $\mu_{*,0}=\delta_0$.
The formula on $[0,1]$ has a bounded holomorphic extension to a
neighborhood of that interval, with a common neighborhood and bound
on each compact temperature interval.  Moreover,
\begin{align}
 \|G_{*,\beta}-G_{*,\beta'}\|_\infty
 \leq\frac{|\beta^2-\beta'^2|}{16\pi}.
 \label{eq:scalar-temperature-continuity}
\end{align}
For every finite positive measure $\sigma$, the Fourier coefficients
\begin{align}
 \widehat{\mathcal D(G_\sigma)}(n)
 &=\bigl(-i\omega_n-\widehat G_\sigma(n)\bigr)^{-1}
 =\frac{i}{\omega_n\left(1+\displaystyle\int_0^\infty
                         \frac{\sigma(dx)}{\omega_n^2+x^2}\right)}
 \label{eq:positive-dyson-map-definition}
\end{align}
define a normalized positive spectral kernel.  For every normalized
positive spectral $G$, the pointwise cube $\beta^2G^3$ belongs to
this source class, and
\begin{align}
 \tfrac14e^{-\beta/2}\|G-G_*\|_{L^2(0,1)}^2
 \leq-\int_0^1(\mathcal D(\beta^2G^3)-G)(G^3-G_*^3)\,dt.
 \label{eq:scalar-coercivity}
\end{align}
No moment assumption on the representing measure of $G$ is needed
for this last assertion.
\end{lemma}
\begin{proof}
\proofstep{Positive measures and the Dyson map}
For $\sigma=\sum_{j=1}^qv_j^2\delta_{x_j}$, form the real skew matrix
with upper entries $K_{0,2j-1}=v_j$, $K_{2j-1,2j}=x_j$, all others zero.
Its orthogonal skew blocks have frequencies $y\geq0$. Their squared
projections of $e_0$ define a probability $\mu$, including the kernel
at $y=0$. By \Cref{fact:schur},
\begin{align}
 \begin{pmatrix}\omega&-x_j\\x_j&\omega\end{pmatrix}^{-1}
 &=\frac1{\omega^2+x_j^2}
       \begin{pmatrix}\omega&x_j\\-x_j&\omega\end{pmatrix},\nonumber\\
 \omega-\sum_j(-v_j,0)
    \begin{pmatrix}\omega&-x_j\\x_j&\omega\end{pmatrix}^{-1}
                    \binom{v_j}{0}
 &=\omega\left(1+\sum_j\frac{v_j^2}{\omega^2+x_j^2}\right).
\end{align}
Consequently spectral decomposition gives
\begin{align}
 [(\omega I-K)^{-1}]_{00}
 &=\frac1{\omega(1+\sum_jv_j^2/(\omega^2+x_j^2))}
   =\omega\int\frac{\mu(dy)}{\omega^2+y^2},\\
 \int y^2\mu(dy)&=(-K^2)_{00}=\sum_jv_j^2,\\
 \int\frac{\mu(dy)}{z+y^2}
 &=\frac1{z(1+\int\sigma(dx)/(z+x^2))}\quad(z>0).
 \label{eq:scalar-dyson-resolvent}
\end{align}
Approximate a general $\sigma$ weakly by finitely supported measures of
the same mass. Their outputs are tight by the second-moment bound; every
limit satisfies \eqref{eq:scalar-dyson-resolvent}. To see that these integrals determine the output, suppose probabilities
$\mu,\mu'$ have the same integral of $(z+y^2)^{-1}$ for every $z>0$.
Differentiation at $z=1$, justified by a bounded integrand for
$z\in[1/2,3/2]$, gives
\begin{align}
 \int(1+y^2)^{-k}\mu(dy)
 &=\frac{(-1)^{k-1}}{(k-1)!}
   \left.\frac{d^{k-1}}{dz^{k-1}}
       \int\frac{\mu(dy)}{z+y^2}\right|_{z=1}
 =\int(1+y^2)^{-k}\mu'(dy)\qquad(k\geq1).
\end{align}
Push both measures forward by $y\mapsto(1+y^2)^{-1}\in(0,1]$, and call
them $\widetilde\mu,\widetilde\mu'$. Their integrals agree on all
polynomials, including the constant polynomial. For $f\in C[0,1]$,
its Bernstein polynomials satisfy
\begin{align}
 p_n(x)&=\sum_{k=0}^nf(k/n)\binom nkx^k(1-x)^{n-k},\nonumber\\
 \|p_n-f\|_\infty
 &\leq\sup_{|x-y|\leq\eta}|f(x)-f(y)|
       +\frac{\|f\|_\infty}{2n\eta^2}\longrightarrow0
       \quad(\eta=n^{-1/4}),\nonumber\\
 \left|\int f\,d\widetilde\mu-\int f\,d\widetilde\mu'\right|
 &\leq2\|f-p_n\|_\infty= o(1).
\end{align}
The polynomial estimate is Chebyshev's inequality for
$\operatorname{Bin}(n,x)/n$, whose variance is at most $1/(4n)$.
Equality of integrals of continuous functions on $[0,1]$ implies
$\widetilde\mu=\widetilde\mu'$, and the inverse map
$x\mapsto\sqrt{x^{-1}-1}$ then gives $\mu=\mu'$.
Monotone convergence also gives
\begin{align}
 \int y^2\mu(dy)
 &=\lim_{z\to\infty}z\left(1-z\int\frac{\mu(dy)}{z+y^2}\right)
 =\lim_{z\to\infty}\frac{z\int\sigma(dx)/(z+x^2)}{1+\int\sigma(dx)/(z+x^2)}
 =\sigma([0,\infty)).
 \label{eq:dyson-second-moment}
\end{align}
This proves existence and weak continuity of the output. Substitution
$z=\omega_n^2$ proves \eqref{eq:positive-dyson-map-definition}.

\proofstep{A fixed point}
For $\beta>0$, minimize over probabilities with second moment at most
$\beta^2/4$ the functional
\begin{align}
 u_n(\mu)&=\omega_n^2\int\frac{\mu(dx)}{\omega_n^2+x^2},
 \qquad \frac1{1+\beta^2/(4\omega_n^2)}\leq u_n(\mu)\leq1,\\
 G_\mu(t)&=\frac12+2\sum_{n\geq0}\frac{u_n(\mu)-1}{\omega_n}\sin(\omega_nt),\\
 \mathcal F_\beta(\mu)
 &=\sum_{n\geq0}(u_n(\mu)-\log u_n(\mu)-1)
       +\frac{\beta^2}{8}\int_0^1G_\mu^4.
 \label{eq:infinite-convex-functional}
\end{align}
The feasible set is weakly compact. Since $1-u_n\leq\beta^2/(4\omega_n^2)$,
the series converge uniformly; dominated convergence makes the
functional continuous. Let $\mu$ minimize it, and let $\nu$ represent
$\mathcal D(\beta^2G_\mu^3)$. It is feasible by
\eqref{eq:dyson-second-moment}. For $v_n=u_n(\nu)$,
\begin{align}
 v_n^{-1}&=1+\frac{\beta^2}{\omega_n}\int_0^1G_\mu^3\sin(\omega_nt)\,dt,\\
 0\leq\left.\frac d{ds}\mathcal F_\beta((1-s)\mu+s\nu)\right|_{s=0}
 &=\sum_{n\geq0}(v_n^{-1}-u_n^{-1})(v_n-u_n)
 =-\sum_{n\geq0}\frac{(v_n-u_n)^2}{u_nv_n}\leq0.
 \label{eq:dyson-descent}
\end{align}
The $O(\omega_n^{-2})$ coordinate bounds justify differentiation. Hence
$u_n=v_n$ and $G_\mu=\mathcal D(\beta^2G_\mu^3)$.
The kernel determines its measure: at $t=1/2+is$ it is the cosine
transform of $\mu(dx)/(2\cosh(x/2))$. The identity theorem and Fourier
uniqueness identify this finite measure; Fourier uniqueness follows by
Gaussian convolution and Fourier inversion. At $\beta=0$ the fixed
point is $G_*=1/2$.

\proofstep{Uniqueness and coercivity}
For positive sources with Fourier coefficients $ib_n,ib'_n$,
$b_n,b'_n\geq0$ for $n\geq0$, \Cref{fact:parseval} gives
\begin{align}
 \int(\mathcal D(\Sigma)-\mathcal D(\Sigma'))(\Sigma-\Sigma')
 &=-2\sum_{n\geq0}\frac{(b_n-b'_n)^2}{(\omega_n+b_n)(\omega_n+b'_n)}\leq0,\\
 \|\mathcal D(\Sigma)-\mathcal D(\Sigma')\|_2
 &\leq\pi^{-2}\|\Sigma-\Sigma'\|_2.
 \label{eq:dyson-antimonotone}
\end{align}
For two fixed points this forces
$\int(G-G')^2(G^2+GG'+(G')^2)=0$, proving uniqueness. Jensen and
\eqref{eq:dyson-second-moment} give
\begin{align}
 G_*(t)&\geq\tfrac12\int e^{-x/2}\mu_{*,\beta}(dx)
 \geq\tfrac12e^{-\frac12(\int x^2\mu_{*,\beta}(dx))^{1/2}}
 =\tfrac12e^{-\beta/4},\\
 -\int(\mathcal D(\beta^2G^3)-G)(G^3-G_*^3)
 &\geq\int(G-G_*)^2(G^2+GG_*+G_*^2)
 \geq\tfrac14e^{-\beta/2}\|G-G_*\|_2^2.
\end{align}
At $\beta=0$ the first inequality in the last line is equality.

\proofstep{Exponential moments}
Write $m_j=\int x^j\mu_{*,\beta}(dx)$ and
$s_j=\int x^j\sigma(dx)$ for the measure of $\beta^2G_*^3$.
Equations \eqref{eq:cubic-positive-source} and
\eqref{eq:scalar-dyson-resolvent} imply
\begin{align}
 s_j&\leq\frac{\beta^2}{4}\EE(X_1+X_2+X_3)^j,
 \qquad X_i\stackrel{\rm iid}{\sim}\mu_{*,\beta},\\
 m_{2k}&=\sum_{j=0}^{k-1}s_{2j}m_{2(k-1-j)},\qquad m_0=1.
 \label{eq:scalar-moment-recursion}
\end{align}
If moments
through $2k-2$ are finite, the source has those moments, and
\begin{align}
 \int\frac{\mu_{*,\beta}(dx)}{1+tx^2}
 -\sum_{j<k}(-t)^jm_{2j}
 &=(-t)^k\int\frac{x^{2k}}{1+tx^2}\mu_{*,\beta}(dx).
\end{align}
Expanding the reciprocal in \eqref{eq:scalar-dyson-resolvent} through
$t^k$ gives a finite limit of the last integral as $t\downarrow0$.
Monotone convergence identifies it with $m_{2k}$ and proves the recursion.
Assuming $m_{2j}\leq\beta^{2j}(2j)!$ for $j<k$, Cauchy--Schwarz and
the multinomial formula yield
\begin{align}
 m_l&\leq\sqrt{m_{l-1}m_{l+1}}\leq\sqrt2\beta^l l!
                   \quad (l<2k-1\text{ odd}),\\
 s_{2j}&\leq\tfrac{\beta^2}{4}(\sqrt2)^3\beta^{2j}(2j)!
                  \binom{2j+2}{2}
         =\tfrac{\sqrt2}{4}\beta^{2j+2}(2j+2)!,\\
 \frac{m_{2k}}{\beta^{2k}(2k)!}
 &\leq\frac{\sqrt2}{4}\sum_{l=1}^k\binom{2k}{2l}^{-1}
 \leq\frac{\sqrt2}{2}<1,\\
 \int e^{x/(2\beta)}\mu_{*,\beta}(dx)
 &\leq1+\sqrt2\sum_{j\geq1}2^{-j}=1+\sqrt2.
\end{align}
The binomial sum is at most $1+(k-1)/\binom{2k}{2}\leq2$.
For $\beta\leq\beta_+$ the same exponential bound holds with
$\beta_+$ in the denominator; dominated differentiation proves the
uniform one-sided holomorphic extension.

\proofstep{Temperature continuity}
Let $u_n,v_n$ be the Fourier ratios at $\beta,\beta'$. Subtracting the
saddle equations gives, by \Cref{fact:bessel} and $\|G_{*,\beta'}^3\|_2\leq1/8$,
\begin{align}
 \|u-v\|_2^2
 &\leq\sum_{n\geq0}\frac{(u_n-v_n)^2}{u_nv_n}
 +\frac{\beta^2}{2}\int(G_{*,\beta}-G_{*,\beta'})
                         (G_{*,\beta}^3-G_{*,\beta'}^3)\nonumber\\
 &=-(\beta^2-\beta'^2)\sum_{n\geq0}\frac{u_n-v_n}{\omega_n}
                         \int G_{*,\beta'}^3\sin(\omega_nt)\,dt
 \leq\frac{|\beta^2-\beta'^2|}{8\sqrt2\pi}\|u-v\|_2,\\
 \|G_{*,\beta}-G_{*,\beta'}\|_\infty
 &\leq2\Bigl(\sum_{n\geq0}\omega_n^{-2}\Bigr)^{1/2}\|u-v\|_2
 \leq\frac{|\beta^2-\beta'^2|}{16\pi}.
\end{align}
The second-moment identity \eqref{eq:saddle-pointwise-bounds} gives
\begin{align}
 |u_n-v_n|
 &\leq(1-u_n)+(1-v_n)
 \leq\frac{\beta^2+\beta'^2}{4\omega_n^2},
\end{align}
which justifies the sums and limits above.
\end{proof}

The second-moment identity \eqref{eq:saddle-pointwise-bounds} also gives the quantitative Fourier bounds
\begin{align}
 0\leq1-\frac{\widehat G_*(n)}{i/\omega_n}
 &=\int\frac{x^2}{\omega_n^2+x^2}\,\mu_{*,\beta}(dx)
 \leq\frac{\beta^2}{4\omega_n^2},\\
 \left|\widehat G_*(n)-\frac i{\omega_n}\right|
 &\leq\frac{\beta^2}{4|\omega_n|^3}.
 \label{eq:saddle-fourier-tail}
\end{align}

\begin{lemma}[Quadratic partition and two-point identities]
\label{lem:quadratic-cut}
Let $\Sigma=G_\sigma$, where $\sigma$ is a positive Borel measure on
$[0,\infty)$ with mass $m<\infty$, and set
\begin{align}
 W(\Sigma)&=\prod_{n\geq0}
      \left(1+\int_0^\infty\frac{\sigma(dx)}{\omega_n^2+x^2}\right).
\end{align}
For $0<u<1$ and $\pi\in\mathcal P_2([2d])$, write
\begin{align}
 \kappa_u(\pi,\mathbf t)
 &=\#\{\{a,b\}\in\pi:a<b,\ t_b<u<t_a\}.
\end{align}
With the $d=0$ integral and product interpreted as one, the identities are
\begin{align}
 W(\Sigma)
 &=\sum_{d\geq0}2^{-d}\sum_{\pi\in\mathcal P_2([2d])}
   (-1)^{\operatorname{cr}(\pi)}
   \int_{1>t_1>\cdots>t_{2d}>0}
       \prod_{\substack{\{a,b\}\in\pi\\a<b}}
               \Sigma(t_a-t_b)\,d\mathbf t,
 \label{eq:quadratic-vacuum-series}\\
 2W(\Sigma)\mathcal D(\Sigma)(u)
 &=\sum_{d\geq0}2^{-d}\sum_{\pi\in\mathcal P_2([2d])}
   \int_{1>t_1>\cdots>t_{2d}>0}
       (-1)^{\operatorname{cr}(\pi)+\kappa_u(\pi,\mathbf t)}
       \prod_{\substack{\{a,b\}\in\pi\\a<b}}
               \Sigma(t_a-t_b)\,d\mathbf t.
 \label{eq:quadratic-two-point-series}
\end{align}
Both series converge absolutely. More precisely,
\begin{align}
 1\leq W(\Sigma)&\leq e^{m/8},
 &2^{-d}\sum_{\pi\in\mathcal P_2([2d])}
   \int_{1>t_1>\cdots>t_{2d}>0}
       \prod_{\substack{\{a,b\}\in\pi\\a<b}}
               |\Sigma(t_a-t_b)|\,d\mathbf t
 &\leq\frac{(m/8)^d}{d!}.
 \label{eq:quadratic-order-majorant}
\end{align}
The two-point identity also holds at $u=0,1$ with the one-sided values
$\mathcal D(\Sigma)(0^+)=\mathcal D(\Sigma)(1^-)=1/2$.
\end{lemma}
\begin{proof}
First let $\sigma=\sum_jv_j^2\delta_{x_j}$ and use the quadratic
Hamiltonian and skew matrix $K$ from the proof of
\Cref{lem:quadratic-dyson-representation}. All traces act on the same
space, including the cavity Majorana and its uncoupled companion.
Let $K_0$ omit the $v_j$ entries and let $y_l$ be the skew-block
frequencies of $K$. By the two-dimensional trace calculation in
\eqref{eq:quadratic-block-traces}, the identity
$\cosh(x/2)=\prod_{n\geq0}(1+x^2/\omega_n^2)$, and the Schur
complement \eqref{eq:schur-fact},
\begin{align}
 \frac{\Tr e^{-H_{\rm quad}}}{\Tr e^{-H_{\rm bath}}}
 &=\frac{\prod_l\cosh(y_l/2)}{\prod_j\cosh(x_j/2)}
 =\prod_{n\geq0}\frac{\det(\omega_nI-K)}{\det(\omega_nI-K_0)}
 \nonumber\\
 &=\prod_{n\geq0}\left(1+\sum_j
                       \frac{v_j^2}{\omega_n^2+x_j^2}\right)
 =W(\Sigma).
 \label{eq:quadratic-determinant-ratio}
\end{align}
Combining this with \eqref{eq:xsd} gives
\begin{align}
 \frac{\Tr(e^{-(1-u)H_{\rm quad}}\psi_{N+1}
                    e^{-uH_{\rm quad}}\psi_{N+1})}
      {\Tr e^{-H_{\rm bath}}}
 &=2W(\Sigma)\mathcal D(\Sigma)(u).
 \label{eq:quadratic-normalized-two-point}
\end{align}

In a diagonal
quadratic Gibbs state $\rho$, a creation or annihilation operator $a$
satisfies $\rho a=r a\rho$, $r>0$. Its anticommutators with linear
Majorana operators $A_j$ are scalars, so
\begin{align}
 aA_2\cdots A_{2d}
 &=\sum_{j=2}^{2d}(-1)^j\{a,A_j\}
                 A_2\cdots\widehat A_j\cdots A_{2d}-A_2\cdots A_{2d}a,\\
 \Tr(\rho Xa)&=r^{-1}\Tr(\rho aX),
 \\
 (1+r^{-1})\Tr(\rho aA_j)&=\{a,A_j\},\\
 \Tr(\rho aA_2\cdots A_{2d})
 &=\sum_{j=2}^{2d}(-1)^j\Tr(\rho aA_j)
                  \Tr(\rho A_2\cdots\widehat A_j\cdots A_{2d}).
 \label{eq:quadratic-wick-recursion}
\end{align}
Linearity in the first factor and induction on $d$ give
\begin{align}
 \Tr(\rho A_1\cdots A_{2d})
 &=\sum_{\pi\in\mathcal P_2([2d])}(-1)^{\operatorname{cr}(\pi)}
         \prod_{\substack{\{a,b\}\in\pi\\a<b}}\Tr(\rho A_aA_b).
 \label{eq:quadratic-full-wick}
\end{align}
For the Majorana vector $\gamma=(\psi_{N+1},\eta_1,\zeta_1,\ldots)$ and the
same real skew matrix $K$,
\begin{align}
 [H_{\rm quad},\gamma_a]&=-i\sum_bK_{ab}\gamma_b,
 &e^{tH_{\rm quad}}\gamma_ae^{-tH_{\rm quad}}
 &=\sum_b(e^{-itK})_{ab}\gamma_b.
 \label{eq:quadratic-linear-evolution}
\end{align}
Thus \eqref{eq:quadratic-full-wick} also applies when each $A_a$ is a
Majorana evolved for its own time.

Set $\Xi=2^{-1/2}\sum_jv_j\eta_j$ and
$\Xi(t)=e^{tH_{\rm bath}}\Xi e^{-tH_{\rm bath}}$. For $0\leq s\leq t\leq1$,
\begin{align}
 H_{\rm quad}-H_{\rm bath}&=i\psi_{N+1}\Xi/\sqrt2,\\
 \frac{\Tr(e^{-H_{\rm bath}}\Xi(t)\Xi(s))}{\Tr e^{-H_{\rm bath}}}
 &=\Sigma(t-s),\\
 (-i/\sqrt2)^{2d}\prod_{j=1}^{2d}\psi_{N+1}\Xi(t_j)
 &=2^{-d}(-1)^{d+d(2d-1)}\psi_{N+1}^{2d}\prod_j\Xi(t_j)
 =2^{-d}\prod_j\Xi(t_j),\\
 \#\{j:t_j>u\}&=2\#\{\{a,b\}\in\pi:t_a,t_b>u\}
                      +\kappa_u(\pi,\mathbf t).
 \label{eq:quadratic-crossing-parity}
\end{align}
With $\Xi(t_j)$ in decreasing time order, Duhamel expansion and
\eqref{eq:quadratic-full-wick} give the two series directly:
\begin{align}
 \frac{\Tr e^{-H_{\rm quad}}}{\Tr e^{-H_{\rm bath}}}
 &=\sum_{d\geq0}2^{-d}\int_{1>t_1>\cdots>t_{2d}>0}
       \frac{\Tr(e^{-H_{\rm bath}}\prod_{j=1}^{2d}\Xi(t_j))}
            {\Tr e^{-H_{\rm bath}}}\,d\mathbf t\nonumber\\
 &=\sum_{d\geq0}2^{-d}\sum_{\pi\in\mathcal P_2([2d])}
       (-1)^{\operatorname{cr}(\pi)}
       \int_{1>t_1>\cdots>t_{2d}>0}
          \prod_{\substack{\{a,b\}\in\pi\\a<b}}\Sigma(t_a-t_b)\,d\mathbf t,
       \nonumber\\
 &\frac{\Tr(e^{-(1-u)H_{\rm quad}}\psi_{N+1}e^{-uH_{\rm quad}}\psi_{N+1})}
      {\Tr e^{-H_{\rm bath}}}\nonumber\\
 &\qquad=\sum_{d\geq0}2^{-d}\sum_{\pi\in\mathcal P_2([2d])}
       \int_{1>t_1>\cdots>t_{2d}>0}
         (-1)^{\operatorname{cr}(\pi)+\kappa_u(\pi,\mathbf t)}
          \prod_{\substack{\{a,b\}\in\pi\\a<b}}\Sigma(t_a-t_b)\,d\mathbf t.
\end{align}
Odd orders vanish because the uncoupled bath has vanishing odd
moments of linear Majorana operators. In the last line moving the external
$\psi_{N+1}$ past the interactions above $u$ contributes
$(-1)^{\#\{j:t_j>u\}}=(-1)^{\kappa_u(\pi,\mathbf t)}$ by
\eqref{eq:quadratic-crossing-parity}. Moreover,
\begin{align}
 0\leq\log W(\Sigma)&\leq m\sum_{n\geq0}\omega_n^{-2}=m/8,\\
 2^{-d}\sum_\pi\int_{t_1>\cdots>t_{2d}}
                         \prod_{\{a,b\}\in\pi}|\Sigma(t_a-t_b)|\,d\mathbf t
 &\leq2^{-d}\frac{(2d-1)!!}{(2d)!}(m/2)^d=\frac{(m/8)^d}{d!}.
\end{align}
Finally choose finitely supported positive measures $\sigma_k$ of mass
$m$ with $\sigma_k\Rightarrow\sigma$. Bounded continuous integrands give
\begin{align}
 G_{\sigma_k}(t)&\longrightarrow G_\sigma(t)\quad(0\leq t\leq1),
 &|G_{\sigma_k}(t)|&\leq m/2,\nonumber\\
 \int\frac{\sigma_k(dx)}{\omega_n^2+x^2}
 &\longrightarrow\int\frac{\sigma(dx)}{\omega_n^2+x^2},
 &0\leq\log\left(1+\int\frac{\sigma_k(dx)}{\omega_n^2+x^2}\right)
 &\leq m/\omega_n^2,\nonumber\\
 \log W(G_{\sigma_k})&\longrightarrow\log W(G_\sigma),
 &\mathcal D(G_{\sigma_k})(t)&\longrightarrow\mathcal D(G_\sigma)(t).
\end{align}
The last convergence uses weak continuity proved in
\Cref{lem:positive-saddle}. For each $d$, dominated convergence applies
to every integral in the two pairing series, and
\begin{align}
 \sup_k\sum_{d>D}2^{-d}\sum_\pi
       \int_{t_1>\cdots>t_{2d}}
            \prod_{\{a,b\}\in\pi}|G_{\sigma_k}(t_a-t_b)|\,d\mathbf t
 &\leq\sum_{d>D}\frac{(m/8)^d}{d!}\longrightarrow0.
\end{align}
Thus both identities pass to $\sigma$. As $u\downarrow0$ or $u\uparrow1$,
\begin{align}
 (-1)^{\kappa_u(\pi,\mathbf t)}&\longrightarrow1
       \quad\text{for almost every }\mathbf t,\nonumber\\
 2W(\Sigma)\mathcal D(\Sigma)(u)&\longrightarrow W(\Sigma),
 &\mathcal D(\Sigma)(0^+)=\mathcal D(\Sigma)(1^-)&=\tfrac12,
\end{align}
by the same summable bound.

\end{proof}

\section{Coincident times and orthogonal reconstruction}\label{app:reconstruction}

\begin{lemma}[Evaluation at coincident times]\label{lem:contact-dictionary}
Let $j_1,\ldots,j_d\in[N]$ and $t_1,\ldots,t_d\in(0,1)$.
For pairwise distinct times let $\pi$ order them decreasingly, and define
\begin{align}
 \mathcal T_J(\mathbf j,\mathbf t)
 &=\operatorname{sgn}(\pi)
 \frac{\Tr\left(e^{(1-t_{\pi(1)})\beta H}
       \psi_{j_{\pi(1)}}e^{(t_{\pi(1)}-t_{\pi(2)})\beta H}
       \cdots\psi_{j_{\pi(d)}}e^{t_{\pi(d)}\beta H}\right)}
      {\Tr e^{\beta H}}.
 \label{eq:explicit-chronological-trace}
\end{align}
For general times, partition $[d]$ into the groups $I_1,\ldots,I_s$
on which $t_a$ is constant. The value representing the Grassmann
product is the independently antisymmetrized value in each group:
\begin{align}
 \left\langle\prod_{a=1}^d\chi_{j_a}(t_a)\right\rangle_N
 &=\lim_{\epsilon\downarrow0}
 \frac{\EE_J\left[\Tr e^{\beta H}\,
       \EE_{\xi}\mathcal T_J(\mathbf j,\mathbf t+\epsilon\boldsymbol\xi)
                 \right]}{\EE_J\Tr e^{\beta H}},
 \qquad \xi_1,\ldots,\xi_d\stackrel{\rm iid}{\sim}{\rm Unif}(-1,1).
 \label{eq:alternating-time-limit}
\end{align}
Every occurrence has its own $\xi_a$, including two occurrences of the
same site and time. Equivalently, replace the ordered product of the
operators in each equal-time group $I=\{a_1<\cdots<a_q\}$ by
\begin{align}
 \frac1{q!}\sum_{\sigma\in S_q}\operatorname{sgn}(\sigma)
                 \psi_{j_{a_{\sigma(1)}}}\cdots\psi_{j_{a_{\sigma(q)}}}.
 \label{eq:equal-time-antisymmetrization}
\end{align}
An \emph{ordered} coincident-time product is a different prescription:
choose distinct fixed offsets $\eta_a$ within each group, substitute
$t_a+\epsilon\eta_a$, and take $\epsilon\downarrow0$ without averaging.
Both prescriptions also apply to products of $Q$ and to their centered
moments by linearity. For times on the thermal circle, including $0$ and
$1$, choose
\begin{align}
 c&\in[0,1)\setminus\{t_a\bmod1:a\in[d]\},
 &k_a&=\lfloor t_a-c\rfloor,
 &\tau_a&=t_a-c-k_a\in(0,1),\nonumber\\
 \left\langle\prod_{a=1}^d\chi_{j_a}(t_a)\right\rangle_N
 &=(-1)^{\sum_ak_a}
       \left\langle\prod_{a=1}^d\chi_{j_a}(\tau_a)\right\rangle_N.
 \label{eq:thermal-endpoint-prescription}
\end{align}
Apply the preceding limits at $\boldsymbol\tau$, with
$0<\epsilon<\frac12\min_a\{\tau_a,1-\tau_a\}$.
\end{lemma}
\begin{proof}
At distinct times use \eqref{eq:chin}. Within an equal-time group of size
$q$, independent continuous offsets give each permutation probability
$1/q!$, hence \eqref{eq:equal-time-antisymmetrization}. This agrees with
the free Grassmann rule: the comparisons in a pairing use disjoint
offsets, and
\begin{align}
 \EE\sgn(\xi_a-\xi_b)&=0,\\
 \lim_{\epsilon\downarrow0}\EE_\xi
 \prod_{\{a,b\}\in\sigma,\ a<b}
          2G_0(t_a+\epsilon\xi_a,t_b+\epsilon\xi_b)
 &=\prod_{\{a,b\}\in\sigma,\ a<b}2G_0(t_a,t_b).
\end{align}
For the interaction expansion, Schatten H\"older and Gaussian
integrability provide the two dominating bounds
\begin{align}
 \left|\Tr(e^{s_0\beta H}X_1\cdots X_qe^{s_q\beta H})\right|
 &\leq\Tr e^{\beta H}
 \quad(s_j\geq0,\ \textstyle\sum_js_j=1,\ \|X_j\|\leq1),\\
 \sum_{q\geq0}\frac1{q!}
   \left(\beta\sqrt{\frac3{8N^3}}\sum_{|E|=4}|J_E|\right)^q
 &=\exp\left(\beta\sqrt{\frac3{8N^3}}\sum_{|E|=4}|J_E|\right)
 \in L^1(d\gamma(J))\quad(N\text{ fixed}).
\end{align}
At each fixed order $q$ of the exponential series, discontinuities occur
only when two of the finitely many times are equal. These sets have
Lebesgue measure zero in the integration variables. The two bounds above
therefore allow the limit to pass through the time integrals, the sum
over $q$, and the Gaussian expectation by dominated convergence. The
denominator is the fixed positive number $\EE\tr e^{\beta H}$. In particular, 
\begin{align}
 \lim_{\epsilon\downarrow0}
 \frac{\Tr(e^{(1-t-\epsilon)\beta H}\psi_i e^{\epsilon\beta H}\psi_i
                    e^{t\beta H})}{\Tr e^{\beta H}}&=1.
\end{align}
For fixed distinct offsets the limiting order is unchanged, and every
resulting matrix product is evaluated with
\begin{align}
 \psi_i^2&=I,\qquad
 \psi_i\psi_j=-\psi_j\psi_i\quad(i\ne j),\nonumber\\
 \left\langle\prod_{r=1}^pQ_{u_rv_r}\right\rangle_N
 &=\frac1{(2N)^p}\sum_{i_1,\ldots,i_p}
       \left\langle\prod_{r=1}^p\chi_{i_r}(u_r)\chi_{i_r}(v_r)\right\rangle_N,\nonumber\\
 &\left\langle\prod_{r=1}^p(Q_{u_rv_r}-G_*(u_r,v_r))\right\rangle_N\nonumber\\
 &\qquad=\sum_{S\subseteq[p]}(-1)^{p-|S|}
       \prod_{r\notin S}G_*(u_r,v_r)
       \left\langle\prod_{r\in S}Q_{u_rv_r}\right\rangle_N.
\end{align}
Apply the chosen limit to each entire summand. The free covariance is
invariant under a simultaneous time translation, and the integrated
quartic interaction is periodic. Translation by $-c$, followed by
$\chi_j(t+1)=-\chi_j(t)$ in each argument, proves
\eqref{eq:thermal-endpoint-prescription}.
\end{proof}

For a function of $d$ time arguments, let $F$ range over the domains and
prescriptions in \eqref{eq:time-cell-definition}--\eqref{eq:time-face-definition}.
The uniform norm used in the hierarchy is
\begin{align}
 \|X\|_\infty
 &=\max_F\sup_{\mathbf t\in F}|X_F(\mathbf t)|,\nonumber\\
 X_{F,\eta}(\mathbf t)
 &=\lim_{\epsilon\downarrow0}X(\mathbf t+\epsilon\boldsymbol\eta),\nonumber\\
 X_{F,\mathrm{alt}}(\mathbf t)
 &=\frac1{\prod_{a=1}^q|I_a|!}
       \sum_{\sigma_a\in S_{I_a},\ a\in[q]}
                         X_{F,\eta^{\boldsymbol\sigma}}(\mathbf t).
 \label{eq:prescribed-uniform-norm}
\end{align}
In the second line the offsets specify the order within every equal-time
group. In the third line one offset choice for each such order suffices;
$X$ already includes the chronological fermionic sign. The limit is
always taken in the entire centered expression. At a fixed maximal
moment order $K$, $d\leq2K$ and there are finitely many such domains and
orders.

An equality involving an integration variable is imposed in the input
before inversion. For example, the repeated pair in the hierarchy means
\begin{align}
 \Bigg[\mathfrak L_\beta^{-1}\operatorname{Alt}_{a,b}
  \int_{[0,1]^2}G_*(a,s)G_*(s,t)G_*(t,b)
       \mu_{p+1}\big((s,t),(s,t),(u_j,v_j)_{j=2}^p\big)\,ds\,dt
                                                        \Bigg](u,v).
 \label{eq:substitution-before-inverse}
\end{align}
In particular, the two occurrences of $(s,t)$ are evaluated by their
prescribed joint limit inside the integral, not by later restricting
an arbitrary $L^2$ representative of the output.

\begin{lemma}[Exact orthogonal reconstruction]\label{lem:orthogonal-reconstruction}
Let $N\geq k$, let $P(Q)$ be a polynomial in finitely many collective
variables, and let $\mathcal G_{k,N}$ be the matrix in \Cref{sec:alg}. Then
\begin{align}
 &\left\langle\chi_{j_1}(t_1)\cdots\chi_{j_{2k}}(t_{2k})P(Q)
                                                    \right\rangle_N
 \nonumber\\
 &=2^k\sum_{\pi,\sigma\in\mathcal P_2([2k])}
     \left(\prod_{\{a,b\}\in\pi}\delta_{j_aj_b}\right)
     (\mathcal G_{k,N}^{-1})_{\pi\sigma}
     (-1)^{\operatorname{cr}(\sigma)}
     \left\langle P(Q)
          \prod_{\{a,b\}\in\sigma,\ a<b}Q_{t_at_b}\right\rangle_N.
 \label{eq:full-orthogonal-reconstruction}
\end{align}
This identity holds for distinct times, for the alternating values in
\Cref{lem:contact-dictionary}, and for ordered limits obtained from
distinct times.
\end{lemma}
\begin{proof}
The free covariance, each $Q$, and the interaction are $O(N)$ invariant.
For
\begin{align}
 T_{j_1,\ldots,j_{2k}}
 &=\langle\chi_{j_1}(t_1)\cdots\chi_{j_{2k}}(t_{2k})P(Q)\rangle_N,
\end{align}
invariance under a sign change at site $b$, and under a permutation
$\tau\in S_N$, gives
\begin{align}
 T_{j_1,\ldots,j_{2k}}
 &=(-1)^{\#\{a:j_a=b\}}T_{j_1,\ldots,j_{2k}},
 &T_{\tau(j_1),\ldots,\tau(j_{2k})}&=T_{j_1,\ldots,j_{2k}}.
 \label{eq:tensor-reflections-permutations}
\end{align}
For $\pi\in\mathcal P_2([2k])$, order its pairs by increasing smaller
element and let $j_a^\pi=l$ when $a$ belongs to its $l$th pair. Define
\begin{align}
 c_\pi&=T_{j_1^\pi,\ldots,j_{2k}^\pi},
 &R_{\mathbf j}&=T_{\mathbf j}-\sum_\pi c_\pi
                             \prod_{\{a,b\}\in\pi}\delta_{j_aj_b}.
\end{align}
The second tensor has the same orthogonal invariance as $T$, since each
Kronecker contraction is invariant. Equations
\eqref{eq:tensor-reflections-permutations} imply
\begin{align}
 \exists b:\#\{a:j_a=b\}\text{ odd}&\quad\Longrightarrow\quad R_{\mathbf j}=0,
 \nonumber\\
 \#\{j_1,\ldots,j_{2k}\}=k,\quad
 \#\{a:j_a=b\}\in\{0,2\}\ (b\in[N])
 &\quad\Longrightarrow\quad R_{\mathbf j}=0.
 \label{eq:orthogonal-remainder-base}
\end{align}
Indeed, in the second line exactly one pairing has all its Kronecker
factors equal to one.

Induct downwards on $\#\{j_1,\ldots,j_{2k}\}$. A nonzero remaining
case has $I=\{a:j_a=b\}$ of size $2r\geq4$. There is an unused site $c$
since the number of used sites is less than $k\leq N$. Let
$j_a^S=c$ for $a\in S\subseteq I$ and $j_a^S=j_a$ otherwise.
Rotate $e_b$ to $(e_b+e_c)/\sqrt2$ while fixing the other used coordinates.
Then
\begin{align}
 R_{\mathbf j}&=2^{-r}\sum_{S\subseteq I}R_{\mathbf j^S},\nonumber\\
 |S|\text{ odd}&\quad\Longrightarrow\quad
 R_{\mathbf j^S}=(-1)^{|S|}R_{\mathbf j^S}=0,\nonumber\\
 0<|S|<2r,\quad |S|\text{ even}
 &\quad\Longrightarrow\quad
 \begin{cases}
 \#\{j_1^S,\ldots,j_{2k}^S\}=\#\{j_1,\ldots,j_{2k}\}+1,\\
 R_{\mathbf j^S}=0,
 \end{cases}\nonumber\\
 R_{\mathbf j^I}&=R_{\mathbf j},\qquad
 R_{\mathbf j}=2^{1-r}R_{\mathbf j},\qquad R_{\mathbf j}=0.
\end{align}
The second line uses the diagonal orthogonal matrix that changes the sign
of $e_c$ only; the third is the induction hypothesis; the fourth uses
the site permutation exchanging $b,c$. Thus $R=0$.

Evaluating the Kronecker tensors at $\mathbf j^\pi$ also proves their
linear independence. In the graph on $[2k]$ whose edges are the pairs
of $\pi$ and $\sigma$, each connected component must carry a single
site label. Different components may receive the same label. Therefore
\begin{align}
 \sum_{\mathbf j\in[N]^{2k}}
     \prod_{\{a,b\}\in\pi}\delta_{j_aj_b}
     \prod_{\{a,b\}\in\sigma}\delta_{j_aj_b}
 &=\prod_{C\in\operatorname{comp}(\pi\cup\sigma)}
                       \left(\sum_{i_C=1}^N1\right)
 =N^{\#\operatorname{comp}(\pi\cup\sigma)},\nonumber\\
 \sum_\pi(\mathcal G_{k,N})_{\sigma\pi}c_\pi
 &=N^{-k}\sum_{\mathbf j}\prod_{\{a,b\}\in\sigma}\delta_{j_aj_b}T_{\mathbf j}
 \nonumber\\
 &=2^k(-1)^{\operatorname{cr}(\sigma)}
       \left\langle P(Q)\prod_{\substack{\{a,b\}\in\sigma\\a<b}}
                                                   Q_{t_at_b}\right\rangle_N.
\end{align}
Regrouping the Grassmann factors into the paired order supplies
$(-1)^{\operatorname{cr}(\sigma)}$, as in \eqref{eq:freewick}.
The Gram matrix is invertible by the linear independence just proved.
Solving for $c_\pi$ gives \eqref{eq:full-orthogonal-reconstruction}.
Its finite sums commute with the ordered limits and averages in
\Cref{lem:contact-dictionary}.
\end{proof}

\section{Physical commutators and source moments}\label{app:bbounds}

We prove the source estimates in \Cref{lem:bbounds}, the ordered moment
inequality in \Cref{lem:ordered-young}, and their consequences for energy
pinching. Throughout,
\begin{align}
 \Phi_N(X)&=\frac{\EE_J\Tr(e^{\beta H}X)}{\EE_J\Tr e^{\beta H}},
 &M_m&=\Phi_N(B^{2m}),\qquad \|B\|\leq1,\nonumber\\
 \|X\|_{2,\Phi_N}^2&=\Phi_N(X^\dagger X),
 &\|X\|_{{\rm row},\Phi_N}^2&=\Phi_N(XX^\dagger).
 \label{eq:physical-two-orientations}
\end{align}
Left multiplication by a contraction decreases the first norm; right
multiplication decreases the second. The state $\Phi_N$ is not assumed
tracial.

\subsection{Exact commutator expansions}

For deterministic piecewise continuous profiles $m_E:\mathbb R\to[-1,1]$,
we use the propagators
\begin{align}
 H_\kappa&=\sqrt{\frac3{8N^3}}\sum_{E\in\binom{[N]}4}
                   \kappa_EJ_E\Gamma_E,
 &\kappa_E&\in[-1,1],\qquad H_{\mathbf1}=H,\nonumber\\
 \partial_tU^m_{[S]}(t,s)
 &=i\beta\sqrt{\frac3{8N^3}}
   \sum_{\substack{E\in\binom{[N]}4\\E\cap S=\varnothing}}
       m_E(t)J_E\Gamma_EU^m_{[S]}(t,s),
 &U^m_{[S]}(s,s)&=I,\nonumber\\
 \mathcal U^m_{[S]}(t,s)X
 &=U^m_{[S]}(t,s)XU^m_{[S]}(t,s)^\dagger,
 &U^m(t,s)&=U^m_{[\varnothing]}(t,s),\nonumber\\
 \mathcal U^m(t,s)&=\mathcal U^m_{[\varnothing]}(t,s),
 &U^m_{[S]}(t,s)^\dagger U^m_{[S]}(t,s)&=I.
 \label{eq:real-profile-evolution}
\end{align}
In particular, $\psi_i(t)=\mathcal U^{\mathbf1}(-t,0)\psi_i$.
For $n\geq0$ and $t\geq0$, set
\begin{align}
 \Delta_n(t)&=\{(s_1,\ldots,s_n)\in\mathbb R^n:
                         t>s_1>\cdots>s_n>0\},
 &\int_{\Delta_n(t)}d\mathbf s&=\frac{t^n}{n!}.
 \label{eq:ordered-simplex-volume}
\end{align}
The zero-dimensional integral is one, including when $t=0$.

\begin{lemma}[Expansion between two sites]\label[lemma]{lem:two-site-expansion}
Let $i\ne j$, $t\geq0$, $d\geq1$, and
\begin{align}
 \mathcal E_d(i,j)
 &=\left\{(E_1,\ldots,E_d)\in\binom{[N]}4^d:
 \begin{array}{l}
 j\in E_1,\quad i\in E_d,\quad i\notin\bigcup_{a<d}E_a,\\
 E_r\cap(\{j\}\cup\bigcup_{a<r-1}E_a)=\varnothing,\\
 E_r\cap E_{r-1}\ne\varnothing\quad(2\leq r\leq d)
 \end{array}\right\},\nonumber\\
 S_r&=\{j\}\cup\bigcup_{a<r}E_a\quad(1\leq r\leq d),\nonumber\\
 S_0&=\varnothing,\qquad s_0=t,\qquad s_{d+1}=0.
 \label{eq:two-site-index-set}
\end{align}
For $\boldsymbol\epsilon\in\{0,1\}^{d+1}$ and $\mathbf s\in\Delta_d(t)$,
write the following matrix product, with the second product in decreasing order:
\begin{align}
 W_{\boldsymbol\epsilon}(\mathbf E,\mathbf s)
 &=\psi_j^{1-\epsilon_0}U^m(t,s_1)
   \left(\prod_{r=1}^d
       \Gamma_{E_r}^{1-\epsilon_r}U^m_{[S_r]}(s_r,s_{r+1})\right)
          \psi_i\nonumber\\*
 &\quad\times\left(\prod_{r=d}^{1}
       U^m_{[S_r]}(s_r,s_{r+1})^\dagger\Gamma_{E_r}^{\epsilon_r}\right)
        U^m(t,s_1)^\dagger\psi_j^{\epsilon_0}.
 \label{eq:two-site-matrix-product}
\end{align}
Then
\begin{align}
 \{\psi_j,\mathcal U^m(t,0)\psi_i\}
 &=\sum_{d=1}^N\left(i\beta\sqrt{\frac3{8N^3}}\right)^d
   \sum_{\mathbf E\in\mathcal E_d(i,j)}
   \int_{\Delta_d(t)}\left(\prod_{r=1}^d m_{E_r}(s_r)J_{E_r}\right)
       \nonumber\\*
 &\quad\times\sum_{\boldsymbol\epsilon\in\{0,1\}^{d+1}}
    (-1)^{\epsilon_1+\cdots+\epsilon_d}
       W_{\boldsymbol\epsilon}(\mathbf E,\mathbf s)\,d\mathbf s,
 \label{eq:explicit-commutator-terms}
\end{align}
Moreover,
\begin{align}
 \mathcal E_d(i,j)&=\varnothing\quad(d>N),
 &\|W_{\boldsymbol\epsilon}(\mathbf E,\mathbf s)\|&=1,\nonumber\\
 \sum_{\boldsymbol\epsilon\in\{0,1\}^{d+1}}
       |(-1)^{\epsilon_1+\cdots+\epsilon_d}|&=2^{d+1},
 &2\sum_{r=0}^d|s_r-s_{r+1}|&=2t.
 \label{eq:commutator-length-and-volume}
\end{align}
For $t<0$, the same formulas apply with
\begin{align}
 \widehat m_E(u)&=-m_E(-u),
 &U^m_{[S]}(t,0)&=U^{\widehat m}_{[S]}(|t|,0).
 \label{eq:negative-time-profile}
\end{align}
\end{lemma}
\begin{proof}
The Clifford relations and Duhamel's formula give
\begin{align}
 \{\psi_j,\mathcal U^m_{[\{j\}]}(t,0)\psi_i\}&=0,\nonumber\\
 \{\psi_j,\mathcal U^m(t,0)\psi_i\}
 &=i\beta\sqrt{\frac3{8N^3}}\sum_{E_1\ni j}\int_0^t
        m_{E_1}(s_1)J_{E_1}\nonumber\\
 &\quad\times\{\psi_j,\mathcal U^m(t,s_1)
       [\Gamma_{E_1},\mathcal U^m_{[\{j\}]}(s_1,0)\psi_i]\}\,ds_1.
 \label{eq:causal-first-step}
\end{align}
For $i\notin S\cup E$, the corresponding identities are
\begin{align}
 [\Gamma_E,\mathcal U^m_{[S\cup E]}(t,0)\psi_i]&=0,\nonumber\\
 [\Gamma_E,\mathcal U^m_{[S]}(t,0)\psi_i]
 &=i\beta\sqrt{\frac3{8N^3}}
   \sum_{\substack{F\cap S=\varnothing\\F\cap E\ne\varnothing}}
       \int_0^t m_F(s)J_F\nonumber\\
 &\quad\times[\Gamma_E,\mathcal U^m_{[S]}(t,s)
        [\Gamma_F,\mathcal U^m_{[S\cup E]}(s,0)\psi_i]]\,ds.
 \label{eq:causal-recursion}
\end{align}
Indeed, conjugation by interactions supported on $[N]\setminus S$
preserves the Clifford algebra generated by that set. Iteration terminates because
\begin{align}
 |E_1\setminus S_1|&=3,\nonumber\\
 E_r\cap S_{r-1}=\varnothing,\quad E_{r-1}\cap S_{r-1}\ne\varnothing
 &\ \Longrightarrow\ E_r\not\subseteq E_{r-1}
 \ \Longrightarrow\ E_r\setminus S_r\ne\varnothing,\nonumber\\
 S_{r+1}&=S_r\cup E_r,
 &|S_{r+1}|&>|S_r|,\nonumber\\
 i\notin\bigcup_{a\leq r}E_a&\ \Longrightarrow\ i\notin S_{r+1}.
\end{align}
The resulting nested brackets expand by
\begin{align}
 [\Gamma_E,X]&=\sum_{\epsilon=0}^1
                  (-1)^\epsilon\Gamma_E^{1-\epsilon}X\Gamma_E^\epsilon,
 &\{\psi_j,X\}&=\sum_{\epsilon=0}^1\psi_j^{1-\epsilon}X\psi_j^\epsilon.
\end{align}
Substitution gives \eqref{eq:two-site-matrix-product} and
\eqref{eq:explicit-commutator-terms}. Finally,
\begin{align}
 \sum_{r=0}^d|s_r-s_{r+1}|
 &=(t-s_1)+\sum_{r=1}^{d-1}(s_r-s_{r+1})+s_d=t.
\end{align}
The last assertion follows by differentiating $U^m_{[S]}(-u,0)$.
\end{proof}

We specify the label constraints used in the next two lemmas. Let $I$
be a finite set of prescribed indices, and let $\sim$ be an equivalence
relation on $I\sqcup([D]\times[4])$. For $\mathbf i\in[N]^I$, set
\begin{align}
 x_a&=i_a\quad(a\in I),
 &x_{(d,h)}&=e_{dh}\quad((d,h)\in[D]\times[4]),\nonumber\\
 \mathcal A_\sim(\mathbf i)
 &=\left\{\mathbf e\in[N]^{D\times[4]}:
 \begin{array}{ll}
 e_{d1}<e_{d2}<e_{d3}<e_{d4}&(d\in[D]),\\
 x_u=x_v&(u\sim v)
 \end{array}\right\},
 &E_d&=\{e_{d1},e_{d2},e_{d3},e_{d4}\}.
 \label{eq:quartet-assignment-set}
\end{align}
For $D'\subseteq[D]$ and $I'\subseteq I$, define the multigraph
$G_\sim(D',I')$ by its vertex set, edge set, and endpoint map:
\begin{align}
 V\bigl(G_\sim(D',I')\bigr)
 &=D'\sqcup\{[u]_\sim:u\in I'\sqcup(D'\times[4])\},\nonumber\\
 E\bigl(G_\sim(D',I')\bigr)&=D'\times[4],
 &\partial(d,h)&=\bigl(d,[(d,h)]_\sim\bigr).
 \label{eq:quartet-constraint-graph}
\end{align}
Thus edges with identical endpoints still have different indices $(d,h)$.
We write $\operatorname{comp}(G)$ for the set of connected components of $G$.

\begin{lemma}[Products and nested commutators]
\label[lemma]{lem:commutator-expansion-structure}
Let $\mathcal T$ be a finite rooted binary tree, with leaf set
$\mathcal L$, internal vertex set $\mathcal B$, and root $o$.
Let $\iota:\mathcal L\to I$ be onto. For $a\in\mathcal L$, choose
$r_a\geq1$, $\varepsilon_{a\nu}\in\{0,1\}$, deterministic
$\kappa^{a\nu}_E\in[-1,1]$, and profiles as in
\eqref{eq:real-profile-evolution}. Define
\begin{align}
 X_a&=\prod_{\nu=1}^{r_a}
       \mathcal U^{m_{a\nu}}(t_{a\nu},0)
       \left(\ad_{\beta H_{\kappa^{a\nu}}}^{\varepsilon_{a\nu}}
                               \psi_{i_{\iota(a)}}\right)
                          &&(a\in\mathcal L),\nonumber\\
 X_a&=\mathcal U^{m_a}(t_a,0)[X_{a_0},X_{a_1}]_{\rm gr}
                          &&(a\in\mathcal B),\nonumber\\
 [Y,Z]_{\rm gr}&=YZ-(-1)^{|Y||Z|}ZY,
 &d_0&=\sum_{a\in\mathcal L}\sum_{\nu=1}^{r_a}\varepsilon_{a\nu},\nonumber\\
 t_a,t_{a\nu}&\geq0,
 &\sum_{a\in\mathcal B}t_a+
       \sum_{a\in\mathcal L}\sum_{\nu=1}^{r_a}t_{a\nu}&\leq T.
 \label{eq:recursive-commutator-expressions}
\end{align}
Here $a_0,a_1$ are the two children of $a$, and $\ad^0$ is the identity.
There is $C\geq1$, depending only on $\mathcal T$ and $(r_a)$, for which
there exist finite sets $\Lambda_n$, relations $\sim_\lambda$ on
$I\sqcup([n+d_0]\times[4])$, and measurable functions
$c_\lambda,m_{\lambda\nu},\tau_{\lambda\nu}$, independent of $J$, such that
\begin{align}
 X_o
 &=\sum_{n=0}^{|\mathcal B|N}
       \left(\beta\sqrt{\frac3{8N^3}}\right)^{n+d_0}
       \sum_{\lambda\in\Lambda_n}
       \sum_{\mathbf e\in\mathcal A_{\sim_\lambda}(\mathbf i)}
       \int_{\Delta_n(CT)}c_\lambda(\mathbf e,\mathbf i,\mathbf u)
                   \left(\prod_{d=1}^{n+d_0}J_{E_d}\right)\nonumber\\*
 &\quad\times U^{m_{\lambda0}}(\tau_{\lambda0},0)
       \psi_{x_{k_{\lambda1}}}U^{m_{\lambda1}}(\tau_{\lambda1},0)
       \cdots\psi_{x_{k_{\lambda\ell_\lambda}}}
           U^{m_{\lambda\ell_\lambda}}(\tau_{\lambda\ell_\lambda},0)
                     \,d\mathbf u,
 \label{eq:general-expanded-word}
\end{align}
The data satisfy
\begin{align}
 |\Lambda_n|&\leq C^{n+1}(n+1)^C,\qquad
       |c_\lambda(\mathbf e,\mathbf i,\mathbf u)|\leq1,\nonumber\\
 k_{\lambda\nu}&\in I\sqcup([n+d_0]\times[4]),\nonumber\\
 \ell_\lambda&\leq\sum_{a\in\mathcal L}r_a+4(n+d_0),\nonumber\\
 \sum_{\nu=0}^{\ell_\lambda}\tau_{\lambda\nu}&\leq CT,
                    \qquad \tau_{\lambda\nu}\geq0,\nonumber\\
 \prod_{\nu=0}^{\ell_\lambda}
      U^{m_{\lambda\nu}}(\tau_{\lambda\nu},0)&=I,\nonumber\\
 \left|\operatorname{comp}
      G_{\sim_\lambda}([n+d_0],I)\right|&=1.
 \label{eq:nested-expansion-bounds}
\end{align}
The bounds hold for every $\mathbf e,\mathbf i,\mathbf u$ in the displayed
sum and integral. Each $m_{\lambda\nu,E}(v)$ belongs to $[-1,1]$ and is
piecewise continuous in $v$. Its dependence on $\mathbf e,\mathbf i,\mathbf u$
is allowed, but not on $J$.

One leaf may instead be the following product, where $p$ is fixed:
\begin{align}
 W&=e^{il_0\beta H}P_1e^{il_1\beta H}\cdots P_se^{il_s\beta H},
 &P_\nu&=\Gamma_{A_\nu},\quad |A_\nu|\leq p,\nonumber\\
 \sum_{\nu=0}^s l_\nu&=0,
 &\sum_{\nu=0}^s|l_\nu|&\leq T.
 \label{eq:closed-real-time-word}
\end{align}
Let $a_*$ be this leaf and let $I$ now be the image of
$\mathcal L\setminus\{a_*\}$ under $\iota$. In place of the corresponding
hypotheses in \eqref{eq:recursive-commutator-expressions}, assume
\begin{align}
 d_0&=\sum_{a\in\mathcal L\setminus\{a_*\}}
                         \sum_{\nu=1}^{r_a}\varepsilon_{a\nu},\nonumber\\
 \sum_{a\in\mathcal B}t_a+
 \sum_{a\in\mathcal L\setminus\{a_*\}}\sum_{\nu=1}^{r_a}t_{a\nu}
       +\sum_{\nu=0}^s|l_\nu|&\leq T,\nonumber\\
 F&=\bigcup_{\nu=1}^s A_\nu,\qquad x_f=f\quad(f\in F).
\end{align}
Use $I\sqcup F$ in \eqref{eq:quartet-assignment-set} and
\eqref{eq:general-expanded-word}. The bounds become
\begin{align}
 |\Lambda_n|&\leq C^{n+1}(1+s+n)^C,\nonumber\\
 \ell_\lambda&\leq\sum_{a\in\mathcal L\setminus\{a_*\}}r_a
                           +ps+4(n+d_0),\nonumber\\
 \sum_{\nu=0}^{\ell_\lambda}\tau_{\lambda\nu}&\leq CT,
                       \qquad\tau_{\lambda\nu}\geq0,\nonumber\\
 \prod_{\nu=0}^{\ell_\lambda}
      U^{m_{\lambda\nu}}(\tau_{\lambda\nu},0)&=I,\nonumber\\
 C'\in\operatorname{comp}G_{\sim_\lambda}([n+d_0],I\sqcup F)
 &\ \Longrightarrow\ C'\cap\{[f]_{\sim_\lambda}:f\in F\}\ne\varnothing,
 \nonumber\\
 \left|\operatorname{comp}G_{\sim_\lambda}([n+d_0],I\sqcup F)\right|
 &\leq |F|.
 \label{eq:word-expansion-bounds}
\end{align}
Also,
\begin{align}
 |c_\lambda(\mathbf e,\mathbf i,\mathbf u)|&\leq1,\qquad
 k_{\lambda\nu}\in(I\sqcup F)\sqcup([n+d_0]\times[4]).
\end{align}
The profiles satisfy the same hypotheses, and $C$ is independent of $s$ and $F$.
For $F=\varnothing$ and $|\mathcal B|\geq1$, take $\Lambda_n=\varnothing$.
\end{lemma}
\begin{proof}
First expand each factor with $\varepsilon_{a\nu}=1$ using
\begin{align}
 [\beta H_\kappa,\psi_i]
 &=2\beta\sqrt{\frac3{8N^3}}\sum_{E\ni i}
                  \kappa_EJ_E\Gamma_E\psi_i,\nonumber\\
 \Gamma_E&=-\psi_{e_1}\psi_{e_2}\psi_{e_3}\psi_{e_4}
                      \quad(e_1<e_2<e_3<e_4),\nonumber\\
 \mathbf1_{i\in E}&=\sum_{h=1}^4\mathbf1_{i=e_h}.
 \label{eq:initial-quartet-expansion}
\end{align}
Each leaf has $\sum_\nu\varepsilon_{a\nu}$ quartet indices,
and every such quartet has an edge to $[\iota(a)]_\sim$. Its graph is
therefore connected. A leaf without such factors has the
single vertex $[\iota(a)]_\sim$.

The algebraic identities used at internal vertices are
\begin{align}
 \mathcal V[Y,Z]_{\rm gr}&=[\mathcal VY,\mathcal VZ]_{\rm gr},\nonumber\\
 [Y,Z_1\cdots Z_l]_{\rm gr}
 &=\sum_{a=1}^l(-1)^{|Y|\sum_{b<a}|Z_b|}
              Z_1\cdots Z_{a-1}[Y,Z_a]_{\rm gr}Z_{a+1}\cdots Z_l,
                 \nonumber\\
 [[X,Y]_{\rm gr},Z]_{\rm gr}
 &=[X,[Y,Z]_{\rm gr}]_{\rm gr}
       -(-1)^{|X||Y|}[Y,[X,Z]_{\rm gr}]_{\rm gr},\nonumber\\
 [\mathcal VX,\mathcal WY]_{\rm gr}
 &=\mathcal V[X,\mathcal V^{-1}\mathcal WY]_{\rm gr},\nonumber\\
 [X,c(J)Y]_{\rm gr}&=c(J)[X,Y]_{\rm gr}.
 \label{eq:graded-tree-identities}
\end{align}
Move the conjugations to the leaves by the first identity. For the
unique path from $o$ to $a$, with composition ordered from root to leaf,
\begin{align}
 \widehat X_a&=\left(\prod_{w\in\operatorname{path}(o,a)\cap\mathcal B}^{o\to a}
                         \mathcal U^{m_w}(t_w,0)\right)X_a.
\end{align}
Choose $a_*$ as the innermost leaf, arbitrarily when $W$ is absent.
Repeated use of the graded Jacobi identity gives
\begin{align}
 X_o&=\sum_{\sigma}c_\sigma
   [\widehat X_{\sigma(1)},[\widehat X_{\sigma(2)},\ldots,
     [\widehat X_{\sigma(v-1)},\widehat X_{a_*}]_{\rm gr}
                         \ldots]_{\rm gr}]_{\rm gr},\nonumber\\*
 v&=|\mathcal L|,\qquad
 \sigma:[v-1]\longrightarrow\mathcal L\setminus\{a_*\}
                      \ \text{bijective},\nonumber\\*
 c_\sigma&\in\mathbb Z,\qquad\sum_\sigma|c_\sigma|\leq C_\mathcal T.
 \label{eq:explicit-bracket-straightening}
\end{align}
This follows by induction on the number of leaves in the argument not
containing $a_*$: the third identity reduces that number in both summands.
For $v=1$, the right side means $\widehat X_{a_*}$.

For the matrix products, concatenation and reversal give
\begin{align}
 U^m(a,0)U^n(b,0)&=U^{\widehat m}(a+b,0),
 &\widehat m_E(u)&=
 \begin{cases}n_E(u),&0\leq u\leq b,\\m_E(u-b),&b<u\leq a+b,\end{cases}
                 \nonumber\\
 U^m(a,0)^\dagger&=U^{\widetilde m}(a,0),
 &\widetilde m_E(u)&=-m_E(a-u).
 \label{eq:profile-concatenation}
\end{align}
Thus each product of propagators below has the form $U^m(\tau,0)$,
with $\tau$ the sum of the nonnegative interval lengths in that product.
For a current summand, write
\begin{align}
 Z&=U_0\psi_{k_1}U_1\cdots\psi_{k_l}U_l,
 &U_\nu&=U^{m_\nu}(\tau_\nu,0),\qquad U_0\cdots U_l=I,\nonumber\\
 Q_a&=U_0\cdots U_{a-1},
 &Z&=\prod_{a=1}^lQ_a\psi_{k_a}Q_a^\dagger,\nonumber\\
 Q_a&=U^{\widehat m_a}\left(\sum_{\nu<a}\tau_\nu,0\right),
 &\sum_{\nu<a}\tau_\nu&\leq\sum_{\nu=0}^l\tau_\nu.
 \label{eq:compact-prefix-factorization}
\end{align}
For $A=U^m(t,0)\psi_jU^m(t,0)^\dagger$, the product rule is the exact identity
\begin{align}
 [A,Z]_{\rm gr}
 &=\sum_{a=1}^l(-1)^{a-1}
     (U_0\psi_{k_1}U_1\cdots\psi_{k_{a-1}}U_{a-1})\nonumber\\
 &\quad\times\{Q_a^\dagger A Q_a,\psi_{k_a}\}
                 (U_a\psi_{k_{a+1}}U_{a+1}\cdots\psi_{k_l}U_l),\nonumber\\
 Q_a^\dagger U^m(t,0)
 &=U^{\widetilde m_a}\left(t+\sum_{\nu<a}\tau_\nu,0\right).
 \label{eq:bracket-with-expanded-word}
\end{align}
Apply \Cref{lem:two-site-expansion} when $j\ne k_a$. When $j=k_a$, use
\begin{align}
 \{\psi_j,\mathcal U^m(t,0)\psi_j\}
 &=\psi_jU^m(t,0)\psi_jU^m(t,0)^\dagger
       +U^m(t,0)\psi_jU^m(t,0)^\dagger\psi_j.
 \label{eq:equal-site-retained-bracket}
\end{align}
Both operations produce the product in \eqref{eq:general-expanded-word}.
In particular, no scalar $J_E$ is duplicated by a bracket.

We now specify the relations $\sim_\lambda$. Order $[4]\times[4]$
lexicographically. For increasingly listed quartets $E,F$,
\begin{align}
 \mathbf1_{E\cap F\ne\varnothing}
 &=\sum_{(h,k)\in[4]^2}\mathbf1_{e_h=f_k}
       \prod_{(h',k')<(h,k)}\mathbf1_{e_{h'}\ne f_{k'}},\nonumber\\
 \mathbf1_{i\in E}&=\sum_{h=1}^4\mathbf1_{i=e_h},
 &1&=\mathbf1_{i=j}+\mathbf1_{i\ne j}.
 \label{eq:intersection-partition}
\end{align}
Each equality selected in these formulas is a generating pair of
$\sim_\lambda$. The other indicators multiply $c_\lambda$. If the
selected factors correspond to $u,v\in I\sqcup([D]\times[4])$, a
positive-length sequence supplies
\begin{align}
 u&\sim_\lambda(d_1,h_0),
 &(d_b,h_b)&\sim_\lambda(d_{b+1},k_{b+1})\quad(1\leq b<n),\nonumber\\
 (d_n,h_n)&\sim_\lambda v.
 \label{eq:nested-expansion-connections}
\end{align}
The zero-length alternative supplies $u\sim_\lambda v$. These relations
join the connected graph of the outer leaf to a component of the previous
graph. Hence, by induction,
\begin{align}
 |\operatorname{comp}G_{\sim_\lambda}([D],I)|&=1
                                      &&\text{without }W,\nonumber\\
 C'\in\operatorname{comp}G_{\sim_\lambda}([D],I\sqcup F)
 &\ \Longrightarrow\ C'\cap\{[f]_{\sim_\lambda}:f\in F\}\ne\varnothing
                                      &&\text{with }W.
\end{align}
For $W$, the initial graph consists of the vertices $F$; the same
induction applies even to vertices not selected by a bracket.

For completeness, index successive outer brackets by $b$. Let
$\tau_{b\nu}$ be the parameters of the current propagators, as in
\eqref{eq:compact-prefix-factorization}, and let $t_{b\mu}$ be those of
the next outer leaf after its finitely many Clifford products are split.
Equations \eqref{eq:two-site-matrix-product} and
\eqref{eq:bracket-with-expanded-word} imply
\begin{align}
 \sum_{\nu=0}^{l_{b+1}}\tau_{b+1,\nu}
 &\leq\sum_{\nu=0}^{l_b}\tau_{b\nu}
       +2\left(\sum_{\nu=0}^{l_b}\tau_{b\nu}
                          +\sum_\mu t_{b\mu}\right)
       +2\sum_\mu t_{b\mu}\nonumber\\
 &\leq8\left(\sum_{\nu=0}^{l_b}\tau_{b\nu}+\sum_\mu t_{b\mu}\right),
             \nonumber\\
 \sum_{\nu=0}^{l_b}\tau_{b\nu}
 &\leq8^b\left(\sum_{\nu=0}^{l_0}\tau_{0\nu}
                      +\sum_{a<b}\sum_\mu t_{a\mu}\right)
 \leq CT.
 \label{eq:nested-duration-bound}
\end{align}
The parameters of the conjugations of the individual Majoranas in
$\widehat X_a$ are bounded by the explicit sum
\begin{align}
 &\sum_{a\in\mathcal L}\sum_{\nu=1}^{r_a}
       (1+4\varepsilon_{a\nu})
       \left(t_{a\nu}+\sum_{w\in\operatorname{path}(o,a)\cap\mathcal B}t_w\right)
       \leq\left(\sum_a r_a+4d_0\right)T.
\end{align}
For the leaf $W$, omit its summand here. Its matrix product instead has
\begin{align}
 \sum_{\nu=0}^s|l_\nu|
       +2\sum_{w\in\operatorname{path}(o,a_*)\cap\mathcal B}t_w
 &\leq 2T.
\end{align}
Thus the constant in \eqref{eq:nested-duration-bound} is independent of $s$.

There are at most $|\mathcal B|$ two-site expansions in each summand.
Their integration variables can be listed in one decreasing order.
Outside $\bigcup_{a\ne b}\{u_a=u_b\}$, for sequence lengths $n_1,\ldots,n_b$ with $\sum_an_a=n$,
\begin{align}
 \left|\left\{(n_1,\ldots,n_b)\in\mathbb Z_{\geq0}^b:
                                     \sum_an_a=n\right\}\right|
 &\leq(n+1)^b,\nonumber\\
 \frac{n!}{\prod_{a=1}^b n_a!}&\leq b^n,\nonumber\\
 \sigma\mathbf u&=(u_{\sigma^{-1}(1)},\ldots,u_{\sigma^{-1}(n)}),\nonumber\\
 \prod_{a=1}^b\Delta_{n_a}(CT)
 &=\bigcup_{\substack{\sigma\in S_n\\
       \sigma^{-1}(1+\sum_{c<a}n_c)<\cdots<
       \sigma^{-1}(\sum_{c\leq a}n_c)\ (a\in[b])}}
                     \sigma\Delta_n(CT).
 \label{eq:common-simplex-decomposition}
\end{align}
Conditional upper endpoints and the additional time inequalities are
indicators in $c_\lambda$. Each coordinate permutation has Jacobian one.
For the single-site leaves, the number of matrix factors and summands obeys
\begin{align}
 \ell_\lambda&\leq\sum_{a\in\mathcal L}r_a+4(n+d_0),\nonumber\\
 |\Lambda_n|
 &\leq C_\mathcal T\,2^{n+d_0+|\mathcal B|}
       4^{2n+2|\mathcal B|+d_0}(n+1)^{|\mathcal B|}
       (|\mathcal B|+1)^n\nonumber\\
 &\quad\times\left[C\left(1+n+d_0+
                    \sum_{a\in\mathcal L}r_a\right)\right]^{|\mathcal B|}
 \leq C^{n+1}(n+1)^C.
\end{align}
The factors bound the two product choices in each bracket, the indices in
\eqref{eq:intersection-partition}, the sequence lengths and coordinate
orders, and the positions in \eqref{eq:bracket-with-expanded-word}.
The factor $2^{d_0}$ replaces each scalar $2$ in
\eqref{eq:initial-quartet-expansion} by two identical summands, so
$|c_\lambda|\leq1$. With the leaf $W$, substitute
\begin{align}
 \sum_{a\in\mathcal L}r_a
 &\longmapsto\sum_{a\in\mathcal L\setminus\{a_*\}}r_a
                              +\sum_{\nu=1}^s|A_\nu|
 \leq\sum_{a\in\mathcal L\setminus\{a_*\}}r_a+ps,\nonumber\\
 |\Lambda_n|&\leq C^{n+1}(1+s+n)^C.
\end{align}
Consequently,
\begin{align}
 \sum_{\lambda\in\Lambda_n}\int_{\Delta_n(CT)}
             \sup_{\mathbf e,\mathbf i}|c_\lambda(\mathbf e,\mathbf i,\mathbf u)|
                         \,d\mathbf u
 &\leq C^{n+1}(1+s+n)^C\frac{(CT)^n}{n!}.
 \label{eq:integrated-coefficient-bound}
\end{align}
Omit $s$ for the version without $W$. The identity $\prod_\nu U_\nu=I$
is preserved by products, conjugation, and each summand in
\eqref{eq:two-site-matrix-product}. Taking adjoints reverses this product
and replaces its profiles by \eqref{eq:profile-concatenation}.
\end{proof}

\subsection{Gaussian integration and moment bounds}

\begin{lemma}[Gaussian derivatives of matrix products]
\label[lemma]{lem:retained-product-derivatives}
Let $r\geq1$ and let all parameters below be deterministic:
\begin{align}
 F(J)&=\prod_{\nu=1}^r
      \left(e^{h_\nu\beta Q_\nu H_\kappa Q_\nu^\dagger}
                   U^{m_\nu}(t_\nu,s_\nu)C_\nu\right),\nonumber\\
 h_\nu&\geq0,
 &\sum_{\nu=1}^r h_\nu&=1,\nonumber\\
 Q_\nu^\dagger Q_\nu&=I,
 &\|C_\nu\|&\leq1,\qquad |\kappa_E|,|m_{\nu,E}(u)|\leq1.
 \label{eq:gaussian-matrix-product}
\end{align}
For $k\geq0$ and $E_1,\ldots,E_k\in\binom{[N]}4$,
\begin{align}
 \left|\Tr\bigl(\partial_{J_{E_1}}\cdots\partial_{J_{E_k}}F(J)\bigr)\right|
 &\leq\Tr(e^{\beta H_\kappa})
       \left(\beta\sqrt{\frac3{8N^3}}\right)^k
       \left(1+\sum_{\nu=1}^r|t_\nu-s_\nu|\right)^k,\nonumber\\
 \frac{\left|\EE_J\Tr\bigl(
          \partial_{J_{E_1}}\cdots\partial_{J_{E_k}}F(J)\bigr)\right|}
      {\EE_J\Tr e^{\beta H_\kappa}}
 &\leq\left(\beta\sqrt{\frac3{8N^3}}\right)^k
       \left(1+\sum_{\nu=1}^r|t_\nu-s_\nu|\right)^k.
 \label{eq:complete-contour-derivative}
\end{align}
For $D\geq0$, the matrix-valued Gaussian identity is
\begin{align}
 \EE_J[J_{E_1}\cdots J_{E_D}F(J)]
 &=\sum_{\substack{\pi\subseteq\binom{[D]}2\\
             A\cap B=\varnothing\ (A\ne B\in\pi)}}
       \left(\prod_{\{a,b\}\in\pi}\delta_{E_aE_b}\right)
       \EE_J\left[\prod_{a\in[D]\setminus\cup\pi}
                             \partial_{J_{E_a}}F(J)\right].
 \label{eq:partial-gaussian-matching}
\end{align}
\end{lemma}
\begin{proof}
Duhamel differentiation gives
\begin{align}
 \partial_{J_E}e^{h\beta QH_\kappa Q^\dagger}
 &=\beta\sqrt{\frac3{8N^3}}\kappa_E
     \int_0^h e^{(h-u)\beta QH_\kappa Q^\dagger}
            Q\Gamma_EQ^\dagger e^{u\beta QH_\kappa Q^\dagger}\,du,
                \nonumber\\
 \partial_{J_E}U^m(t,s)
 &=i\beta\sqrt{\frac3{8N^3}}
     \int_s^t m_E(u)U^m(t,u)\Gamma_EU^m(u,s)\,du.
 \label{eq:masked-gibbs-derivative}
\end{align}
For nonnegative $a_j$ and contractions $Y_j$, Schatten H\"older yields
\begin{align}
 \left|\Tr\prod_{j=1}^l
                   (e^{a_j\beta Q_jH_\kappa Q_j^\dagger}Y_j)\right|
 &\leq\prod_{j=1}^l
           \|e^{a_j\beta Q_jH_\kappa Q_j^\dagger}\|_{1/a_j}
       \prod_{j=1}^l\|Y_j\|\nonumber\\
 &=\Tr(e^{\beta H_\kappa})\prod_{j=1}^l\|Y_j\|
                \qquad\left(\sum_{j=1}^l a_j=1\right).
\end{align}
Here $a_j=0$ uses the operator norm. For $k$ labelled derivatives,
\begin{align}
 \sum_{\sigma\in S_k}\int_{\Delta_k(h)}d\mathbf u&=h^k,
 &\sum_{\sigma\in S_k}\int_{\Delta_k(|t-s|)}d\mathbf u&=|t-s|^k.
\end{align}
Consequently,
\begin{align}
 \left|\Tr(\partial_{J_{E_1}}\cdots\partial_{J_{E_k}}F)\right|
 &\leq\Tr(e^{\beta H_\kappa})
       \left(\beta\sqrt{\frac3{8N^3}}\right)^k\nonumber\\
 &\quad\times\sum_{\substack{\alpha,\delta\in\mathbb Z_{\geq0}^r\\
                                    |\alpha|+|\delta|=k}}
       \frac{k!}{\prod_\nu\alpha_\nu!\delta_\nu!}
                   \prod_{\nu=1}^r h_\nu^{\alpha_\nu}
                                      |t_\nu-s_\nu|^{\delta_\nu}\nonumber\\
 &=\Tr(e^{\beta H_\kappa})
       \left(\beta\sqrt{\frac3{8N^3}}\right)^k
       \left(1+\sum_\nu|t_\nu-s_\nu|\right)^k.
\end{align}
This proves both derivative bounds. Repeated use of
\begin{align}
 \EE[J_Ef(J)]&=\EE[\partial_{J_E}f(J)],
 &\partial_{J_E}J_F&=\delta_{EF}
\end{align}
gives \eqref{eq:partial-gaussian-matching} by induction on $D$.
At fixed $N$, all entries and derivatives in that induction are bounded by
\begin{align}
 C_{N,D,\mathbf h,\mathbf t,\mathbf s}
       \left(1+\sum_E|J_E|\right)^D
       \exp\left(\beta\sqrt{\frac3{8N^3}}\sum_E|J_E|\right)
 &\in L^1(d\gamma(J)).
\end{align}
Thus the identities apply to the full matrix product, including every
factor $U^{m_\nu}(t_\nu,s_\nu)$.
\end{proof}

\begin{lemma}[Counting quartet labels]\label[lemma]{lem:causal-count}
Let $I$ be finite, let $D\geq0$, and use
\eqref{eq:quartet-assignment-set}--\eqref{eq:quartet-constraint-graph}.
Suppose
\begin{align}
 [D]&=D_1 \mathbin{\dot{\cup}} \cdots \mathbin{\dot{\cup}} D_k,
 &I_a&\subseteq I,\qquad c_a\in\mathbb Z_{\geq0},\nonumber\\*
 |\operatorname{comp}G_\sim(D_a,I_a)|&\leq c_a,
 &C'\in\operatorname{comp}G_\sim(D_a,I_a)
 &\ \Longrightarrow\ C'\cap\{[i]_\sim:i\in I_a\}\ne\varnothing,
                       \nonumber\\*
 \pi&\subseteq\binom{[D]}2,
 &A\cap B&=\varnothing\quad(A\ne B\in\pi).
 \label{eq:quartet-count-hypotheses}
\end{align}
For every $\mathbf i\in[N]^I$,
\begin{align}
 &\left(\frac{3\beta^2}{8N^3}\right)^{D-|\pi|}
   \sum_{\mathbf e\in\mathcal A_\sim(\mathbf i)}
        \prod_{\{a,b\}\in\pi}\prod_{h=1}^4\mathbf1_{e_{ah}=e_{bh}}
        \nonumber\\*
 &\qquad\leq
   \left(\frac{3\beta^2}{8}\right)^{D-|\pi|}
       N^{-\left\lceil\frac12\sum_{a=1}^k
                         (|\{i_u:u\in I_a\}|-c_a)\right\rceil}.
 \label{eq:incidence-gain}
\end{align}
More generally, let $I=[L]\sqcup F$, fix distinct numerical labels
$(j_f)_{f\in F}$, and suppose
\begin{align}
 \sum_{a=1}^k\mathbf1_{u\in I_a}&\leq2\quad(u\in[L]).
 \label{eq:averaged-variable-multiplicity}
\end{align}
Then, with $(|F|+L)^L=1$ when $L=0$,
\begin{align}
 &\frac1{N^L}\left(\frac{3\beta^2}{8N^3}\right)^{D-|\pi|}
   \sum_{\mathbf i\in[N]^L}
   \sum_{\mathbf e\in\mathcal A_\sim(\mathbf i,\mathbf j)}
        \prod_{\{a,b\}\in\pi}\prod_{h=1}^4\mathbf1_{e_{ah}=e_{bh}}
        \nonumber\\*
 &\qquad\leq (|F|+L)^L
       \left(\frac{3\beta^2}{8}\right)^{D-|\pi|}
       N^{-\left\lceil\frac12\sum_{a=1}^k(|I_a|-c_a)\right\rceil}.
 \label{eq:averaged-incidence-gain}
\end{align}
\end{lemma}
\begin{proof}
Complete $\pi$ by its singleton blocks:
\begin{align}
 \widehat\pi&=\pi\cup\{\{d\}:d\in[D]\setminus\cup\pi\},
 &b&=|\widehat\pi|=D-|\pi|.
\end{align}
Let $\approx$ be the equivalence relation generated by $\sim$ and
\begin{align}
 (a,h)&\approx(b,h)
       &&(\{a,b\}\in\pi,\ h\in[4]),\nonumber\\
 u&\approx v
       &&(u,v\in I,\ i_u=i_v).
\end{align}
An inconsistent assignment contributes zero:
\begin{align}
 &\bigl(\exists u,v\in I:\ u\approx v,\ i_u\ne i_v\bigr)
 \ \text{or}\ 
   \bigl(\exists d,\ h\ne h':\ (d,h)\approx(d,h')\bigr)
              \nonumber\\
 &\qquad\Longrightarrow\quad
 \sum_{\mathbf e\in\mathcal A_\sim(\mathbf i)}
        \prod_{\{a,b\}\in\pi}\prod_{h=1}^4\mathbf1_{e_{ah}=e_{bh}}=0.
\end{align}
Otherwise put
\begin{align}
 \mathcal S&=(I\sqcup([D]\times[4]))/\!\approx,
 &\mathcal R&=\{[u]_\approx:u\in I\},\nonumber\\
 f&=|\mathcal S\setminus\mathcal R|,
 &k_a&=|\{i_u:u\in I_a\}|,\nonumber\\
 \sum_{\mathbf e\in\mathcal A_\sim(\mathbf i)}
        \prod_{\{a,b\}\in\pi}\prod_{h=1}^4\mathbf1_{e_{ah}=e_{bh}}
 &\leq\sum_{x\in[N]^{\mathcal S\setminus\mathcal R}}1=N^f.
 \label{eq:free-class-assignment-count}
\end{align}
No distinctness condition is imposed between different classes in the
last sum.

Identify $\mathcal R$ to one vertex $*$. For $B\in\widehat\pi$,
$h\in[4]$, and any $d\in B$, define
\begin{align}
 \mathcal V&=\widehat\pi\sqcup(\mathcal S\setminus\mathcal R)\sqcup\{*\},
 &\mathcal E&=\widehat\pi\times[4],\nonumber\\
 \xi(B,h)&=
 \begin{cases}
 *,&[(d,h)]_\approx\in\mathcal R,\\
 [(d,h)]_\approx,&[(d,h)]_\approx\notin\mathcal R,
 \end{cases}
 &\mathsf D_{v,(B,h)}&=\mathbf1_{v=B}-\mathbf1_{v=\xi(B,h)}.
\end{align}
The hypotheses imply that this graph is connected. Hence its incidence
matrix satisfies
\begin{align}
 |\mathcal E|&=4b,
 &|\mathcal V|&=b+f+1,\nonumber\\
 \operatorname{rank}\mathsf D&=b+f,
 &\dim\ker\mathsf D&=3b-f.
 \label{eq:cycle-dimension}
\end{align}
A spanning tree supplies $b+f$ independent columns; every column has
coordinate sum zero.

For each $a$, the subgraph before the identification to $*$ has
\begin{align}
 \mathcal Q_a&=\{B\in\widehat\pi:B\cap D_a\ne\varnothing\},
 &\mathcal E_a&=\mathcal Q_a\times[4],\nonumber\\
 \mathcal V_a^0&=\mathcal Q_a\sqcup
                 \{[u]_\approx:u\in I_a\sqcup(D_a\times[4])\},
 &\partial_a^0(B,h)&=(B,[(d,h)]_\approx)\quad(d\in B\cap D_a),\nonumber\\
 |\operatorname{comp}(\mathcal V_a^0,\mathcal E_a)|&\leq c_a,
 &|\mathcal V_a^0\cap\mathcal R|&\geq k_a.
\end{align}
The quotient cannot increase the number of components. If $\mathsf D_a^0$
and $\mathsf D_a$ are the incidence matrices before and after identifying
$\mathcal V_a^0\cap\mathcal R$ to $*$, then
\begin{align}
 \dim\ker\mathsf D_a
 &=\dim\ker\mathsf D_a^0+|\mathcal V_a^0\cap\mathcal R|
                      -|\operatorname{comp}(\mathcal V_a^0,\mathcal E_a)|
 \geq k_a-c_a.
 \label{eq:subgraph-cycle-gain}
\end{align}
An empty subgraph contributes zero. Let
\begin{align}
 P&=\operatorname{proj}_{\ker\mathsf D},\nonumber\\
 P_a&=\operatorname{proj}_{\{z\in\mathbb R^{\mathcal E}:
                 \mathsf Dz=0,\ z_e=0\ (e\notin\mathcal E_a)\}},
 &R_a&=\operatorname{diag}_{e\in\mathcal E}\mathbf1_{e\in\mathcal E_a}.
\end{align}
Then
\begin{align}
 \Tr P_a&\geq k_a-c_a,
 &PP_a=P_aP&=P_a,
 &0\preceq P_a&\preceq R_a,\nonumber\\
 \sum_{a=1}^k\mathbf1_{(B,h)\in\mathcal E_a}
 &=\sum_{a=1}^k\mathbf1_{B\cap D_a\ne\varnothing}
 \leq |B|\leq2,
 &\sum_aR_a&\preceq2I,\nonumber\\
 \sum_a(k_a-c_a)
 &\leq\sum_a\Tr P_a
   =\Tr\left(P\sum_aP_aP\right)\nonumber\\
 &\leq2\Tr P=2(3b-f),\nonumber\\
 \left(\frac{3\beta^2}{8N^3}\right)^bN^f
 &\leq\left(\frac{3\beta^2}{8}\right)^b
                 N^{-\lceil\frac12\sum_a(k_a-c_a)\rceil}.
 \label{eq:incidence-normalized-assignment}
\end{align}
Together with \eqref{eq:free-class-assignment-count}, this proves
\eqref{eq:incidence-gain}.

For \eqref{eq:averaged-incidence-gain}, let $\operatorname{Part}(I)$
be the set of partitions of $I=[L]\sqcup F$, and set $x_u=i_u$ for
$u\in[L]$ and $x_f=j_f$ for $f\in F$. For
$\mathcal Q\in\operatorname{Part}(I)$ with $|C'\cap F|\leq1$ for
$C'\in\mathcal Q$,
\begin{align}
 K&=|\{C'\in\mathcal Q:C'\cap F=\varnothing\}|,\nonumber\\
 \left|\left\{\mathcal Q\in\operatorname{Part}(I):
              |C'\cap F|\leq1\quad(C'\in\mathcal Q)\right\}\right|
 &\leq(|F|+L)^L,\nonumber\\
 \left|\left\{\mathbf i\in[N]^L:
          \{x^{-1}(z):z\in x(I)\}=\mathcal Q\right\}\right|
 &\leq N^K.
\end{align}
For a class $C'$ avoiding $F$, write
\begin{align}
 d&=|C'|,
 &n_a&=|C'\cap I_a|,
 &h&=|\{a:n_a>0\}|,\nonumber\\
 \sum_a n_a&\leq2d,
 &\sum_a(n_a-1)_+
 &\leq
 \begin{cases}
 0,&h=0,\\
 d-1,&h=1,\\
 2d-h\leq2(d-1),&h\geq2.
 \end{cases}
\end{align}
For a class meeting $F$, put $d=|C'\cap[L]|$; its reduction of
$\sum_a|I_a|$ is at most $2d$. Summing gives
\begin{align}
 \sum_a(|I_a|-k_a)&\leq2(L-K),
 \label{eq:owner-collision-loss}\\
 N^{-L}N^K N^{-\lceil\frac12\sum_a(k_a-c_a)\rceil}
 &\leq N^{-\lceil\frac12\sum_a(|I_a|-c_a)\rceil}.
\end{align}
To count the partitions, assign each element of $[L]$ either its class's
fixed element in $F$ or its least element in $[L]$. There are at most
$(|F|+L)^L$ resulting maps. Applying \eqref{eq:incidence-gain} in each
class of assignments proves the averaged estimate.
\end{proof}

\begin{lemma}[Summation of Gaussian expectations]\label[lemma]{lem:causal-moments}
Fix $C_0\geq1$, integers $d_0,k,L\geq0$, and $r\geq0$.
Let $I=[L]\sqcup F$, with distinct fixed labels $(j_f)_{f\in F}$.
For each $n\geq0$, let $\Lambda_n$ be finite, and for
$\lambda\in\Lambda_n$ set $D=n+d_0$. Assume
\begin{align}
 |\Lambda_n|&\leq C_0^{n+1}(1+s+n)^{C_0},
 &|c_\lambda(\mathbf e,\mathbf i,\mathbf u)|&\leq1,
 \label{eq:gaussian-summation-data}\\
 F_\lambda(J;\mathbf e,\mathbf i,\mathbf u)
 &=\prod_{\nu=1}^{q_\lambda}
      \left(e^{h_{\lambda\nu}\beta Q_{\lambda\nu}H_\kappa
                                      Q_{\lambda\nu}^\dagger}
        U^{m_{\lambda\nu}}(t_{\lambda\nu},s_{\lambda\nu})C_{\lambda\nu}\right),
                    \nonumber\\
 h_{\lambda\nu}&\geq0,
 &\sum_{\nu=1}^{q_\lambda}h_{\lambda\nu}&=1,\nonumber\\
 Q_{\lambda\nu}^\dagger Q_{\lambda\nu}&=I,
 &\|C_{\lambda\nu}\|&\leq1,\nonumber\\
 \sum_{\nu=1}^{q_\lambda}|t_{\lambda\nu}-s_{\lambda\nu}|&\leq C_0T
                  &&(\mathbf u\in\Delta_n(C_0T)).
 \label{eq:gaussian-summation-products}
\end{align}
All parameters and coefficient functions are measurable and independent
of $J$ at fixed $\mathbf e,\mathbf i,\mathbf u$; the profiles and masks satisfy
\eqref{eq:real-profile-evolution}. For each $\lambda$, suppose
$\sim_\lambda,(D_{a\lambda},I_{a\lambda},c_{a\lambda})_{a=1}^k$
satisfy \eqref{eq:quartet-count-hypotheses} and
\eqref{eq:averaged-variable-multiplicity}, and
\begin{align}
 \sum_{a=1}^k(|I_{a\lambda}|-c_{a\lambda})&\geq2r.
 \label{eq:gaussian-summation-gain}
\end{align}
There is $C=C(\beta,C_0,d_0,k,L)$ such that
\begin{align}
 &\frac1{N^L\EE_J\Tr e^{\beta H_\kappa}}
   \sum_{\mathbf i\in[N]^L}\sum_{n\geq0}
       \left(\beta\sqrt{\frac3{8N^3}}\right)^{n+d_0}
   \sum_{\lambda\in\Lambda_n}
   \sum_{\mathbf e\in\mathcal A_{\sim_\lambda}(\mathbf i,\mathbf j)}
          \nonumber\\*
 &\quad\times\int_{\Delta_n(C_0T)}
     |c_\lambda(\mathbf e,\mathbf i,\mathbf u)|
     \left|\EE_J\Tr\left[
          \left(\prod_{d=1}^{n+d_0}J_{E_d}\right)
                F_\lambda(J;\mathbf e,\mathbf i,\mathbf u)\right]\right|
                    d\mathbf u\nonumber\\*
 &\qquad\leq C(1+s)^C(1+|F|)^L(1+T)^C
                           e^{C(1+T)^2}N^{-\lceil r\rceil}.
 \label{eq:causal-time-envelope}
\end{align}
The constants are independent of $N,s,F,T$ and $q_\lambda$.
\end{lemma}
\begin{proof}
For $D=n+d_0$ and $\ell=|\pi|$, the Gaussian identity and derivative
bound give
\begin{align}
 b&=D-\ell,
 &\left(\beta\sqrt{\frac3{8N^3}}\right)^{D+D-2\ell}
     &=\left(\frac{3\beta^2}{8N^3}\right)^b,
 \label{eq:gaussian-coefficient-count}
\end{align}
By \eqref{eq:partial-gaussian-matching} and
\eqref{eq:complete-contour-derivative},
\begin{align}
 &\frac{\left(\beta\sqrt{3/(8N^3)}\right)^D}
             {\EE_J\Tr e^{\beta H_\kappa}}
       \left|\EE_J\Tr\left[J_{E_1}\cdots J_{E_D}F_\lambda\right]\right|
                  \nonumber\\
 &\qquad\leq
   \sum_{\substack{\pi\subseteq\binom{[D]}2\\
                       A\cap B=\varnothing\ (A\ne B\in\pi)}}
       \left(\frac{3\beta^2}{8N^3}\right)^{D-|\pi|}
       (1+C_0T)^{D-2|\pi|}
           \prod_{\{a,b\}\in\pi}\prod_{h=1}^4\mathbf1_{e_{ah}=e_{bh}}.
\end{align}
Apply \eqref{eq:averaged-incidence-gain} to each $\pi$. Since
\begin{align}
 \left|\left\{\pi\subseteq\binom{[D]}2:
       |\pi|=\ell,\ A\cap B=\varnothing\ (A\ne B\in\pi)\right\}\right|
 &=\frac{D!}{2^\ell\ell!(D-2\ell)!},\nonumber\\
 \sum_{\ell=0}^{\lfloor D/2\rfloor}
       \frac{D!x^{D-2\ell}}{2^\ell\ell!(D-2\ell)!}
 &=\EE_{g\sim\mathcal N(0,1)}(x+g)^D
 \leq\EE(x+|g|)^D\quad(x\geq0),
\end{align}
the left side of \eqref{eq:causal-time-envelope} is at most
\begin{align}
 &C(1+|F|)^L N^{-\lceil r\rceil}
   \sum_{n\geq0}\frac{(CT)^n}{n!}(1+s+n)^C
          \EE[C(1+T)+|g|]^{n+d_0}\nonumber\\*
 &\quad\leq C(1+s)^C(1+|F|)^L N^{-\lceil r\rceil}
   \EE\left[(1+T+|g|)^{d_0}
       \sum_{n\geq0}(n+1)^C
           \frac{[CT(C(1+T)+|g|)]^n}{n!}\right]\nonumber\\*
 &\quad\leq C(1+s)^C(1+|F|)^L N^{-\lceil r\rceil}
   \EE(1+T+|g|)^C e^{CT(C(1+T)+|g|)}\nonumber\\*
 &\quad\leq C(1+s)^C(1+|F|)^L(1+T)^C
                         e^{C(1+T)^2}N^{-\lceil r\rceil}.
 \label{eq:gaussian-majorant-sum}
\end{align}
The constants can increase from line to line. The last two inequalities use
\begin{align}
 \sum_{n\geq0}(n+1)^C\frac{x^n}{n!}
 &\leq C'(1+x)^{\lceil C\rceil}e^x,\nonumber\\
 -\frac{u^2}{2}+au&=-\frac{(u-a)^2}{2}+\frac{a^2}{2},\nonumber\\
 \EE(1+T+|g|)^Ce^{a|g|}
 &\leq C'(1+T+a)^Ce^{a^2/2}\quad(a\geq0).
\end{align}
Tonelli's theorem applies to these nonnegative majorants. For the
expansions in \Cref{lem:commutator-expansion-structure}, only finitely
many $n$ occur at fixed $N$. Their original expectations and Gaussian
integrations by parts are therefore justified by
\Cref{lem:retained-product-derivatives}.
\end{proof}

\begin{corollary}[High moments of physical commutators]
\label[corollary]{cor:physical-commutator-moments}
Let $B,B_0,\ldots,B_j,A_1,\ldots,A_h$ be Hermitian operators of the form
\begin{align}
 A&=b_0I+\sum_{\mu=1}^r\frac{b_\mu}{N}
        \sum_{i=1}^N\prod_{\nu=1}^{2d_\mu}\psi_i(t_{\mu\nu}),
 &\psi_i(t)&=e^{-it\beta H}\psi_i e^{it\beta H},\nonumber\\
 r&\in\{0,\ldots,K\},\quad d_\mu\in\{1,\ldots,K\},
 &|b_0|+\sum_{\mu=1}^r|b_\mu|&\leq K,\nonumber\\
 |t_{\mu\nu}|&\leq T,
 &A&=A^\dagger,
 \label{eq:empirical-observable-class}
\end{align}
where $K\geq1$ and $h\geq0$ are fixed integers, and put $V=A_1\cdots A_h$.
For fixed integers $q,j\geq1$,
\begin{align}
 &\Phi_N\left[V^\dagger
   \left((\ad_{B_j}\cdots\ad_{B_1}B_0)^\dagger
                   (\ad_{B_j}\cdots\ad_{B_1}B_0)\right)^qV\right]
   \leq C(1+T)^Ce^{C(1+T)^2}N^{-jq},
 \label{eq:nested-empirical-high-moment}\\
 &\Phi_N\left[V^\dagger
   \left((\ad_B^jW)^\dagger\ad_B^jW\right)^qV\right]
   \leq C(1+s)^C(1+T)^Ce^{C(1+T)^2}N^{-jq},
 \label{eq:nested-anchored-high-moment}
\end{align}
where $W$ satisfies \eqref{eq:closed-real-time-word}. For fixed
$j,h\geq0$ and $q\geq1$,
\begin{align}
 &\Phi_N\left[B^h
   \left((\ad_B^j[B,[-\beta H,B]])^\dagger
                    \ad_B^j[B,[-\beta H,B]]\right)^qB^h\right]
   \leq C(1+T)^Ce^{C(1+T)^2}N^{-(j+1)q}.
 \label{eq:curvature-high-moment}
\end{align}
The three inequalities also hold after replacing each displayed
nested commutator by its adjoint.
The constants depend on $\beta,K,h,j,q$ and the fixed degree $p$ in
\eqref{eq:closed-real-time-word}, but not on $N,s,T$.
\end{corollary}
\begin{proof}
Each occurrence of an empirical sum has a separate summation index:
\begin{align}
 \prod_{a=1}^l\left(\frac1N\sum_{i=1}^NY_a(i)\right)
 &=\frac1{N^l}\sum_{i_1,\ldots,i_l=1}^N
                              Y_1(i_1)\cdots Y_l(i_l),\nonumber\\
 [cI,X]&=[X,cI]=0.
\end{align}
For the $2q$ nested expressions in
\eqref{eq:nested-empirical-high-moment}, take disjoint $I_a$ with
\begin{align}
 |I_a|&=j+1,
 &c_{a\lambda}&=1,
 &\frac12\sum_{a=1}^{2q}(|I_a|-c_{a\lambda})&=jq,\nonumber\\
 \sum_{a=1}^{2q}\mathbf1_{u\in I_a}&=1
                        &&\left(u\in\bigcup_a I_a\right).
\end{align}
Apply \Cref{lem:commutator-expansion-structure} separately to these
expressions and keep their quartet-position sets $D_{a\lambda}$ disjoint.
Multiplying their sums and using \eqref{eq:common-simplex-decomposition}
gives \eqref{eq:gaussian-summation-data}--\eqref{eq:gaussian-summation-products}.
The additional factors are $e^{\beta H},V,V^\dagger$; their propagator
parameters have sum at most $C T$. The scalar coefficients and their
normalized empirical sums satisfy
\begin{align}
 \|A\|&\leq |b_0|+\sum_{\mu=1}^r|b_\mu|\leq K,
 &N^{-l}\sum_{\mathbf i\in[N]^l}1&=1.
\end{align}
Consequently \Cref{lem:causal-moments} proves the first estimate.

For \eqref{eq:nested-anchored-high-moment}, include all distinct static
sites of $W$, not only sites chosen in a product rule:
\begin{align}
 F&=\bigcup_{\nu=1}^s A_\nu,
 &|F|&\leq ps,\nonumber\\
 I_a&=\{i_{a1},\ldots,i_{aj}\}\sqcup F,
 &c_{a\lambda}&\leq|F|,\nonumber\\
 \frac12\sum_{a=1}^{2q}(|I_a|-c_{a\lambda})&\geq jq,
 &(1+|F|)^{2qj}&\leq(1+ps)^{2qj}.
\end{align}
The component condition is \eqref{eq:word-expansion-bounds}. All factors
in \eqref{eq:causal-time-envelope} depending on $F$ are therefore
polynomial in $1+s$. This proves the second estimate. If $F=\varnothing$,
then $W=I$ and $\ad_B^jW=0$.

Finally,
\begin{align}
 [\beta H,\psi_i(t)]
 &=e^{-it\beta H}[\beta H,\psi_i]e^{it\beta H},\nonumber\\
 [\beta H,\psi_i]
 &=2\beta\sqrt{\frac3{8N^3}}\sum_{E\ni i}J_E\Gamma_E\psi_i,
 \label{eq:energy-derivative-star}\\
 [\beta H,\prod_{\nu=1}^{2d}\psi_i(t_\nu)]
 &=\sum_{\mu=1}^{2d}
      \left(\prod_{\nu<\mu}\psi_i(t_\nu)\right)
      e^{-it_\mu\beta H}[\beta H,\psi_i]e^{it_\mu\beta H}
      \left(\prod_{\nu>\mu}\psi_i(t_\nu)\right).
\end{align}
Thus each copy of $\ad_B^j[B,[-\beta H,B]]$ has one factor with
$\varepsilon_{a\nu}=1$ in \eqref{eq:recursive-commutator-expressions}, and
\begin{align}
 d_0&=2q,
 &|I_a|&=j+2,
 &c_{a\lambda}&=1,\nonumber\\
 \frac12\sum_{a=1}^{2q}(|I_a|-c_{a\lambda})&=(j+1)q.
\end{align}
The same Gaussian summation proves \eqref{eq:curvature-high-moment}.
For the reversed products,
\begin{align}
 (U_0\psi_{k_1}U_1\cdots\psi_{k_l}U_l)^\dagger
 &=U_l^\dagger\psi_{k_l}U_{l-1}^\dagger\cdots\psi_{k_1}U_0^\dagger,\nonumber\\
 \sum_{\nu=0}^l\tau_\nu&=\sum_{\nu=0}^l\tau_{l-\nu}.
\end{align}
The same relations $\sim_\lambda$ and bounds apply.
\end{proof}

\subsection{Curvature and commutators of powers}

\begin{fact}[Young's inequality]\label{fact:young}
For $x,y\geq0$, $0<\alpha<1$, and $a,\varepsilon>0$,
\begin{align}
 x^\alpha y^{1-\alpha}&\leq\alpha x+(1-\alpha)y,\nonumber\\
 a x^\alpha&\leq\varepsilon x+
 (1-\alpha)\alpha^{\alpha/(1-\alpha)}
       \varepsilon^{-\alpha/(1-\alpha)}a^{1/(1-\alpha)}.
 \label{eq:young-fact}
\end{align}
\end{fact}
\begin{proof}
Concavity of $\log$ proves the first inequality for positive $x,y$;
continuity includes zero. For the second, apply the first to
\begin{align}
 a x^\alpha
 &=\left(\frac{\varepsilon x}{\alpha}\right)^\alpha
   \left[a^{1/(1-\alpha)}
          \left(\frac\alpha\varepsilon\right)^{\alpha/(1-\alpha)}
   \right]^{1-\alpha}.
\end{align}
\end{proof}

\begin{lemma}[Physical source estimates]\label[lemma]{lem:bbounds}
Let $B$ be a fixed Hermitian empirical observable as in
\Cref{cor:physical-commutator-moments}, with $\|B\|\leq1$.
For $c=1/8$ and every fixed integer $m\geq1$,
\begin{align}
 0\leq\Phi_N([B^m,[-\beta H,B^m]])
 &\leq C_m(N^{-c}M_{m-1}+N^{-mc}).
 \label{eq:source-curvature}
\end{align}
For the word $W$ in \eqref{eq:closed-real-time-word},
\begin{align}
 \|[B^m,W]\|_{2,\Phi_N}^2+
 \|[B^m,W]\|_{{\rm row},\Phi_N}^2
 &\leq C_m(1+s)^{C_m}(1+T)^{C_m}e^{C_m(1+T)^2}
       (N^{-c}M_{m-1}+N^{-mc}).
 \label{eq:source-power-commutator}
\end{align}
The same estimate holds with a bounded Hermitian empirical observable
$C$ in place of $W$, with its fixed number of factors and times in
the prefactor. The right-hand sides use only moments of $B$.
\end{lemma}
\begin{proof}
For any matrix $D$, $q\geq1$, and $\varepsilon>0$, spectral calculus and
Cauchy--Schwarz in $\Phi_N$ give
\begin{align}
 D^\dagger D&\preceq\varepsilon I+
                 \varepsilon^{1-q}(D^\dagger D)^q,\nonumber\\
 |\Phi_N(B^aDB^b)|
 &\leq\sqrt{M_a}\left[
       \varepsilon M_b+
       \varepsilon^{1-q}\Phi_N(B^b(D^\dagger D)^qB^b)
                    \right]^{1/2}.
 \label{eq:source-operator-young}
\end{align}
The product rule gives
\begin{align}
 [B^m,[-\beta H,B^m]]
 &=\sum_{l,r=0}^{m-1}
       B^{l+r}[B,[-\beta H,B]]B^{2m-2-l-r},\nonumber\\
 B^aDB^b&=B^{a-1}DB^{b+1}+B^{a-1}[B,D]B^b\quad(a\geq1),\nonumber\\
 B^aDB^b&=B^{a+1}DB^{b-1}-B^a[B,D]B^{b-1}\quad(b\geq1).
\end{align}
Move a factor only from a side longer than $m-1$. The pair consisting
of the total outside degree and the absolute degree difference decreases
lexicographically in each resulting term. Hence
\begin{align}
 [B^m,[-\beta H,B^m]]
 &=\sum_{\substack{a,b,j\geq0\\a,b\leq m-1\\a+b+j=2m-2}}
     c_{abj}B^a\ad_B^j[B,[-\beta H,B]]B^b,
 &\sum_{a,b,j}|c_{abj}|&\leq C_m.
 \label{eq:balanced-source-curvature}
\end{align}
Use \eqref{eq:source-operator-young} with $q=2m$ and
$\varepsilon=N^{-2(j+1)c}$. By \eqref{eq:curvature-high-moment},
\begin{align}
 |\Phi_N(B^a\ad_B^j[B,[-\beta H,B]]B^b)|
 &\leq N^{-(j+1)c}\sqrt{M_aM_b}
       +C_mN^{-(j+1)(m-c(2m-1))}\nonumber\\
 &\leq N^{-(j+1)c}\sqrt{M_aM_b}+C_mN^{-mc}.
\end{align}
For $m\geq2$, scalar moment interpolation and \Cref{fact:young} give
\begin{align}
 \sqrt{M_aM_b}&\leq M_{m-1}^{(a+b)/(2m-2)},\nonumber\\
 N^{-(j+1)c}M_{m-1}^{1-j/(2m-2)}
 &=\bigl(N^{-c}M_{m-1}\bigr)^{1-j/(2m-2)}
       \bigl(N^{-(2m-1)c}\bigr)^{j/(2m-2)}\nonumber\\
 &\leq N^{-c}M_{m-1}+N^{-mc}.
\end{align}
The endpoints $j=0,2m-2$ follow directly; for $m=1$, use $M_0=1$.
Positivity follows in an eigenbasis $-\beta H\ket a=E_a\ket a$:
\begin{align}
 \Tr(e^{\beta H}[F,[-\beta H,F]])
 &=\sum_{a,b}(E_b-E_a)(e^{-E_a}-e^{-E_b})|F_{ab}|^2\geq0,
 \qquad F=F^\dagger.
 \label{eq:source-curvature-positive}
\end{align}
This proves \eqref{eq:source-curvature}.

For the commutator of a power, the two exact expansions are
\begin{align}
 [B^m,W]&=\sum_{j=1}^m\binom mj\ad_B^j(W)B^{m-j},
 \label{eq:power-commutator-binomial}\\
 [B^m,W]&=\sum_{j=1}^m(-1)^{j-1}\binom mj B^{m-j}\ad_B^j(W).
\end{align}
Use the first expansion for the column norm and the second for the
row norm. With $q=2m$ and cutoff $N^{-jc}$,
\eqref{eq:nested-anchored-high-moment} implies
\begin{align}
 &\|\ad_B^j(W)B^{m-j}\|_{2,\Phi_N}^2
    +\|B^{m-j}\ad_B^j(W)\|_{{\rm row},\Phi_N}^2\nonumber\\
 &\quad\leq2N^{-jc}M_{m-j}
   +C_m(1+s)^{C_m}(1+T)^{C_m}e^{C_m(1+T)^2}
                       N^{-j(2m-c(2m-1))}.
\end{align}
The latter exponent is at least $mc$. For $m>1$ and $1\leq j<m$,
\begin{align}
 M_{m-j}&\leq M_{m-1}^{(m-j)/(m-1)},\nonumber\\
 N^{-jc}M_{m-j}
 &\leq\bigl(N^{-c}M_{m-1}\bigr)^{(m-j)/(m-1)}
       \bigl(N^{-mc}\bigr)^{(j-1)/(m-1)}\nonumber\\
 &\leq N^{-c}M_{m-1}+N^{-mc}.
 \label{eq:lower-degree-interpolation}
\end{align}
For $j=m$ use $M_0=1$. Finally, in either norm,
\begin{align}
 \left\|\sum_{j=1}^m X_j\right\|^2&\leq m\sum_{j=1}^m\|X_j\|^2.
\end{align}
This proves \eqref{eq:source-power-commutator}. For empirical $C$,
replace \eqref{eq:nested-anchored-high-moment} by
\eqref{eq:nested-empirical-high-moment} in the same calculation.
\end{proof}

\begin{corollary}[Products with nonzero total real time]
\label[corollary]{cor:source-general-real-word}
The bound \eqref{eq:source-power-commutator} holds for
\begin{align}
 W&=e^{il_0\beta H}P_1e^{il_1\beta H}\cdots P_se^{il_s\beta H},
 &\sum_a|l_a|&\leq T,
\end{align}
without requiring $\sum_a l_a=0$.
For every Hermitian $F$ and $t\in\mathbb R$,
\begin{align}
 \|[e^{it\beta H},F]\|_{2,\Phi_N}^2+
 \|[e^{it\beta H},F]\|_{{\rm row},\Phi_N}^2
 &\leq C(|t|+t^2)\Phi_N([F,[-\beta H,F]]).
 \label{eq:whole-power-motion}
\end{align}
\end{corollary}
\begin{proof}
For real $x,t$,
\begin{align}
 |e^{itx}-1|^2&\leq\min\{4,t^2x^2\},\nonumber\\
 |x|\leq1&\ \Longrightarrow\
 |e^{itx}-1|^2\leq Ct^2x\tanh(x/2),\nonumber\\
 |x|\geq1&\ \Longrightarrow\
 |e^{itx}-1|^2\leq2|tx|\leq C|t|x\tanh(x/2).
\end{align}
Symmetrizing the energy-basis sums and using
\eqref{eq:source-curvature-positive} proves
\eqref{eq:whole-power-motion}.

Put $t=\sum_a l_a$ and
$V=e^{il_0\beta H}P_1e^{il_1\beta H}\cdots P_se^{i(l_s-t)\beta H}$.
Then $W=Ve^{it\beta H}$ and $V$ has zero total time and absolute
duration at most $2T$. The identities
\begin{align}
 [Ve^{it\beta H},F]
 &=[V,F]e^{it\beta H}+V[e^{it\beta H},F],\nonumber\\
 [Ve^{it\beta H},F]
 &=e^{it\beta H}[e^{-it\beta H}Ve^{it\beta H},F]
   +[e^{it\beta H},F]e^{-it\beta H}Ve^{it\beta H}
\end{align}
give the two norm estimates
\begin{align}
 \|[Ve^{it\beta H},F]\|_{2,\Phi_N}^2
 &\leq2\|[V,F]\|_{2,\Phi_N}^2
       +2\|[e^{it\beta H},F]\|_{2,\Phi_N}^2,\nonumber\\
 \|[Ve^{it\beta H},F]\|_{{\rm row},\Phi_N}^2
 &\leq2\|[e^{-it\beta H}Ve^{it\beta H},F]\|_{{\rm row},\Phi_N}^2
       +2\|[e^{it\beta H},F]\|_{{\rm row},\Phi_N}^2.
 \label{eq:partial-word-two-orientations}
\end{align}
Here $e^{it\beta H}$ commutes with the Gibbs density. The conjugated
word has duration at most $4T$. Apply \Cref{lem:bbounds} with $F=B^m$.
\end{proof}

\subsection{Ordered products of empirical observables}

\begin{lemma}[A scalar sum of squares]\label[lemma]{lem:scalar-moment-squares}
For $\alpha\in\mathbb Z_{\geq0}^q$ with $|\alpha|=2m$, both polynomials
\begin{align}
 \sum_{j=1}^q\frac{\alpha_j}{2m}x_j^{2m}\pm x^\alpha
 \label{eq:scalar-moment-polynomials}
\end{align}
are finite sums of squares of real homogeneous polynomials of degree
$m$. The representations depend only on $\alpha$.
\end{lemma}
\begin{proof}
For the minus sign, write
\begin{align}
 F_\alpha(x)&=\sum_{j=1}^q\frac{\alpha_j}{2m}x_j^{2m}-x^\alpha.
 \label{eq:ag-sos-polynomial}
\end{align}
Whenever $\gamma+\delta=\alpha$ and $|\gamma|=|\delta|=m$,
\begin{align}
 F_\alpha&=\tfrac12F_{2\gamma}+\tfrac12F_{2\delta}
                 +\tfrac12(x^\gamma-x^\delta)^2.
 \label{eq:ag-sos-recursion}
\end{align}
Thus it suffices to treat even multi-indices. At each even nonpure
$\eta$, fix distinct $i,j$ with $\eta_i,\eta_j>0$ and put
\begin{align}
 \gamma_\eta&=\eta/2+e_i-e_j,
 &\delta_\eta&=\eta/2-e_i+e_j.
\end{align}
Starting at $\alpha$, let $\alpha_n$ move from $\eta$ to
$2\gamma_\eta$ or $2\delta_\eta$, each with probability $1/2$,
and stop at the first pure index, at time $\tau$. Then
\begin{align}
 \EE[\|\alpha_{n+1}\|_2^2-\|\alpha_n\|_2^2\mid\alpha_n]
 &=8\mathbf1_{\{n<\tau\}},\nonumber\\
 8\EE(\tau\wedge n)
 &=\EE\|\alpha_{\tau\wedge n}\|_2^2-\|\alpha\|_2^2
 \leq(2m)^2,\nonumber\\
 \EE\tau&\leq(2m)^2/8.
\end{align}
Iterating \eqref{eq:ag-sos-recursion} gives
\begin{align}
 F_\alpha(x)
 &=\EE F_{\alpha_{\tau\wedge n}}(x)
   +\frac12\EE\sum_{h<\tau\wedge n}
        (x^{\gamma_{\alpha_h}}-x^{\delta_{\alpha_h}})^2,\nonumber\\
 |\EE F_{\alpha_{\tau\wedge n}}(x)|
 &\leq\max_{|\eta|=2m}|F_\eta(x)|\Pr(\tau>n)\longrightarrow0,\nonumber\\
 F_\alpha(x)
 &=\frac12\sum_{\substack{|\eta|=2m\\\eta\text{ even, nonpure}}}
    \EE\#\{h<\tau:\alpha_h=\eta\}
        (x^{\gamma_\eta}-x^{\delta_\eta})^2.
 \label{eq:ag-sos-finite-representation}
\end{align}
The final sum is finite and has nonnegative finite coefficients.
For the plus sign, change the sign of an odd coordinate of $\alpha$.
If all coordinates are even, use directly
\begin{align}
 \sum_j\frac{\alpha_j}{2m}x_j^{2m}+x^\alpha
 &=\sum_j\frac{\alpha_j}{2m}(x_j^m)^2+(x^{\alpha/2})^2.
\end{align}
\end{proof}

\begin{lemma}[Ordered moment inequality]\label[lemma]{lem:ordered-young}
Let $B_1,\ldots,B_q$ be a fixed family of Hermitian empirical
observables as in \Cref{cor:physical-commutator-moments}. For every
ordered word of length $2m$,
\begin{align}
 |\Phi_N(B_{i_1}\cdots B_{i_{2m}})|
 &\leq C_{m,q}\sum_{j=1}^q\Phi_N(B_j^{2m})+C N^{-m/4}.
 \label{eq:ordered-young}
\end{align}
The first constant depends only on $m,q$. The additive constant also
depends on the fixed coefficient bounds and real times; for times in
$[-T,T]$, it is at most $C(1+T)^Ce^{C(1+T)^2}$.
\end{lemma}
\begin{proof}
For $m=1$, Cauchy--Schwarz gives
\begin{align}
 |\Phi_N(B_iB_j)|
 &\leq\sqrt{\Phi_N(B_i^2)\Phi_N(B_j^2)}
 \leq\tfrac12\Phi_N(B_i^2)+\tfrac12\Phi_N(B_j^2).
\end{align}
Suppose the result holds at all smaller positive orders.
Two words with the same commuting monomial differ by adjacent exchanges,
so their difference is a finite sum of $U[B_i,B_j]V$. The identities
\begin{align}
 (UB_i)DV&=UD(B_iV)+U[B_i,D]V,\nonumber\\
 UD(B_iV)&=(UB_i)DV-U[B_i,D]V
\end{align}
move a factor from a side longer than $m-1$. In each resulting term,
$(|U|+|V|,\,\lvert|U|-|V|\rvert)$ decreases lexicographically. The
process stops with
\begin{align}
 D&=\ad_{B_{j_r}}\cdots\ad_{B_{j_1}}B_{j_0},
 &1\leq r&\leq2m-1,\nonumber\\
 u=|U|&\leq m-1,
 &v=|V|&\leq m-1,
 &u+v&=2m-r-1.
 \label{eq:balanced-ordered-remainder}
\end{align}
The number and coefficients of these terms depend only on $m,q$.

For $0\leq l<m$, moment interpolation and the induction hypothesis imply
\begin{align}
 \sum_i\Phi_N(B_i^{2l})
 &\leq q^{1-l/m}\left(\sum_i\Phi_N(B_i^{2m})\right)^{l/m},\nonumber\\
 \Phi_N(UU^\dagger)
 &\leq C\left(\sum_i\Phi_N(B_i^{2u})+N^{-u/4}\right)
 \leq C\left(\sum_i\Phi_N(B_i^{2m})+N^{-m/4}\right)^{u/m},\nonumber\\
 \Phi_N(V^\dagger V)
 &\leq C\left(\sum_i\Phi_N(B_i^{2m})+N^{-m/4}\right)^{v/m}.
\end{align}
Empty words have expectation one. Take the high moment at order $2m$:
\begin{align}
 D^\dagger D
 &\preceq N^{-r/2}I+N^{r(2m-1)/2}(D^\dagger D)^{2m},\nonumber\\
 |\Phi_N(UDV)|
 &\leq\Phi_N(UU^\dagger)^{1/2}
   \left[N^{-r/2}\Phi_N(V^\dagger V)
      +N^{r(2m-1)/2}\Phi_N(V^\dagger(D^\dagger D)^{2m}V)
   \right]^{1/2}\nonumber\\
 &\leq C N^{-r/4}
     \left(\sum_i\Phi_N(B_i^{2m})+N^{-m/4}\right)^{1-(r+1)/(2m)}
     +C N^{-r(2m+1)/4}.
 \label{eq:ordered-remainder-cutoff}
\end{align}
The last term uses \eqref{eq:nested-empirical-high-moment} and the
bounded operator norm of $U$. No factor has been moved around the state.
For $1\leq r<2m-1$, \Cref{fact:young} yields
\begin{align}
 &C N^{-r/4}
   \left(\sum_i\Phi_N(B_i^{2m})+N^{-m/4}\right)^{1-(r+1)/(2m)}
 \nonumber\\
 &\quad\leq\varepsilon\sum_i\Phi_N(B_i^{2m})
             +\varepsilon N^{-m/4}
             +C_\varepsilon N^{-rm/(2(r+1))},\nonumber\\
 \frac{rm}{2(r+1)}&\geq\frac m4,
 &\frac{r(2m+1)}4&\geq\frac m4,\nonumber\\
 |\Phi_N(UDV)|
 &\leq\varepsilon\sum_i\Phi_N(B_i^{2m})
                  +C_\varepsilon N^{-m/4}.
 \label{eq:ordered-remainder-young}
\end{align}
For $r=2m-1$, the exponent of the sum in
\eqref{eq:ordered-remainder-cutoff} is zero, and the same estimate
follows directly.

Let $W=B_{i_1}\cdots B_{i_{2m}}$ and let $\alpha$ record its multiplicities.
In \Cref{lem:scalar-moment-squares}, replace a degree-$m$ monomial
$x^\gamma$ by $B_1^{\gamma_1}\cdots B_q^{\gamma_q}$ and each square
by $P^\dagger P$. The difference between the lifted identity and its
commuting polynomial consists of the ordering errors just bounded.
Thus, for each sign,
\begin{align}
 \sum_i\frac{\alpha_i}{2m}B_i^{2m}
       \mp\frac{W+W^\dagger}{2}
 &=\sum_\nu P_\nu^\dagger P_\nu
          +\sum_\mu c_\mu U_\mu D_\mu V_\mu,
 &\sum_\mu|c_\mu|&\leq C_{m,q},\nonumber\\
 \pm\Re\Phi_N(W)
 &\leq\sum_i\frac{\alpha_i}{2m}\Phi_N(B_i^{2m})
          +\sum_\mu|c_\mu|\,|\Phi_N(U_\mu D_\mu V_\mu)|.
\end{align}
Also $(W-W^\dagger)/(2i)$ has zero commuting polynomial, so it is a
sum of the same ordering errors. Fixing $\varepsilon$ according to
their total coefficient gives
\begin{align}
 |\Phi_N(W)|
 &\leq|\Re\Phi_N(W)|+|\Im\Phi_N(W)|\nonumber\\
 &\leq C_{m,q}\sum_i\Phi_N(B_i^{2m})+C N^{-m/4}.
\end{align}
All dependence on the physical coefficients and times occurs in the
additive constant. Its time dependence follows from
\eqref{eq:causal-time-envelope}, since the proof takes only finitely
many sums, products, and fixed powers of that bound.
\end{proof}

\subsection{Transport through positive-heat factors}

\begin{lemma}[Source moments on a positive half-heat word]
\label[lemma]{lem:positive-heat-source}
Let $P_1,\ldots,P_s$ be static Clifford unitaries of bounded degree and
\begin{align}
 R(\mathbf z)&=e^{z_0\beta H}P_1e^{z_1\beta H}\cdots P_se^{z_s\beta H},
 &\sum_{a=0}^s z_a&=\tfrac12.
\end{align}
For each fixed $m\geq1$ and $0<\eta<1$,
\begin{align}
 \frac{\EE\|B^mR(\mathbf u)\|_{\rm HS}^2}{\EE\Tr e^{\beta H}}
 &\leq C_me^{C_ms}(1+s)^{K_m}(M_m+N^{-m\eta}),
 \qquad u_a\geq0,\quad\sum_a u_a=\tfrac12.
 \label{eq:positive-heat-source}
\end{align}
The constants may depend on $m,\eta$ and the fixed physical data, but
not on $s$, the site labels, or the individual $u_a$. The result also
holds with the source on the right and with a fixed number of
fixed-duration real-time local factors.
\end{lemma}
\begin{proof}
At $\mathbf z=e_b/2+i\mathbf t$, $\sum_a t_a=0$, factor
\begin{align}
 R(e_b/2+i\mathbf t)&=W_Le^{\beta H/2}W_R,
 &W_LW_L^\dagger=W_RW_R^\dagger&=I.
\end{align}
Insert opposite phases next to the heat factor to make $W_L,W_R$
zero-total-time words. Their combined absolute duration is at most
$2\sum_a|t_a|$. The binomial expansion gives
\begin{align}
 B^mW_L&=\sum_{j=0}^m\binom mj\ad_B^j(W_L)B^{m-j},\nonumber\\
 \frac{\EE\|B^mR(e_b/2+i\mathbf t)\|_{\rm HS}^2}{\EE\Tr e^{\beta H}}
 &\leq(m+1)\sum_{j=0}^m\binom mj^2
    \Phi_N\left[B^{m-j}(\ad_B^jW_L)^\dagger
                              (\ad_B^jW_L)B^{m-j}\right].
\end{align}
The $j=0$ term is $M_m$. For $j\geq1$, use the cutoff $N^{-j\eta}$
and \eqref{eq:nested-anchored-high-moment}, choosing $q$ with
$q(1-\eta)+\eta\geq m\eta$:
\begin{align}
 &\Phi_N\left[B^{m-j}(\ad_B^jW_L)^\dagger
                              (\ad_B^jW_L)B^{m-j}\right]\nonumber\\
 &\quad\leq N^{-j\eta}M_{m-j}
   +C_m(1+s)^{K_m}e^{C_m(1+\sum_a|t_a|)^2}
                   N^{-j(q(1-\eta)+\eta)},\nonumber\\
 N^{-j\eta}M_{m-j}
 &\leq N^{-j\eta}M_m^{1-j/m}
 \leq\varepsilon M_m+C_{m,\varepsilon}N^{-m\eta},\nonumber\\
 \frac{\EE\|B^mR(e_b/2+i\mathbf t)\|_{\rm HS}^2}{\EE\Tr e^{\beta H}}
 &\leq C_mM_m+C_m(1+s)^{K_m}
                  e^{C_m(1+\sum_a|t_a|)^2}N^{-m\eta}.
 \label{eq:half-heat-real-boundary}
\end{align}
The endpoint $j=m$ uses $M_0=1$.

Fix $u_a>0$, $\sum_a u_a=1/2$, and choose $\lambda>0$ below. Define the
holomorphic function on $\sum_a z_a=1/2$ by
\begin{align}
 F(\mathbf z)
 &=\frac{\EE\Tr[R(\overline{\mathbf z})^\dagger B^{2m}R(\mathbf z)]}
        {\EE\Tr e^{\beta H}}
      \exp\left[-\lambda\sum_a\frac{z_a(1/2-z_a)}{u_a}\right].
 \label{eq:half-heat-gaussian-window}
\end{align}
Gaussian exponential integrability at fixed $N$ justifies dominated
differentiation, as in \Cref{lem:retained-product-derivatives}.
Schatten H\"older bounds the trace ratio by one when $\Re z_a\geq0$.
For $z_a=x_a+it_a$ in that domain,
\begin{align}
 \Re\sum_a\frac{z_a(1/2-z_a)}{u_a}
 &=\sum_a\frac{x_a(1/2-x_a)+t_a^2}{u_a}\geq0,\nonumber\\
 \left(\sum_a|t_a|\right)^2&\leq\tfrac12\sum_a t_a^2/u_a,\nonumber\\
 \left|\exp\left[-\lambda\sum_a\frac{z_a(1/2-z_a)}{u_a}\right]\right|
 &=e^{-\lambda\sum_a t_a^2/u_a}
       \qquad(\mathbf z=e_b/2+i\mathbf t).
\end{align}
Cauchy--Schwarz applied to $B^mR(e_b/2\pm i\mathbf t)$, followed by
\eqref{eq:half-heat-real-boundary}, shows that a fixed sufficiently
large $\lambda$, independent of $s$, satisfies
\begin{align}
 \sup_{\sum_a t_a=0}|F(e_b/2+i\mathbf t)|
 &\leq C_m\bigl(M_m+(1+s)^{K_m}N^{-m\eta}\bigr).
\end{align}
The function $F$ is bounded on the whole domain $\Re z_a\geq0$.
Apply \Cref{fact:three-lines} on the lines
$(1-z)\mathbf x+z\mathbf y+i\mathbf t$ and iterate over convex
combinations of the vertices. Since $\mathbf u=\sum_b(2u_b)e_b/2$,
\begin{align}
 |F(\mathbf u)|
 &\leq\prod_{b=0}^s
     \left(\sup_{\sum_a t_a=0}|F(e_b/2+i\mathbf t)|\right)^{2u_b}
 \leq C_m\bigl(M_m+(1+s)^{K_m}N^{-m\eta}\bigr),\nonumber\\
 \frac{\EE\|B^mR(\mathbf u)\|_{\rm HS}^2}{\EE\Tr e^{\beta H}}
 &=e^{\lambda\sum_a(1/2-u_a)}F(\mathbf u)
  =e^{\lambda s/2}F(\mathbf u).
\end{align}
This proves \eqref{eq:positive-heat-source}. Zero coordinates follow
by continuity. The right-source bound follows from
$\|R B^m\|_{\rm HS}=\|B^mR^\dagger\|_{\rm HS}$. The additional
fixed-duration local factors change only the fixed boundary constants.
\end{proof}

\subsection{Physical estimates after energy pinching}

For the rounded state, write
$\|X\|_{{\rm row},\rho(u)}^2=\Tr(\rho(u)XX^\dagger)$.

\begin{lemma}[Powers and mixed products after pinching]
\label[lemma]{lem:mixed-source-pinching}
Choose $\delta_E$ and $c_0$ as in \eqref{eq:energy-width-choice}.
For every fixed integer $p\geq1$,
\begin{align}
 \EE_{J,u}\|B^p-(\Pi_uB)^p\|_{2,\rho(u)}^2
 &\leq C_p(N^{-c_0}M_{p-1}+N^{-pc_0}).
 \label{eq:all-power-pinching-input}
\end{align}
Let $C$ be a Hermitian empirical contraction with times in $[-T,T]$;
the source $B$ keeps its fixed times. For $0<c_1\leq c_0/4$ and every fixed $m\geq1$,
\begin{align}
 \EE_{J,u}\|[\Pi_uC,(\Pi_uB)^p]\|_{2,\rho(u)}^2
 &\leq C_p(1+T)^{C_p}e^{C_p(1+T)^2}
           (N^{-c_0}M_{p-1}+N^{-pc_0}),
 \label{eq:pinched-power-commutator}\\
 \EE_{J,u}\|CB^m-(\Pi_uC)(\Pi_uB)^m\|_{2,\rho(u)}^2
 &\leq C_m(1+T)^{C_m}e^{C_m(1+T)^2}
           (N^{-c_1}M_{m-1}+N^{-mc_1}).
 \label{eq:mixed-source-pinching}
\end{align}
The mixed-product bound also holds for the row norm.
\end{lemma}
\begin{proof}
Equation \eqref{eq:all-power-pinching-input} is
\Cref{lem:all-power-pinching}; its proof uses only
\eqref{eq:source-curvature} and the elementary pinching calculations in
\Cref{app:measurement-calculus}, not the conclusions below.

Fix $J,u$. The pinching map is an orthogonal projection in the weighted
Hilbert--Schmidt norm and its image commutes with $\rho(u)$. Its
bimodule property and Schwarz inequality give
\begin{align}
 [\Pi_uF,\Pi_uC]
 &=\Pi_u[F,C]-\Pi_u[F-\Pi_uF,C-\Pi_uC],
 \label{eq:pinched-commutator-bimodule}\\
 \|\Pi_u(XY)\|_{2,\rho(u)}
 &\leq\min\{\|X\|\|Y\|_{2,\rho(u)},
                  \|Y\|\|X\|_{{\rm row},\rho(u)}\},\nonumber\\
 \|B^p-\Pi_u(B^p)\|_{2,\rho(u)}
   +\|(\Pi_uB)^p-\Pi_u(B^p)\|_{2,\rho(u)}
 &\leq2\|B^p-(\Pi_uB)^p\|_{2,\rho(u)}.
\end{align}
For Hermitian $F$ and $\|C\|\leq1$, the second term in the first line
has norm at most $4\|F-\Pi_uF\|_{2,\rho(u)}$. Consequently,
\begin{align}
 \|[\Pi_uC,(\Pi_uB)^p]\|_{2,\rho(u)}
 &\leq\|[C,B^p]\|_{2,\rho(u)}
         +6\|B^p-(\Pi_uB)^p\|_{2,\rho(u)}.
\end{align}
Square, average, and apply \eqref{eq:gibbs-as-rounded-average},
\Cref{lem:bbounds}, and \eqref{eq:all-power-pinching-input}. Since
$c_0<c=1/8$, this proves \eqref{eq:pinched-power-commutator}.

Choose $m+2$ distinct fixed numbers $z_j\in[-1,1]$. The invertible
Vandermonde matrix gives constants $\lambda_j$ satisfying
\begin{align}
 \sum_{j=0}^{m+1}\lambda_jz_j^r
 &=\frac{2^{m+1}}{m+1}\delta_{r1},
 \qquad 0\leq r\leq m+1,\nonumber\\
 \frac1{m+1}\sum_{r=0}^m B^rCB^{m-r}
 &=h^{-1}\sum_{j=0}^{m+1}\lambda_j
                    \left(\frac{B+hz_jC}{2}\right)^{m+1}.
 \label{eq:shrinking-polarization}
\end{align}
The identity also holds with $B,C$ replaced by $\Pi_uB,\Pi_uC$.
For $0<h\leq1$, \Cref{lem:ordered-young} gives
\begin{align}
 \Phi_N\left[\left(\frac{B+hz_jC}{2}\right)^{2m}\right]
 &\leq C_m(M_m+h^{2m})
        +C_m(1+T)^{C_m}e^{C_m(1+T)^2}N^{-m/4}.
\end{align}
Apply \eqref{eq:all-power-pinching-input} at power $m+1$ to each
$(B+hz_jC)/2$, whose norm and coefficients are uniformly bounded.
With $h=N^{-c_0/4}$, this yields
\begin{align}
 &\EE_{J,u}\left\|
  \frac1{m+1}\sum_{r=0}^m
    \bigl[B^rCB^{m-r}-(\Pi_uB)^r(\Pi_uC)(\Pi_uB)^{m-r}\bigr]
                   \right\|_{2,\rho(u)}^2\nonumber\\
 &\quad\leq C_m(1+T)^{C_m}e^{C_m(1+T)^2}h^{-2}
     \left[N^{-c_0}(M_m+h^{2m}+N^{-m/4})+N^{-(m+1)c_0}\right]
 \nonumber\\
 &\quad\leq C_m(1+T)^{C_m}e^{C_m(1+T)^2}
     \left[N^{-c_0/2}M_{m-1}+N^{-mc_0/2}\right].
 \label{eq:symmetrized-mixed-pinching}
\end{align}
Both symmetrized products are Hermitian, so this also bounds the row norm.

The physical ordering error has the exact expansion
\begin{align}
 \frac1{m+1}\sum_{r=0}^m B^rCB^{m-r}-CB^m
 &=\sum_{j=1}^m\frac1{j+1}\binom mj\ad_B^j(C)B^{m-j}.
 \label{eq:symmetrized-product-remainder}
\end{align}
The calculation in the proof of \eqref{eq:source-power-commutator},
with cutoff $N^{-jc_0}$, bounds its squared column norm by
$C_m(1+T)^{C_m}e^{C_m(1+T)^2}(N^{-c_0}M_{m-1}+N^{-mc_0})$.
Its adjoint differs from it by $[C,B^m]$, so
\Cref{lem:bbounds} gives the same row bound.

For the pinched ordering error, use a common eigenbasis of
$\Pi_uB$ and $\rho(u)$. The scalar multipliers satisfy
\begin{align}
 \left|y^m-\frac1{m+1}\sum_{r=0}^m x^ry^{m-r}\right|
 &\leq C_m|x-y|(|x|+|y|)^{m-1},\nonumber\\
 |x^p-y^p|&\geq c_p|x-y|(|x|+|y|)^{p-1}
                         \qquad(p\text{ odd}),\nonumber\\
 \left|y^m-\frac1{m+1}\sum_{r=0}^m x^ry^{m-r}\right|
 &\leq C_m|x^m-y^m|\qquad(m\text{ odd}),\nonumber\\
 \left|y^m-\frac1{m+1}\sum_{r=0}^m x^ry^{m-r}\right|^2
 &\leq C_m|x^{m-1}-y^{m-1}|\,|x^{m+1}-y^{m+1}|
                                      \qquad(m\text{ even}).
 \label{eq:pinched-ordering-multipliers}
\end{align}
The first bound follows by factoring $x^r-y^r$. For the second, the
ratio is positive and continuous on $|x|+|y|=1$, with a positive
continuous value on $x=y$.

For odd $m$, apply \eqref{eq:pinched-power-commutator} directly.
For even $m\geq2$, Cauchy--Schwarz over the matrix entries and $J,u$
bounds the squared norm by the same time prefactor times
\begin{align}
 &\left[(N^{-c_0}M_{m-2}+N^{-(m-1)c_0})
        (N^{-c_0}M_m+N^{-(m+1)c_0})\right]^{1/2}\nonumber\\
 &\quad\leq N^{-c_0}M_{m-1}^{1-1/(2m-2)}
       +N^{-(m+2)c_0/2}M_{m-1}^{(m-2)/(2m-2)}\nonumber\\
 &\qquad+N^{-mc_0/2}M_{m-1}^{1/2}+N^{-mc_0}\nonumber\\
 &\quad\leq C_m\left(N^{-c_0/2}M_{m-1}+N^{-mc_0/2}\right).
\end{align}
The last step is \Cref{fact:young}; $M_0=1$ covers $m=2$.
The pinched products commute with $\rho(u)$, so their row and column
norms coincide. Combining the physical ordering error, the pinched
ordering error, and \eqref{eq:symmetrized-mixed-pinching} proves
\eqref{eq:mixed-source-pinching}, in fact with $c_0/2$ in place of $c_1$.
\end{proof}

\begin{lemma}[Source-dependent measurement bounds]
\label[lemma]{lem:measurement-physical-input}
Let $C,C'$ be Hermitian empirical contractions with times in $[-T,T]$,
and keep the source $B$ at its fixed times. Choose $\delta_E,c_0$ as
in \eqref{eq:energy-width-choice} and $0<c_1\leq c_0/4$.
Writing $\rho(u)=\sum_jw_jP_j(u)$, set
\begin{align}
 \|X\|_{{\rm h},u}^2
 &=\sum_{j,k}\frac{w_jw_k}{w_j+w_k}
                         \|P_k(u)XP_j(u)\|_{\rm HS}^2.
 \label{eq:harmonic-gibbs-norm}
\end{align}
For every fixed $m\geq1$,
\begin{align}
 &\EE_{J,u}\Bigg[
   \|[\Pi_uC,(\Pi_uB)^m]\|_{2,\rho(u)}^2
   +\frac1N\sum_i\|[\psi_i,\Pi_uC](\Pi_uB)^m\|_{{\rm h},u}^2
   +\frac1N\sum_i\|[\psi_i,(\Pi_uB)^m]\|_{{\rm h},u}^2
                 \Bigg]\nonumber\\
 &\qquad\leq C_m(1+T)^{C_m}e^{C_m(1+T)^2}
                     (N^{-c_1}M_{m-1}+N^{-mc_1}).
 \label{eq:harmonic-rooted-source-bound}
\end{align}
This also holds with each commutator product replaced by its adjoint.
Moreover,
\begin{align}
 &\EE_{J,u}\Big[
   \|[\Pi_uC,\Pi_uC'](\Pi_uB)^m\|_{2,\rho(u)}^2
   +\|(C-\Pi_uC)(\Pi_uB)^m\|_{2,\rho(u)}^2\Big]\nonumber\\
 &\qquad\leq C_m(1+T)^{C_m}e^{C_m(1+T)^2}
                     (N^{-c_1}M_{m-1}+N^{-mc_1}).
 \label{eq:weighted-pinched-pair-input}
\end{align}
\end{lemma}
\begin{proof}
The harmonic weights and $\psi_i^2=I$ imply
\begin{align}
 \|X\|_{{\rm h},u}^2
 &\leq\min\{\Tr(\rho(u)X^\dagger X),\Tr(\rho(u)XX^\dagger)\},\nonumber\\
 \|X^\dagger\|_{{\rm h},u}&=\|X\|_{{\rm h},u},\nonumber\\
 \|\psi_iX\|_{{\rm h},u}^2&\leq\Tr(\rho(u)X^\dagger X),
 &\|X\psi_i\|_{{\rm h},u}^2&\leq\Tr(\rho(u)XX^\dagger),\nonumber\\
 \|[\psi_i,X]\|_{{\rm h},u}^2
 &\leq2\|X\|_{2,\rho(u)}^2+2\|X\|_{{\rm row},\rho(u)}^2.
 \label{eq:harmonic-elementary-bounds}
\end{align}
A block-diagonal contraction decreases either orientation of the
harmonic norm, since it acts inside each matrix block.

The first term of \eqref{eq:harmonic-rooted-source-bound} is
\eqref{eq:pinched-power-commutator}. For the last term,
\begin{align}
 [\psi_i,(\Pi_uB)^m]
 &=[\psi_i,B^m]+[\psi_i,(\Pi_uB)^m-B^m],\nonumber\\
 \EE_{J,u}\|[\psi_i,(\Pi_uB)^m]\|_{{\rm h},u}^2
 &\leq2\|[\psi_i,B^m]\|_{2,\Phi_N}^2
       +8\EE_{J,u}\|(\Pi_uB)^m-B^m\|_{2,\rho(u)}^2.
\end{align}
Apply \Cref{lem:bbounds,lem:mixed-source-pinching}.

For the weighted probe, \eqref{eq:nested-anchored-high-moment} with
$W=\psi_i$ and empirical observable $C$ gives
\begin{align}
 \Phi_N\left[B^m
     ([\psi_i,C]^\dagger[\psi_i,C])^{2m}B^m\right]
 &\leq C_m(1+T)^{C_m}e^{C_m(1+T)^2}N^{-2m},\nonumber\\
 \|[\psi_i,C]B^m\|_{2,\Phi_N}^2
 &\leq N^{-c_1}M_m
       +C_m(1+T)^{C_m}e^{C_m(1+T)^2}N^{-2m+c_1(2m-1)}\nonumber\\
 &\leq C_m(1+T)^{C_m}e^{C_m(1+T)^2}
                     (N^{-c_1}M_{m-1}+N^{-mc_1}).
 \label{eq:weighted-physical-probe}
\end{align}
Now use the exact identities
\begin{align}
 [\psi_i,\Pi_uC](\Pi_uB)^m
 &=[\psi_i,(\Pi_uC)(\Pi_uB)^m]
                  -(\Pi_uC)[\psi_i,(\Pi_uB)^m],\nonumber\\
 [\psi_i,(\Pi_uC)(\Pi_uB)^m]
 &=[\psi_i,CB^m]
       +[\psi_i,(\Pi_uC)(\Pi_uB)^m-CB^m],\nonumber\\
 [\psi_i,CB^m]&=[\psi_i,C]B^m+C[\psi_i,B^m].
\end{align}
The mixed-product difference is bounded in both weighted norms by
\Cref{lem:mixed-source-pinching}. The other terms use
\eqref{eq:weighted-physical-probe}, \Cref{lem:bbounds}, and left
multiplication by the contraction $C$ in the column norm. Apply
\eqref{eq:harmonic-elementary-bounds} and average over $i$ to prove
\eqref{eq:harmonic-rooted-source-bound}. Adjoint invariance of the
harmonic norm proves the stated reversed version.

For the last two estimates, the same spectral cutoff gives
\begin{align}
 \|[C,C']B^m\|_{2,\Phi_N}^2
 &\leq N^{-c_1}M_m
      +N^{c_1(2m-1)}\Phi_N\left[B^m
               ([C,C']^\dagger[C,C'])^{2m}B^m\right]\nonumber\\
 &\leq C_m(1+T)^{C_m}e^{C_m(1+T)^2}
                    (N^{-c_1}M_{m-1}+N^{-mc_1}),\nonumber\\
 [CB^m,C']&=C[B^m,C']+[C,C']B^m,\nonumber\\
 [\Pi_u(CB^m),\Pi_uC']
 &=\Pi_u[CB^m,C']
    -\Pi_u[CB^m-\Pi_u(CB^m),C'-\Pi_uC'].
\end{align}
Because $(\Pi_uC)(\Pi_uB)^m$ lies in the image of $\Pi_u$,
\begin{align}
 \|CB^m-\Pi_u(CB^m)\|_{2,\rho(u)}
 &\leq\|CB^m-(\Pi_uC)(\Pi_uB)^m\|_{2,\rho(u)},\nonumber\\
 \|CB^m-\Pi_u(CB^m)\|_{{\rm row},\rho(u)}
 &\leq\|CB^m-(\Pi_uC)(\Pi_uB)^m\|_{{\rm row},\rho(u)}.
\end{align}
The pinching product inequality in the preceding proof bounds both
terms of the last commutator. Replace $\Pi_u(CB^m)$ by
$(\Pi_uC)(\Pi_uB)^m$ using \Cref{lem:mixed-source-pinching}, and use
\begin{align}
 [\Pi_uC,\Pi_uC'](\Pi_uB)^m
 &=[(\Pi_uC)(\Pi_uB)^m,\Pi_uC']
       -(\Pi_uC)[(\Pi_uB)^m,\Pi_uC'].
\end{align}
All factors in this identity are block diagonal, so multiplication by
$\Pi_uC$ or $\Pi_uC'$ is contractive in both weighted norms.
Finally,
\begin{align}
 (C-\Pi_uC)(\Pi_uB)^m
 &=C((\Pi_uB)^m-B^m)
       +CB^m-(\Pi_uC)(\Pi_uB)^m,\nonumber\\
 \|(C-\Pi_uC)(\Pi_uB)^m\|_{2,\rho(u)}^2
 &\leq2\|(\Pi_uB)^m-B^m\|_{2,\rho(u)}^2
       +2\|CB^m-(\Pi_uC)(\Pi_uB)^m\|_{2,\rho(u)}^2.
\end{align}
Average and apply \Cref{lem:mixed-source-pinching}. This proves
\eqref{eq:weighted-pinched-pair-input}.
\end{proof}

\section{Concentration of the chronological collective moments}
\label{app:mubound}
The physical moment bound concerns ordinary products of empirical
operators. The moments $\mu_p$ instead use fermionic chronological
ordering. We first give the exact reordering, keeping all repeated
site labels, and then analytically continue the resulting scalar bounds to imaginary
time.

\subsection{Reordering while retaining the scalar centers}

\begin{lemma}[Centered chronological products]
\label{lem:centered-chronological-real}
Fix $\beta\geq0$ and $T<\infty$. For $p\geq1$, let
$t_1,\ldots,t_{2p}\in[-T,T]$ and let $\pi\in S_{2p}$ satisfy
\begin{align}
 \pi^{-1}(2j-1)&<\pi^{-1}(2j)\qquad(j\in[p]).
\end{align}
For $S\subseteq[p]$, let $\pi_S$ be the list obtained from
$\pi(1),\ldots,\pi(2p)$ by retaining only indices in
$\{2j-1,2j:j\in S\}$. Its sign is measured relative to the increasing
list of these indices. Set
\begin{align}
 \psi_i(t)&=e^{-it\beta H}\psi_i e^{it\beta H},
 &h_j&=2G_*(i(t_{2j-1}-t_{2j})),\\
 W_S((i_j)_{j\in S})
 &=\operatorname{sgn}(\pi_S)
       \prod_{k=1}^{2|S|}
       \psi_{i_{\lceil\pi_S(k)/2\rceil}}(t_{\pi_S(k)}),
 &W_\varnothing&=I.
\end{align}
The product is in the displayed order. There is
$\theta_0=\theta_0(\beta,T)>0$, independent of $p$, such that
\begin{align}
 &\left|\Phi_N\left(
   \sum_{S\subseteq[p]}(-1)^{p-|S|}
       \left(\prod_{j\notin S}h_j\right)
       N^{-|S|}\sum_{(i_j)_{j\in S}\in[N]^S}
                           W_S((i_j)_{j\in S})\right)\right|
 \leq C_{\beta,T,p}N^{-\theta_0p}.
 \label{eq:chronological-centered-sum}
\end{align}
Every site sum includes repeated numerical values. The constant is
uniform over the displayed times and permutations; the times themselves
need not be distinct.
\end{lemma}
\begin{proof}
Set
\begin{align}
 Y_j&=\frac1N\sum_i\psi_i(t_{2j-1})\psi_i(t_{2j})-h_jI.
 \label{eq:chronological-ordinary-centers}
\end{align}
For fixed site indices, put $w_a=\psi_{i_{\lceil a/2\rceil}}(t_a)$
and use commuting idempotents $z_j^2=z_j$. Define
\begin{align}
 x_a&=(1-z_{\lceil a/2\rceil})I+z_{\lceil a/2\rceil}w_a,\\
 \epsilon_\pi(\mathbf z)
 &=\prod_{\substack{a<b\\\pi^{-1}(a)>\pi^{-1}(b)}}
                    (1-2z_{\lceil a/2\rceil}z_{\lceil b/2\rceil}),\\
 \left.\epsilon_\pi(\mathbf z)\prod_kx_{\pi(k)}\right|_{\mathbf z=\mathbf1_S}
 &=W_S.
 \label{eq:chronological-subset-polynomial}
\end{align}
Apply, for every $j$, the commuting linear operation
\begin{align}
 f&\longmapsto f|_{z_j=1}-h_jf|_{z_j=0}.
 \label{eq:subset-center-operation}
\end{align}
After $N^{-p}\sum_{i_1,\ldots,i_p}$ this is exactly the centered subset
sum. Unused site sums supply the missing factors of $N$.

For different pairs $j,k$, and a homogeneous expression $T$ independent
of $z_j$, the exact exchanges are
\begin{align}
 x_ax_b&=(1-2z_jz_k)x_bx_a+z_jz_k\{w_a,w_b\},
 \label{eq:optional-graded-swap}\\
 x_aT&=(1-z_j+(-1)^{|T|}z_j)Tx_a+z_j[w_a,T]_{\rm gr},
 \label{eq:optional-tree-swap}\\
 x_{2j-1}x_{2j}&\longmapsto w_{2j-1}w_{2j}-h_jI,
 &z_jT&\longmapsto T|_{z_j=1}.
 \label{eq:optional-center-operation}
\end{align}
Here $[X,Y]_{\rm gr}=XY-(-1)^{|X||Y|}YX$. The first two identities
are verified at $z_j,z_k\in\{0,1\}$; the last follows from
\eqref{eq:subset-center-operation}.

For each summand keep disjoint sets $C,F\subseteq[p]$: $C$ records
pairs already replaced by $w_{2j-1}w_{2j}-h_jI$, at the left of the
product; $F$ records pairs required to be present. If $c(\mathbf z)$ is
that summand's coefficient, this means
\begin{align}
 j\in F&\quad\Longrightarrow\quad z_jc(\mathbf z)=c(\mathbf z),
 &j&=\min([p]\setminus(C\cup F))
 \quad\text{at the next step}.
\end{align}
Use \eqref{eq:optional-graded-swap}--\eqref{eq:optional-tree-swap} to
move $x_{2j-1}$ immediately after the factors indexed by $C$, and then
move $x_{2j}$ immediately after $x_{2j-1}$. Apply each identity to both
of its summands. The updates are
\begin{align}
 \text{adjacent factors }x_{2j-1}x_{2j}:\qquad
 &x_{2j-1}x_{2j}\longmapsto w_{2j-1}w_{2j}-h_jI,
 &C&\longmapsto C\cup\{j\},\nonumber\\
 \text{term containing }z_jz_k\{w_a,w_b\}:\qquad
 &F\longmapsto F\cup\{j,k\},\nonumber\\
 \text{term containing }z_j[w_a,T]_{\rm gr}:\qquad
 &F\longmapsto F\cup\{j\}\cup
           \{\lceil b/2\rceil:b\in\mathcal I(T)\}.
\end{align}
Here $\mathcal I(T)$ is the set of original occurrence indices in $T$.
After either commutator update, restart with the least remaining index.
No factor indexed by $C$ is moved again. Every update increases
$|C\cup F|$ by at least one, and arranging the two selected occurrences
uses at most $4p$ exchanges. Therefore
\begin{align}
 \#\{\text{updates along one sequence of summands}\}&\leq p,
 &\#\{\text{exchanges along that sequence}\}&\leq4p^2.
\end{align}
The expansion is consequently finite for each fixed $p$.

The sets $\mathcal I(T)$ of original occurrences in the remaining
single factors and brackets stay disjoint:
\begin{align}
 \mathcal I([T,T']_{\rm gr})&=\mathcal I(T)\cup \mathcal I(T'),
 &\bigcup_T\mathcal I(T)&=\{\text{remaining occurrences}\}.
 \label{eq:chronological-occurrence-partition}
\end{align}
For each bracket let $v(T)$ count its distinct formal pairs, let
$R=\sum_T(v(T)-1)$, and let $F$ be the marked pairs. A new bracket
adds the previously unmarked moving pair to the crossed expression.
It therefore increases $R$ by one and marks one or two new pairs:
\begin{align}
 \Delta R&=1,\qquad\Delta|F|\leq2,
 &|F|&\leq2R,\qquad\ell:=p-|F|,\qquad p\leq\ell+2R.
 \label{eq:centered-pair-rank}
\end{align}
This includes crossing a single occurrence of an already marked pair.
An intact adjacent pair passes $k$ other occurrences with sign
$(-1)^{2k}=1$. Thus its deletion does not change the final sign.
Applying the center operations gives a finite exact sum
\begin{align}
 \pm UZ,\qquad U&=Y_{j_1}\cdots Y_{j_\ell},
 &Z&=N^{-|F|}\sum_{(i_j)_{j\in F}}T_1\cdots T_q,
 \qquad p\leq\ell+2R.
 \label{eq:chronological-reordered-terms}
\end{align}
The site sums in $U$ and $Z$ are independent formal sums, with no
restriction on numerical coincidences.

Each pair has two occurrences, so its formal site index enters at most
two bracket factors. In $Z^\dagger Z$ the two copies have separate site
sums and total $\sum_T(v(T)-1)=2R$. Hence
\Cref{lem:causal-count,lem:causal-moments}, including their equality
partitions, give
\begin{align}
 \Phi_N(ZZ^\dagger)+\Phi_N(Z^\dagger Z)&\leq C_{\beta,T,p}N^{-R}.
 \label{eq:chronological-bracket-norm}
\end{align}
No equal-site anticommutator is discarded. Each $Y_j$ is a simultaneous
real-time translate of the source at $t_{2j-1}-t_{2j}$; stationarity
preserves its Hermitian quadrature moments. Thus
\Cref{thm:mbound,lem:ordered-young}, uniformly on the bounded time
interval, give a common $c>0$ with
\begin{align}
 \Phi_N(UU^\dagger)+\Phi_N(U^\dagger U)&\leq C_{\beta,T,p}N^{-c\ell},\\
 |\Phi_N(UZ)|&\leq\sqrt{\Phi_N(UU^\dagger)\Phi_N(Z^\dagger Z)}
 \leq C_{\beta,T,p}N^{-(c\ell+R)/2}
 \leq C_{\beta,T,p}N^{-\min\{c/2,1/4\}p}.
\end{align}
Empty products have norm one. Summing the finite expansion proves the
claim with $\theta_0=\min\{c/2,1/4\}$.
\end{proof}

\subsection{An analytic neighborhood independent of the number of pairs}

\begin{fact}[Two-constants inequality]
\label{fact:two-constants}
Let $D\subset\mathbb C$ be a bounded connected domain, and let
$E\subset\partial D$. Suppose that the harmonic measure $h$ of $E$
is continuous on $\overline D$, with boundary values $1$ on $E$ and
$0$ on $\partial D\setminus E$. If $u$ is subharmonic and bounded
above on $D$, and
\begin{align}
 \limsup_{D\ni z\to\xi}u(z)
 &\leq
 \begin{cases}
 a,&\xi\in E,\\
 b,&\xi\in\partial D\setminus E,
 \end{cases}
\end{align}
then
\begin{align}
 u(z)&\leq h(z)a+(1-h(z))b,\qquad z\in D.
 \label{eq:two-constants-subharmonic}
\end{align}
In particular, if $f$ is bounded and holomorphic on $D$ and its
boundary upper limits in modulus are at most $A>0$ on $E$ and
$B>0$ on the rest of the boundary, then
\begin{align}
 |f(z)|&\leq A^{h(z)}B^{1-h(z)}.
 \label{eq:two-constants-holomorphic}
\end{align}
These hypotheses apply when $D=\Omega\setminus E$, where $\Omega$
is a bounded rectangle or a neighborhood
$\{z:\operatorname{dist}(z,[0,1])<\eta\}$ and $E$ is a nondegenerate
closed line segment contained in $\Omega$. In these cases
$0<h(z)<1$ throughout $D$, so $\min_K h>0$ for each compact
$K\subset D$.
\end{fact}
\begin{proof}
The function $u-ah-b(1-h)$ is subharmonic and has nonpositive
boundary upper limits. The maximum principle gives
\eqref{eq:two-constants-subharmonic}; applying it to $u=\log|f|$
gives \eqref{eq:two-constants-holomorphic}. Zeros of $f$ are allowed,
with $\log0=-\infty$.

For the stated slit domains, the outer boundary and the interior of
each side of the slit are regular for the Dirichlet problem. At a
slit endpoint, a square-root change of coordinate opens the slit
into a straight boundary, giving the same regularity. Thus the
boundary data $1$ on $E$ and $0$ on $\partial\Omega$ have a continuous
harmonic extension. The strong maximum principle gives $0<h<1$,
and continuity gives the positive minimum on $K$.
\end{proof}

\begin{lemma}[A fixed-width continuation]
\label{lem:chronological-analytic-neighborhood}
For each $\beta\geq0$ there are $\eta_\beta>0$ and $C_\beta<\infty$
with the following property. Let $N$ be even, let $i_1,\ldots,i_q\in[N]$
satisfy
\begin{align}
 \#\{k:i_k=i\}&\in2\mathbb Z\quad(i\in[N]),
 &r&=\#\{i_1,\ldots,i_q\},
\end{align}
and choose $b\in\{0,\ldots,q\}$ and real $u_0,\ldots,u_q$ with
\begin{align}
 u_j&\geq0,
 &\sum_{j=0}^q u_j&=1,
 &z_j(s)&=\mathbf1_{j=b}+s(u_j-\mathbf1_{j=b}).
 \label{eq:vertex-to-heat-line}
\end{align}
Then the function
\begin{align}
 f_N(s)&=\frac{\EE_J\tr\left[
       e^{z_0(s)\beta H}\psi_{i_1}e^{z_1(s)\beta H}
       \cdots\psi_{i_q}e^{z_q(s)\beta H}\right]}
                   {\EE_J\tr e^{\beta H}}
 \label{eq:chronological-line-function}
\end{align}
is entire and satisfies
\begin{align}
 \sup_{\operatorname{dist}(s,[0,1])<\eta_\beta}|f_N(s)|
 &\leq C_\beta^{r+1}.
 \label{eq:chronological-line-uniform-bound}
\end{align}
The constants are independent of $N,q,r,b$ and $\mathbf u$. The empty
word has $f_N=1$.
\end{lemma}
\begin{proof}
Let $S=\{i_1,\ldots,i_q\}$, $P_j=\psi_{i_1}\cdots\psi_{i_j}$,
and $H_j=P_jHP_j^\dagger$, with $P_0=I$. Even multiplicities give
$P_q=\epsilon I$, $\epsilon\in\{\pm1\}$, and
\begin{align}
 P_j\Gamma_EP_j^\dagger
 &=(-1)^{\sum_{k\leq j}\mathbf1_{i_k\in E}}\Gamma_E,\\
 H_j-H&=\sqrt{\frac3{8N^3}}\sum_{E\cap S\ne\varnothing}
       \left((-1)^{\sum_{k\leq j}\mathbf1_{i_k\in E}}-1\right)J_E\Gamma_E,\\
 \tr\left(e^{z_0\beta H}\prod_{j=1}^q(\psi_{i_j}e^{z_j\beta H})\right)
 &=\epsilon\tr\prod_{j=0}^q e^{z_j\beta H_j},\qquad
 \tr e^{\beta H_j}=\tr e^{\beta H}.
 \label{eq:chronological-sign-conjugates}
\end{align}

\proofstep{A bound when the coefficients sum to zero}
Take
\begin{align}
 W(s)&=\prod_{j=1}^ke^{sc_j\beta H_{a_j}},
 &\sum_jc_j&=0,\qquad\sum_j|c_j|\leq T.
 \label{eq:zero-sum-sector-product}
\end{align}
Its symmetrized degree-$d$ coefficients satisfy
\begin{align}
 W^{(d)}(0)&=(\beta\sqrt{3/(8N^3)})^d
               \sum_{E_1,\ldots,E_d}J_{E_1}\cdots J_{E_d}A_{E_1,\ldots,E_d},\\
 \|A_{E_1,\ldots,E_d}\|
 &\leq\sum_{n_1+\cdots+n_k=d}\frac{d!}{n_1!\cdots n_k!}
                       \prod_j|c_j|^{n_j}
 =\left(\sum_j|c_j|\right)^d\leq T^d.
 \label{eq:zero-sum-coefficient-bound}
\end{align}
A component $\mathcal C$ of its quartet-intersection graph disjoint
from $S$ has the same sign in every $H_{a_j}$ and commutes with the
other components. Restricting to the variables of this monomial gives
\begin{align}
 \prod_j\exp\left(sc_j\beta\sqrt{\frac3{8N^3}}
                             \sum_{E\in\mathcal C}J_E\Gamma_E\right)=I.
\end{align}
Thus every nonzero coefficient has all components meeting $S$; this
cancellation precedes absolute values.

Let $R$ retain positive-real-part exponentials of the $H_j$ with
real lengths summing to one, total imaginary variation at most one,
and fixed Clifford contractions. By \eqref{eq:complete-contour-derivative},
\begin{align}
 \frac{|\EE\tr[(\partial_{J_{F_1}}\cdots\partial_{J_{F_k}}R)A]|}
      {\EE\tr e^{\beta H}}
 &\leq(2\beta\sqrt{3/(8N^3)})^k\|A\|
       \quad(A\text{ independent of }J).
 \label{eq:retained-conjugate-derivatives}
\end{align}
After choosing $\ell$ Gaussian pairs, $b=d-\ell$ quartet indices
remain. Root a spanning tree of each component at one of the $r$ sites
in $S$. Each quartet chooses that site or a position in a parent
quartet, then at most $N^3/6$ other labels. Hence
\begin{align}
 \#\{\text{assignments of the }b\text{ indices}\}
 &\leq(r+4b)^b(N^3/6)^b,
 \label{eq:marked-forest-count}\\
 \#\{\pi:|\pi|=\ell\}&=\frac{d!}{2^\ell\ell!(d-2\ell)!}\leq2^dd^\ell,\\
 \frac{|\EE\tr(RW^{(d)}(0))|}{\EE\tr e^{\beta H}}
 &\leq T^d\sum_{\ell\leq d/2}2^dd^\ell2^{d-2\ell}
       \left(\frac{3\beta^2}{8N^3}\right)^{d-\ell}
       (r+4(d-\ell))^{d-\ell}(N^3/6)^{d-\ell}\nonumber\\
 &\leq[C_\beta T(r+d)]^d.
 \label{eq:marked-sector-taylor}
\end{align}
Parent choices already enumerate the trees; assigning along each tree
adds no factorial. Repeated numerical quartets are allowed. Since
$(r+d)^d\leq e^d(r+1)\cdots(r+d)$,
\begin{align}
 \sum_{d\geq0}\frac{[C_\beta T(r+d)]^d}{d!}
 &\leq(1-eC_\beta T)^{-r-1},
 \label{eq:marked-sector-taylor-sum}\\
 eC_\beta T\leq\tfrac12
 &\quad\Longrightarrow\quad
 \frac{|\EE\tr(RW(1))|}{\EE\tr e^{\beta H}}\leq2^{r+1}.
 \label{eq:zero-sum-product-bound}
\end{align}
The term $d=0$ is bounded by one. At fixed $N$, Gaussian exponential
integrability makes the expression entire, so the Taylor series is its
actual value.

\proofstep{The neighborhood}
Choose $\eta_\beta\leq1/4$ with $6eC_\beta\eta_\beta\leq1/2$.
For $|s|<\eta_\beta$, rotate the $b$th exponential first and retain
$e^{\beta H_b}$. The remaining coefficients sum to zero and have
absolute sum at most $2|s|$, so \eqref{eq:zero-sum-product-bound} applies.
In the rectangle $0\leq\Re s\leq1+\eta_\beta$, $|\Im s|<\eta_\beta$,
\begin{align}
 \Re z_j(s)&\geq0\quad(j\ne b),
 &\Re z_b(s)&\geq-\eta_\beta,\qquad
 V:=\sum_j|\Im z_j(s)|\leq2\eta_\beta.
\end{align}
When all real parts are nonnegative, Schatten H\"older bounds the
normalized trace by one. Otherwise rotate the negative interval first,
write its length as $-a$, and include following positive intervals with
total real length $a$, splitting the last if necessary. Their
coefficients $z'_j$ have purely imaginary sum $w$, and
\begin{align}
 \prod_je^{z'_j\beta H_{a_j}}
 &=\left(\prod_je^{z'_j\beta H_{a_j}}\right)e^{-w\beta H}e^{w\beta H},\\
 \sum_jz'_j-w&=0,
 &\sum_j|z'_j|+|w|&\leq2(a+V)\leq6\eta_\beta.
 \label{eq:negative-heat-removal}
\end{align}
The remaining factors, including $e^{w\beta H}$, have nonnegative
real lengths summing to one and imaginary variation at most $V\leq1$.
They form an admissible $R$, so \eqref{eq:zero-sum-product-bound}
again gives $2^{r+1}$. This rectangle and the initial disk cover
$\operatorname{dist}(s,[0,1])<\eta_\beta$. The constants do not depend
on the number of factors. Entirety also follows directly by Gaussian
exponential integrability in \eqref{eq:chronological-line-function}.
\end{proof}

\begin{proof}[Proof of \Cref{lem:collective-concentration}]
First take distinct times, reversing pairs by antisymmetry so that
$t_{2j-1}>t_{2j}$. Let $\pi$ order all times decreasingly and put
$\tau_k=t_{\pi(k)}$. With $\pi_S$ as in
\Cref{lem:centered-chronological-real}, define
\begin{align}
 F_p(s)&=2^{-p}\sum_{S\subseteq[p]}(-1)^{p-|S|}
       \left(\prod_{j\notin S}2G_*(s(t_{2j-1}-t_{2j}))\right)
       N^{-|S|}\sum_{(i_j)_{j\in S}}\operatorname{sgn}(\pi_S)\nonumber\\
 &\quad\times\Phi_N\left[\prod_{a\in\pi_S}^{\text{in list order}}
       e^{-st_a\beta H}\psi_{i_{\lceil a/2\rceil}}e^{st_a\beta H}\right].
 \label{eq:collective-line-definition}
\end{align}
Expansion of the centered product and \eqref{eq:chin} give
\begin{align}
 F_p(1)&=\mu_p((t_1,t_2),\ldots,(t_{2p-1},t_{2p})),
 \label{eq:collective-line-target}\\
 \sup_{|t|\leq T_0}|F_p(it)|&\leq C_{\beta,p}N^{-\theta_0p}
                        \qquad(T_0>0\text{ fixed}),
 \label{eq:collective-real-vertex}
\end{align}
since $F_p(it)$ is exactly the real-time subset expression of
\Cref{lem:centered-chronological-real}, multiplied by $2^{-p}$.

For analyticity, keep the original positions, inserting identities at
deleted pairs. Multiplication of the exponentials shows that their
lengths lie on \eqref{eq:vertex-to-heat-line} with $b=0$ and
\begin{align}
 u_0&=1-\tau_1,
 &u_k&=\tau_k-\tau_{k+1}\ (1\leq k<2p),
 &u_{2p}&=\tau_{2p},\qquad\sum_ku_k=1.
\end{align}
Every site in a retained word occurs evenly, and at most $|S|$ sites
occur. Thus \Cref{lem:chronological-analytic-neighborhood} bounds its
trace by $C_\beta^{|S|+1}$. The scalar factors have the same common
neighborhood by \eqref{eq:saddle-exponential-moment}, because
$\operatorname{dist}(sd,[0,1])\leq\operatorname{dist}(s,[0,1])$ for
$0\leq d\leq1$. For a fixed sufficiently small $\eta>0$,
\begin{align}
 \Omega_\eta&=\{s:\operatorname{dist}(s,[0,1])<\eta\},\\
 \sup_{s\in\Omega_\eta}|F_p(s)|
 &\leq2^{-p}\sum_{S\subseteq[p]}(2C_\beta)^{p-|S|}C_\beta^{|S|+1}
 \leq C_\beta^{p+1}.
 \label{eq:collective-common-domain}
\end{align}
Take $\eta/2\leq T_0$ and set
\begin{align}
 E&=i[-\eta/2,\eta/2],\qquad
 D=\Omega_\eta\setminus E.
\end{align}
Let $h$ be the harmonic measure of both sides of $E$ in $D$,
with boundary values $1$ on $E$ and $0$ on $\partial\Omega_\eta$.
By \Cref{fact:two-constants}, $\omega=h(1)>0$. The bounds
\eqref{eq:collective-real-vertex} and
\eqref{eq:collective-common-domain} give the boundary hypotheses
of that fact for $\log|F_p|$, and hence
\begin{align}
 |F_p(1)|&\leq(C_\beta^{p+1})^{1-\omega}
                   (C_{\beta,p}N^{-\theta_0p})^\omega
 \leq C_{\beta,p}N^{-\theta p},
 &\theta&=\omega\theta_0>0.
 \label{eq:collective-heat-continuation}
\end{align}
The domain, and hence $\omega$, is independent of $p,N$ and the time
vector. A fixed chain of disks from $E$ to $1$, with three-circle
estimates, gives an effective positive lower bound.

The estimate is uniform up to every equality or endpoint of the time
variables. Apply the ordered limits or finite averages in
\Cref{lem:contact-dictionary} to the whole centered expression; both
preserve the same bound. Reversing the initially reoriented pairs changes
only signs. This proves the asserted supremum estimate. The argument
also includes $\beta=0$, where $F_p$ is constant.
\end{proof}

\section{Operator, covariance, and numerical estimates}
\label{app:ladder}
This appendix proves \Cref{lem:full-bilocal-inversion}. We first give
the argument excluding a first singular temperature, and then prove
the operator, covariance, and curvature estimates that it uses.
We work on antisymmetric kernels in $L^2([0,1]^2)$,
antiperiodically extended in each argument. Convolution includes
integration over $[0,1]$.

\subsection{Excluding a singularity of the ladder operator}
\label{sec:ladder-proof}
The concentration estimate in \Cref{lem:collective-concentration}
was proved without inverting the ladder operator. At a fixed temperature where the inverse exists, the
finite hierarchy therefore gives its leading covariance coefficient.
We use this coefficient only at such regular temperatures to exclude
a first singularity. Define
\begin{align}
 \mathcal J_\beta X&=G_* *X*G_*,
 &(\mathcal M_\beta X)(u,v)&=3\beta^2G_*(u,v)^2X(u,v),\\
 \mathcal S_\beta&=I-\mathcal M_\beta^{1/2}
                         \mathcal J_\beta\mathcal M_\beta^{1/2},
 &\mathcal C_\beta&=2\mathfrak L_\beta^{-1}\mathcal J_\beta
          \quad(\beta\text{ regular}).
\end{align}
The operator estimates, covariance calculation, and physical
curvature bounds used below are proved in the following subsections.

\begin{proof}[Proof of \Cref{lem:full-bilocal-inversion}]
At $\beta=0$ the operator is the identity.  Fix a compact interval
$[0,\beta_+]$.  By \Cref{lem:ladder-structure}, $\mathfrak L_\beta$ is
similar, at positive temperature, to a norm-continuous self-adjoint
compact perturbation $\mathcal S_\beta$ of the identity.  Moreover,
$\mathcal S_0=I$.  Suppose there is a singular temperature in this
interval, and let $\beta_*>0$ be the first one.  Then
$\mathcal S_\beta>0$ for $0\leq\beta<\beta_*$.  For a sequence
$\beta_n\uparrow\beta_*$, put
\begin{align*}
 t_n=\|\mathcal S_{\beta_n}^{-1}\|_{2\to2}.
\end{align*}
The numbers $t_n$ tend to infinity.  The spectral argument in
\Cref{lem:critical-covariance} gives a subsequence such that
\begin{align}\label{eq:critical-positive-limit}
 t_n^{-1}\mathcal C_{\beta_n}\longrightarrow\mathcal A_*
 \quad\hbox{in operator norm},\qquad
 \mathcal A_*\geq0,\quad \mathcal A_*\ne0,
 \quad \operatorname{ran}\mathcal A_*\subseteq\ker\mathfrak L_{\beta_*}.
\end{align}
Here $\mathcal C_\beta$ is the limiting covariance at a regular
 temperature, defined in \Cref{lem:regular-covariance}.  Both legs of
$\mathcal A_*$ lie in the nullspace, since $\mathcal A_*$ is self-adjoint.
Only finitely many center frequencies occur in this nullspace, and none
is $0$ or $\pm1$.  Choose an integer $P$ bounding their absolute values.

Choose $\delta>0$ smaller than the uniform curvature radius in
\Cref{lem:short-curvature}, and also smaller than
$1/16$ and $1/[100(P+1)]$.  Let $\chi$ be a nonzero nonnegative smooth
function supported in $(4\delta,5\delta)$, and set
\begin{align*}
 \varphi(s)=\chi(s)+\chi(1-s),\qquad h=\varphi''.
\end{align*}
With $\widehat h(p)=\int_0^1h(s)e^{-2\pi ips}\,ds$,
\begin{align}\label{eq:negative-center-frequency-multipliers}
 \widehat h(0)=0,\qquad
 \widehat h(p)=-(2\pi p)^2\,2\int_0^1\chi(s)\cos(2\pi ps)\,ds<0
 \quad(1\leq |p|\leq P).
\end{align}
The strict inequality holds because the cosine is positive on the
support of $\chi$.

For $\eta\in C_c^\infty(0,\delta)$, let
\begin{align}\label{eq:relative-projector}
 (P_p\eta)(r)=\frac12\left[\eta(r)+e^{2\pi ipr}\eta(1-r)\right].
\end{align}
By \Cref{lem:physical-covariance-test}, the following expression is
bounded uniformly in $n$:
\begin{align*}
 \sum_{p\in\mathbb Z}\widehat h(-p)
 \langle P_p\eta,(\mathcal C_{\beta_n})_pP_p\eta\rangle.
\end{align*}
The test is continuous in operator norm.  Divide by $t_n$ and use
\eqref{eq:critical-positive-limit} to obtain
\begin{align*}
 0=\sum_{0<|p|\leq P}\widehat h(-p)
       \langle P_p\eta,(\mathcal A_*)_pP_p\eta\rangle.
\end{align*}
Every quadratic form is nonnegative and every displayed multiplier is
strictly negative.  Thus each quadratic form is zero, for every such
$\eta$.

The orthogonal projections of the vectors $P_p\eta$ into the critical
space span that space.  Indeed, a critical vector orthogonal to all of
them vanishes on $(0,\delta)$, because $P_p$ is the orthogonal projection
onto the relative functions in sector $p$.  It then vanishes identically
by \Cref{lem:critical-analyticity}.  The critical space is finite
dimensional, so no closure issue remains.  Positivity now implies
$(\mathcal A_*)_p=0$ in every sector.  This contradicts
$\mathcal A_*\ne0$.

There is therefore no singular temperature in $[0,\beta_+]$.  Since the
interval was arbitrary, the $L^2$ inverse exists at every finite
$\beta$.  Norm continuity makes its norm locally bounded.  The
pointwise estimate in \Cref{lem:ladder-contact} gives
\eqref{eq:analysis-invertibility}, including the prescribed coincident-time
values.  Finally, \Cref{lem:ladder-certificate} makes the bound effective.
\end{proof}

\subsection{The operator and the covariance}

\begin{lemma}[Self-adjoint form and center frequencies]
\label{lem:ladder-structure}
Define
\begin{align*}
 \mathcal J_\beta X=G_* *X*G_*,\qquad
 (\mathcal M_\beta X)(u,v)=3\beta^2G_*(u,v)^2X(u,v),\qquad
 \mathcal S_\beta=I-\mathcal M_\beta^{1/2}
                  \mathcal J_\beta\mathcal M_\beta^{1/2}.
\end{align*}
Then $\mathcal J_\beta$ is compact and self-adjoint, and
$\mathcal S_\beta$ is a norm-continuous self-adjoint compact perturbation
of $I$.  For $\beta>0$,
\begin{align}\label{eq:bilocal-similarity}
 \mathfrak L_\beta
 =\mathcal M_\beta^{-1/2}\mathcal S_\beta\mathcal M_\beta^{1/2}.
\end{align}
In center-frequency sectors $p=0,\pm1$, one has
$\mathcal S_\beta\geq I$.  In every other nonzero sector,
\begin{align}\label{eq:high-center-frequency-bound}
 \|I-\mathfrak L_\beta\|_p,
 \ \|I-\mathcal S_\beta\|_p
 \leq\frac{3\beta^2}{4\pi^2(2|p|-1)}.
\end{align}
Consequently only finitely many center sectors can be singular on a
compact temperature interval.
\end{lemma}
\begin{proof}
The bounded kernel $G_*(u,s)G_*(t,v)$ is symmetric under simultaneous
exchange of input and output, since both factors change sign. Thus
$\mathcal J_\beta$ is self-adjoint and Hilbert--Schmidt. The scalar
bounds give $\mathcal M_\beta\geq(3\beta^2/4)e^{-\beta/2}I$ for
$\beta>0$; multiplication verifies \eqref{eq:bilocal-similarity}, and
\eqref{eq:scalar-temperature-continuity} gives norm continuity.

For $e_r(t)=e^{-2\pi irt}$, $r\in\mathbb Z+1/2$, put
\begin{align}
 a_r&=2\pi r\int\frac{\mu_{*,\beta}(dx)}{(2\pi r)^2+x^2},
 &\sgn a_r&=\sgn r,\qquad|a_r|\leq(2\pi|r|)^{-1},\\
 \mathcal J_\beta(e_r\otimes e_s-e_s\otimes e_r)
 &=a_ra_s(e_r\otimes e_s-e_s\otimes e_r).
\end{align}
Left and right convolution supply $ia_r$ and $-ia_s$, respectively.
Multiplication by $G_*^2$ preserves the center frequency $p=r+s$.
Distinct half-integers with $p=0,\pm1$ have opposite signs, so
$\mathcal J_{\beta,p}\leq0$ and $\mathcal S_{\beta,p}\geq I$.
For every $p\ne0$,
\begin{align}
 |r|\,|p-r|&\geq(2|p|-1)/4,
 &\|\mathcal J_{\beta,p}\|&\leq[\pi^2(2|p|-1)]^{-1}.
\end{align}
Together with $\|\mathcal M_\beta\|\leq3\beta^2/4$, this proves
\eqref{eq:high-center-frequency-bound} and the final claim.
\end{proof}

\begin{lemma}[Uniform representatives and coincident-time evaluations]
\label{lem:ladder-contact}
At any temperature where the $L^2$ inverse exists,
\begin{align}\label{eq:analysis-contact-inverse}
 \|\mathfrak L_\beta^{-1}X\|_\infty
 \leq\left(1+\frac{3\beta^2}{16}
       +\frac{9\beta^4}{256}\|\mathfrak L_\beta^{-1}\|_{2\to2}\right)
       \|X\|_\infty.
\end{align}
\end{lemma}
\begin{proof}
For $K=I-\mathfrak L_\beta$ the kernel is bounded by $3\beta^2/16$.
On the unit-measure square,
\begin{align}
 \|K\|_{\infty\to\infty},\ \|K\|_{\infty\to2},\ \|K\|_{2\to\infty}
 &\leq3\beta^2/16,\\
 \mathfrak L_\beta^{-1}&=I+K+K\mathfrak L_\beta^{-1}K.
 \label{eq:ab-smooth}
\end{align}
This proves \eqref{eq:analysis-contact-inverse}. If $\ell_{u,v}$ and
$c_{s,t}$ are a row and column of $K$, the last term has the specified
kernel
\begin{align}
 R(u,v;s,t)&=\ell_{u,v}\mathfrak L_\beta^{-1}c_{s,t},
 &|R(u,v;s,t)|&\leq(3\beta^2/16)^2\|\mathfrak L_\beta^{-1}\|_{2\to2}.
\end{align}
For ordered limits $x_\epsilon\to x$, $y_\epsilon\to y$, let $\ell_x,c_y$
denote the limits for those orders, rather than independently assigned
diagonal values. Here $x,y$ are pairs of time arguments. Translation of the jump of $G_*$ changes
its $L^2$ norm by $O(\sqrt\epsilon)$. The smooth part is uniformly
continuous on each of its one-sided intervals. Thus
\begin{align}
 \|\ell_{x_\epsilon}-\ell_x\|_2+
                  \|c_{y_\epsilon}-c_y\|_2&\longrightarrow0,\nonumber\\
 |R(x_\epsilon;y_\epsilon)-R(x;y)|
 &\leq\|\mathfrak L_\beta^{-1}\|_{2\to2}
   \left(\|\ell_{x_\epsilon}-\ell_x\|_2\|c_{y_\epsilon}\|_2
              +\|\ell_x\|_2\|c_{y_\epsilon}-c_y\|_2\right)
 \longrightarrow0,\nonumber\\
 (\mathfrak L_\beta^{-1}X)_{F,\eta}
 &=X_{F,\eta}+(KX)_{F,\eta}+(RX)_{F,\eta}.
 \label{eq:inverse-prescribed-limits}
\end{align}
In the last line each term is the limit of the entire expression, using
the same offsets $\eta$; bounded convergence applies to the integrals.
A finite average over the orders in an equal-time group gives its
alternating value by \eqref{eq:prescribed-uniform-norm}. Each ordered
limit satisfies the same bound, and the averaging weights sum to one.
This proves on every domain $F$,
\begin{align}
 \sup_{\mathbf t\in F}|(\mathfrak L_\beta^{-1}X)_F(\mathbf t)|
 &\leq\left(1+\frac{3\beta^2}{16}
       +\frac{9\beta^4}{256}\|\mathfrak L_\beta^{-1}\|_{2\to2}\right)
       \max_{F'}\sup_{\mathbf t\in F'}|X_{F'}(\mathbf t)|.
 \label{eq:facewise-inverse-bound}
\end{align}
This includes each ordered limit and each finite alternating average.
Repeated arguments involving an integration variable are evaluated in
$X$ before applying the inverse, as in
\eqref{eq:substitution-before-inverse}. An equality involving
an integration variable is imposed in the input or column first,
exactly as in \eqref{eq:substitution-before-inverse}.
\end{proof}

\begin{lemma}[Exact equation for collective moments]
\label{lem:exact-hierarchy}
Equation \eqref{eq:explicit-collective-hierarchy} is exact, with the coincident-time evaluation
prescriptions of \Cref{lem:contact-dictionary}.  Its order-one terms use
only $\mu_{p+1},\mu_{p+2},\mu_{p+3}$, and its explicit $N^{-1}$ terms
use only $\mu_{p-1},\mu_{p-2}$.
\end{lemma}
\begin{proof}
Apply the free Wick integration-by-parts identity termwise to the
interaction series. For finitely many independent fields it is
\begin{align}
 \langle\chi_i(a)F\rangle_0
 &=2\sum_cG_0(a,c)\langle\partial_{\chi_i(c)}F\rangle_0,\\
 \partial_{\chi_i(a)}\left(\frac{N\beta^2}{8}\int Q_{st}^4\,ds\,dt\right)
 &=\frac{\beta^2}{2}\int Q_{ac}^3\chi_i(c)\,dc.
\end{align}
The second identity is interpreted inside the time integrals of each
finite-order term; the two endpoints contribute equally. For a polynomial
$P$ in finitely many independent coordinates $Q_{cd}$, $c<d$, the
left derivative of $\chi_i(b)P$ and summation over $i$ give the bounded-kernel
identity
\begin{align}
 \langle Q_{ab}P\rangle_N
 &=G_0(a,b)\langle P\rangle_N
   +\beta^2\int G_0(a,s)\langle Q_{st}^3Q_{tb}P\rangle_N\,ds\,dt
   \nonumber\\
 &\quad+\frac1N\sum_{c<d}
 \left\langle\bigl(G_0(a,c)Q_{db}-G_0(a,d)Q_{cb}\bigr)
                         \frac{\partial P}{\partial Q_{cd}}\right\rangle_N.
 \label{eq:exact-master-identity}
\end{align}
All sums and limits are justified by the absolute interaction-series
bound in \Cref{lem:contact-dictionary}. Division is only by the continuum
normalization $\EE\tr e^{\beta H}>0$; no inverse of a sampled covariance
or nonzero finite-grid partition function is assumed.

Use $G_*=G_0+\beta^2G_0*G_*^{\circ3}*G_*$ and the bounded convolution
identity
\begin{align}
 (I-\beta^2G_0*G_*^{\circ3})^{-1}
 &=I+\beta^2G_* *G_*^{\circ3},\\
 (I-\beta^2G_0*G_*^{\circ3})^{-1}G_0&=G_*.
\end{align}
Their Fourier multipliers verify both equalities. Subtract the scalar
saddle, set $Q=G_*+T$, and apply this inverse to the first argument of
\eqref{eq:exact-master-identity}. The linear term is
$\mathfrak L_\beta T$ and the remaining polynomial is
\begin{align}
 \beta^2G_* *\left[(3G_*T^{\circ2}+T^{\circ3})*G_*
 +(3G_*^2T+3G_*T^{\circ2}+T^{\circ3})*T\right].
 \label{eq:exact-nonlinearity}
\end{align}
Insert $P=\prod_{j=2}^pT_{u_jv_j}$ and antisymmetrize the output pair.
The five nonlinear terms have orders $p+1,p+2,p+3$; differentiating
$P$ removes one pair and the remaining $Q=G_*+T$ gives orders $p-2,p-1$.
\end{proof}

\begin{lemma}[Covariance at a regular temperature]
\label{lem:regular-covariance}
Call $\beta$ regular when $\mathfrak L_\beta$ has an $L^2$ inverse.  At
each fixed regular temperature,
\begin{align}\label{eq:ab-cov}
 \mu_1=O_\beta(N^{-1}),\qquad
 N(\mu_2-\mu_1\mu_1)=\mathcal C_\beta+O_\beta(N^{-1}),\qquad
 \mathcal C_\beta=2\mathfrak L_\beta^{-1}\mathcal J_\beta.
\end{align}
The second equality identifies a four-point kernel with its integral
operator using full-square integration against antisymmetric tests.
\end{lemma}
\begin{proof}
At a fixed regular temperature choose $\theta K>2$.
The finite hierarchy calculation in \eqref{eq:hierarchy-fixed-point}
uses only this inverse and gives
\begin{align}
 \boldsymbol\mu&=e_0+N^{-1}\left(\sum_{r<K}A_0^r\right)A_1e_0+O_\beta(N^{-2}).
\end{align}
Only row $2$ of $A_1e_0$ is nonzero. Its forcing kernel is
$G_*(a,c)G_*(d,b)-G_*(a,d)G_*(c,b)$, which acts on an antisymmetric
test as $2\mathcal J_\beta$. The operator $A_0$ can move it to lower
rows, but not back to row $2$. Consequently
\begin{align}
 \mu_1&=O_\beta(N^{-1}),
 &\mu_2&=2N^{-1}\mathfrak L_\beta^{-1}\mathcal J_\beta+O_\beta(N^{-2}).
\end{align}
Subtracting $\mu_1\mu_1=O_\beta(N^{-2})$ proves the assertion. No
remainder estimate uniform near a singular temperature is used.
\end{proof}

\begin{lemma}[The normalized critical covariance]
\label{lem:critical-covariance}
At a first singular temperature $\beta_*>0$, the subsequence and the
positive nonzero operator in \eqref{eq:critical-positive-limit} exist.
\end{lemma}
\begin{proof}
Because $\mathcal S_{\beta_*}-I$ is compact, its nullspace is finite
dimensional and isolated from the remaining spectrum. A resolvent
contour around zero and norm continuity give spectral projections
$E_n\to E_*$ in norm, of the same finite rank for large $n$.
On $I-E_n$, the inverse is uniformly bounded; on $E_n$, it is positive.
Identifying the finite-dimensional ranges by the projections and taking
a subsequence yields
\begin{align}
 t_n^{-1}\mathcal S_{\beta_n}^{-1}&\longrightarrow R_*,
 &R_*&\succeq0,\quad\|R_*\|=1,\quad
             \operatorname{ran}R_*\subseteq\ker\mathcal S_{\beta_*}.
\end{align}
The norm is one because the complementary inverse divided by $t_n$
tends to zero. Direct multiplication gives
\begin{align}
 \mathcal C_\beta
 &=2\mathcal M_\beta^{-1/2}(\mathcal S_\beta^{-1}-I)\mathcal M_\beta^{-1/2}.
 \label{eq:covariance-symmetrized-resolvent}
\end{align}
The multiplication operators and their inverses are norm-continuous
near $\beta_*>0$. Hence
$\mathcal A_*=2\mathcal M_{\beta_*}^{-1/2}R_*\mathcal M_{\beta_*}^{-1/2}$
is the required nonzero positive limit, with range in
$\ker\mathfrak L_{\beta_*}$ by similarity. All operators preserve center
frequency; \Cref{lem:ladder-structure} leaves only the stated finite
set of nonzero sectors.
\end{proof}

\subsection{A physical bound on short-time tests}

\begin{lemma}[Thermal curvature and a second-derivative test]
\label{lem:thermal-curvature}
For the positive annealed functional
\begin{align*}
 \Phi_N(X)=\frac{\EE_J\Tr(e^{\beta H}X)}{\EE_J\Tr e^{\beta H}},
 \qquad
 \mathcal Q_N(U)=\Phi_N([U^*,[-\beta H,U]]),
\end{align*}
one has $\mathcal Q_N(U)\geq0$.  For any disorder-dependent matrices
$U,V$ for which the expressions are integrable, and any
$\varphi\in C_c^\infty(0,1)$,
\begin{align}\label{eq:ab-fsum}
 \left|\int_0^1\varphi''(s)
       \Phi_N(e^{-s\beta H}Ue^{s\beta H}V)\,ds\right|
 \leq\|\varphi\|_\infty
            \sqrt{\mathcal Q_N(U)\mathcal Q_N(V)}.
\end{align}
\end{lemma}
\begin{proof}
Diagonalize $\beta H$ at each sample, with eigenvalues $\lambda_a$.
Expansion of the double commutator gives
\begin{align*}
 \mathcal Q_N(U)
 =\frac{\EE_J\sum_{a,b}(\lambda_b-\lambda_a)
              (e^{\lambda_b}-e^{\lambda_a})|U_{ab}|^2}
             {\EE_J\Tr e^{\beta H}}.
\end{align*}
These weights are nonnegative and symmetric in $a,b$.  Twice integrating
by parts, with no boundary terms, gives
\begin{align*}
 \left|\int_0^1\varphi''(s)e^{sx}\,ds\right|
 =\left|x^2\int_0^1\varphi(s)e^{sx}\,ds\right|
 \leq\|\varphi\|_\infty x(e^x-1).
\end{align*}
Multiply by $e^{\lambda_a}$, set $x=\lambda_b-\lambda_a$, and apply
Cauchy--Schwarz to the joint sum over $a,b$ and the disorder.  Symmetry
of the weights identifies the second squared factor with
$\mathcal Q_N(V)$.
\end{proof}

\begin{lemma}[Short-relative-time curvature]\label{lem:short-curvature}
Let
\begin{align*}
 B_z=\frac1{2N}\sum_i e^{-z\beta H}\psi_i e^{z\beta H}\psi_i.
\end{align*}
For every finite $\beta_+$ there are effective $\delta>0$ and $C<\infty$
such that, uniformly for $0\leq\beta\leq\beta_+$ and $0<r<\delta$,
\begin{align}\label{eq:ab-short}
 0\leq\mathcal Q_N(B_r)\leq C/N.
\end{align}
The reversed family satisfies the same bound. More explicitly, for all
$f,g\in L^1(0,\delta)$,
\begin{align}
 0\leq\mathcal Q_N\!\left(\int_0^\delta[f(r)B_r+g(r)B_r^*] \,dr\right)
 &\leq\frac CN(\|f\|_1+\|g\|_1)^2.
 \label{eq:short-curvature-integrals}
\end{align}
\end{lemma}
\begin{proof}
Use independent complex parameters and the holomorphic reversed family
\begin{align}
 B_z^{\rm rev}&=\frac1{2N}\sum_i\psi_i e^{z\beta H}\psi_i e^{-z\beta H},\\
 \left.\partial_z^d B_z\right|_0
 &=\frac{(-1)^d}{2N}\sum_i\ad_{\beta H}^d(\psi_i)\psi_i.
\end{align}
For integers $d_1,d_2\geq0$, put $d=d_1+d_2$. Differentiation gives
\begin{align}
 \left.\partial_z^{d_1}B_z^{\rm rev}\right|_{z=0}
 &=\frac1{2N}\sum_i\psi_i\ad_{\beta H}^{d_1}(\psi_i),\\
 &\left.\partial_z^{d_1}\partial_w^{d_2}
     \Phi_N([B_z^{\rm rev},[-\beta H,B_w]])\right|_{z=w=0}\nonumber\\
 &\quad=\frac{(-1)^{d_2+1}}{4N^2}\sum_{i,j}
 \Phi_N\!\left(
 \left[\psi_i\ad_{\beta H}^{d_1}(\psi_i),
 \left[\beta H,\ad_{\beta H}^{d_2}(\psi_j)\psi_j\right]\right]\right),
 \label{eq:short-curvature-derivatives}\\
 B_0=B_0^{\rm rev}&=\tfrac12I,\qquad
 \left.\partial_z^{d_1}\partial_w^{d_2}
     \Phi_N([B_z^{\rm rev},[-\beta H,B_w]])\right|_{z=w=0}=0
     \quad(d_1d_2=0).
\end{align}
For $d_1,d_2\geq1$, expand each occurrence of $\beta H$. There are
$D=d+1$ Gaussian factors, since
\begin{align}
 \ad_{\beta H}^{h}(\psi_i)
 &=\left(\beta\sqrt{\frac3{8N^3}}\right)^h
   \sum_{E_1,\ldots,E_h\in\binom{[N]}4}
   \left(\prod_{a=1}^hJ_{E_a}\right)
   \ad_{\Gamma_{E_h}}\cdots\ad_{\Gamma_{E_1}}(\psi_i),\\
 \bigl\|\ad_{\Gamma_{E_h}}\cdots\ad_{\Gamma_{E_1}}(\psi_i)\bigr\|
 &\leq2^h,\qquad \|[U,V]\|\leq2\|U\|\|V\|.
\end{align}
In a nonzero nested commutator, every new interaction meets the preceding
support:
\begin{align}
 \supp\!\left(\ad_{\Gamma_{E_a}}\cdots\ad_{\Gamma_{E_1}}(\psi_i)\right)
 &\subseteq\{i\}\cup\bigcup_{q\leq a}E_q,\\
 \ad_{\Gamma_{E_a}}\cdots\ad_{\Gamma_{E_1}}(\psi_i)\ne0
 &\quad\Longrightarrow\quad
 E_a\cap\left(\{i\}\cup\bigcup_{q<a}E_q\right)\ne\varnothing.
 \label{eq:short-curvature-support}
\end{align}
Both matrices in the outer commutator of
\eqref{eq:short-curvature-derivatives} are even. If their supports are
disjoint they commute. Thus every nonzero summand has a connected graph
of intersecting quartets containing both original sites $i,j$.

Fix a partial Gaussian pairing $\pi$ with $\ell$ pairs and put
$b=D-\ell$. Let $\mathcal A_{i,j}(\pi)$ be the assignments of these $b$
quartet variables for which the expanded commutator is nonzero. Choose
a spanning tree of the intersection graph, adding the root $\{i\}$.
There are at most $(b+1)^b$ choices of parents and $16^b$ choices of
positions witnessing the tree intersections. Assigning the labels along
the tree leaves $3b$ free positions. If $j\ne i$, at least one of those
positions must carry $j$. Consequently
\begin{align}
 \#\mathcal A_{i,j}(\pi)
 &\leq[16(b+1)]^b\,3bN^{3b-1}\qquad(i\ne j),\\
 \#\mathcal A_{i,i}(\pi)
 &\leq[16(b+1)]^bN^{3b},\\
 \frac1{N^2}\sum_{i,j}\#\mathcal A_{i,j}(\pi)
 &\leq[16(b+1)]^b(3b+1)N^{3b-1}.
 \label{eq:short-curvature-label-count}
\end{align}
Different quartet variables may have the same numerical labels; the
count does not exclude such assignments.

Apply \eqref{eq:partial-gaussian-matching}. The unpaired factors
differentiate only $e^{\beta H}$, since the remaining Clifford matrices
are independent of $J$. For any such matrix $X$,
\begin{align}
 \frac{\left|\EE_J\Tr\!\left[
   (\partial_{J_{F_1}}\cdots\partial_{J_{F_q}}e^{\beta H})X\right]\right|}
      {\EE_J\Tr e^{\beta H}}
 &\leq\left(\beta\sqrt{\frac3{8N^3}}\right)^q\|X\|,\\
 \#\{\pi:|\pi|=\ell\}
 &=\frac{D!}{2^\ell\ell!(D-2\ell)!}\leq2^DD^\ell,\\
 \left(\beta\sqrt{\frac3{8N^3}}\right)^{D+D-2\ell}
 &=\left(\frac{3\beta^2}{8N^3}\right)^b,
 \qquad b+\ell=D.
\end{align}
The first line is \eqref{eq:complete-contour-derivative} with total heat
one and no real-time intervals. Combining the bounds gives, uniformly
for $0\leq\beta\leq\beta_+$,
\begin{align}
 &\left|\left.\partial_z^{d_1}\partial_w^{d_2}
   \Phi_N([B_z^{\rm rev},[-\beta H,B_w]])\right|_{z=w=0}\right|\nonumber\\
 &\quad\leq\frac{2^{D+1}}{4N}
   \sum_{\ell=0}^{\lfloor D/2\rfloor}
      2^DD^\ell\left(\frac{3\beta_+^2}{8}\right)^{D-\ell}
      [16(D-\ell+1)]^{D-\ell}(3D+1)\nonumber\\
 &\quad\leq\frac1N[32(1+\beta_+^2)(d+2)]^{d+3}.
 \label{eq:short-curvature-taylor-coefficients}
\end{align}
For fixed $N$, the integrand is holomorphic in $z,w$. On every compact
set it has an integrable bound, because
\begin{align}
 &\left|\Tr(e^{\beta H}[B_z^{\rm rev},[-\beta H,B_w]])\right|\nonumber\\
 &\quad\leq\Tr(e^{\beta H})\|\beta H\|
                   e^{2(|z|+|w|)\|\beta H\|},\\
 \|\beta H\|&\leq\beta_+\sqrt{\frac3{8N^3}}\sum_E|J_E|,\\
 \EE_J\exp\!\left(A\sum_E|J_E|\right)&<\infty
             \qquad(A<\infty,\ N\text{ fixed}).
\end{align}
Thus expectation and differentiation commute. The Taylor series and
\eqref{eq:short-curvature-taylor-coefficients} imply
\begin{align}
 &\sup_{|z|,|w|\leq\delta}
        |\Phi_N([B_z^{\rm rev},[-\beta H,B_w]])|\nonumber\\
 &\quad\leq\frac1N\sum_{d\geq0}\frac{(2\delta)^d}{d!}
                      [32(1+\beta_+^2)(d+2)]^{d+3}\nonumber\\
 &\quad\leq\frac{C_{\beta_+}}N
        \sum_{d\geq0}[64e(1+\beta_+^2)\delta]^d(d+2)^3
 \leq\frac{C_{\beta_+}}N,
 \qquad 0<\delta\leq\frac1{128e(1+\beta_+^2)}.
 \label{eq:short-curvature-complex-bound}
\end{align}
Here $\sum_{d_1+d_2=d}(d_1!d_2!)^{-1}=2^d/d!$ and
$(d+2)^d/d!\leq e^{d+2}$. For real $r$,
$B_r^{\rm rev}=B_r^*$, so \Cref{lem:thermal-curvature} proves
\eqref{eq:ab-short}.

The symmetric nonnegative energy weights in that lemma also give
\begin{align}
 \mathcal Q_N(U^*)&=\mathcal Q_N(U),\\
 |\Phi_N([U^*,[-\beta H,V]])|
 &\leq\sqrt{\mathcal Q_N(U)\mathcal Q_N(V)},\\
 \mathcal Q_N\!\left(\int_0^\delta
          [f(r)B_r+g(r)B_r^*]\,dr\right)^{1/2}
 &\leq\int_0^\delta
       \left(|f(r)|\mathcal Q_N(B_r)^{1/2}
             +|g(r)|\mathcal Q_N(B_r^*)^{1/2}\right)dr\nonumber\\
 &\leq\sqrt{C/N}\,(\|f\|_1+\|g\|_1).
\end{align}
The last inequality holds for all $f,g\in L^1(0,\delta)$ and proves the
assertion about integrals.
\end{proof}

\begin{lemma}[The bounded physical covariance test]
\label{lem:physical-covariance-test}
Fix the short-time interval in \Cref{lem:short-curvature}.  For
$\eta,\xi\in C_c^\infty(0,\delta)$ and
$\varphi\in C_c^\infty(3\delta,1-3\delta)$, the functional
\begin{align}\label{eq:explicit-center-frequency-covariance-test}
 \mathcal R_{\eta,\xi,\varphi''}(C)
 =\sum_{p\in\mathbb Z}\widehat{\varphi''}(-p)
                    \langle P_p\eta,C_pP_p\xi\rangle
\end{align}
is continuous in operator norm.  Its value on $\mathcal C_\beta$ is
locally uniformly bounded over regular real temperatures.
\end{lemma}
\begin{proof}
For $r,t\in(0,\delta)$ and $s\in\operatorname{supp}\varphi$, the ordering
$s+r>s>t>0$ gives the exact identity
\begin{align}
 \Phi_N(e^{-s\beta H}B_re^{s\beta H}B_t)
 &=\frac1{4N^2}\sum_{i,j}\Phi_N(\psi_i(s+r)\psi_i(s)\psi_j(t)\psi_j(0)),\\
 \left|\int\varphi''(s)\Phi_N(e^{-s\beta H}B_re^{s\beta H}B_t)\,ds\right|
 &\leq C\|\varphi\|_\infty/N.
\end{align}
Here $\psi_i(u)=e^{-u\beta H}\psi_i e^{u\beta H}$; the bound is
\Cref{lem:thermal-curvature,lem:short-curvature}. It includes $i=j$
and the reversed pairs. Subtracting deterministic or finite-$N$
stationary centers changes only terms constant in $s$, annihilated by
$\int\varphi''$.

Integrate against $\overline\eta(r)\xi(t)$, multiply by $N$, and use
\Cref{lem:regular-covariance} at the fixed regular temperature.
In center sector $p$, writing a bilocal as $e^{2\pi ipb}f(a-b)$ gives
\begin{align}
 f(1-r)&=e^{-2\pi ipr}f(r),\\
 |\mathcal R_{\eta,\xi,\varphi''}(\mathcal C_\beta)|
 &\leq C\|\varphi\|_\infty\|\eta\|_1\|\xi\|_1.
\end{align}
The $+1$ projection of the unitary involution
$\eta(r)\mapsto e^{2\pi ipr}\eta(1-r)$ is $P_p$.
Fourier expansion in $s$ therefore gives precisely
\eqref{eq:explicit-center-frequency-covariance-test}. It is justified
first on finite Fourier sums and then by
\begin{align}
 |\mathcal R_{\eta,\xi,h}(C)|
 &\leq\|C\|\,\|\eta\|_2\|\xi\|_2\sum_p|\widehat h(p)|<\infty
 \qquad(h\text{ smooth}).
\end{align}
This also proves operator-norm continuity. The physical bound is uniform
on compact temperature intervals; the large-$N$ limit was taken only
at each fixed regular temperature.
\end{proof}

\subsection{Analyticity of critical vectors}

\begin{lemma}[A critical vector is determined locally]
\label{lem:critical-analyticity}
In any fixed center sector, a vector in $\ker\mathfrak L_\beta$ is
analytic as a function of the relative time on the one-sided branch
$[0,1]$.  In particular, its restriction to any nonempty open
subinterval of $(0,1)$ is injective on the nullspace.
\end{lemma}
\begin{proof}
Write $X(a,b)=e^{i\nu b}f(a-b)$, $\nu=2\pi p$, and
$G_{*,p}(r)=e^{i\nu r}G_*(r)$. The null equation gives
\begin{align}
 f&=3\beta^2G_* *(G_*^2f)*G_{*,p}.
\end{align}
The outer Fourier multipliers decay as inverse frequencies, with a
fixed shift in the second, so $f\in H^2$ and $f,f'$ are continuous.
Using $DG_*=I+k$, $k=\beta^2G_*^{\circ3}*G_*$, yields
\begin{align}
 f''-i\nu f'&=3\beta^2\{G_*^2f+S_p*(G_*^2f)\},
 &S_p&=k+e^{i\nu\cdot}k+k*(e^{i\nu\cdot}k).
 \label{eq:critical-ode}
\end{align}
The derivative of $G_*$ includes its jump of size one at zero;
\eqref{eq:critical-ode} therefore holds in the sense of distributions.
We construct a holomorphic solution locally and then identify it with
$f$, rather than assuming that $f$ is analytic.

By \eqref{eq:saddle-exponential-moment}, the positive-time expression for
$G_*$ is bounded and holomorphic on a complex neighborhood of $[0,1]$.
This property is preserved by antiperiodic convolution. Indeed, if
$A_+,B_+$ are the holomorphic extensions of two positive-time branches,
then, for $0<r<1$,
\begin{align}
 (A*B)(r)
 &=\int_0^r A_+(r-s)B_+(s)\,ds
      -\int_r^1 A_+(1+r-s)B_+(s)\,ds,\\
 (A*B)_+(z)
 &=z\int_0^1A_+(z(1-t))B_+(zt)\,dt\nonumber\\
 &\quad-(1-z)\int_0^1
    A_+(1-(1-z)t)B_+(z+(1-z)t)\,dt.
 \label{eq:analytic-antiperiodic-convolution}
\end{align}
Every argument of $A_+$ or $B_+$ in the second formula remains in their
common neighborhood when $z$ is sufficiently close to $[0,1]$.
The formula is holomorphic there by dominated differentiation.
Apply it to the convolutions defining $k$ and $S_p$ in
\eqref{eq:critical-ode}. There are holomorphic functions $S_+,S_-$,
on neighborhoods of $[0,1]$ and $[-1,0]$, respectively, satisfying
\begin{align}
 S_+(r)&=S_p(r)\quad(0<r<1),\\
 S_-(z)&=-S_+(z+1),\\
 \sup_{\operatorname{dist}(z,[0,1])<a}
      (|G_*(z)|^2+|S_+(z)|)
   +\sup_{\operatorname{dist}(z,[-1,0])<a}|S_-(z)|
 &\leq C_{\beta,p}
 \label{eq:critical-analytic-coefficients}
\end{align}
for some $a>0$. The notation $G_*(z)$ here always uses the
positive-time branch.

Fix $r_0\in[0,1]$. For $|z-r_0|<a/2$, define
\begin{align}
 F_{r_0}(z)
 &=\int_0^{r_0}S_+(z-s)G_*(s)^2f(s)\,ds
    +\int_{r_0}^1S_-(z-s)G_*(s)^2f(s)\,ds,\\
 |F_{r_0}(z)|&\leq C_{\beta,p}\|G_*^2f\|_1.
 \label{eq:critical-fixed-integrals}
\end{align}
This is holomorphic, since the factors depending on $s$ are integrable
and the two kernels satisfy \eqref{eq:critical-analytic-coefficients}.
For real $r\in[0,1]$ with $|r-r_0|<a/2$, splitting the same integral
at $r$ instead of $r_0$ gives
\begin{align}
 (S_p*(G_*^2f))(r)
 &=F_{r_0}(r)
   +\int_{r_0}^r[S_+(r-s)-S_-(r-s)]G_*(s)^2f(s)\,ds,\\
 f''(r)
 &=i\nu f'(r)+3\beta^2G_*(r)^2f(r)+3\beta^2F_{r_0}(r)\nonumber\\
 &\quad+3\beta^2\int_{r_0}^r
        [S_+(r-s)-S_-(r-s)]G_*(s)^2f(s)\,ds.
 \label{eq:critical-local-integral-equation}
\end{align}
The integrals are oriented when $r<r_0$. Since $f\in H^2$, the
right side is continuous; thus $f$ is $C^2$ on the real interval and
the second equality holds pointwise, with one-sided values at its ends.

On the disk $|z-r_0|<\rho<a/4$, define holomorphic functions recursively,
using straight line segments for all integrals:
\begin{align}
 y_0(z)&=f(r_0),\qquad v_0(z)=f'(r_0),\\
 y_{n+1}(z)&=f(r_0)+\int_{r_0}^zv_n(s)\,ds,\\
 v_{n+1}(z)&=f'(r_0)+\int_{r_0}^z
     \Bigg[i\nu v_n(s)+3\beta^2G_*(s)^2y_n(s)+3\beta^2F_{r_0}(s)
     \nonumber\\
 &\hspace{33mm}+3\beta^2\int_{r_0}^s
     [S_+(s-t)-S_-(s-t)]G_*(t)^2y_n(t)\,dt\Bigg]ds.
 \label{eq:critical-holomorphic-iteration}
\end{align}
Take supremum norms on this disk. For $n\geq1$,
\begin{align}
 \|y_{n+1}-y_n\|_\infty
 &\leq\rho\|v_n-v_{n-1}\|_\infty,\\
 \|v_{n+1}-v_n\|_\infty
 &\leq|\nu|\rho\|v_n-v_{n-1}\|_\infty
       +C_{\beta,p}(\rho+\rho^2)\|y_n-y_{n-1}\|_\infty,\\
 \max\{\|y_{n+1}-y_n\|_\infty,\|v_{n+1}-v_n\|_\infty\}
 &\leq\tfrac12\max\{\|y_n-y_{n-1}\|_\infty,\|v_n-v_{n-1}\|_\infty\},
 \label{eq:critical-iteration-contraction}
\end{align}
where $\rho$ is chosen so that
$(1+|\nu|)\rho+C_{\beta,p}(\rho+\rho^2)\leq1/2$.
The first differences are bounded by \eqref{eq:critical-fixed-integrals},
so the geometric estimate proves uniform convergence to holomorphic
functions $y,v$. Passing to the integrals gives
\begin{align}
 y'&=v,\qquad y(r_0)=f(r_0),\qquad v(r_0)=f'(r_0).
\end{align}
On $[0,1]\cap[r_0-\rho,r_0+\rho]$, the pair $(f,f')$ satisfies the same
integral equations by \eqref{eq:critical-local-integral-equation}.
The difference estimate used in \eqref{eq:critical-iteration-contraction}
therefore gives
\begin{align}
 \max\{\|y-f\|_\infty,\|v-f'\|_\infty\}
 &\leq\tfrac12\max\{\|y-f\|_\infty,\|v-f'\|_\infty\}.
\end{align}
Hence $y=f$ and $v=f'$ on this interval. The radius $\rho$ is independent
of $r_0$.
The local holomorphic functions agree on overlapping disks by the
identity theorem, including the disks at $0$ and $1$. Finally,
\begin{align}
 f|_I=0\quad\text{for a nonempty open }I\subset(0,1)
 \quad&\Longrightarrow\quad f\equiv0\text{ on }[0,1]
 \quad\Longrightarrow\quad X=0.
\end{align}
This proves both assertions.
\end{proof}

\subsection{Certification and numerical stability}

\begin{lemma}[Effective scalar and bilocal approximation]
\label{lem:ladder-certificate}
At each fixed temperature, the scalar kernel
can be approximated to uniform one-sided error $\varepsilon$ in time
polynomial in $\varepsilon^{-1}$.  Once invertibility has been proved,
a finite refinement search certifies a bound for
$\|\mathfrak L_\beta^{-1}\|_{2\to2}$.  With this certificate, the inverse
operations and coincident-time evaluations used by any fixed truncated hierarchy can be
computed to accuracy $\varepsilon$ in time polynomial in
$\varepsilon^{-1}$.
\end{lemma}
\begin{proof}
By \Cref{lem:saddle}, with Euclidean gradient residual $R_M$,
\begin{align}
 \|\widetilde G_*-G_*\|_\infty&\leq C_\beta(M^{-2}+R_M).
 \label{eq:certificate-effective-saddle-error}
\end{align}
Keep the free jump exact. Let $P_h$ be the orthogonal projection onto
antisymmetric step kernels on a mesh of width $h$, and form the finite
compression $K_{\rm fin}=P_h\widetilde K P_h$, with $K=I-\mathfrak L_\beta$.
For the kernel $K(u,v;s,t)$, fix the order of $u,v,s,t$.
Its smooth factors have uniformly bounded first derivatives; the
possible changes of formula occur at $u=s$, $t=v$, and $s=t$.
The union of the sets $|u-s|\leq h$, $|t-v|\leq h$, and $|s-t|\leq h$
has four-dimensional volume at most $6h$. Consequently
\begin{align}
 \|K-K_{\rm fin}\|_{2\to2}
 &\leq\|K-K_{\rm fin}\|_{\rm HS}
 \leq\varepsilon_K:=C_\beta(M^{-2}+R_M+\sqrt h),\\
 \min\{1,\sigma_{\min}(L_{\rm fin})\}&>\varepsilon_K
 \quad\Longrightarrow\quad
 \|\mathfrak L_\beta^{-1}\|_{2\to2}
 \leq\bigl(\min\{1,\sigma_{\min}(L_{\rm fin})\}-\varepsilon_K\bigr)^{-1}.
\end{align}
Here $L_{\rm fin}=I-K_{\rm fin}$ on the finite space and the approximate
operator is $I$ on its orthogonal complement. Matrix-entry and singular-value
rounding errors are included in $\varepsilon_K$. Rational interval
arithmetic certifies the strict inequality. Refinement terminates
because the true inverse exists and $\varepsilon_K\to0$.

For a row $\ell_x$ and column $c_y$ of $K$, where $x,y$ are pairs of
time variables on their specified branches, translating a jump changes
an $L^2$ norm by $O(\sqrt h)$. Thus
\begin{align}
 \|\ell_x-\ell_{x'}\|_2&\leq C_\beta\|x-x'\|_\infty^{1/2},
 &\|c_y-c_{y'}\|_2&\leq C_\beta\|y-y'\|_\infty^{1/2},\\
 R(x;y)&=\ell_x\mathfrak L_\beta^{-1}c_y,\\
 |R(x;y)-R(x';y')|
 &\leq C_\beta\bigl(\|x-x'\|_\infty^{1/2}
                        +\|y-y'\|_\infty^{1/2}\bigr).
 \label{eq:inverse-kernel-holder}
\end{align}
The last constant includes the certified inverse norm. Separate
representatives are used at the specified equality faces. Projecting
rows and columns and using the finite inverse approximates $R$ uniformly:
\begin{align}
 \|\widetilde R-R\|_\infty
 &\leq C_\beta(M^{-2}+R_M+\sqrt h).
\end{align}
This follows by telescoping the row, the middle inverse, and the column;
the middle error is the resolvent identity. It also gives a uniform
supremum-norm bound for the finite inverse, using $I+K_{\rm fin}+
K_{\rm fin}L_{\rm fin}^{-1}K_{\rm fin}$ and the bounded row and column
$L^2$ norms.

We next prove the regularity estimates used in quadrature. Throughout,
$F$ ranges over the ordered domains and prescribed equalities in
\eqref{eq:time-cell-definition}--\eqref{eq:time-face-definition}.
For products, use the same ordered limit in both factors and take any
finite average only after multiplication. Products and finite sums satisfy
\begin{align}
 \|XY\|_{1/2,\mathrm{pw}}
 &\leq\|X\|_{1/2,\mathrm{pw}}\|Y\|_{1/2,\mathrm{pw}},\\
 \left\|\sum_{j=1}^qc_jX_j\right\|_{1/2,\mathrm{pw}}
 &\leq\sum_{j=1}^q|c_j|\,\|X_j\|_{1/2,\mathrm{pw}}.
 \label{eq:holder-products-and-sums}
\end{align}
Every substitution in \eqref{eq:explicit-collective-hierarchy} selects,
repeats, or permutes time coordinates. With $x_0=0$ and $x_{a+1}=1$,
such a substitution has the form
\begin{align}
 \vartheta(x)&=(x_{i_1},\ldots,x_{i_q}),
       \qquad i_j\in\{0,\ldots,a+1\},\\
 \|\vartheta(x)-\vartheta(x')\|_\infty
 &\leq\|x-x'\|_\infty,\\
 |X(\vartheta(x))-X(\vartheta(x'))|
 &\leq\|X\|_{1/2,\mathrm{pw}}\|x-x'\|_\infty^{1/2}
       \qquad(x,x'\in F).
 \label{eq:holder-coordinate-substitutions}
\end{align}
The order of the image coordinates is fixed on $F$. Repeated coordinates
are evaluated using the specified equal-time value of the entire input.
An ordered limit preserves the bound, and an average of such limits
also preserves it because its weights are nonnegative and sum to one.

For integration, let $x,x'\in F\subseteq[0,1]^a$ and
$y\in[0,1]^k$. Write the integrand as $f(x,y)$, including any indicator
that specifies an order of its arguments. Set
\begin{align}
 d&=\|x-x'\|_\infty,\qquad
 E_{x,x'}=\bigcup_{i=1}^a\bigcup_{j=1}^k
                 \{y\in[0,1]^k:|y_j-x_i|\leq d\},\\
 |E_{x,x'}|&\leq2akd.
\end{align}
Outside $E_{x,x'}$, the arguments $(x,y)$ and $(x',y)$ have the same
order. Equalities between integration variables have measure zero;
prescribed equalities are first parametrized by their free coordinates.
It follows that
\begin{align}
 \left|\int_{[0,1]^k}f(x,y)\,dy\right|&\leq\|f\|_\infty,\\
 &\left|\int_{[0,1]^k}f(x,y)\,dy
            -\int_{[0,1]^k}f(x',y)\,dy\right|\nonumber\\
 &\quad\leq\int_{[0,1]^k\setminus E_{x,x'}}
                 |f(x,y)-f(x',y)|\,dy
            +2|E_{x,x'}|\,\|f\|_\infty\nonumber\\
 &\quad\leq(d^{1/2}+4akd)\|f\|_{1/2,\mathrm{pw}}
       \leq(1+4ak)d^{1/2}\|f\|_{1/2,\mathrm{pw}}.
 \label{eq:holder-parameter-integral}
\end{align}
Dominated convergence gives the same statement at each prescribed
ordered limit; finite averages are taken afterwards.

The known factors $G_*$ have bounded first derivatives on each
one-sided interval, by \eqref{eq:saddle-exponential-moment}. Apply
\eqref{eq:holder-products-and-sums}--\eqref{eq:holder-parameter-integral}
to the explicit kernel of $K$, and use
\eqref{eq:inverse-kernel-holder} for $R$. Since
$\mathfrak L_\beta^{-1}=I+K+R$,
\begin{align}
 \|KX\|_{1/2,\mathrm{pw}}+\|RX\|_{1/2,\mathrm{pw}}
 &\leq C_{\beta,a}\|X\|_{1/2,\mathrm{pw}},\\
 \|\mathfrak L_\beta^{-1}X\|_{1/2,\mathrm{pw}}
 &\leq C_{\beta,a}\|X\|_{1/2,\mathrm{pw}}.
 \label{eq:hierarchy-holder-stability}
\end{align}
Here $a$ bounds the number of time coordinates, including the passive
ones. For the inverse on two variables alone the constant is $C_\beta$.
For a vector of inputs, take the maximum of its component norms.
At fixed truncation order $K$, all the operations defining $A_0,A_1$
therefore obey
\begin{align}
 \|e_0\|_{1/2,\mathrm{pw}}&=1,\qquad
 \|A_iX\|_{1/2,\mathrm{pw}}
       \leq C_{\beta,K}\|X\|_{1/2,\mathrm{pw}}\quad(i=0,1),\\
 \|T_l\cdots T_1e_0\|_{1/2,\mathrm{pw}}
 &\leq C_{\beta,K}^{\,l}
       \qquad(T_j\in\{A_0,A_1\}).
 \label{eq:exact-iterate-holder-bound}
\end{align}
The same bounds apply between the elementary operations inside each
block. Expanding the finite iteration into such products proves a bound
$C_{\beta,K,j}$ for every exact intermediate expression, independently
of $N$, since all powers of $N^{-1}$ are at most one. This argument
starts from $e_0$ and does not require regularity of the unknown moments.

For completeness, the quadrature can be specified without crossing any
of the inequalities that determine the time order. For fixed external
coordinates $x$, list the distinct numbers
\begin{align}
 \{b_0,\ldots,b_R\}
 &=\{0,1,x_1,\ldots,x_a\}
       \cup\{qh: q\in\mathbb Z,\ 0<qh<1\},\\
 0=b_0&<b_1<\cdots<b_R=1,\qquad
 b_l-b_{l-1}\leq h,\qquad
 R\leq\lceil h^{-1}\rceil+a+1.
\end{align}
Assign each of the $k$ integration coordinates to one of the intervals
$(b_{l-1},b_l)$, and specify their order within each interval. Let $Q$ be
one such region and $r_l$ its number of coordinates in the $l$th interval.
These regions partition $[0,1]^k$ up to a set of measure zero, and
\begin{align}
 \sum_{l=1}^Rr_l&=k,\qquad
 |Q|=\prod_{l=1}^R\frac{(b_l-b_{l-1})^{r_l}}{r_l!},\\
 \operatorname{diam}_\infty Q&\leq h,\qquad
 \#\{Q\}\leq k!R^k,\qquad \sum_Q|Q|=1.
 \label{eq:explicit-ordered-quadrature-regions}
\end{align}
For rational grid coordinates $x$ and rational $h$, the volumes and an
interior point $y_Q$ are computable rational numbers. Every comparison
$y_i<y_j$ or $y_i<x_j$ is constant in $Q$. Consequently
\begin{align}
 \left|\int_Q f(x,y)\,dy-|Q|f(x,y_Q)\right|
 &\leq |Q|h^{1/2}\|f\|_{1/2,\mathrm{pw}},\\
 \left|\int_{[0,1]^k}f(x,y)\,dy
              -\sum_Q|Q|f(x,y_Q)\right|
 &\leq h^{1/2}\|f\|_{1/2,\mathrm{pw}},\\
 \left|\sum_Q|Q|f(x,y_Q)\right|&\leq\|f\|_\infty.
 \label{eq:certificate-positive-quadrature}
\end{align}
On a prescribed equality face, apply this construction to its free
coordinates. For reconstruction in the external coordinates, choose a
grid representative in the same ordered domain and within distance $h$:
\begin{align}
 x_{\rm grid}&\in F,\qquad
       \|x-x_{\rm grid}\|_\infty\leq h,\\
 |X_F(x)-X_F(x_{\rm grid})|
 &\leq h^{1/2}\|X\|_{1/2,\mathrm{pw}}.
 \label{eq:certificate-external-reconstruction}
\end{align}
Such representatives are obtained by placing the distinct coordinates
in each mesh interval at equally spaced interior points, keeping equal
coordinates equal and leaving coordinates fixed at $0$ or $1$ unchanged. In fixed dimension their number is polynomial in
$h^{-1}$. The identity term of $I+K+R$ is evaluated separately and is
not replaced by an integral.

Use the already certified approximations of $K,R$ at these quadrature
points. For each exact linear block or elementary operation $T$ and its
computed version $\widetilde T$, the preceding estimates give
\begin{align}
 \|\widetilde T X\|_\infty&\leq C_{\beta,K}\|X\|_\infty,\\
 \|(\widetilde T-T)X\|_\infty
 &\leq C_{\beta,K}(M^{-2}+R_M+h^{1/2})
                         \|X\|_{1/2,\mathrm{pw}}.
 \label{eq:certificate-operation-error}
\end{align}
For the integral terms, the kernel replacement error is multiplied by
$\sum_Q|Q|=1$, not by the number of quadrature points. For a product of
$l$ operations, use the exact identity
\begin{align}
 \widetilde T_l\cdots\widetilde T_1e_0-T_l\cdots T_1e_0
 &=\sum_{r=1}^l\widetilde T_l\cdots\widetilde T_{r+1}
       (\widetilde T_r-T_r)T_{r-1}\cdots T_1e_0,\\
 \|\widetilde T_l\cdots\widetilde T_1e_0-T_l\cdots T_1e_0\|_\infty
 &\leq C_{\beta,K,l}(M^{-2}+R_M+h^{1/2}).
 \label{eq:certificate-composition-error}
\end{align}
Only the exact inputs $T_{r-1}\cdots T_1e_0$ require the bound
\eqref{eq:exact-iterate-holder-bound}. Taking $R_M\leq M^{-2}$ and
$h=M^{-4}$ makes the final error $O_{\beta,K,j}(M^{-2})$.

There are polynomially many stored values, matrix entries, and
quadrature regions, because their dimensions are fixed:
\begin{align}
 \dim\operatorname{ran}P_h&\leq C h^{-2}=C M^8,\\
 \#\{Q\}&\leq C_{a,k}h^{-k}=C_{a,k}M^{4k},\\
 \|L_{\rm fin}^{-1}\|_{2\to2}+\|L_{\rm fin}\|_{2\to2}
 &\leq C_\beta
       \qquad\text{for all sufficiently fine certified meshes}.
\end{align}
The initial certificate is obtained at fixed $\beta$, independently
of the requested accuracy. Increasing the number of arithmetic bits
to $C_{\beta,K,j}\log(M+2)$ bounds the total rounding error by $M^{-2}$;
the constant accounts for the polynomial number of operations and the
certified inverse bound. Finally choose
$M\geq C_{\beta,K,j}\varepsilon^{-1/2}$, increasing the constant to
cover the error bound. This gives polynomial bit cost in
$\varepsilon^{-1}$.
\end{proof}

\section{Scalar pinching and Gaussian measurement calculations}
\label{app:measurement-calculus}
\subsection{Scalar inequalities and energy pinching}

\begin{lemma}[Two scalar inequalities]\label{lem:odd-scalar-moments}
Let $Y$ be a real random variable with a finite $2m$th moment, where
$m$ is odd, and let $b=\EE Y$.  Then
\begin{align}
 0\leq\EE Y^{2m}-b^{2m}
 &\leq C_m\operatorname{Var}(Y^m),
 \label{eq:scalar-odd-variance}\\
 \EE|Y^m-b^m|^2
 &\leq C_m\bigl(\EE Y^{2m}-b^{2m}\bigr).
 \label{eq:scalar-power-jensen-remainder}
\end{align}
Consequently the first right side may be replaced by
$C_m\EE|Y^m-y^m|^2$ for any fixed real $y$.
\end{lemma}
\begin{proof}
For odd $m$, homogeneity gives the scalar bounds
\begin{align}
 (x-y)(x^{2m-1}-y^{2m-1})&\leq C_m(x^m-y^m)^2,
 \label{eq:odd-power-compression}\\
 |x^m-y^m|^2&\leq C_m[x^{2m}-y^{2m}-2my^{2m-1}(x-y)].
 \label{eq:scalar-pointwise-power-remainder}
\end{align}
Indeed, at $y=1$ the ratios are continuous except at a removable
singularity $x=1$, with limits $(2m-1)/m^2$ and $m/(2m-1)$;
both tend to one at infinity. Oddness and strict convexity exclude
other denominator zeros; $y=0$ is immediate.
For an independent copy $Y'$ and $b=\EE Y$, convexity gives
\begin{align}
 0\leq\EE Y^{2m}-b^{2m}
 &\leq2m\EE[(Y-b)Y^{2m-1}]\nonumber\\
 &=m\EE[(Y-Y')(Y^{2m-1}-(Y')^{2m-1})]
 \leq C_m\operatorname{Var}(Y^m).
\end{align}
Integrating the second scalar bound at $y=b$ proves the other claim;
$\operatorname{Var}(Y^m)\leq\EE|Y^m-y^m|^2$ proves the final assertion.
\end{proof}

\begin{lemma}[Pinching odd powers]\label{lem:source-power-pinching}
Choose $\delta_E=N^{-\varepsilon}$ with $0<2\varepsilon<c$ and set
$c_0=(c-2\varepsilon)/4$, as in \eqref{eq:energy-width-choice}.
For every fixed odd $m\geq1$,
\begin{align}
 0\leq M_m-\EE_{J,u}\Tr\bigl(\rho(u)(\Pi_uB)^{2m}\bigr)
 &\leq C_m\bigl(N^{-c_0}M_{m-1}+N^{-mc_0}\bigr),
 \label{eq:pinching-loss-final}\\
 \EE_{J,u}\|B^m-(\Pi_uB)^m\|_{2,\rho(u)}^2
 &\leq C_m\bigl(N^{-c_0}M_{m-1}+N^{-mc_0}\bigr).
 \label{eq:pinched-power-error}
\end{align}
\end{lemma}
\begin{proof}
Fix $J,u$. In a block eigenbasis $v$ of $\Pi_u(B^m)$, write
$\rho(u)v=w_vv$ and apply Jensen and \Cref{lem:odd-scalar-moments}:
\begin{align}
 \Tr\rho(u)[B^{2m}-(\Pi_uB)^{2m}]
 &\leq\sum_vw_v[\langle v,B^{2m}v\rangle-\langle v,Bv\rangle^{2m}]\nonumber\\
 &\leq C_m\sum_vw_v[\langle v,B^{2m}v\rangle-\langle v,B^mv\rangle^2]
 =C_m\|B^m-\Pi_u(B^m)\|_{2,\rho(u)}^2.
 \label{eq:pinching-by-power-variance}
\end{align}
In a block eigenbasis of $\Pi_uB$, the other scalar inequality gives
\begin{align}
 0&\leq\Tr\rho(u)[B^{2m}-(\Pi_uB)^{2m}],\\
 \|B^m-(\Pi_uB)^m\|_{2,\rho(u)}^2
 &\leq C_m\Tr\rho(u)[B^{2m}-(\Pi_uB)^{2m}].
 \label{eq:pinching-power-jensen}
\end{align}
No operator convexity of powers is used.

In an eigenbasis of $-\beta H$, let $x_{ab}=|E_a-E_b|$.
The shifts separating $E_a,E_b$ have length $\min(\delta_E,x_{ab})$.
Since $e^{\beta\widetilde H(u)}\leq e^{\delta_E}e^{\beta H}$ and
$e^{\delta_E}-1\geq\delta_E$, the exact joint law implies
\begin{align}
 \EE_{J,u}\|F-\Pi_uF\|_{2,\rho(u)}^2
 &\leq\frac{e}{\EE\Tr e^{\beta H}}\EE\sum_{a,b}e^{-E_a}
                    \min(1,x_{ab}/\delta_E)|F_{ab}|^2,
 \label{eq:quadratic-pinching-multiplier}\\
 \Phi_N([F,[-\beta H,F]])
 &=\frac{\EE\sum_{a,b}(e^{-E_a}+e^{-E_b})
               x_{ab}\tanh(x_{ab}/2)|F_{ab}|^2}{\EE\Tr e^{\beta H}}.
 \label{eq:pinching-curvature-spectral}
\end{align}
For $x\geq0$, $\delta_E\leq1$,
$\min(1,x/\delta_E)\leq C\delta_E^{-1}\sqrt{x\tanh(x/2)}$.
Symmetrize the first sum, apply Cauchy--Schwarz, and set $F=B^m$:
\begin{align}
 \EE_{J,u}\|B^m-\Pi_u(B^m)\|_{2,\rho(u)}^2
 &\leq C_mN^\varepsilon
           \sqrt{M_m(N^{-c}M_{m-1}+N^{-mc})}\nonumber\\
 &\leq C_mN^{-2c_0}M_{m-1}
       +C_mN^{\varepsilon-mc/2}\sqrt{M_m}\nonumber\\
 &\leq C_m(N^{-2c_0}M_{m-1}+N^{-2mc_0}).
 \label{eq:pinching-loss-odd}
\end{align}
Indeed, $2c_0=c/2-\varepsilon$, $M_m\leq M_{m-1}$, and
\begin{align}
 N^{\varepsilon-mc/2}\sqrt{M_m}
 &\leq N^{-2c_0}M_{m-1}
       +\tfrac14N^{-c(m-1)-2c_0}\nonumber\\
 &\leq N^{-2c_0}M_{m-1}+\tfrac14N^{-2mc_0}.
\end{align}
Average \eqref{eq:pinching-by-power-variance} and
\eqref{eq:pinching-power-jensen}, and use
\eqref{eq:gibbs-as-rounded-average}. This proves both claims with the
stronger exponent $2c_0$, which will also be used for even powers below.
\end{proof}

\begin{lemma}[Pinching at every fixed power]\label{lem:all-power-pinching}
For $c_0$ chosen in \eqref{eq:energy-width-choice},
\eqref{eq:all-power-pinching-input} holds for every integer $p\geq1$.
\end{lemma}
\begin{proof}
Only even $p$ remains. The elementary upper bound for $|x^p-y^p|$
and lower bounds for the adjacent odd divided differences give
\begin{align}
 |x^p-y^p|^2
 &\leq C_p|x^{p-1}-y^{p-1}|\,|x^{p+1}-y^{p+1}|\nonumber\\
 &\leq C_p(s^2|x^{p-1}-y^{p-1}|^2+s^{-2}|x^{p+1}-y^{p+1}|^2).
 \label{eq:adjacent-odd-powers}
\end{align}
In a block eigenbasis $v$ of $\Pi_uB$, with eigenvalues $b_v$, the
squared norm is the sum of the spectral integrals of $(x^p-b_v^p)^2$
with weights $w_v$. Use the stronger odd-power estimates with exponent
$2c_0$ proved in \Cref{lem:source-power-pinching}, and take
$s^2=N^{-c_0}$. Then
\begin{align}
 \EE_{J,u}\|B^p-(\Pi_uB)^p\|_{2,\rho(u)}^2
 &\leq C_p\bigl(N^{-3c_0}M_{p-2}+N^{-(2p-1)c_0}
                  +N^{-c_0}M_p+N^{-(2p+1)c_0}\bigr)\nonumber\\
 &\leq C_p(N^{-c_0}M_{p-1}+N^{-pc_0}).
\end{align}
Here $M_p\leq M_{p-1}$. For $p>2$, moment interpolation and
\Cref{fact:young} give
\begin{align}
 N^{-3c_0}M_{p-2}
 &\leq N^{-3c_0}M_{p-1}^{(p-2)/(p-1)}\nonumber\\
 &=\bigl(N^{-c_0}M_{p-1}\bigr)^{(p-2)/(p-1)}
       \bigl(N^{-(2p-1)c_0}\bigr)^{1/(p-1)}\nonumber\\
 &\leq N^{-c_0}M_{p-1}+N^{-(2p-1)c_0}
 \leq N^{-c_0}M_{p-1}+N^{-pc_0}.
\end{align}
For $p=2$, use $M_0=1$ and $N^{-3c_0}\leq N^{-2c_0}$.
\end{proof}

\subsection{Integrated Gaussian identities}

\begin{lemma}[Gaussian Kraus calculations]\label{lem:kraus-calculus}
Let $\rho=\sum_jw_jP_j$ be scalar on its spectral blocks, and let $F$
be a block-diagonal Hermitian contraction. For any matrix $X$,
\begin{align}
 \int_{\mathbb R}K_x(F)^2\,dx&=I,\\
 \int_{\mathbb R}\|[K_x(F),X]\|_{2,\rho}^2\,dx
 &\leq\frac1{4\sigma^2}\|[F,X]\|_{2,\rho}^2,
 \label{eq:kraus-commutator-bound}\\
 \int_{\mathbb R}\|(F^m-x^m)K_x(F)\|_{2,\rho}^2\,dx
 &\leq C_m\left(\sigma^2\Tr(\rho F^{2m-2})+\sigma^{2m}\right).
 \label{eq:power-readout-bound}
\end{align}
For a Hermitian unitary $\psi$ and block-diagonal $X$, the harmonic
norm in \eqref{eq:harmonic-gibbs-norm} satisfies
\begin{align}
 \int\|[\psi,K_x(F)]X\|_{\rm h}^2\,dx
 &\leq\frac1{2\sigma^2}\|[\psi,F]X\|_{\rm h}^2
                      +\frac3{8\sigma^4}\|[F,X]\|_{2,\rho}^2.
 \label{eq:one-kraus-harmonic-integrated}
\end{align}
All reversed orientations satisfy the analogous statements.
\end{lemma}
\begin{proof}
In a common eigenbasis of $F,\rho$, the first three assertions follow
from
\begin{align}
 \int K_x(\lambda)^2\,dx&=1,\\
 \int|K_x(\lambda)-K_x(\mu)|^2\,dx
 &=2(1-e^{-(\lambda-\mu)^2/(8\sigma^2)})
 \leq(\lambda-\mu)^2/(4\sigma^2),\\
 |\lambda^m-x^m|^2
 &\leq C_m(|\lambda|^{2m-2}|x-\lambda|^2+|x-\lambda|^{2m}).
\end{align}
For the last line integrate against $x\sim\mathcal N(\lambda,\sigma^2)$.
For the harmonic estimate, Fourier Plancherel turns its left side into
$\EE_t\|[\psi,e^{-itF}]X\|_{\rm h}^2$, where
$\EE t^2=(4\sigma^2)^{-1}$, $\EE t^4=3/(16\sigma^4)$.
Duhamel and block-unitary invariance give
\begin{align}
 [\psi,e^{-itF}]X
 &=-i\int_0^t e^{-isF}[\psi,F]e^{-i(t-s)F}X\,ds,\\
 \|[e^{-ivF},X]\|_{2,\rho}&\leq|v|\|[F,X]\|_{2,\rho},\\
 \|[\psi,e^{-itF}]X\|_{\rm h}
 &\leq|t|\|[\psi,F]X\|_{\rm h}+t^2\|[F,X]\|_{2,\rho}.
\end{align}
For the second term move the entire $X$ once, use
\eqref{eq:harmonic-elementary-bounds} and $\|[\psi,F]\|\leq2$.
Squaring by $(a+b)^2\leq2a^2+2b^2$ and averaging gives the two stated
constants. Reversal proves the other orientations.
\end{proof}

\begin{lemma}[Sequential source and variance estimates]
\label{lem:sequential-measurement}
Use the forward/reverse instrument of \Cref{sec:concentration}, with the source
measured first in the forward order. For some $b>0$ independent of
fixed odd $m$, its posterior satisfies
\begin{align}
 0\leq M_m-\EE|a|^{2m}
 &\leq C_m(N^{-b}M_{m-1}+N^{-mb}),
 \label{eq:posterior-moment-gap}\\
 \frac12\EE_{J,u}\int\Bigl(
 \|O_{\mathbf x}(\Pi_uB)^m-a^mO_{\mathbf x}\|_{2,\rho(u)}^2
 &+\|(\Pi_uB)^mO_{\mathbf x}-a^mO_{\mathbf x}\|_{2,\rho(u)}^2
                   \Bigr)\,d\mathbf x\nonumber\\
 &\leq C_m(N^{-b}M_{m-1}+N^{-mb}).
 \label{eq:posterior-power-jensen}
\end{align}
For every measured probe $C_\ell$, both conditional variance
orientations obey
\begin{align}
 \EE\bigl[|a|^{2m}\operatorname{Var}_{\rho_x}(\Pi_uC_\ell)\bigr]
 &\leq C_m(N^{-b}M_{m-1}+N^{-mb}).
 \label{eq:posterior-grid-variance}
\end{align}
Finally,
\begin{align}
 &\frac1{2N}\EE_{J,u}\int |a|^{2m}\sum_i
       \left(\|[\psi_i,O_{\mathbf x}]\|_{{\rm h},u}^2
                    +\|[\psi_i,O_{\mathbf x}^\dagger]\|_{{\rm h},u}^2\right)
                     d\mathbf x\nonumber\\
 &\qquad\leq C_m(N^{-b}M_{m-1}+N^{-mb}).
 \label{eq:source-weighted-harmonic-bound}
\end{align}
\end{lemma}
\begin{proof}
Write $K_0=K_{x_0}(\Pi_uB)$, $K_\ell=K_{x_\ell}(\Pi_uC_\ell)$,
and $O=K_L\cdots K_0$. When $X$ is independent of $x$, completeness
and $[K_x(F),\rho]=0$ imply
\begin{align}
 \int\|K_x(F)X\|_{2,\rho}^2\,dx
 &=\int\|XK_x(F)\|_{2,\rho}^2\,dx=\|X\|_{2,\rho}^2.
 \label{eq:integrated-kraus-isometry}
\end{align}
The two exact telescopes
\begin{align}
 O(\Pi_uB)^m-x_0^mO
 &=K_L\cdots K_1((\Pi_uB)^m-x_0^m)K_0,\\
 [(\Pi_uB)^m,O]
 &=\sum_{\ell=1}^LK_L\cdots K_{\ell+1}
               [(\Pi_uB)^m,K_\ell]K_{\ell-1}\cdots K_0
 \label{eq:product-commutator}
\end{align}
and \Cref{lem:kraus-calculus,lem:measurement-physical-input} give
\begin{align}
 &\EE_{J,u}\int\bigl(\|O(\Pi_uB)^m-x_0^mO\|_{2,\rho(u)}^2
              +\|(\Pi_uB)^mO-x_0^mO\|_{2,\rho(u)}^2\bigr)d\mathbf x
 \nonumber\\
 &\quad\leq C_m(\sigma^2M_{m-1}+\sigma^{2m})
 +C_mL^2\sigma^{-2}(1+T_N)^{C_m}e^{C_m(1+T_N)^2}
                       (N^{-c_1}M_{m-1}+N^{-mc_1})\nonumber\\
 &\quad\leq C_m(N^{-b}M_{m-1}+N^{-mb}).
 \label{eq:power-intertwining-readout}
\end{align}
The source noise term uses scalar Jensen for the pinched even moment;
the other variables integrate by \eqref{eq:integrated-kraus-isometry}.
The scales of \Cref{sec:concentration} allow $b=c_1/32$, including the pinching loss.
For any block-diagonal Hermitian $F$ and real $y$, the two measurement
orders give the exact identity
\begin{align}
 p_u(\mathbf x)\Tr\left[\rho_x(F-yI)^2\right]
 &=\frac12\Tr\left[\rho(u)(OO^\dagger+O^\dagger O)(F-yI)^2\right]
 \nonumber\\
 &=\frac12\left(\|(F-yI)O\|_{2,\rho(u)}^2
                  +\|O(F-yI)\|_{2,\rho(u)}^2\right).
 \label{eq:two-order-spectral-error}
\end{align}
Conditionally on $(J,u,\mathbf x)$, let $Y$ have the spectral distribution
of $\Pi_uB$ in $\rho_x$. State preservation and the preceding identity give
\begin{align}
 \EE[Y\mid J,u,\mathbf x]&=a,
 &\EE Y^{2m}&=\EE_{J,u}\Tr[\rho(u)(\Pi_uB)^{2m}].\nonumber
\end{align}
Applying \eqref{eq:two-order-spectral-error} with $F=(\Pi_uB)^m$ gives
\begin{align}
 &\EE|Y^m-x_0^m|^2\nonumber\\
 &\quad=\frac12\EE_{J,u}\int\left(
 \|O(\Pi_uB)^m-x_0^mO\|_{2,\rho(u)}^2
 +\|(\Pi_uB)^mO-x_0^mO\|_{2,\rho(u)}^2\right)d\mathbf x,\nonumber\\
 &0\leq\EE_{J,u}\Tr[\rho(u)(\Pi_uB)^{2m}]-\EE|a|^{2m}
 \leq C_m\EE|Y^m-x_0^m|^2,\nonumber\\
 &\EE|Y^m-a^m|^2
 \leq C_m\left(\EE_{J,u}\Tr[\rho(u)(\Pi_uB)^{2m}]-\EE|a|^{2m}\right).
 \label{eq:measurement-conditional-scalar-comparison}
\end{align}
The last two lines apply \Cref{lem:odd-scalar-moments} conditionally,
then average. Add \eqref{eq:pinching-loss-final} and apply
\eqref{eq:power-intertwining-readout} to obtain
\eqref{eq:posterior-moment-gap}. Taking $F=(\Pi_uB)^m$ and $y=a^m$ in
\eqref{eq:two-order-spectral-error} identifies the last left side with
the half-sum in \eqref{eq:posterior-power-jensen}, proving that estimate.

For the harmonic estimate first use the outcome-independent
$X=(\Pi_uB)^m$ and expand
\begin{align}
 [\psi_i,O]X
 &=\sum_{\ell=0}^LK_L\cdots K_{\ell+1}[\psi_i,K_\ell]
                                   XK_{\ell-1}\cdots K_0\nonumber\\
 &\quad+\sum_{\ell=0}^LK_L\cdots K_{\ell+1}[\psi_i,K_\ell]
                                   [K_{\ell-1}\cdots K_0,X].
 \label{eq:fixed-source-sequential-telescope}
\end{align}
For the first sum use \eqref{eq:one-kraus-harmonic-integrated}; expand
the second commutator and use
\begin{align}
 \int\|[\psi_i,K_x(F)]Y\|_{{\rm h},u}^2\,dx
 &\leq4\|Y\|_{2,\rho(u)}^2\qquad(Y\text{ block diagonal}).
\end{align}
This follows by splitting the commutator and applying
\eqref{eq:harmonic-elementary-bounds}. Integrate all outside factors
block by block. At most $(L+1)^2$ terms give a squared cost
$C(L+1)^4\sigma^{-4}$ times \eqref{eq:harmonic-rooted-source-bound}.
Replace the fixed source by its posterior mean using
\begin{align}
 a^m[\psi_i,O]
 &=[\psi_i,O](\Pi_uB)^m
   +[\psi_i,a^mO-O(\Pi_uB)^m]+O[\psi_i,(\Pi_uB)^m].
 \label{eq:posterior-source-commutator-identity}
\end{align}
The middle term uses the power error without any inverse noise factor;
the last uses the physical commutator after integration. Thus the
harmonic left side is bounded by
\begin{align}
 C_m(N^{-b}M_{m-1}+N^{-mb})
 +C_m(L+1)^4\sigma^{-4}(1+T_N)^{C_m}e^{C_m(1+T_N)^2}
                         (N^{-c_1}M_{m-1}+N^{-mc_1}).
\end{align}
Reverse all factors for $O^\dagger$ and decrease the common $b$ to
obtain \eqref{eq:source-weighted-harmonic-bound}.

For a measured $R=\Pi_uC_\ell$, clip only its readout to
$y_\ell=\max(-1,\min(1,x_\ell))$. Then
\begin{align}
 \int(R-y_\ell)^2K_{x_\ell}(R)^2\,dx_\ell&\leq\sigma^2I,\\
 a^m(R-y_\ell)O
 &=(R-y_\ell)O(\Pi_uB)^m
                +(R-y_\ell)(a^mO-O(\Pi_uB)^m).
 \label{eq:variance-source-telescope}
\end{align}
The second term costs at most four times the power error. In the first,
move $R$ to its own Kraus factor and move $X=(\Pi_uB)^m$ once using
\eqref{eq:fixed-source-sequential-telescope}. The resulting commutators
are $[R,\Pi_uC_j]X$ and $[\Pi_uC_j,X]$; their bounds are
\eqref{eq:weighted-pinched-pair-input} and
\eqref{eq:harmonic-rooted-source-bound}. The same Fourier calculation
as in \Cref{lem:kraus-calculus}, in the block Hilbert norm, costs at most
$C(L+1)^4\sigma^{-4}$. Therefore the variance is bounded by the preceding
display plus $C_m\sigma^2M_m$. Minimizing over scalar centers gives
\eqref{eq:posterior-grid-variance}, in either order. Every completeness
identity was applied only after separating its outcome-dependent factor.
\end{proof}

\subsection{Products and continuous real times}

\begin{lemma}[Conditional product approximation]\label{lem:posterior-product}
Use the empirical operator $C(t)$ in \eqref{eq:measured-probes}.
For $|s_j|,|t_j|,|s_j-t_j|\leq T_N$ and fixed or logarithmically growing $r$, let
\begin{align}
 T_j&=\frac1N\sum_i
     (e^{-is_j\beta H}\psi_i e^{is_j\beta H})
     (e^{-it_j\beta H}\psi_i e^{it_j\beta H}).
\end{align}
For the same measurement law as above and some $b>0$ independent of
fixed $m$,
\begin{align}
 &\left|\EE_{J,u}\Tr\left[\rho(u)(\Pi_uB)^{2m}T_1\cdots T_r\right]
      -\EE\left[|a|^{2m}\prod_{j=1}^r2G(i(s_j-t_j))\right]\right|
 \nonumber\\
 &\qquad\leq C_m(1+r)^{C_m}
                  (N^{-b}M_{m-1}+N^{-mb})+C_mr^2N^{-L_0}.
 \label{eq:continuous-conditional-product}
\end{align}
A reversed factor uses $G(-it)=G(1+it)$. The grid can be chosen once
for all these factors with $(\log(N+2))^{C_{m,L_0}}$ real points.
\end{lemma}
\begin{proof}
\proofstep{One interpolation per conditional variance}
Differentiating a propagated Majorana $d$ times and applying
\eqref{eq:partial-gaussian-matching} to its squared norm gives
\begin{align}
 \sup_{|t|\leq T}\bigl(\|C^{(d)}(t)\|_{2,\Phi_N}
                    +\|C^{(d)}(t)\|_{{\rm row},\Phi_N}\bigr)
 &\leq[C_\beta(1+T)d]^d.
 \label{eq:physical-coordinate-derivatives}
\end{align}
Indeed, after $\ell$ pairs there are $b=2d-\ell$ rooted quartet
indices, at most $[C(b+1)N^3]^b$ assignments, and at most
$2^{2d}(2d)^\ell$ partial pairings. The remaining derivatives cost
$[C(1+T)]^{2d-2\ell}$ by \eqref{eq:complete-contour-derivative};
$N^{-3b}$ cancels the label count. Taking the square root proves the
bound, including coincidences.

Taylor's formula and $d!\geq(d/e)^d$ give a Hilbert-valued analytic
disk of radius $c_\beta/(1+T)$ at each real time. Cover $[-T,T]$ by
intervals of one tenth that radius and interpolate at Chebyshev nodes
of degree $D=C_{L_0}\log(N+2)$. The fixed-ellipse contour remainder
and the harmonic-sum bound for the Lagrange coefficients give
\begin{align}
 C(t)&=\sum_j\lambda_j(t)C(t_j)+R_t,
 &\sum_j|\lambda_j(t)|&\leq C\log(D+1),\\
 \|R_t\|_{2,\Phi_N}+\|R_t\|_{{\rm row},\Phi_N}&\leq N^{-L_0},
 &\#\{t_j\}&\leq C_\beta(T(1+T)+1)(D+1).
\end{align}
Thus the grid is polylogarithmic at $T=T_N$. In the direct-integral
Hilbert space of posterior states, apply interpolation to just one
variance:
\begin{align}
 \bigl(\EE|a|^{2m}\operatorname{Var}_{\rho_x}(\Pi_uC(t))\bigr)^{1/2}
 &\leq\sum_j|\lambda_j(t)|
       \bigl(\EE|a|^{2m}\operatorname{Var}_{\rho_x}(\Pi_uC(t_j))\bigr)^{1/2}
       +N^{-L_0}.
 \label{eq:one-factor-posterior-interpolation}
\end{align}
The remainder uses $|a|\leq1$, contractivity of pinching, and
$\EE_x\rho_x=\rho(u)$. Apply \eqref{eq:posterior-grid-variance}
to the two Hermitian quadratures, and to their reversed orientations.

\proofstep{Products and the source weight}
For contractions $F_j$ and $f_j=\Tr(\rho_xF_j)$,
\begin{align}
 \prod_{j=1}^rF_j-\prod_{j=1}^rf_jI
 &=\sum_{j=1}^rF_1\cdots F_{j-1}(F_j-f_jI)\prod_{k>j}f_k.
 \label{eq:posterior-product-telescope}
\end{align}
Cauchy--Schwarz bounds each expected summand by that factor's standard
deviation. Comparing $(\Pi_uB)^mO$ with $a^mO$, and also the reverse
amplitudes, gives
\begin{align}
 &\left|\EE_{J,u}\Tr[\rho(u)(\Pi_uB)^{2m}\prod_jF_j]
       -\EE[|a|^{2m}\Tr(\rho_x\prod_jF_j)]\right|\nonumber\\
 &\quad\leq2\sqrt{M_m}\left[\frac12\EE_{J,u}\int
       \bigl(\|O(\Pi_uB)^m-a^mO\|_{2,\rho(u)}^2
            +\|(\Pi_uB)^mO-a^mO\|_{2,\rho(u)}^2\bigr)d\mathbf x\right]^{1/2}.
 \label{eq:source-state-replacement}
\end{align}
This is $\|XX^\dagger-YY^\dagger\|_1\leq
(\|X\|_2+\|Y\|_2)\|X-Y\|_2$. Combining the two displays costs
$r\sqrt{M_m}$ times the largest conditional standard deviation, with
only one interpolation coefficient mass, not its $r$th power.

Replace each $T_j$ by $\Pi_uT_j$ in turn. For a block-diagonal
contraction prefix $P$, error $E=T_j-\Pi_uT_j$, and contraction
suffix $U$,
\begin{align}
 |\Tr(\rho(u)(\Pi_uB)^{2m}PEU)|
 &\leq\Tr(\rho(u)(\Pi_uB)^{2m})^{1/2}
                  \|(\Pi_uB)^mPE\|_{{\rm row},\rho(u)}.
\end{align}
Move $(\Pi_uB)^m$ through $P$ using at most $r$ commutators, then use
\eqref{eq:harmonic-rooted-source-bound} and
\eqref{eq:weighted-pinched-pair-input}. The suffix contracts in the
row norm; summing costs $Cr^2$ times the same source-weighted error.
For a block-diagonal $X$,
\begin{align}
 \|e^{-it\beta H}Xe^{it\beta H}-X\|
 &\leq2|t|\delta_E\|X\|,
\end{align}
because $[\beta\widetilde H(u),X]=0$. Hence rounded evolution removes
the common-time translations at cost $CT_N\delta_EM_m$. The resulting
means are the characteristic functions of $\nu_x$. The tensor estimate
\eqref{eq:conditional-tensor-db} replaces their product by that of
$\mathcal P\nu_x$ at cost $r\EE[|a|^{2m}D]$, without changing $a$.
Finally,
\begin{align}
 \sqrt{M_m(N^{-c}M_{m-1}+N^{-mc})}
 &\leq C_m(N^{-c/2}M_{m-1}+N^{-mc/2}).
 \label{eq:productive-square-root}
\end{align}
Use $M_m\leq M_{m-1}\leq1$ and Young's inequality for this last step.
Absorb the single polylogarithmic factor and the fixed rounding rate
by decreasing $b$. This proves \eqref{eq:continuous-conditional-product}
with the same grid and instrument for every factor.
\end{proof}

\section{Changing the cavity source and bounding the omitted terms}
\label{app:cavity-errors}
The estimates here concern the Hamiltonians and source in
\eqref{eq:cavity-hamiltonian-split}--\eqref{eq:cavity-four-correlators}.
All traces use the same ambient even Clifford algebra. The normalization
of an empirical source is $1/N$ throughout, including after addition of
the cavity Majorana.

\subsection{Moving and pinching the source}

\begin{lemma}[Moving the source and rounding the spectrum]
\label{lem:cavity-source-rounding}
Use $T,U_b,V_b$ from \eqref{eq:vt} and \eqref{eq:cavity-boundary-unitaries}, with
$2d+2$ static factors of degree at most four. For each fixed odd $m$,
there is $a_1>0$, independent of $m$, such that
\begin{align}
 &\left|\Phi_N(U_bB^{2m}V_b)-\Phi_N(B^{2m}U_bV_b)\right|\nonumber\\*
 &\quad+\left|\Phi_N(B^{2m}U_bV_b)
 -\EE_{J,u}\Tr\left[\rho(u)(\Pi_uB)^{2m}U_bV_b\right]\right|\nonumber\\*
 &\qquad\leq C_m(1+d)^{C_m}(1+T)^{C_m}e^{C_m(1+T)^2}
 (N^{-a_1}M_{m-1}+N^{-ma_1}).
 \label{eq:cavity-source-rounding-error}
\end{align}
\end{lemma}
\begin{proof}
For $R\in\{U_b,V_b,U_bV_b\}$, write its intervals explicitly as
\begin{align}
 R&=e^{il_0\beta H}P_1e^{il_1\beta H}\cdots P_he^{il_h\beta H},
 &h&\leq2d+2,\nonumber\\
 t&=\sum_{a=0}^h l_a,
 &\sum_{a=0}^h|l_a|&\leq T,\nonumber\\
 R&=(Re^{-it\beta H})e^{it\beta H},
 &\sum_{a=0}^{h-1}l_a+(l_h-t)&=0,\nonumber\\
 |t|&\leq T,
 &\sum_{a=0}^{h-1}|l_a|+|l_h-t|&\leq2T.
\end{align}
Apply \eqref{eq:source-power-commutator} to the first factor and
\eqref{eq:whole-power-motion} to the second, in the two orientations
\eqref{eq:partial-word-two-orientations}. With
$\|X\|_{2,\Phi_N}^2=\Phi_N(X^\dagger X)$, this gives
\begin{align}
 \|[R,B^m]\|_{2,\Phi_N}^2+\|[R^\dagger,B^m]\|_{2,\Phi_N}^2
 &\leq C_m(1+d)^{C_m}(1+T)^{C_m}e^{C_m(1+T)^2}
 (N^{-c}M_{m-1}+N^{-mc}).
 \label{eq:cavity-compact-commutator-bound}
\end{align}
The exact identities needed to move the source are
\begin{align}
 U_bB^{2m}V_b-B^{2m}U_bV_b
 &=[U_b,B^m]B^mV_b+B^m[U_b,B^m]V_b,\nonumber\\
 B^mV_b&=V_bB^m+[B^m,V_b],\nonumber\\
 [U_b,B^m]V_b&=[U_bV_b,B^m]-U_b[V_b,B^m].
 \label{eq:cavity-source-cut-relocation}
\end{align}
Cauchy--Schwarz in the positive state $\Phi_N$ and unitarity imply
\begin{align}
 \|B^mV_b\|_{2,\Phi_N}
 &\leq\sqrt{M_m}+\|[V_b,B^m]\|_{2,\Phi_N},\nonumber\\
 &\left|\Phi_N(U_bB^{2m}V_b)-\Phi_N(B^{2m}U_bV_b)\right|\nonumber\\*
 &\quad\leq\|[U_b^\dagger,B^m]\|_{2,\Phi_N}
 \left(\sqrt{M_m}+\|[V_b,B^m]\|_{2,\Phi_N}\right)\nonumber\\*
 &\qquad+\sqrt{M_m}\left(\|[U_bV_b,B^m]\|_{2,\Phi_N}
 +\|[V_b,B^m]\|_{2,\Phi_N}\right).
 \label{eq:cavity-source-relocation-bound}
\end{align}

For the pinching step, use the exact marginal
\eqref{eq:gibbs-as-rounded-average} and
\begin{align}
 B^{2m}-(\Pi_uB)^{2m}
 &=(B^m-(\Pi_uB)^m)B^m+(\Pi_uB)^m(B^m-(\Pi_uB)^m),\nonumber\\
 B^mU_bV_b&=U_bV_bB^m+[B^m,U_bV_b],\nonumber\\
 [\rho(u),(\Pi_uB)^m]&=0,\nonumber\\
 &\left|\Phi_N(B^{2m}U_bV_b)
 -\EE_{J,u}\Tr\left[\rho(u)(\Pi_uB)^{2m}U_bV_b\right]\right|\nonumber\\*
 &\quad\leq
 \left(\EE_{J,u}\|B^m-(\Pi_uB)^m\|_{2,\rho(u)}^2\right)^{1/2}
 \left(2\sqrt{M_m}+\|[B^m,U_bV_b]\|_{2,\Phi_N}\right).
 \label{eq:cavity-source-rounded-once}
\end{align}
For the second summand in the first line, cyclicity moves $(\Pi_uB)^m$
through $\rho(u)$ before Cauchy--Schwarz is applied. Its squared norm
is at most $M_m$ by \eqref{eq:mm1}. The other source difference is
bounded by \eqref{eq:mm2}. Finally, for any $a>0$,
\begin{align}
 \sqrt{M_m(N^{-a}M_{m-1}+N^{-ma})}
 &\leq N^{-a/2}\sqrt{M_mM_{m-1}}+N^{-ma/2}\sqrt{M_m}\nonumber\\
 &\leq N^{-a/2}M_{m-1}+N^{-ma/2}.
 \label{eq:cavity-square-root-reduction}
\end{align}
Apply this with $a=\min(c,c_0)$; the products of two error norms are
bounded by their squared norms. Choosing $a_1=\min(c,c_0)/2$ proves
the claim, after increasing $C_m$.
\end{proof}

\subsection{Pairing Majoranas at different real times}

\begin{lemma}[Pairing the evolved Majoranas]
\label{lem:cavity-majorana-pairing}
Use the disjoint assignments, cyclic order, and positions $p_i,q_i$
of \Cref{lem:cavity-coefficient-comparison}. There is $a_2>0$,
independent of every fixed odd $m$, such that
\begin{align}
 &\left|\EE_{J,u}\Tr\left[\rho(u)(\Pi_uB)^{2m}U_bV_b\right]\right.\nonumber\\*
 &\qquad\left.-\epsilon_{A,\pi}\EE_{J,u}\Tr\left[\rho(u)(\Pi_uB)^{2m}
 \prod_{i\in(A\cap[N])\cup\bigcup_{e\in\pi}T_e}
 \psi_i(-\alpha_{p_i})\psi_i(-\alpha_{q_i})\right]\right|\nonumber\\*
 &\quad\leq C_m(1+d)^{C_m}(1+T)^{C_m}e^{C_m(1+T)^2}
 (N^{-a_2}M_{m-1}+N^{-ma_2}).
 \label{eq:cavity-majorana-pairing-error}
\end{align}
The product is ordered by increasing $q_i$, then increasing $i$.
The coefficient of $(1+T)^2$ in the exponential is independent of $d$.
\end{lemma}
\begin{proof}
Expand each static factor in \eqref{eq:uv} in canonical Majorana order.
The cavity factors obey
\begin{align}
 e^{-it\beta H}\psi_{N+1}e^{it\beta H}&=\psi_{N+1},
 &\{\psi_{N+1},\psi_i(t)\}&=0\quad(i\leq N).
\end{align}
They therefore cancel exactly. The remaining $r=3d+|A\cap[N]|$
distinct sites each occur twice. Starting with the rightmost remaining
factor $V$, let $U$ be the earlier factor at the same site. The exact
identity
\begin{align}
 PUX_1\cdots X_hV
 &=(-1)^hPX_1\cdots X_hUV\nonumber\\*
 &\quad+\sum_{j=1}^h(-1)^{j-1}
 PX_1\cdots X_{j-1}\{U,X_j\}X_{j+1}\cdots X_hV
 \label{eq:cavity-rightmost-pair-extraction}
\end{align}
places this pair on the right. Repeat only on the first summand.
The extracted pairs have the stated order, and the accumulated sign
is $\epsilon_{A,\pi}$: setting all real times to zero makes every
extracted pair equal to $I$, while the original static product is
\eqref{eq:cavity-clifford-signs}.

There are at most $r(r-1)$ error terms. Each has the form
\begin{align}
 PDSR,\qquad D=\{\psi_i(s),\psi_j(t)\},\quad i\ne j,
 \qquad PP^\dagger=SS^\dagger=RR^\dagger=I,
 \qquad\|D\|\leq2.
\end{align}
Here $P,S$ retain the order of subsequences of the original factors,
and $R$ consists of previously extracted pairs. Since
$\rho(u)(\Pi_uB)^{2m}\succeq0$, Cauchy--Schwarz cancels the entire suffix:
\begin{align}
 &\left|\Tr\left[\rho(u)(\Pi_uB)^{2m}PDSR\right]\right|^2\nonumber\\*
 &\quad\leq\Tr\left[\rho(u)(\Pi_uB)^{2m}\right]
 \Tr\left[\rho(u)(\Pi_uB)^m
 PDSRR^\dagger S^\dagger D^\dagger P^\dagger(\Pi_uB)^m\right]\nonumber\\*
 &\quad=\Tr\left[\rho(u)(\Pi_uB)^{2m}\right]
 \Tr\left[\rho(u)(\Pi_uB)^mPDD^\dagger P^\dagger(\Pi_uB)^m\right],
 \label{eq:cavity-tail-cancellation}\\
 (PDD^\dagger P^\dagger)^q&=P(DD^\dagger)^qP^\dagger.\nonumber
\end{align}
For any subsequence, \eqref{eq:uv} gives
\begin{align}
 |\alpha_{j_1}|+\sum_{h=1}^{k-1}
 |\alpha_{j_{h+1}}-\alpha_{j_h}|+|\alpha_{j_k}|
 &\leq3T\qquad(j_1<\cdots<j_k),\nonumber\\
 |\alpha_p-\alpha_q|&\leq T.
 \label{eq:cavity-subsequence-duration}
\end{align}
Thus $P$ occurs only once on each side of the $q$th power; neither $R$
nor $S$ remains. The two distinct sites in each of the $2q$
anticommutators give exponent $q$ in \Cref{lem:causal-count}.
\Cref{lem:causal-moments} therefore yields
\begin{align}
 \Phi_N\left(B^mP(DD^\dagger)^qP^\dagger B^m\right)
 &\leq C_{m,q}(1+d)^{C_{m,q}}(1+T)^{C_{m,q}}
 e^{C_{m,q}(1+T)^2}N^{-q}.
 \label{eq:cavity-anticommutator-moments}
\end{align}
This application has $O(d+1)$ static factors but total real-time
duration $C_q(1+T)$, not $C_qdT$.

Take a fixed $0<\eta\leq1/4$ and $q=2m+2$. Functional calculus gives
\begin{align}
 PDD^\dagger P^\dagger
 &\preceq N^{-\eta}I+N^{\eta(q-1)}P(DD^\dagger)^qP^\dagger,\nonumber\\
 -q+\eta(q-1)&\leq-(m+1)\eta,\nonumber\\
 \Phi_N(B^mPDD^\dagger P^\dagger B^m)
 &\leq N^{-\eta}M_m+
 C_m(1+d)^{C_m}(1+T)^{C_m}e^{C_m(1+T)^2}N^{-(m+1)\eta},
 \label{eq:cavity-local-error-source}\\
 \EE_{J,u}\|D^\dagger P^\dagger(\Pi_uB)^m\|_{2,\rho(u)}^2
 &\leq2\Phi_N(B^mPDD^\dagger P^\dagger B^m)
 +8\EE_{J,u}\|B^m-(\Pi_uB)^m\|_{2,\rho(u)}^2.
 \label{eq:cavity-pairing-pinched-source}
\end{align}
The last line uses the exact marginal \eqref{eq:gibbs-as-rounded-average}
and $\|D^\dagger P^\dagger\|\leq2$. Average
\eqref{eq:cavity-tail-cancellation}, apply Cauchy--Schwarz, and then use
\eqref{eq:mm1}--\eqref{eq:mm2} and \eqref{eq:cavity-square-root-reduction}:
\begin{align}
 \EE_{J,u}\left|\Tr\left[\rho(u)(\Pi_uB)^{2m}PDSR\right]\right|
 &\leq\sqrt{M_m}
 \left(\EE_{J,u}\|D^\dagger P^\dagger(\Pi_uB)^m\|_{2,\rho(u)}^2\right)^{1/2}
 \nonumber\\
 &\leq C_m(1+d)^{C_m}(1+T)^{C_m}e^{C_m(1+T)^2}
 (N^{-a_2}M_{m-1}+N^{-ma_2}),
\end{align}
where $a_2=\min(\eta,c_0)/2$ suffices. Summing at most
$r(r-1)\leq(3d+4)^2$ errors proves the lemma.
\end{proof}

\subsection{Averaging the paired site labels}

\begin{lemma}[Replacing distinct labels by independent labels]
\label{lem:cavity-site-average}
For distinct $i_1,\ldots,i_r\in[N]$ and arbitrary real times
$s_\ell,t_\ell$, keeping the displayed product order fixed,
\begin{align}
 &\left|\EE_{J,u}\Tr\left[\rho(u)(\Pi_uB)^{2m}
 \prod_{\ell=1}^r\psi_{i_\ell}(s_\ell)\psi_{i_\ell}(t_\ell)\right]\right.
 \nonumber\\*
 &\qquad\left.-\EE_{J,u}\Tr\left[\rho(u)(\Pi_uB)^{2m}
 \prod_{\ell=1}^r\left(\frac1N\sum_{j=1}^N
 \psi_j(s_\ell)\psi_j(t_\ell)\right)\right]\right|
 \leq\frac{r(r-1)}N M_m.
 \label{eq:cavity-independent-label-error}
\end{align}
For $r=0$ both products are $I$ and the difference is zero.
\end{lemma}
\begin{proof}
The joint law of $J,u$ is permutation invariant, and the rounded state
and $\Pi_uB$ transform by the same unitary as $H$. Consequently
\begin{align}
 &\EE_{J,u}\Tr\left[\rho(u)(\Pi_uB)^{2m}
 \prod_{\ell=1}^r\psi_{i_\ell}(s_\ell)\psi_{i_\ell}(t_\ell)\right]
 \nonumber\\*
 &\quad=\frac{(N-r)!}{N!}
 \sum_{\substack{j_1,\ldots,j_r\in[N]\\j_\ell\ne j_k\ (\ell\ne k)}}
 \EE_{J,u}\Tr\left[\rho(u)(\Pi_uB)^{2m}
 \prod_{\ell=1}^r\psi_{j_\ell}(s_\ell)\psi_{j_\ell}(t_\ell)\right].
 \label{eq:cavity-distinct-site-average}
\end{align}
For independent uniform indices $I_1,\ldots,I_r$, conditioning on
all labels being distinct gives exactly the average on the right.
Moreover,
\begin{align}
 \Pr[\exists\ell<k:I_\ell=I_k]&\leq\binom r2\frac1N,\nonumber\\
 \left|\Tr\left[\rho(u)(\Pi_uB)^{2m}
 \prod_{\ell=1}^r\psi_{I_\ell}(s_\ell)\psi_{I_\ell}(t_\ell)\right]\right|
 &\leq\Tr[\rho(u)(\Pi_uB)^{2m}],\nonumber\\
 2\Pr[\exists\ell<k:I_\ell=I_k]\,
 \EE_{J,u}\Tr[\rho(u)(\Pi_uB)^{2m}]
 &\leq\frac{r(r-1)}N M_m.
\end{align}
The last line bounds the difference between the conditional and
unconditional averages. Averaging independent indices inside the ordered
product gives
\begin{align}
 \frac1{N^r}\sum_{j_1,\ldots,j_r}
 \prod_{\ell=1}^r\psi_{j_\ell}(s_\ell)\psi_{j_\ell}(t_\ell)
 =\prod_{\ell=1}^r\left(\frac1N\sum_{j=1}^N
 \psi_j(s_\ell)\psi_j(t_\ell)\right),
\end{align}
which proves \eqref{eq:cavity-independent-label-error}.
\end{proof}

\subsection{Source powers under deletion of one Majorana}

\begin{lemma}[Deleting the cavity interactions]\label{lem:cavity-source-change}
There is $b>0$, independent of the fixed source order $m$, such that
\begin{align}
 \sup_{0\leq t\leq1}\left|U_A(B^+;t)-U_A(B;t)\right|
 &\leq C_m\left(N^{-b}M_{m-1}+N^{-mb}\right)
 \label{eq:cavity-source-restoration}
\end{align}
for each of the four fixed sets $A$ used in the cavity comparison.
\end{lemma}
\begin{proof}
Put $H_\lambda=H+\lambda(H^+-H)$ and let $B_\lambda$ be the source
of \eqref{eq:cavity-full-source} with this Hamiltonian. Thus
$B_0=B$ and $B_1=B^+$. Each individual enlarged source in
\eqref{eq:cavity-full-source} has norm at most $1+1/N$, uniformly in
$\lambda$. By a common fixed rescaling of the coefficients defining
$B$ and every $B_\lambda$, we may therefore assume
$\sup_{0\leq\lambda\leq1}\|B_\lambda\|\leq1$.
Use its own Gibbs normalization:
\begin{align}
 Z_\lambda&=\EE\tr e^{\beta H_\lambda},
 &M_j(\lambda)&=Z_\lambda^{-1}\EE\tr(e^{\beta H_\lambda}B_\lambda^{2j}).
 \label{eq:cavity-own-profile-moments}
\end{align}
For $\psi_i(t)=e^{-it\beta H_\lambda}\psi_i e^{it\beta H_\lambda}$,
Duhamel gives the oriented integral
\begin{align}
 \partial_\lambda\psi_i(t)
 &=-i\int_0^t e^{-i(t-s)\beta H_\lambda}
 [\beta(H^+-H),e^{-is\beta H_\lambda}\psi_i e^{is\beta H_\lambda}]
                      e^{i(t-s)\beta H_\lambda}\,ds.
 \label{eq:cavity-source-derivative}
\end{align}
Each displayed interaction contains $N+1$. For $i\ne N+1$, the path
expansion connects it to the averaged site $i$; $i=N+1$ already costs
$N^{-1}$. Each further empirical commutator adds an averaged site.
The joint count in \Cref{lem:causal-count,lem:causal-moments} therefore
gives, for a fixed positive half-heat word $R$,
\begin{align}
 \frac{\EE\tr\left[R^\dagger B_\lambda^h
 ((\ad_{B_\lambda}^j\partial_\lambda B_\lambda)^\dagger
                  \ad_{B_\lambda}^j\partial_\lambda B_\lambda)^q
 B_\lambda^hR\right]}{Z_\lambda}
 &\leq C_{h,j,q,R}N^{-(j+1)q}.
 \label{eq:cavity-derivative-moments}
\end{align}
The same bound holds for a fixed-quartet directional derivative
\begin{align}
 \left.\frac d{d\epsilon}B(H_\lambda+\epsilon\Gamma_E)\right|_{\epsilon=0},
 \qquad E\ni N+1,
 \label{eq:cavity-directional-source}
\end{align}
and for $B_1-B_\lambda$ or $B_0-B_\lambda$, by integrating the
Duhamel formula with its bounded-mask propagators retained.
More generally, for a product $D$ of such derivatives or differences
with nested depths $r_1,\ldots,r_k$, the same joint count gives
\begin{align}
 w&=\sum_{l=1}^k(r_l+1),
 &\frac{\EE\tr[R^\dagger B_\lambda^h(D^\dagger D)^qB_\lambda^hR]}{Z_\lambda}
 &\leq C_{h,q,D,R}N^{-wq}.
 \label{eq:cavity-difference-high-moments}
\end{align}
This uses joint Gaussian pairings, not a product of separate thermal
norm bounds. All constants are uniform in $\lambda\in[0,1]$, and the
reference remains $H_\lambda$.

\proofstep{Comparing the normalized moments}
Balance the factors in each derivative of $B_\lambda^{2j}$ using
$B_\lambda D=DB_\lambda+[B_\lambda,D]$. Each term has a derivative
of nested depth $r$, with outside degrees $a,b\leq j$ and
$a+b=2j-1-r$. Split its heat into two halves and apply
\eqref{eq:source-operator-young},
\eqref{eq:cavity-derivative-moments}, and
\Cref{lem:positive-heat-source}. A cutoff $N^{-(r+1)\eta}$ and
sufficiently large fixed moment order give
\begin{align}
 |\text{normalized term}|
 &\leq C_jN^{-(r+1)\eta/2}
       \sqrt{(M_a(\lambda)+N^{-a\eta})(M_b(\lambda)+N^{-b\eta})}
             +C_jN^{-j\eta/4}\nonumber\\
 &\leq C_jM_j(\lambda)+C_jN^{-j\eta/4}.
 \label{eq:cavity-balanced-first-variation}
\end{align}
For the last step use $M_l\leq M_j^{l/j}$ and Young's inequality:
$N^{-(r+1)\eta/2}M_j^{1-(r+1)/(2j)}
\leq C_j(M_j+N^{-j\eta})$. The mixed error terms satisfy the same
bound, since $a+b=2j-1-r$.

The complete derivative is
\begin{align}
 (Z_\lambda M_j(\lambda))'
 &=\sum_{h=0}^{2j-1}\EE\tr(e^{\beta H_\lambda}
             B_\lambda^h(\partial_\lambda B_\lambda)B_\lambda^{2j-1-h})
 \nonumber\\
 &\quad+\beta\sqrt{\frac3{8N^3}}\sum_{E\ni N+1}\int_0^1
       \EE J_E\tr(e^{(1-u)\beta H_\lambda}\Gamma_E
                            e^{u\beta H_\lambda}B_\lambda^{2j})\,du.
 \label{eq:cavity-moment-derivative}
\end{align}
Integrate by parts in $J_E$. A derivative on the two heat factors has
weights $1-u,u$, summing to one, and is bounded by
\Cref{lem:positive-heat-source}; on $B_\lambda$ it is
$\lambda\sqrt{3/(8N^3)}$ times \eqref{eq:cavity-directional-source},
controlled above. Since $(3/(8N^3))\binom N3\leq1/16$, the same
calculation for the denominator yields
\begin{align}
 \frac{Z_\lambda'}{Z_\lambda}
 &=\frac{3\beta^2\lambda}{8N^3Z_\lambda}\sum_{E\ni N+1}\int_0^1
       \EE\tr(e^{(1-u)\beta H_\lambda}\Gamma_Ee^{u\beta H_\lambda}\Gamma_E)\,du,
 &0\leq\frac{Z_\lambda'}{Z_\lambda}&\leq\frac{\beta^2\lambda}{16}.
 \label{eq:cavity-partition-ratio}
\end{align}
Positivity is immediate in the energy basis. For a common $\eta_1>0$,
\begin{align}
 |M_j'(\lambda)|&\leq C_j(M_j(\lambda)+N^{-j\eta_1}),
 &|\partial_\lambda\log(M_j(\lambda)+N^{-j\eta_1})|&\leq C_j,\\
 M_j(\lambda)&\leq C_j(M_j(0)+N^{-j\eta_1}),
 &M_j(0)&\leq C_j(M_j(\lambda)+N^{-j\eta_1}),\\
 1&\leq Z_1/Z_0\leq e^{\beta^2/32}.
 \label{eq:cavity-own-state-comparison}
\end{align}

\proofstep{Restoring the source}
Expand $B_0^m-B_1^m$ about $B_1$ and move its remaining powers to the
right. Every term is $DB_1^h$ with $w+h=m$, $w\geq1$ in
\eqref{eq:cavity-difference-high-moments}. Operator Young and positive
half-heat transport give, for a common $\eta_2>0$,
\begin{align}
 N^{-w\eta_2}(M_h(1)+N^{-h\eta_2})+C_mN^{-m\eta_2}
 &\leq C_m(N^{-\eta_2}M_{m-1}(1)+N^{-m\eta_2}),\\
 \frac{\EE\|(B_0^m-B_1^m)R\|_{\rm HS}^2}{\EE\Tr e^{\beta H_1}}
 &\leq C_m(N^{-\eta_2}M_{m-1}(1)+N^{-m\eta_2}).
 \label{eq:cavity-power-difference}
\end{align}
Use $B_0^m=B_1^m+(B_0^m-B_1^m)$ for the other source norm.
In the correlator, write
\begin{align}
 B_1^{2m}-B_0^{2m}
 &=(B_1^m-B_0^m)B_1^m+B_0^m(B_1^m-B_0^m).
\end{align}
Split the heat at cumulative length $1/2$ and apply Hilbert--Schmidt
Cauchy--Schwarz. Now use \eqref{eq:cavity-own-state-comparison} and
\eqref{eq:productive-square-root}, decreasing the common exponent,
to obtain \eqref{eq:cavity-source-restoration}. Both halves keep their
$H_1$ Gibbs factors until the own-state comparison is applied.
\end{proof}

\subsection{Repeated site labels and the factorial tail}

\begin{lemma}[Omitted cavity terms]\label{lem:cavity-omitted}
For each fixed odd $m$ and each fixed $L_0>0$, one may choose
$P_N=\lceil C_{m,L_0}\log(N+2)\rceil$ so that the part of
\eqref{eq:exact-star-gaussian-expansion} with $d>P_N$, or with triples
that are not mutually disjoint or meet $A\cap[N]$, has absolute value
at most
\begin{align}
 C_m\left(N^{-b}M_{m-1}+N^{-mb}+N^{-L_0}\right)
 \label{eq:cavity-collision-input}
\end{align}
for a common $b>0$. No assumption of distinct labels is used in
bounding the omitted coefficients.
\end{lemma}
\begin{proof}
Split each coefficient at cumulative heat $1/2$ and use
Cauchy--Schwarz and \Cref{lem:positive-heat-source}, retaining all
$2d+2$ Clifford insertions and repeated labels:
\begin{align}
 |c_{A,\pi}(\mathbf T;\mathbf t,t)|
 &\leq C_me^{C_md}(1+d)^{C_m}(M_m+N^{-m\eta}).
 \label{eq:cavity-coefficient-source-bound}
\end{align}
The source is independent of the cavity Gaussians. Their total
order-$2d$ pairing and integration weight is
\begin{align}
 \left(\frac{3\beta^2}{8N^3}\binom N3\right)^d
          \frac{(2d-1)!!}{(2d)!}
 &\leq\frac{(\beta^2/32)^d}{d!}.
 \label{eq:cavity-factorial-envelope}
\end{align}
A uniform triple contains a fixed site with probability $3/N$, and
intersects another independent triple with probability at most $9/N$.
The union bound therefore gives
\begin{align}
 \Pr[\exists e:T_e\cap(A\cap[N])\ne\varnothing
         \text{ or }\exists e<f:T_e\cap T_f\ne\varnothing]
 &\leq\min\{1,C(d+4)^2/N\}.
 \label{eq:cavity-label-collision-count}
\end{align}
Multiplying these bounds and summing yields
\begin{align}
 |\text{sum of intersecting assignments}|
 &\leq\frac{C_m}{N}(M_m+N^{-m\eta})
           \sum_{d\geq0}\frac{C_m^d(1+d)^{C_m+2}}{d!}\nonumber\\
 &\leq C_m(N^{-b}M_{m-1}+N^{-mb}).
 \label{eq:cavity-summed-collisions}
\end{align}
Choose $b\leq\min(1,\eta)$ and use $M_m\leq M_{m-1}$.
Finally $d!\geq(d/e)^d$ gives
\begin{align}
 \sum_{d>P_N}\frac{C_m^d(1+d)^{C_m}}{d!}&\leq N^{-L_0}
 \label{eq:cavity-physical-tail}
\end{align}
for the stated $P_N$, after increasing its constant and the fixed
threshold. Since $M_m\leq1$, this bounds the omitted tail and justifies
all regroupings of the absolutely convergent series.
\end{proof}

\subsection{Summing the coefficient comparison}

\begin{lemma}[Summing the retained cavity coefficients]
\label{lem:cavity-coefficient-sum}
With the choices in \Cref{lem:cavity-coefficient-comparison}, there is
$b>0$, independent of the fixed odd $m$, such that
\begin{align}
 &\left|U_A(B;t)-\sum_{d\geq0}\left(\frac{\beta^2}{16}\right)^d
 \sum_{\pi\in\mathcal P_2([2d])}
 \int_{1>t_1>\cdots>t_{2d}>0}
 \ell_{A,\pi}(\mathbf t,t)\,d\mathbf t\right|\nonumber\\
 &\qquad\leq C_m(N^{-b}M_{m-1}+N^{-mb})
 \label{eq:cavity-summed-coefficient-error}
\end{align}
for the four external sets $A$ in \eqref{eq:cavity-four-role-formulas},
uniformly in $0\leq t\leq1$. The scalar series converges absolutely.
\end{lemma}
\begin{proof}
For $d\leq P_N$ and sufficiently large $N$, sequentially choosing the
disjoint triples gives
\begin{align}
 |\mathcal R_{A,\pi}|
 &=\prod_{j=0}^{d-1}\binom{N-|A\cap[N]|-3j}{3},\nonumber\\
 \left(\frac{3\beta^2}{8N^3}\right)^d|\mathcal R_{A,\pi}|
 &=\left(\frac{\beta^2}{16}\right)^d
 \prod_{\ell=0}^{3d-1}\left(1-\frac{|A\cap[N]|+\ell}{N}\right),\nonumber\\
 0\leq1-\prod_{\ell=0}^{3d-1}
 \left(1-\frac{|A\cap[N]|+\ell}{N}\right)
 &\leq\sum_{\ell=0}^{3d-1}\frac{|A\cap[N]|+\ell}{N}
 \leq\frac{C(d+1)^2}{N},\nonumber\\
 \frac{(2d-1)!!}{(2d)!}\left(\frac{\beta^2}{16}\right)^d
 &=\frac{(\beta^2/32)^d}{d!}.
 \label{eq:cavity-duhamel-summability}
\end{align}
Since $|2G(z)|\leq1$ for $0\leq\Re z\leq1$,
\begin{align}
 |\ell_{A,\pi}(\mathbf t,t)|&\leq\EE|a|^{2m}\leq M_m,\nonumber\\
 \sum_{d\geq0}\left(\frac{\beta^2}{16}\right)^d
 \sum_\pi\int_{1>t_1>\cdots>t_{2d}>0}
 |\ell_{A,\pi}(\mathbf t,t)|\,d\mathbf t
 &\leq M_m\sum_{d\geq0}\frac{(\beta^2/32)^d}{d!}<\infty,\nonumber\\
 \frac{CM_m}{N}\sum_{d\geq0}\frac{(\beta^2/32)^d(d+1)^2}{d!}
 &\leq\frac{C_\beta}{N}M_{m-1}.
 \label{eq:cavity-counting-error-sum}
\end{align}
The last line bounds the error in replacing the exact count by its
leading value. Summing \eqref{eq:cavity-positive-heat-comparison} over
$d\leq P_N$, the pairings, assignments, and ordered time simplices costs
at most
\begin{align}
 &C_m\sum_{d\geq0}\frac{(\beta^2/32)^de^{C_md}(1+d)^{C_m}}{d!}
 \left[N^{-3a_0/4}M_{m-1}+N^{-(m-1/4)a_0}+N^{-L_0}\right]\nonumber\\
 &\qquad\leq C_m\left[N^{-3a_0/4}M_{m-1}
 +N^{-(m-1/4)a_0}+N^{-L_0}\right].
 \label{eq:cavity-comparison-error-sum}
\end{align}
The physical omitted terms are bounded by \Cref{lem:cavity-omitted}.
For the scalar omitted terms, increase the constant in $P_N$ so that
\begin{align}
 M_m\sum_{d>P_N}\frac{(\beta^2/32)^d}{d!}&\leq C_mN^{-L_0}.
\end{align}
Choose $b$ no larger than the exponent of \Cref{lem:cavity-omitted}
and than $\min(a_0/2,1)$. Since $L_0>ma_0+10$,
\begin{align}
 N^{-3a_0/4}M_{m-1}+N^{-(m-1/4)a_0}
 +N^{-L_0}+N^{-1}M_{m-1}
 &\leq C_m(N^{-b}M_{m-1}+N^{-mb}).
\end{align}
This proves the claim. All bounds are uniform in the external insertion
time, so the prescribed endpoint limits follow by dominated convergence.
\end{proof}

\section{Recovery from a positive spectral kernel}
\label{app:recovery}
\begin{proof}[Proof of \Cref{lem:heat-to-real-recovery}]
Write
\begin{align}
 k_x(z)&=\frac{e^{-zx}+e^{-(1-z)x}}{2(1+e^{-x})},
 &G(z)&=\int k_x(z)\mu(dx),
 &G_*(z)&=\int k_x(z)\mu_{*,\beta}(dx).
 \label{eq:spectral-kernel-exponential-form}
\end{align}
We use only that both measures are probabilities and that the target
has an exponential moment.  More explicitly, \Cref{lem:positive-saddle}
and \eqref{eq:saddle-exponential-moment} give
\begin{align}
 \mu([0,\infty))&=\mu_{*,\beta}([0,\infty))=1,
 &\lambda&=\frac1{2\max\{\beta,1\}},\nonumber\\*
 \int e^{\lambda x}\mu_{*,\beta}(dx)&\leq1+\sqrt2.
 \label{eq:recovery-target-exponential-moment}
\end{align}
At $\beta=0$, the last inequality follows from
$\mu_{*,0}=\delta_0$.  Put $\epsilon=\|G-G_*\|_2>0$.

For $0<t\leq1/2$, differentiation under the integral is justified
by the bounded derivative of $k_x$ and yields
\begin{align}
 |G'(t)|
 &\leq\frac12\int x e^{-tx}\mu(dx)
 \leq\frac1{2et},
 &|G_*'(t)|&\leq\frac1{2et}.
\end{align}
For $0<h\leq1/4$, $h\leq t\leq1/4$, and $0<s\leq h/2$,
\begin{align}
 |G(t)-G_*(t)|
 &\leq\frac1{2s}\int_{t-s}^{t+s}|G(v)-G_*(v)|\,dv
       +\frac C{2s h}\int_{t-s}^{t+s}|v-t|\,dv\nonumber\\*
 &\leq\frac\epsilon{\sqrt{2s}}+\frac{Cs}{h},\nonumber\\*
 \sup_{h\leq t\leq1/4}|G(t)-G_*(t)|
 &\leq C\epsilon^{2/3}h^{-1/3}.
 \label{eq:real-interval-recovery}
\end{align}
For the last line, take $s=(\epsilon h)^{2/3}$ if $s\leq h/2$;
otherwise $\epsilon^{2/3}h^{-1/3}>1/2$ and use
$|G(t)-G_*(t)|\leq1$.

Choose
$0<\ell\leq\min\{1/4,(1-e^{-1})\lambda/8\}$ and, for
$R\geq4/\ell$, set $q=2\lfloor\ell R/2\rfloor$.
Then $q$ is even and $\ell R/2\leq q\leq\ell R$.
The binomial identity gives a nonnegative integral:
\begin{align}
 2\sum_{j=0}^q(-1)^j\binom qjG(j/R)
 &=\int(1-e^{-x/R})^q
          \frac{1+e^{-(1-q/R)x}}{1+e^{-x}}\mu(dx)\nonumber\\*
 &\geq(1-e^{-1})^q\mu([R,\infty)).
 \label{eq:tail-finite-difference}
\end{align}
For the target, the ratio in this integrand is at most two, and
\eqref{eq:recovery-target-exponential-moment} gives
\begin{align}
 &\frac{2\sum_{j=0}^q(-1)^j\binom qjG_*(j/R)}
              {(1-e^{-1})^q}\nonumber\\*
 &\qquad\leq\frac{2\int x^q\mu_{*,\beta}(dx)}
                       {[R(1-e^{-1})]^q}
 \leq\frac{2(1+\sqrt2)q!}{[\lambda R(1-e^{-1})]^q}\nonumber\\*
 &\qquad\leq2(1+\sqrt2)
         \left(\frac\ell{\lambda(1-e^{-1})}\right)^q
 \leq2(1+\sqrt2)8^{-q}.
 \label{eq:recovery-target-finite-difference}
\end{align}
Since $G(0)=G_*(0)=1/2$, the $j=0$ term cancels in their
difference.  All remaining arguments lie in $[1/R,1/4]$, so
\eqref{eq:real-interval-recovery} implies
\begin{align}
 \mu([R,\infty))
 &\leq C8^{-q}
    +\frac2{(1-e^{-1})^q}
       \sum_{j=1}^q\binom qj|G(j/R)-G_*(j/R)|\nonumber\\*
 &\leq Ce^{-cR}
       +C\epsilon^{2/3}R^{1/3}
                         \left(\frac2{1-e^{-1}}\right)^q\nonumber\\*
 &\leq Ce^{-cR}+C\epsilon^{2/3}e^{C_1R}.
 \label{eq:spectral-tail-recovery}
\end{align}
Here $c,C_1>0$ depend only on $\beta$.  Increasing $C$ makes
the last bound valid for every $R\geq0$.

Fix $0<d<\min\{c,\lambda\}$ and
$0<\kappa<1/[3(C_1+d)]$.  For sufficiently small $\epsilon$,
let $L=\kappa\log(1/\epsilon)$.  Integrating
\eqref{eq:spectral-tail-recovery} gives
\begin{align}
 \int_{[0,L)}e^{dx}\mu(dx)
 &\leq1+d\int_0^Le^{ds}\mu([s,\infty))\,ds\nonumber\\*
 &\leq1+Cd\int_0^Le^{-(c-d)s}\,ds
       +Cd\epsilon^{2/3}\int_0^Le^{(C_1+d)s}\,ds\nonumber\\*
 &\leq C+C\epsilon^{2/3-\kappa(C_1+d)}\leq C,
 \label{eq:truncated-uniform-exponential-moment}\\
 \mu([L,\infty))
 &\leq C\epsilon^{c\kappa}
            +C\epsilon^{2/3-C_1\kappa}
 \leq C(\epsilon^{c\kappa}+\epsilon^{1/3}).
 \label{eq:power-small-tail}
\end{align}
Define $G_L(z)=\int_{[0,L)}k_x(z)\mu(dx)$, without
renormalizing.  For $-d\leq\Re z\leq1+d$,
\begin{align}
 |k_x(z)|&\leq\tfrac12e^{dx},
 &|G_L(z)|+|G_*(z)|&\leq C,
\end{align}
by \eqref{eq:truncated-uniform-exponential-moment} and
\eqref{eq:recovery-target-exponential-moment}.  Both functions are
holomorphic in the open strip.  On a fixed real interval,
\begin{align}
 \sup_{\ell/2\leq t\leq\ell}|G_L(t)-G_*(t)|
 &\leq\sup_{\ell/2\leq t\leq\ell}|G(t)-G_*(t)|
       +\sup_{\ell/2\leq t\leq\ell}
                    \int_{[L,\infty)}k_x(t)\mu(dx)\nonumber\\*
 &\leq C\epsilon^{2/3}+e^{-\ell L/2}
 \leq C\epsilon^\theta,
 \label{eq:truncated-real-bound}
\end{align}
where $\theta=\min\{2/3,\ell\kappa/2\}>0$.

In the fixed domain
\begin{align}
 \Omega
 &=\{z:-d/2<\Re z<1+d/2,\ |\Im z|<T_0+1\}
       \setminus[\ell/2,\ell],
\end{align}
let $h(z)$ be the harmonic measure of the removed interval.  It is
positive in $\Omega$, and hence
$\omega_0=\min_{|t|\leq T_0}h(it)>0$.
Apply \Cref{fact:two-constants} to $G_L-G_*$, using
\eqref{eq:truncated-real-bound} on both sides of the removed
interval and the uniform bound on the outer boundary:
\begin{align}
 |G_L(it)-G_*(it)|
 &\leq(C\epsilon^\theta)^{h(it)}C^{1-h(it)}
 \leq C\epsilon^{\theta\omega_0}
       \qquad(|t|\leq T_0),\nonumber\\*
 \sup_{|t|\leq T_0}|G(it)-G_*(it)|
 &\leq\sup_{|t|\leq T_0}|G_L(it)-G_*(it)|
                          +\tfrac12\mu([L,\infty))\nonumber\\*
 &\leq C\bigl(\epsilon^{\theta\omega_0}
                  +\epsilon^{c\kappa}+\epsilon^{1/3}\bigr).
\end{align}
The second line uses $|k_x(it)|\leq1/2$.  Taking
$\zeta=\min\{\theta\omega_0,c\kappa,1/3\}>0$ proves the result
for sufficiently small positive $\epsilon$.  For larger $\epsilon$,
increase $C$ and use $|G(it)-G_*(it)|\leq1$.

If $\epsilon=0$, continuity on $(0,1)$ and the identity theorem
give $G(z)=G_*(z)$ on $0<\Re z<1$.  Dominated convergence then gives
\begin{align}
 G(it)&=\lim_{s\downarrow0}G(s+it)
       =\lim_{s\downarrow0}G_*(s+it)=G_*(it),
\end{align}
since $|k_x(s+it)|\leq1/2$ for $0\leq s\leq1$.
\end{proof}

\section{Classical cavity estimates and reconstruction}
\label{sec:pspin-appendix}

This section proves the estimates used in Section~\ref{sec:pspin-main}.
Throughout, the interaction orders and coefficients are fixed,
$0\le\beta\le b<\beta_*$, and the number of replicas and all truncation orders
are fixed independently of $N$. Constants may depend on these data but not on
physical site labels. We use the exact coefficients in
\eqref{eq:pspin-model}.

\subsection{Gaussian differentiation of normalized expectations}
\label{sec:pspin-gaussian}

Let $H_J=\sum_eJ_e\phi_e$ on a finite configuration space with a positive
reference measure. For a deterministic function $f$ of $n$ replicas,
differentiating the numerator and $Z^{-n}$ gives
\begin{equation}
 \partial_{J_e}\langle f\rangle_{\beta,J}
 =\beta\left\langle f\left(
       \sum_{a=1}^{n}\phi_e(\sigma^a)-n\phi_e(\sigma^{n+1})
                       \right)\right\rangle_{\beta,J}.
 \label{eq:pspin-quotient-derivative}
\end{equation}
The last term is represented by one additional independent Gibbs sample.
This identity applies to repeated derivative indices.

Consider first an interpolation
$H_t=H_0+\sqrt t\,H_1$ with independent Gaussian fields. Write $\Delta$ for
the covariance of $H_1$, and assume $\Delta(\sigma,\sigma)$ is constant in
$\sigma$. For $t>0$, differentiate in $t$, apply Gaussian integration by parts
to $H_1$, and then use \eqref{eq:pspin-quotient-derivative}. Before simplifying
by replica exchangeability, this gives
\begin{align}
 \nu_t'(f)=\frac{\beta^2}{2}\nu_t\!\Bigg[f\Bigg(&
      \sum_{a,b\le n}\Delta_{ab}-n\Delta_{n+1,n+1}
      -(n-1)\sum_{a\le n}\Delta_{a,n+1} \notag\\[-2pt]
      &-(n+1)\sum_{a\le n}\Delta_{a,n+2}
       +n(n+1)\Delta_{n+1,n+2}\Bigg)\Bigg].
 \label{eq:pspin-interpolation-unsimplified}
\end{align}
The $n$ old diagonal terms cancel the new diagonal term. The two unused
replicas have the same joint distribution with the first $n$, so
\begin{equation}
 \nu_t'(f)=\beta^2\nu_t(\mathscr D_n^\Delta f),
 \qquad
 \mathscr D_n^\Delta f=f\left[
    \sum_{a<b\le n}\Delta_{ab}-n\sum_{a\le n}\Delta_{a,n+1}
            +\frac{n(n+1)}2\Delta_{n+1,n+2}\right].
 \label{eq:pspin-interpolation-identity}
\end{equation}
For $H_t=\sqrt{1-t}\,H_0+\sqrt t\,H_1$, the same calculation gives
\eqref{eq:pspin-interpolation-identity} with
$\Delta=K_1-K_0$, provided its diagonal is scalar. Positivity of $\Delta$ is
not required in this second use.

The absolute coefficient sum in the brackets is $2n^2$. Hence for $f\ge0$,
\begin{equation}
 |\nu_t'(f)|\le2\beta^2n^2\|\Delta\|_\infty\nu_t(f),
 \qquad
 \nu_t(f)\le e^{2b^2n^2\|\Delta\|_\infty}\nu_u(f),
 \quad t,u\in[0,1].
 \label{eq:pspin-comparison}
\end{equation}
The second inequality follows by Gr\"onwall in either direction. It compares
positive replica functions without comparing partition functions separately.

At fixed $N$, the configuration space is finite and all feature functions are
bounded. Gaussian exponential moments justify differentiation and integration
by parts in the open interpolation interval. The right-hand side of
\eqref{eq:pspin-interpolation-identity}, and each fixed iterate of it, is
continuous and bounded up to the endpoints. Integration followed by induction
therefore gives the endpoint derivatives used in finite Taylor expansions.
No complex-temperature extension is used.

\subsection{The statistical input and finite-size normalization}
\label{sec:pspin-normalization}

The ordered model associated with \eqref{eq:pspin-model} is
\begin{equation}
 H_N^{\mathrm{ord}}(\sigma)=
   \sum_{p\in\mathcal P}\frac{\lambda_p}{\sqrt{2N^{p-1}}}
     \sum_{i_1,\ldots,i_p=1}^{N}
           J_{p;i_1\ldots i_p}\sigma_{i_1}\cdots\sigma_{i_p},
 \qquad
 \mathbb E H_N^{\mathrm{ord}}(\sigma)H_N^{\mathrm{ord}}(\tau)
       =N\xi(R(\sigma,\tau)).
 \label{eq:pspin-ordered-model}
\end{equation}
The ordered coordinates are independent standard Gaussians. Repeated positions
are included.

In the pure case, the Gaussian field in Theorem~3.1 of \cite{CHL21} has
covariance $NR^p$. For $\lambda_p=1$, our Gibbs law at inverse temperature
$\beta$ corresponds to their parameter $\beta/\sqrt2$. Their self-overlap compensation is constant
for Ising spins and cancels from normalized expectations. The theorem yields
\eqref{eq:pspin-overlap}, uniformly on each fixed interval below the transition.
For mixtures, Theorems~3--4 and Remark~3 of \cite{Che17} give the corresponding
statements when the smallest nonzero order is at least three. More explicitly,
put $v=\sum_p\lambda_p^2$, $c_p=\lambda_p^2/v$, and use inverse temperature
$\beta\sqrt v$ in the convention with covariance
$N\sum_pc_pR^p/2$. This matches the covariance of the Gaussian field in our
Gibbs exponent. Odd interaction orders are allowed.

We next transfer the overlap estimate to the distinct-site model without
changing the target Hamiltonian. Let $K_{\mathrm{ord}}$ be the covariance in
\eqref{eq:pspin-ordered-model} and $K_{\mathrm{dis}}$ the covariance in
\eqref{eq:pspin-model}. The contribution of ordered tuples with all positions
distinct is exactly $K_{\mathrm{dis}}$: their $p!$ permutations give the factor
$p!$ in $a_{p,N}^2$. Therefore
\begin{align}
 \|K_{\mathrm{ord}}-K_{\mathrm{dis}}\|_\infty
 &\le \frac12\sum_p\lambda_p^2N
                \left(1-\frac{(N)_p}{N^p}\right)
 \le\frac14\sum_p\lambda_p^2p(p-1),
 \label{eq:pspin-covariance-comparison}\\
 (N)_p&=N(N-1)\cdots(N-p+1).\notag
\end{align}
The second inequality is the union bound for a repetition in $p$ independent
uniform labels. Both covariances have scalar diagonals; their diagonal values
need not be equal. Apply \eqref{eq:pspin-comparison} along an interpolation of
independent copies of the two fields to $|R_{12}|^{2r}$. This proves
\eqref{eq:pspin-overlap} for the exact distinct-site normalization, with a
constant factor.

For the same interpolation, differentiating a single logarithmic partition
function gives
\begin{equation}
 \frac{d}{dt}\mathbb E\log Z_t
    =\frac{\beta^2}{2}\mathbb E\left\langle
          \Delta(\sigma^1,\sigma^1)-\Delta(\sigma^1,\sigma^2)
                                     \right\rangle_t.
 \label{eq:pspin-pressure-comparison}
\end{equation}
Thus the expected log partition functions differ by $O(1)$. Their limiting
pressures agree, and hence so do the static boundaries defined in
Section~\ref{sec:pspin-model}. In particular, the ordered model's annealed
pressure is $\log2+\beta^2\xi(1)/2$, while the distinct-site model has the same
limit.

The finite constants in the overlap theorem do not need to be known to run
the algorithm. They determine a sufficiently-large-$N$ threshold. The cutoffs
\eqref{eq:pspin-error-choices}--\eqref{eq:pspin-cavity-choices} depend on
exponents, not these constants. Likewise, $\beta<\beta_*$ is a promise on the
fixed parameters; the algorithm is not required to compute $\beta_*$.

\subsection{A cavity proof of the overlap-moment upgrade}
\label{sec:pspin-overlap-upgrade}

For completeness, we give the classical cavity induction behind the
all-moment input. This also permits one to use the published exclusion of
macroscopic overlaps in place of the full moment theorem. Assume that the
ordered model satisfies, uniformly for $0\le\beta\le b$, the estimate
\begin{equation}
 \nu_\beta(\mathbf1_{\{|R_{12}|>\varepsilon\}})
        \le C_\varepsilon e^{-c_\varepsilon N}
 \quad\text{for every fixed }\varepsilon>0.
 \label{eq:pspin-macro-overlap}
\end{equation}
We show that it implies \eqref{eq:pspin-overlap}.

Write $\epsilon_a=\sigma_N^a$ and
$R_{ab}^-=N^{-1}\sum_{i<N}\sigma_i^a\sigma_i^b$. Delete every ordered tensor
feature that contains site $N$, retaining the ambient normalization. Its
cavity covariance is exactly
\begin{equation}
 \Delta_{ab}=N\left[\xi\left(R_{ab}^-+\frac{\epsilon_a\epsilon_b}{N}\right)
                                  -\xi(R_{ab}^-)\right].
 \label{eq:pspin-upgrade-covariance}
\end{equation}
Since $\xi'(q)=O(q^2)$ and $\xi''$ is bounded on $[-1,1]$,
\begin{equation}
 |\Delta_{ab}|\le C\left(|R_{ab}^-|^2+N^{-1}\right)\le C.
 \label{eq:pspin-upgrade-local-bound}
\end{equation}
Interpolate the deleted features with $\sqrt t$. Equation~\eqref{eq:pspin-comparison}
then bounds every nonnegative replica function along the interpolation by a
constant times its full-coupling expectation.

Set $M_j=\nu_\beta(|R_{12}|^j)$ and assume inductively that
$M_{2m}\le C_mN^{-m}$. Site exchangeability and
$R_{12}=R_{12}^-+\epsilon_1\epsilon_2/N$ give
\begin{equation}
 M_{2m+2}
 =\nu_1(\epsilon_1\epsilon_2R_{12}^{2m+1})
 =\nu_1(\epsilon_1\epsilon_2(R_{12}^-)^{2m+1})+O(N^{-m-1}).
 \label{eq:pspin-upgrade-exchange}
\end{equation}
Indeed, the difference is bounded by
$C_mN^{-1}\nu_1(|R_{12}|^{2m}+|R_{12}^-|^{2m})$, and
$|R_{12}^-|^{2m}\le C_m(|R_{12}|^{2m}+N^{-2m})$.

The function $g=\epsilon_1\epsilon_2(R_{12}^-)^{2m+1}$ has expectation zero
at $t=0$. Integrate \eqref{eq:pspin-interpolation-identity}, use
\eqref{eq:pspin-upgrade-local-bound}, and compare absolute values with the
full state. H\"older's inequality and replica exchangeability give
\begin{equation}
 \nu_1\!\left(|R_{12}^-|^{2m+1}|R_{ab}^-|^2\right)
     \le \nu_1(|R_{12}^-|^{2m+3})
 \quad(a\ne b).
 \label{eq:pspin-upgrade-holder}
\end{equation}
Also $N^{-1}\nu_1(|R_{12}^-|^{2m+1})=O(N^{-m-1})$, since
$|R_{12}^-|\le1$. It follows that
\begin{equation}
 M_{2m+2}\le C_m'M_{2m+3}+C_m''N^{-m-1}.
 \label{eq:pspin-upgrade-induction}
\end{equation}
Choose $\varepsilon>0$ with $C_m'\varepsilon\le1/2$. By
\eqref{eq:pspin-macro-overlap},
\begin{equation}
 M_{2m+3}\le\varepsilon M_{2m+2}
                 +C_\varepsilon e^{-c_\varepsilon N}.
 \label{eq:pspin-upgrade-absorb}
\end{equation}
Absorb the first term and start the induction at $M_0=1$. This proves
\eqref{eq:pspin-overlap}. All constants are uniform on the fixed interval.

If the macroscopic estimate is initially given at each fixed positive
temperature, its uniformity on a compact interval follows directly. For each
configuration, $H_N(\sigma)$ is Gaussian with variance $O(N)$. A union bound
over $2^N$ configurations gives
$\Pr(\max_\sigma|H_N(\sigma)|>CN)\le e^{-cN}$ for sufficiently large fixed
$C$. On the complementary event, the ratio of two-replica densities at
inverse temperatures differing by $d$ is at most $e^{4CN|d|}$. A fixed
exponential overlap bound therefore persists in a neighborhood of that
temperature. Near zero, the uniform-spin estimate
$\Pr(|R_{12}|>\varepsilon)\le2e^{-N\varepsilon^2/2}$ and the same density
comparison give the needed neighborhood. A finite cover yields uniform
constants on $[0,b]$. Transfer to the distinct-site model then follows from
Section~\ref{sec:pspin-normalization}.

\subsection{Global and cavity covariance moments}
\label{sec:pspin-covariance-proof}

Let $z_j\in\{-1,1\}$ and $U=\sum_{j=1}^{M}z_j$. The elementary symmetric
polynomials obey Newton's identities
\begin{equation}
 q e_q(z)=\sum_{j=1}^{q}(-1)^{j-1}e_{q-j}(z)s_j,
 \qquad
 s_j=\sum_{h=1}^{M}z_h^j=
 \begin{cases}U,&j\text{ odd},\\M,&j\text{ even},\end{cases}
 \qquad e_0=1.
 \label{eq:pspin-newton}
\end{equation}
Assign weight one to $U$ and weight two to $M$. Induction shows that every
monomial $U^uM^v$ in $e_q$ has $u+2v\le q$. For $M\le N$ this gives
\begin{equation}
 |e_q(z)|\le C_qN^{q/2}\left(1+\frac{|U|}{\sqrt N}\right)^q.
 \label{eq:pspin-elementary-bound}
\end{equation}
The number of terms and their coefficients depend only on fixed $q$.

For the distinct-site model and $z_j=\sigma_j^a\sigma_j^b$,
\begin{equation}
 K_{ab}=\sum_pa_{p,N}^2e_p(z),
 \qquad
 |K_{ab}|\le C\sum_pN^{1-p/2}(1+\sqrt N|R_{ab}|)^p.
 \label{eq:pspin-global-bound}
\end{equation}
For every fixed positive integer $r$, sufficiently high even moments in
\eqref{eq:pspin-overlap} bound the $r$th moment of the right-hand side by
$C_rN^{-r\eta}$. This proves
\eqref{eq:pspin-covariance-moments}. H\"older then bounds any product of a
fixed number of off-diagonal covariances, even when their replica pairs
share labels.

For the local covariance in \eqref{eq:pspin-local-covariance}, use instead
\begin{equation}
 \Delta_i^{ab}=
       \sum_pa_{p,N}^2z_i e_{p-1}((z_j)_{j\ne i}),
 \qquad
 \left|\frac1N\sum_{j\ne i}z_j-R_{ab}\right|=\frac1N.
 \label{eq:pspin-local-elementary}
\end{equation}
Equations~\eqref{eq:pspin-elementary-bound} and~\eqref{eq:pspin-overlap}
therefore yield
\begin{equation}
 \nu_1(|\Delta_i^{ab}|^r)\le C_rN^{-r\kappa},\qquad a\ne b.
 \label{eq:pspin-local-full-moment}
\end{equation}
Since $\|\Delta_i\|_\infty\le\Lambda_{i,N}\le C$, apply
\eqref{eq:pspin-comparison} to the nonnegative function
$|\Delta_i^{ab}|^r$. It follows that
\begin{equation}
 \sup_{0\le t\le1}\nu_t(|\Delta_i^{ab}|^r)
       \le C_{r,b}N^{-r\kappa}.
 \label{eq:pspin-local-path-moment}
\end{equation}
All these bounds are uniform in the chosen site and in $0\le\beta\le b$.

\subsection{Proof of active-site concentration}
\label{sec:pspin-support-proof}

\begin{proof}[Proof of Lemma~\ref{lem:pspin-support}]
Flip spin $i$ in every replica and change the sign of each Gaussian coordinate
whose interaction contains $i$. The disorder law is unchanged, and the Gibbs
weights transform by this relabeling. A monomial changes by
$(-1)^{|B_i|}$. This proves the exact zero whenever an active replica subset
has odd size. Ordinary permutations of the physical sites also preserve the
joint law.

Suppose now that every $B_j$ is even. For a nonempty even subset $B$ of the
replica labels, set
$R_B=N^{-1}\sum_i\sigma_i^B$, and set $R_\varnothing=1$. Conditional
independence of replicas gives
\begin{align}
 \nu_\beta(R_B^{2r})
 &=N^{-2r}\sum_{j_1,\ldots,j_{2r}}
    \mathbb E_J\!\left[
      \langle\sigma_{j_1}\cdots\sigma_{j_{2r}}\rangle_{\beta,J}^{|B|}
                 \right] \notag\\
 &\le \nu_\beta(R_{12}^{2r})\le C_rN^{-r}.
 \label{eq:pspin-multioverlap}
\end{align}
Indeed, $0\le x^{|B|}\le x^2$ for $x\in[-1,1]$ and nonempty even $B$.
All coincidences among the summation labels are included.

Permutation symmetry expresses the expectation of $f$ as the normalized sum
over distinct physical labels. Inclusion--exclusion on their equality
partitions gives
\begin{equation}
 \nu_\beta(f)=\frac1{(N)_s}
   \sum_{\pi\in\operatorname{Part}([s])}
    \mu(\pi)N^{|\pi|}
    \nu_\beta\!\left[
       \prod_{C\in\pi}R_{\triangle_{j\in C}B_j}
                  \right],
 \qquad
 \mu(\pi)=\prod_{C\in\pi}(-1)^{|C|-1}(|C|-1)!.
 \label{eq:pspin-distinct-inversion}
\end{equation}
Here $\operatorname{Part}([s])$ is the set of partitions of $[s]$, and
$\triangle$ denotes symmetric difference. To verify the identity, fix a
physical label tuple. The sum of $\mu(\pi)$ over partitions whose blocks have
equal labels is one when all labels are distinct and zero otherwise. Summing
the unconstrained label of each block then gives its displayed multi-overlap.

For a term in \eqref{eq:pspin-distinct-inversion}, let $z$ be the number of
blocks with empty symmetric difference and let $q=|\pi|-z$. Every other
symmetric difference is nonempty and even. H\"older and
\eqref{eq:pspin-multioverlap}, using a sufficiently high fixed even moment,
bound the product expectation by $C N^{-q/2}$. An empty-difference block
cannot be a singleton, so $s\ge q+2z$. For $N\ge2s$, the full term is at most
\begin{equation}
 C_sN^{|\pi|-s-q/2}
   =C_sN^{q/2+z-s}\le C_sN^{-s/2}.
 \label{eq:pspin-support-count}
\end{equation}
The number of partitions is fixed. This proves the claimed estimate for all
sufficiently large $N$; increasing the constant covers the remaining sizes
where the prescribed distinct sites exist.
\end{proof}

\subsection{The local cavity remainder}
\label{sec:pspin-local-proof}

\begin{proof}[Proof of Lemma~\ref{lem:pspin-local-remainder}]
Use \eqref{eq:pspin-interpolation-identity} with the covariance
$\Delta_i$ of the deleted interactions. The operator is independent of $t$,
and its $j$th application uses $n+2j$ declared replicas. Hence
\eqref{eq:pspin-local-derivative} holds at every fixed order, including at the
endpoints by Section~\ref{sec:pspin-gaussian}.

Taylor's formula about one, evaluated at zero, reads
\begin{equation}
 0=\nu_0(f)=\sum_{k=0}^{L-1}\frac{(-1)^k}{k!}\nu_1^{(k)}(f)
       +\frac{(-1)^L}{(L-1)!}\int_0^1t^{L-1}\nu_t^{(L)}(f)\,dt.
 \label{eq:pspin-backward-taylor}
\end{equation}
Solving for the $k=0$ term proves \eqref{eq:pspin-cavity-row}, with exact
remainder
\begin{equation}
 r_L(f)=\frac{(-1)^{L+1}}{(L-1)!}
                         \int_0^1t^{L-1}\nu_t^{(L)}(f)\,dt.
 \label{eq:pspin-integral-remainder}
\end{equation}
Every term of $\mathscr D_i^Lf$, before expanding the interactions, contains
$L$ off-diagonal cavity covariances. The replica coefficient mass is at most
\begin{equation}
 \prod_{j=0}^{L-1}2(n+2j)^2.
 \label{eq:pspin-replica-mass}
\end{equation}
H\"older and \eqref{eq:pspin-local-path-moment} bound each product expectation
by $C N^{-L\kappa}$. The original monomial has absolute value one, so the
same bound holds with it inserted. Integrating proves
\eqref{eq:pspin-local-error}. No independence of the cavity covariance
factors is needed.
\end{proof}

\subsection{Collision counting and the finite recursion}
\label{sec:pspin-recursion-proof}

We first prove \eqref{eq:pspin-row-mass}. Fix a term of order $k<L$, its
replica pairs, and its interaction orders $p_1,\ldots,p_k$. Choose
$T_j=E_j\setminus\{i\}$ independently and uniformly from the
$(p_j-1)$-subsets of $[N]\setminus\{i\}$. If the old support has size $s$,
the union bound gives
\begin{equation}
 \Pr(\text{a collision})
 \le\frac{
       (s-1)\sum_j(p_j-1)
         +\sum_{j<h}(p_j-1)(p_h-1)}{N-1}.
 \label{eq:pspin-collision-probability}
\end{equation}
The first term counts contacts with old active sites; the second counts
intersections between new sets. There is no restriction that the interaction
labels be distinct: any repeated label is among the collision assignments.

The sum of interaction weights for each incident label is
$\Lambda_{i,N}\le\Lambda:=\tfrac12\sum_pp\lambda_p^2$. After summing over
replica terms and over $k$, the total row mass is bounded by
\begin{equation}
 \sum_{k=1}^{L-1}\frac{b^{2k}}{k!}\Lambda^k
                         \prod_{j=0}^{k-1}2(n+2j)^2.
 \label{eq:pspin-total-row-bound}
\end{equation}
All factors are fixed. Multiplying the fixed-order summands by
\eqref{eq:pspin-collision-probability} proves the second bound in
\eqref{eq:pspin-row-mass}. For a collision-free assignment, every new
non-cavity site belongs to exactly one off-diagonal replica pair. It remains
active after reduction, while every old site other than $i$ is unchanged.
This proves \eqref{eq:pspin-support-increase}.

\begin{proof}[Proof of Lemma~\ref{lem:pspin-recursion}]
Between two collision rows, every update increases support by at least one.
Every internal node has support below $S$. A path has at most $R$ collision
rows, so $(R+1)(S+1)$ is a bound on its number of row expansions, including
updates leading to terminal nodes. Each row adds at most $2(L-1)$ declared
replicas. This proves \eqref{eq:pspin-depth}. In particular, there is a fixed
replica range on which all row constants can be chosen uniformly.

A row has at most $L-1$ incident interaction choices. There are
$O(N^{p-1})$ choices for an interaction of order $p$ through the selected site.
The fixed number of replica-pair choices and the fixed depth therefore give
at most
\begin{equation}
 C N^{(P-1)(L-1)T+O(1)}
 \label{eq:pspin-tree-cost}
\end{equation}
arithmetic operations for one initial pattern. Reducing monomials and testing
collisions involve only a fixed number of slots at each node. Every row
coefficient is an explicit product of $a_{p,N}^2$, integer replica
coefficients, reciprocal fixed factorials, and powers of $\beta$. Along a path
the degree in $\beta$ is at most $2(L-1)T$.

For the error, substitute the exact identity
\eqref{eq:pspin-cavity-row} down the implemented finite tree, leaving the true
expectation at each terminal node. Subtract the tree evaluated by the
algorithm. Constant and gauge-zero terminals have zero error. Other errors
come from local remainders, support terminals, or collision terminals.

Choose $C$ large enough that a collision-free row has absolute coefficient
mass at most $C$ and a collision row at most $C/N$ for every reachable internal
state. For any fixed sequence of collision-free and collision statuses of
length at most $T$, summing products of absolute coefficients over its
numerical labels costs at most $C^T N^{-r}$, where $r$ is the number of
collisions in the sequence. There are at most $2^{T+1}$ such status sequences.
Thus the number of physical assignments has already been included in the row
mass and gives no further power of $N$.

A support terminal has at most $S-1+(P-1)(L-1)$ active sites: its parent has
support below $S$ and one row introduces at most $(P-1)(L-1)$ sites. Its
replica count also has the fixed bound \eqref{eq:pspin-depth}. Consequently
Lemma~\ref{lem:pspin-support} applies uniformly and gives true expectation
$O(N^{-S/2})$. A collision terminal has bounded true expectation and path
weight with at least $R$ collision factors. Finally, each local remainder is
$O(N^{-L\kappa})$ by Lemma~\ref{lem:pspin-local-remainder}. Summing the
weighted errors over the fixed status sequences and possible depths proves
\eqref{eq:pspin-evaluator-error}.
\end{proof}

In particular, for initial support at most $s_0$ and replica count at most
$n_0$, the choices
\begin{equation}
 S\ge\max(s_0+1,2Q+2),\qquad R\ge Q+1,\qquad
 L\ge2,\quad L\kappa\ge Q+1
 \label{eq:pspin-local-cutoffs}
\end{equation}
give error $O(N^{-Q-1})$. The constants may depend on the cutoffs, but all
cutoffs remain fixed for each requested precision exponent.

\subsection{The finite triangular hierarchy}
\label{sec:pspin-triangular}

We record the algebra used when cavity equations are collected into a finite
moment vector. It also applies to a finite centered-moment hierarchy after
inversion of its leading response operator.

\begin{lemma}[Finite hierarchy truncation]
\label{lem:pspin-triangular}
Let $m,e,r$ lie in a normed vector space and suppose
\begin{equation}
 (I-U-N^{-1}V)m=e+r,
 \qquad U^K=0,\quad Ue=0,
 \qquad \|U\|+\|V\|+\|m\|\le C.
 \label{eq:pspin-abstract-hierarchy}
\end{equation}
Set $B=(I-U)^{-1}V=(\sum_{j<K}U^j)V$. For fixed $K,R$ and sufficiently
large $N$,
\begin{equation}
 \left\|m-\sum_{j=0}^{R-1}N^{-j}B^je\right\|
       \le C_K\|r\|+C_K(C_K/N)^R.
 \label{eq:pspin-abstract-error}
\end{equation}
\end{lemma}

\begin{proof}
Multiplication by $(I-U)^{-1}$ gives
$m=e+N^{-1}Bm+(I-U)^{-1}r$. Iterating exactly $R$ times yields
\begin{equation}
 m-\sum_{j=0}^{R-1}N^{-j}B^je
   =N^{-R}B^Rm+
        \sum_{j=0}^{R-1}N^{-j}B^j(I-U)^{-1}r.
 \label{eq:pspin-abstract-telescope}
\end{equation}
Both $\|(I-U)^{-1}\|$ and $\|B\|$ are bounded by constants depending only on
$K,C$. For sufficiently large $N$, the sum of norms on the right is bounded
as in \eqref{eq:pspin-abstract-error}.
\end{proof}

If the blocks are computed numerically, a finite product is controlled by the
identity
\begin{equation}
 \widetilde T_d\cdots\widetilde T_1-T_d\cdots T_1
 =\sum_{j=1}^{d}\widetilde T_d\cdots\widetilde T_{j+1}
      (\widetilde T_j-T_j)T_{j-1}\cdots T_1.
 \label{eq:pspin-numerical-telescope}
\end{equation}
Uniform stability of the computed blocks and consistency error $\epsilon$
on the finitely many exact intermediate inputs therefore add $O(\epsilon)$
to \eqref{eq:pspin-abstract-error} at fixed truncation orders.

For the classical recursion, collision-free rows are the upward part and
collision rows have total mass $O(1/N)$. The support cutoff makes the upward
expansion nilpotent. Since rows introduce new replicas, the implementation is
the finite unfolding proved in Section~\ref{sec:pspin-recursion-proof}, with
the collision counter retained. We do not apply
Lemma~\ref{lem:pspin-triangular} to an unspecified infinite replica space.

\subsection{Hermite reconstruction and the complete tail}
\label{sec:pspin-hermite-proof}

\begin{proof}[Proof of Lemma~\ref{lem:pspin-hermite}]
Repeated use of \eqref{eq:pspin-quotient-derivative} gives
\eqref{eq:pspin-disorder-derivative}. At the $j$th step the total coefficient
mass increases by at most $2j$, proving the bound $2^dd!$ for
\eqref{eq:pspin-insertion}. Gaussian integration by parts in each coordinate
then gives \eqref{eq:pspin-exact-coefficient}. At fixed $N$ and derivative
order, all Gibbs derivatives are bounded by this finite insertion formula, so
all Gaussian boundary terms vanish.

Squaring a derivative uses two disjoint groups of $d+1$ replicas. Expand both
insertions before summing over $e_1,\ldots,e_d$. Each label occurs once in
each group, so its sum produces
\begin{equation}
 \sum_ea_e^2\sigma_e^u\sigma_e^v=K_{uv},
 \qquad u\text{ in the first group},\quad v\text{ in the second}.
 \label{eq:pspin-cross-contraction}
\end{equation}
In particular $u\ne v$. Every expanded term has $d$ such covariances and two
bounded target monomials. Their total coefficient mass is at most
$4^d(d!)^2\beta^{2d}$. H\"older and
\eqref{eq:pspin-covariance-moments} prove
\eqref{eq:pspin-gaussian-sobolev}, including repeated interaction labels and
shared replica pairs.

Here is a direct tail argument. For a multi-index $v$ of size $r$, integration
by parts gives
\begin{equation}
 \mathbb E_J[(\partial^vF_A)h_\gamma]
       =\sqrt{\frac{(\gamma+v)!}{\gamma!}}\,c_{A,\gamma+v}.
 \label{eq:pspin-bessel-coefficient}
\end{equation}
Apply Bessel's inequality to $\partial^vF_A$ and sum over $v$ with
multiplicity $r!/v!$, corresponding to ordered derivative lists. For each
$\alpha$, the coefficient obtained is
\begin{equation}
 \sum_{\substack{v\le\alpha\\|v|=r}}
       \frac{r!}{v!}\frac{\alpha!}{(\alpha-v)!}
       =(|\alpha|)_r.
 \label{eq:pspin-derivative-multiplicity}
\end{equation}
This counts ordered selections of $r$ objects from groups of sizes
$\alpha_e$. All summands are nonnegative. Thus
\begin{equation}
 \mathbb E_J\|D^rF_A\|_{\mathrm{HS}}^2
     \ge\sum_{j\ge r}(j)_r\sum_{|\alpha|=j}|c_{A,\alpha}|^2.
 \label{eq:pspin-bessel-tail}
\end{equation}
For $r=D+1$, every factor $(j)_r$ on the right is at least $(D+1)!$.
Equation~\eqref{eq:pspin-gaussian-sobolev} proves
\eqref{eq:pspin-hermite-tail}.

The Hermite products are complete in Gaussian $L^2$. One direct argument is
that a function orthogonal to all polynomials has an entire Gaussian-weighted
Laplace transform whose Taylor coefficients vanish. Cauchy--Schwarz against
Gaussian exponential moments justifies the entire extension and
termwise differentiation. The transform is therefore zero; on imaginary
arguments, uniqueness of the Fourier transform gives the zero function.
Consequently Parseval identifies the discarded coefficient sum with the
squared approximation error. No convergence of a differentiated infinite
Hermite series has been assumed.

The cavity error $O(N^{-Q})$ applied to each of the fixed number of monomials
in \eqref{eq:pspin-insertion} proves
\eqref{eq:pspin-coefficient-error}. Finally,
\begin{equation}
 \sum_ea_e^2
    =\frac12\sum_p\lambda_p^2N\frac{(N)_p}{N^p}\le C N.
 \label{eq:pspin-feature-squares}
\end{equation}
The multinomial theorem proves \eqref{eq:pspin-multinomial}. Summing the
squared coefficient errors and the orthogonal tail gives
\eqref{eq:pspin-total-l2}.
\end{proof}

\subsection{Counting, finite input, and bit complexity}
\label{sec:pspin-bits}

We supply the numerical details used in the proof of
Theorem~\ref{thm:pspin-main}. First, for $M_p=\binom Np$,
$Q_p=\sum_{E\in\binom{[N]}p}J_{p,E}^2$ has moment generating function
\begin{equation}
 \mathbb E e^{tQ_p}=(1-2t)^{-M_p/2},\qquad 0<t<1/2.
 \label{eq:pspin-chi-mgf}
\end{equation}
For example, evaluating Chernoff's bound at $t=1/4$ shows that
$\Pr(Q_p>C_1M_p)\le e^{-c_1M_p}$ for a sufficiently large fixed $C_1$.
Since $a_{p,N}^2M_p\le\lambda_p^2N/2$, a union bound over the fixed set of
orders proves
\begin{equation}
 \Pr\!\left(\sum_ea_e^2J_e^2>C_0N\right)
       \le C e^{-c_0N^{p_*}}.
 \label{eq:pspin-energy-event}
\end{equation}
This is the only coupling-norm event used in
\eqref{eq:pspin-counting-cs}--\eqref{eq:pspin-counting-ms}.

Fix a coordinate box $|J_e|\le B_N$, where
$B_N=C\lceil\log_2(N+2)\rceil$. The Gaussian tail bound and
$|\mathcal I_N|=O(N^P)$ give
\begin{equation}
 \Pr(\max_e|J_e|>B_N)
        \le C N^P e^{-B_N^2/2}.
 \label{eq:pspin-input-box}
\end{equation}
This is smaller than any prescribed inverse polynomial for sufficiently large
$N$. The finite input records rounded coordinates and an overflow flag. On
overflow the algorithm returns a designated bounded output; that event is
included in its failure probability.

At the fixed cutoffs, every ideal approximation is a fixed-degree polynomial
in the couplings and in $\beta$. The degree in $\beta$ is at most
$D+2(L-1)T$. Its coefficients are generated by a polynomial number of
operations with fixed factorials, powers of $N$, and the fixed mixture
parameters. On the input box, their magnitudes and first derivatives are
polynomially bounded. The same is true for all intermediate expressions: a
row expands finitely many products and polynomially many sums, and
multiplicative depth is bounded by the fixed recursion depth.

The two factors $\sqrt{\alpha!}$ in
\eqref{eq:pspin-reconstruction} can be combined. The contribution of a
multi-index $\alpha$ is evaluated as
\begin{equation}
 \frac{\beta^{|\alpha|}\prod_ea_e^{\alpha_e}}{\alpha!}
       \widehat\nu_\beta(\mathcal F_{A;\mathbf e})
       \prod_e\operatorname{He}_{\alpha_e}(J_e).
 \label{eq:pspin-unnormalized-evaluation}
\end{equation}
Thus no Hermite-normalization square root is required. The remaining scalar
square roots in $a_{p,N}$ can be approximated by bisection to inverse-polynomial
absolute accuracy. Fixed-order Hermite polynomials have explicit coefficients
or can be generated by
$\operatorname{He}_{j+1}(x)=x\operatorname{He}_j(x)-j\operatorname{He}_{j-1}(x)$.

More formally, let $N^A$ bound the number of arithmetic operations and let
$N^B$ bound the magnitudes and sensitivities of the fixed-depth products on a
slightly enlarged input box. Such exponents follow explicitly from the
enumeration in \eqref{eq:pspin-arithmetic-cost}. Rounding inputs and scalar
operations to $N^{-B_{\mathrm{bit}}}$, with a sufficiently large fixed
$B_{\mathrm{bit}}$, bounds the accumulated scalar error by an arbitrarily
small fixed inverse power of $N$. Allocate per-coordinate error at most
$\tfrac14 C_\ell^{-1/2}N^{-c-\ell/2}$ for the observable vector. Increase
$B_{\mathrm{bit}}$ to include the polynomial number of coefficients and
coupling factors in \eqref{eq:pspin-counting-output}; then its numerical error
is at most $\tfrac14N^{-c}$. Both allocations require only $O(\log N)$
fractional bits. Intermediate integer parts also have $O(\log N)$ bits.

The counting expression is evaluated by representing each $P_e(J;u)$ as a
fixed-degree polynomial in $u$ and integrating each power exactly in real
arithmetic. The term $N\log2$ is evaluated to the same absolute precision.
Clipping is applied only to the final observable outputs, not to these
polynomials. Exponentiation of an accurate logarithm produces a relative
approximation to the partition function with polynomial output length, for
example as a mantissa and binary exponent. On the input box,
$\max_\sigma|H_N(\sigma)|\le B_N\sum_e|a_e|=N^{O(1)}$, so the logarithm
has polynomial magnitude.

For clarity, the physical target is stable under rounding as well. If each
coupling changes by at most $N^{-B_{\mathrm{bit}}}$, then
\begin{equation}
 \|\Delta H_N\|_\infty
     \le N^{-B_{\mathrm{bit}}}\sum_e|a_e|
     \le C N^{(P+1)/2-B_{\mathrm{bit}}}.
 \label{eq:pspin-input-hamiltonian-error}
\end{equation}
Interpolation between $H_N$ and $H_N+\Delta H_N$ gives
\begin{equation}
 |\Delta\langle O\rangle_\beta|
       \le2b\|O\|_\infty\|\Delta H_N\|_\infty,
 \qquad
 |\Delta\log Z_N|\le b\|\Delta H_N\|_\infty.
 \label{eq:pspin-physical-rounding}
\end{equation}
Indeed, the derivative of the first quantity is a Gibbs covariance with
$\beta\Delta H_N$, and that of the second is its Gibbs mean. These estimates
also handle sufficiently accurate fixed-parameter approximations. A
coefficient-normalized degree-$\ell$ observable has
$\|O\|_\infty\le C_\ell N^{\ell/2}$, or one may control the full vector
as above.

The ideal mean-square estimates always concern the exact Gaussian disorder.
The numerical estimates just proved are deterministic on the input box.
Combining the two on that box and then adding its failure probability proves
the bit-complexity statement; it does not assert an unconditional mean-square
numerical bound from a boxwise calculation.

\subsection{Ordered interactions and exact-variance normalization}
\label{sec:pspin-other-models}

\begin{corollary}[Other interaction conventions]
\label{cor:pspin-conventions}
Theorem~\ref{thm:pspin-main} also holds for the ordered model
\eqref{eq:pspin-ordered-model}, including repeated positions, and for the
distinct-site model with coefficients
\begin{equation}
 a^*_{p,N}=\lambda_p\sqrt{\frac{N}{2\binom Np}}.
 \label{eq:pspin-exact-variance}
\end{equation}
The static boundary is the same in each convention. Each algorithm uses the
exact coefficients of its target model.
\end{corollary}

\begin{proof}
For \eqref{eq:pspin-exact-variance}, the exact coefficient relation is
\begin{equation}
 a_{p,N}=\sqrt{\rho_{p,N}}\,a^*_{p,N},
 \qquad \rho_{p,N}=\frac{(N)_p}{N^p}.
 \label{eq:pspin-exact-factor}
\end{equation}
The two distinct-site covariances differ in supremum norm by at most
$\tfrac12\sum_p\lambda_p^2N(1-\rho_{p,N})=O(1)$. Equations
\eqref{eq:pspin-comparison} and~\eqref{eq:pspin-pressure-comparison} transfer
the overlap estimate and identify the limiting pressure. Newton's bounds
apply because $(a^*_{p,N})^2=O(N^{1-p})$ for all sufficiently large $N$.
The incident weight is now exactly
\begin{equation}
 \sum_p(a^*_{p,N})^2\binom{N-1}{p-1}
       =\frac12\sum_pp\lambda_p^2.
 \label{eq:pspin-exact-incident}
\end{equation}
Thus all local remainder, collision, reconstruction, and finite-bit estimates
hold with $a^*_{p,N}$ retained in every formula. In particular, we have not
replaced \eqref{eq:pspin-exact-factor} by one at finite $N$.

For the ordered model, the global covariance is exactly $N\xi(R)$, so its
moment estimate follows directly from \eqref{eq:pspin-overlap}.
Delete every tensor coordinate containing the chosen site $i$ at least once.
The remaining Hamiltonian is independent of that spin, and the cavity
covariance is \eqref{eq:pspin-upgrade-covariance}, with $i$ in place of $N$.
Expanding each power gives
\begin{equation}
 \Delta_i^{ab}=\frac12\sum_p\lambda_p^2
    \sum_{j=1}^{p}\binom pj N^{1-j}
       (\sigma_i^a\sigma_i^b)^j(R_{ab}^-)^{p-j},
 \qquad R_{ab}^-=\frac1N\sum_{h\ne i}\sigma_h^a\sigma_h^b.
 \label{eq:pspin-ordered-local}
\end{equation}
Its $j$th summand has moments at scale
$N^{1-j-(p-j)/2}=N^{-(p+j-2)/2}$. The slowest is $j=1$, giving the same
$N^{-\kappa}$ estimate. The total incident weight is bounded by
$\tfrac12\sum_pp\lambda_p^2$, because
$N^p-(N-1)^p\le pN^{p-1}$. The local comparison and remainder proof therefore
apply unchanged. Some deleted features contain the cavity site an even number
of times and are independent of that spin after parity reduction. Deleting
these independent Gaussian features is still a valid interpolation; they are
retained in \eqref{eq:pspin-ordered-local}.

For collision counting, call an assignment collision-free only if every
incident tuple contains $i$ exactly once and all other slots are distinct
and avoid the old support. Such an assignment obeys
\eqref{eq:pspin-support-increase}. To bound the remaining weight, designate
one occurrence of $i$ in every incident tuple. There are at most
$\prod_{j=1}^{k}p_j$ designations. The other
$d=\sum_j(p_j-1)$ slots have at most $N^d$ assignments, each of weight at
most $C N^{-d}$. A non-collision-free assignment has a witness: a remaining
slot equals $i$ or an old active site, or two remaining slots are equal.
There are only a fixed number of possible witnesses at fixed $s,L,P$, and
each leaves at most $N^{d-1}$ assignments. Overcounting by the designations
is harmless for this absolute bound. Thus all collision weights sum to
$O(1/N)$, proving the same row estimates and finite recursion.

Gauge symmetry now flips the sign of coordinates in which a site has odd
slot multiplicity, so the active-site proof is unchanged. In the Hermite
insertion, $\sigma_e$ is the product over the ordered slots; it is reduced
modulo two only when evaluating its replica monomials. There are still
$O(N^P)$ interaction coordinates and $\sum_ea_e^2=O(N)$, so the Hermite,
counting, and bit estimates hold. Counting includes every ordered coordinate,
including those whose spin monomial is constant. Such a random constant
energy shift cancels from normalized observables but contributes exactly to
\eqref{eq:pspin-counting-output}. Its monomial expectation is one. This
completes the proof in both conventions.
\end{proof}

\section{SK cavity estimates and graph reconstruction}
\label{sec:sk-appendix}

This section proves the estimates and gives the exact graph evaluator used
in Section~\ref{sec:sk-main}. Throughout, $0<\beta<1$, the observable
degree $\ell$ is fixed, and the edge weights are those in
\eqref{eq:sk-edge-weights}. All probability estimates concern the original
Gaussian disorder. Numerical approximation is treated separately in
Section~\ref{sec:sk-bits}.

\subsection{Bounds for the omitted graphs}
\label{sec:sk-graph-bounds}

We first bound the total squared graph mass at a parameter below one.
For $0\le x<1$, the parity identity gives
\begin{align}
 \sum_{\partial G=A}x^{|G|}
  &=2^{-N}\sum_\sigma\sigma_A
                    \prod_{i<j}(1+x\sigma_i\sigma_j)
       \notag\\
  &\le 2^{-N}\sum_\sigma\prod_{i<j}(1+x\sigma_i\sigma_j)
   \le \mathbb E_\sigma e^{x(U^2-N)/2},
   \qquad U=\sum_i\sigma_i.
 \label{eq:sk-positive-graph-bound}
\end{align}
The product is nonnegative, and $1+y\le e^y$ applies to both $y=x$
and $y=-x$. For a standard Gaussian $g$ independent of the spins,
\begin{equation}
 \mathbb E_\sigma e^{x(U^2-N)/2}
  =e^{-Nx/2}\mathbb E_g\cosh(\sqrt x\,g)^N
  \le e^{-Nx/2}\mathbb E_g e^{Nxg^2/2}
  =\frac{e^{-Nx/2}}{\sqrt{1-Nx}}
 \label{eq:sk-gaussian-graph-bound}
\end{equation}
when $Nx<1$. The inequality $\cosh z\le e^{z^2/2}$ follows by
integrating $\tanh z\le z$ for $z\ge0$ and using evenness. The scalar
factor $e^{-Nx/2}$ in the equality is retained. In particular,
\begin{equation}
 \sum_{\partial G=A}(\lambda/N)^{|G|}
      \le(1-\lambda)^{-1/2},\qquad 0\le\lambda<1.
 \label{eq:sk-total-graph-mass}
\end{equation}
This holds uniformly in $N$ and $A$, including the empty boundary.

\begin{proof}[Proof of Lemma~\ref{lem:sk-truncation}]
By \eqref{eq:sk-orthogonality}, $q_N\le\beta^2/N$, and
\eqref{eq:sk-total-graph-mass} at $\lambda=(\beta')^2$,
\begin{align}
 \mathbb E_J\left|\sum_{\substack{\partial G=A\\|G|>k}}w(G)\right|^2
 &=\sum_{\substack{\partial G=A\\|G|>k}}q_N^{|G|}
       \notag\\
 &\le\rho^k\sum_{\partial G=A}((\beta')^2/N)^{|G|}
 \le (1-(\beta')^2)^{-1/2}\rho^k.
 \label{eq:sk-edge-tail-proof}
\end{align}
This proves \eqref{eq:sk-edge-tail}.

For the excess bound, fix $|A|=r=2h$, an edge count $m\ge1$, and an
excess $d\ge0$. There are $v=m-h-d$ active unmarked vertices; a term
with $v<0$ is absent. Choose their labels in at most $N^v/v!$ ways.
For a specified degree list, pair $2m$ labeled half-edges in
$(2m-1)!!$ ways. Every simple graph with that degree list is represented
by exactly $\prod_i d_i!$ pairings. Allowing pairings with loops or
parallel edges only increases the numerator count. Hence the number of
simple graphs on those labels is at most
$(2m-1)!!/\prod_i d_i!$.

For integers $s\ge0$,
\begin{equation}
 \frac1{(2s+2)!}\le\frac{12^{-s}}{2s!},\qquad
 \frac1{(2s+1)!}\le\frac{6^{-s}}{s!}.
 \label{eq:sk-degree-factorials}
\end{equation}
The statements hold at $s=0$. The ratios of consecutive left-hand
denominators are at least $12(s+1)$ and $6(s+1)$, respectively, which
proves them by induction. Summing the degree lists in
\eqref{eq:sk-degrees} with $\sum_i s_i=d$ gives
\begin{equation}
 \sum_{\sum_i s_i=d}\frac1{\prod_i d_i!}
       \le 2^{-v}\frac{(v/12+r/6)^d}{d!}.
 \label{eq:sk-degree-list-sum}
\end{equation}
This is the multinomial identity applied to $v$ weights $1/12$ and
$r$ weights $1/6$.

Multiplying the graph count by $(\beta^2/N)^m$, the factorial part is
\begin{align}
 \frac{(2m)!}{2^{m+v}m!v!}
  &=\frac{\binom{2m}{m}}{4^m}
                  2^{h+d}\frac{m!}{(m-h-d)!}
       \notag\\
  &\le C_h m^{h-1/2}(2m)^d.
 \label{eq:sk-pairing-factorial}
\end{align}
Here $m!/(m-j)!\le m^j$ and
$\binom{2m}{m}/4^m\le C m^{-1/2}$. For the latter bound, use
\[
 \int_0^{\pi/2}\cos^{2m}u\,du
    =\frac\pi2\frac{\binom{2m}{m}}{4^m},
 \qquad \cos u\le e^{-u^2/2}\quad(0\le u<\pi/2),
\]
and compare the integral with a Gaussian integral. Since $v\le m$ and
$r\le2m$ whenever a nonempty graph exists, equations
\eqref{eq:sk-degree-list-sum}--\eqref{eq:sk-pairing-factorial} yield
\begin{align}
 \sum_{\substack{\partial G=A,\ |G|=m\\\delta(G)=d}}q_N^m
 &\le \beta^{2m}N^{-h-d}
       \frac{(2m)!}{2^{m+v}m!v!}
       \frac{(v/12+r/6)^d}{d!}
       \notag\\
 &\le C_r\beta^{2m}N^{-h}m^{h-1/2}
                      \frac{(C_rm^2/N)^d}{d!}.
 \label{eq:sk-excess-count-proof}
\end{align}
The empty graph is treated
separately and has excess zero.

For $x\ge0$ and an integer $B\ge1$,
$\sum_{d\ge B}x^d/d!\le x^Be^x$. Sum
\eqref{eq:sk-excess-count-proof} over $d\ge B$ and $1\le m\le k$.
The factors involving $m^2/N$ are bounded by their values at $k$, and
\[
 \sum_{m\ge1}\beta^{2m}m^{h-1/2}<\infty
\]
for every fixed $\beta<1$. This proves \eqref{eq:sk-excess-tail}.
The two omitted graph classes are disjoint, so their sums are
orthogonal by \eqref{eq:sk-orthogonality}.
\end{proof}

\subsection{Exact core and subset calculations}
\label{sec:sk-core-proof}

We give the evaluator in Lemma~\ref{lem:sk-colorful}. Fix $A$ of size
$r=2h$, cutoffs $k,B$, a support size $s\ge1$, and a coloring
$\chi:[N]\to[s]$. All calculations below are exact for arbitrary signed
edge weights. No positivity of an individual graph weight is used.

\paragraph{The core.}
Retain every vertex in $A$ and every unmarked vertex of degree at least
four. Each retained unmarked vertex contributes at least one to the
excess, so their number $b$ is at most $B-1$. Every remaining active
vertex has degree two. A component with no retained vertex is a simple
cycle and is put aside. In a component with a retained vertex, suppress
the interiors of its maximal degree-two chains. This gives a uniquely
determined core multigraph, with loops and parallel edges allowed.

Suppressing a chain reduces the edge count and the number of unmarked
vertices by the same amount. Removing a free cycle does the same.
Thus the core has $r+b$ vertices and $q=h+b+\delta$ edges, as in
\eqref{eq:sk-core-size}. Its marked vertices have positive odd degree,
and its unmarked vertices have even degree at least four. Enumerate
all symmetric nonnegative integer adjacency-multiplicity arrays with
these constraints and $q-b-h<B$. A diagonal entry counts a loop once
in $q$ and twice in the degree. The bound $q\le h+2(B-1)$ makes the
number of arrays finite at fixed $B,\ell$.

Give the $b$ unmarked core vertices labeled slots. For each array, sum
over injective assignments of these slots to $[N]\setminus A$. If
any two core vertices have the same color, this embedding contributes
zero. Otherwise let $\mathcal R$ be the core colors. Every chain
interior is restricted to colors in $[s]\setminus\mathcal R$;
in particular, it avoids every physical core vertex.

\paragraph{Paths and core loops.}
For a core endpoint $u$, define $F_u(S,x)$ by
\eqref{eq:sk-path-recursion}. Induction on $|S|$ shows that this is
exactly the sum of paths from $u$ to $x$ whose vertices after $u$
have color set $S$. In the induction step, deleting the last edge
uniquely determines the previous endpoint and removes the color
$\chi(x)$. That color was absent from the shorter path, so the
vertices remain distinct. For distinct core endpoints $u,v$, set
\begin{equation}
 K_{uv}(S)=\sum_x F_u(S,x)t_{xv},
 \qquad K_{uv}(\varnothing)=t_{uv}.
 \label{eq:sk-chain-kernel}
\end{equation}
This is the sum of chains with those endpoints and interior colors
exactly $S$.

For a core loop at $u$, use
\begin{equation}
 L_u(S)=
 \begin{cases}
  \displaystyle\frac12\sum_xF_u(S,x)t_{xu},&|S|\ge2,\\
  0,&|S|<2.
 \end{cases}
 \label{eq:sk-loop-kernel}
\end{equation}
The restriction $|S|\ge2$ excludes a loop of length one and a
length-two traversal of the same edge twice. A simple cycle through
$u$ has exactly two orientations, giving the factor $1/2$.
For a fixed $u$, the path recurrence costs $O(2^sN^2)$ arithmetic
operations. Computing all needed endpoints remains
$2^{O(s)}N^{O(1)}$.

\paragraph{Free cycles.}
For a color set $S$ with $|S|\ge3$, let $a=\min S$. Evaluate the same
path recurrence using only the colors $S\setminus\{a\}$ after the
starting vertex, and put
\begin{equation}
 C(S)=\frac12\sum_{u:\chi(u)=a}\sum_x
                        F_u(S\setminus\{a\},x)t_{xu}.
 \label{eq:sk-free-cycle}
\end{equation}
A colorful cycle has a unique vertex of minimum color and two
orientations. Thus \eqref{eq:sk-free-cycle} counts each undirected
cycle exactly once. Set $C(S)=0$ when $|S|<3$.
The weights of collections of disjoint cycles satisfy
\begin{equation}
 Z_{\mathrm{cyc}}(\varnothing)=1,\qquad
 Z_{\mathrm{cyc}}(T)=
  \sum_{\substack{S\subseteq T,\ \min T\in S\\|S|\ge3}}
                  C(S)Z_{\mathrm{cyc}}(T\setminus S)
 \quad(T\ne\varnothing).
 \label{eq:sk-cycle-collections}
\end{equation}
The component containing the minimum color of $T$ is unique. Induction
therefore proves the formula without any further factorial for cycle
permutations. Disjoint color sets ensure vertex-disjoint components.
Once the cycle kernels are known, the subset recurrence costs at most
$3^s$ operations over all $T$.

\paragraph{Joining chains.}
For arrays on subsets of a color set, use disjoint-subset convolution
\begin{equation}
 (F*G)(S)=\sum_{T\subseteq S}F(T)G(S\setminus T).
 \label{eq:sk-subset-convolution}
\end{equation}
Its unit is the array equal to one at $\varnothing$ and zero elsewhere.
Convolution enforces disjoint interiors among different chains.

For $d\ge1$ parallel core edges between distinct endpoints $u,v$,
orient every chain from the smaller endpoint to the larger endpoint.
Define $K^+_{uv}(\varnothing)=0$ and $K^+_{uv}(S)=K_{uv}(S)$ for
nonempty $S$. The combined array is
\begin{equation}
 \frac{(K^+_{uv})^{*d}
                 +d\,t_{uv}(K^+_{uv})^{*(d-1)}}{d!}.
 \label{eq:sk-parallel-chains}
\end{equation}
For $d=0$, use the convolution unit. At most one chain may be the
direct edge $\{u,v\}$: allowing two direct chains would repeat an
original edge. The first term in \eqref{eq:sk-parallel-chains} uses
no direct chain, and the second uses exactly one. All remaining
chains have nonempty disjoint interiors and are distinct. Ordering
an unordered collection gives exactly $d!$ presentations; if the
direct edge is present there are $d$ choices for its position and
$(d-1)!$ orders for the other chains. This proves the formula.

For $d$ core loops at a single vertex, use $L_u^{*d}/d!$. The individual
kernels already divide out the two loop orientations. Their disjoint
nonempty interiors make the loops distinct, and $d!$ removes their
permutations.

Convolve all core-edge arrays with $Z_{\mathrm{cyc}}$, evaluated on
exactly the colors in $[s]\setminus\mathcal R$. If the chain interiors
use $j$ colors and the free cycles use $t$ colors, then
\begin{equation}
 |V(G)|=r+b+j+t=s,\qquad
 |G|=q+j+t=q+s-r-b.
 \label{eq:sk-reconstructed-size}
\end{equation}
Use this formula to impose $|G|\le k$. The core already imposes
$\delta(G)<B$. When the core is empty, the free-cycle recurrence
alone gives the contribution. The empty graph is evaluated separately
as the constant term of \eqref{eq:sk-estimator}.

Finally, divide the contribution of each labeled core array and its
embeddings by $b!$. A physical graph has exactly $b!$ bijections between
its unmarked branching vertices and the labeled slots. All labeled
adjacency arrays were enumerated, so this remains true even if the core
has automorphisms. No additional automorphism factor is required.

\begin{proof}[Proof of Lemma~\ref{lem:sk-colorful}]
The suppressed core and its maximal chains are uniquely determined by
a retained graph. The path, loop, and free-cycle recurrences enumerate
exactly these decorations. The factors in
\eqref{eq:sk-loop-kernel}, \eqref{eq:sk-free-cycle}, and
\eqref{eq:sk-parallel-chains}, together with $b!$, remove all orientation
and labeling multiplicities. Distinct color sets prevent intersections
between interiors and between free components. Core colors are excluded
from every interior, repeated direct edges are excluded by
\eqref{eq:sk-parallel-chains}, and backtracking loops are excluded by
\eqref{eq:sk-loop-kernel}. Thus every surviving graph is simple, and
every retained colorful simple graph has coefficient one.

There are finitely many cores at fixed $B,\ell$, at most $N^{B-1}$
branch embeddings for each, and a fixed number of core edges.
All path and subset calculations cost
$2^{O_{B,\ell}(s)}N^{O(1)}$. This proves
\eqref{eq:sk-colorful-cost}.
\end{proof}

\subsection{Deterministic average-balanced colorings}
\label{sec:sk-color-proof}

We prove Lemma~\ref{lem:sk-color-family}. For a coloring $\chi$, write
$I_\chi(V)=\mathbf1_{\{\chi|_V\ \mathrm{injective}\}}$ for $|V|=s$.
The success probability of a uniform coloring is $p_s=s!/s^s$, and
\begin{equation}
 p_s\ge e^{-s}.
 \label{eq:sk-color-probability}
\end{equation}
Indeed, $\log(s!)\ge\int_1^s\log x\,dx=s\log s-s+1$.
For $m$ independent uniform colorings, the variance of the normalized
Bernoulli average at each fixed $V$ gives
\begin{equation}
 \mathbb E_{\mathcal C}Q_s(\mathcal C)
          =\frac{p_s^{-1}-1}{m}.
 \label{eq:sk-random-color-mean}
\end{equation}
We construct a deterministic family with no larger discrepancy by
choosing colors one vertex at a time.

Suppose $j$ complete colorings $\chi_1,\ldots,\chi_j$ have been chosen
and the next coloring $\chi$ is partially assigned. Define
\begin{align}
 S_0(\chi)&=\sum_{|V|=s}\mathbb E[I_\chi(V)\mid\chi\text{ partial}],
       \label{eq:sk-single-color-count}\\
 S_i(\chi)&=\sum_{|V|=s}I_{\chi_i}(V)
                    \mathbb E[I_\chi(V)\mid\chi\text{ partial}],
                        \qquad 1\le i\le j.
       \label{eq:sk-joint-color-count}
\end{align}
The expectations complete the unassigned colors independently and
uniformly. Up to a positive factor and an additive term independent of
the current coloring, the conditional expectation of the final
$Q_s$ is
\begin{equation}
 \Psi(\chi)=2\sum_{i=1}^jS_i(\chi)
                   +[1-2(j+1)p_s]S_0(\chi).
 \label{eq:sk-color-potential}
\end{equation}
To verify this, fix $V$, let $a_V=\sum_{i\le j}I_{\chi_i}(V)$, let
$I$ be the current indicator, and let $Y$ count successes among the
$m-j-1$ future colorings. In
$\mathbb E(a_V+I+Y-mp_s)^2$, use $I^2=I$ and
$\mathbb EY=(m-j-1)p_s$. The coefficient of $\mathbb EI$ is
$2a_V+1-2(j+1)p_s$. Summing over $V$ proves
\eqref{eq:sk-color-potential}.

Both counts can be evaluated without listing $\binom Ns$ sets.
Suppose the assigned color classes of $\chi$ have sizes
$n_1,\ldots,n_s$ and $u$ vertices remain unassigned. Then
\begin{equation}
 S_0(\chi)=\sum_{a=0}^s e_a(n_1,\ldots,n_s)
                                  \frac{(u)_{s-a}}{s^{s-a}},
 \label{eq:sk-single-count-formula}
\end{equation}
where $e_a$ is the elementary symmetric polynomial and
$(u)_b=u(u-1)\cdots(u-b+1)$, with $(u)_0=1$ and $(u)_b=0$ for
$b>u$. The factor $e_a$ chooses assigned vertices with different
colors. The remaining vertices are chosen in $\binom u{s-a}$ ways
and occupy the missing colors bijectively with probability
$(s-a)!/s^{s-a}$, giving the displayed falling factorial.
The values $e_a$ are computed by multiplying
$\prod_{b=1}^s(1+n_bz)$ and retaining its coefficients.

For a fixed earlier coloring $\chi_i$, let $n_{ab}$ be the number of
assigned vertices with old color $a$ and new color $b$, and let $u_a$
be the number of unassigned vertices with old color $a$. Then
\begin{equation}
 S_i(\chi)=\operatorname{per}
                \left(n_{ab}+\frac{u_a}{s}\right)_{a,b=1}^s.
 \label{eq:sk-joint-count-formula}
\end{equation}
A set colorful under $\chi_i$ chooses exactly one vertex in each old
color class. These classes are disjoint, so the new colors of the
selected unassigned vertices are independent. A permutation in the
permanent imposes distinct new colors for the different classes.
This proves \eqref{eq:sk-joint-count-formula}.

For any $s\times s$ matrix $M$, the exact subset formula is
\begin{equation}
 \operatorname{per}(M)
   =\sum_{T\subseteq[s]}(-1)^{s-|T|}
                          \prod_{a=1}^s\left(\sum_{b\in T}M_{ab}\right).
 \label{eq:sk-permanent-formula}
\end{equation}
On expansion, a row-to-column map with image $U$ has coefficient
$\sum_{T\supseteq U}(-1)^{s-|T|}$, which vanishes unless $U=[s]$.
A surjective map between two $s$-sets is a permutation, and its
coefficient is one. The formula takes $O(s2^s)$ rational operations.

For the next unassigned vertex, evaluate \eqref{eq:sk-color-potential}
for each of its $s$ colors and choose the first minimizer. The average
of these values is the previous conditional potential. Each choice
therefore does not increase the conditional expectation. Continue for
all $Nm$ assignments. Equation~\eqref{eq:sk-random-color-mean} implies
\eqref{eq:sk-color-family-bound} for the completed deterministic list.

A conservative operation count is
$m^2N^2\operatorname{poly}(s)2^s$. Class sizes are integers at most
$N$. The formulas contain products of at most $s$ such counts,
powers of $s$ in the denominators, and the factor $p_s$.
Their numerator and denominator bit lengths are polynomial in
$s\log(N+2)+s\log(s+1)+\log(m+1)$.
This proves Lemma~\ref{lem:sk-color-family} in both arithmetic and bit
complexity. When $s=1$ every coloring is exact; a repeated list of the
requested length is allowed. The empty graph requires no coloring.

\subsection{From average color balance to simultaneous error}
\label{sec:sk-color-error-proof}

We prove \eqref{eq:sk-color-error}. For brevity, write
$d_s(V)=d_{\mathcal C_s}(V)$. The coefficient of a nonempty retained
graph $G$ in \eqref{eq:sk-estimator} is
$1+d_{|V(G)|}(V(G))$. The coloring lists are deterministic and are
chosen independently of the edge disorder. Therefore
\eqref{eq:sk-orthogonality} gives the exact identity
\begin{align}
 &\mathbb E_J\left[
   \sum_{A\in\mathcal A_\ell}|\widehat T_A-T_A^{k,B}|^2
       +R_\ell|\widehat T_\varnothing-T_\varnothing^{k,B}|^2
                       \right]
       \notag\\
 &\quad=
 \sum_{\substack{G\ne\varnothing,\ |G|\le k,\ \delta(G)<B\\
                 \partial G\in\mathcal A_\ell}}
    \left(1+R_\ell\mathbf1_{\{\partial G=\varnothing\}}\right)
                  q_N^{|G|}d_{|V(G)|}(V(G))^2.
 \label{eq:sk-color-exact-error}
\end{align}
Fix a support size $s$. The sum of the coefficient before $d_s(V)^2$
over all retained graphs supported on a particular $s$-set $V$ is
independent of that set. Indeed, relabeling vertices preserves the
whole target family, both cutoffs, and the extra empty-boundary
weight. Call the common value $W_s$. The contribution of size $s$
to \eqref{eq:sk-color-exact-error} is
\begin{equation}
 W_s\sum_{|V|=s}d_s(V)^2
       =W_s\binom Ns Q_s(\mathcal C_s).
 \label{eq:sk-support-symmetry}
\end{equation}
By \eqref{eq:sk-total-graph-mass} with $\lambda=\beta^2$,
\begin{equation}
 \sum_{s\ge1}W_s\binom Ns
   \le\sum_{A\in\mathcal A_\ell}\sum_{\partial G=A}q_N^{|G|}
            +R_\ell\sum_{\partial G=\varnothing}q_N^{|G|}
   \le2C_\beta R_\ell.
 \label{eq:sk-weighted-variance-mass}
\end{equation}
Equations~\eqref{eq:sk-support-symmetry} and
\eqref{eq:sk-weighted-variance-mass} prove \eqref{eq:sk-color-error}.
There is no additional factor for the number of support sizes: their
variance masses have already been summed. A discrepancy guarantee
for every individual $V$ is not needed.

\subsection{A quantitative lower tail for the denominator}
\label{sec:sk-denominator-proof}

We first record the Gaussian concentration fact used in the argument,
including a proof so that no model-specific concentration theorem is
required.

\begin{lemma}[Gaussian concentration and distance to a set]
\label{lem:sk-gaussian-concentration}
If $F$ is a real $1$-Lipschitz function on standard Gaussian space, then
for $t\ge0$,
\begin{equation}
 \Pr(F-\mathbb EF\ge t)\le e^{-t^2/2},\qquad
 \Pr(F-\mathbb EF\le-t)\le e^{-t^2/2}.
 \label{eq:sk-gaussian-concentration}
\end{equation}
If a nonempty measurable set $E$ has Gaussian probability at least
$p>0$, its Euclidean distance function satisfies
\begin{equation}
 \Pr\!\left(d_E>\sqrt{2\log(1/p)}+t\right)\le e^{-t^2/2}.
 \label{eq:sk-set-distance}
\end{equation}
\end{lemma}

\begin{proof}
Let $\gamma$ be standard Gaussian measure and
$P_tf(x)=\mathbb E_g f(e^{-t}x+\sqrt{1-e^{-2t}}\,g)$ its
Ornstein--Uhlenbeck semigroup. For smooth positive bounded $f$ bounded
away from zero, Gaussian integration by parts gives
\begin{equation}
 -\frac{d}{dt}\int(P_tf)\log(P_tf)\,d\gamma
           =\int\frac{|\nabla P_tf|^2}{P_tf}\,d\gamma.
 \label{eq:sk-entropy-derivative}
\end{equation}
Indeed, the generator is $\Delta-x\cdot\nabla$ and its integration by
parts formula is
$\int u(\Delta-x\cdot\nabla)v\,d\gamma=-\int\nabla u\cdot\nabla v\,d\gamma$.
Moreover, $\nabla P_tf=e^{-t}P_t\nabla f$ and weighted
Cauchy--Schwarz gives
\[
 \frac{|P_t\nabla f|^2}{P_tf}
             \le P_t\left(\frac{|\nabla f|^2}{f}\right).
\]
Integrate \eqref{eq:sk-entropy-derivative} from zero to infinity and
use Gaussian invariance of $P_t$. Since $P_tf$ tends to $\int f\,d\gamma$,
\begin{equation}
 \operatorname{Ent}_\gamma(f)
 :=\int f\log f\,d\gamma
        -\left(\int f\,d\gamma\right)\log\left(\int f\,d\gamma\right)
 \le\frac12\int\frac{|\nabla f|^2}{f}\,d\gamma.
 \label{eq:sk-gaussian-log-sobolev}
\end{equation}
For $f=e^{\lambda F}$, write $\psi(\lambda)=\log\mathbb Ee^{\lambda F}$.
The last inequality implies
$\lambda\psi'(\lambda)-\psi(\lambda)\le\lambda^2/2$
when $|\nabla F|\le1$. Integrating the derivative of
$\psi(\lambda)/\lambda$ from zero proves
$\log\mathbb E e^{\lambda(F-\mathbb EF)}\le\lambda^2/2$.
Chernoff's inequality at $\lambda=t$ gives the upper tail; applying
it to $-F$ gives the lower tail. Clipping and smoothing extend the
argument to every Lipschitz $F$. The bound
$|F(x)|\le |F(0)|+|x|$ supplies the Gaussian exponential integrability
needed for the limiting steps.

The function $d_E$ is $1$-Lipschitz and vanishes on $E$.
The lower tail in \eqref{eq:sk-gaussian-concentration} gives
$p\le\Pr(d_E=0)\le e^{-(\mathbb Ed_E)^2/2}$, hence
$\mathbb Ed_E\le\sqrt{2\log(1/p)}$. Apply the upper tail to $d_E$.
Replacing $E$ by its closure, if necessary, does not change the
required conclusion.
\end{proof}

\begin{proof}[Proof of Lemma~\ref{lem:sk-denominator}]
Put $\lambda=\beta^2<1$, $M=\binom N2$, and
$W=Z_N/\mathbb E_JZ_N$. Since $Z_N$ is the ordinary spin sum,
\begin{equation}
 \mathbb E_JZ_N=2^N e^{\lambda M/(2N)},\qquad
 W=2^{-N}\sum_\sigma
       \exp\left(\frac\beta{\sqrt N}\sum_{i<j}J_{ij}\sigma_i\sigma_j
                           -\frac{\lambda M}{2N}\right).
 \label{eq:sk-annealed-normalization}
\end{equation}
For two independent uniform spin configurations, their coordinatewise
products are independent uniform signs. Let $U$ be their sum.
A direct Gaussian average gives
\begin{equation}
 \mathbb E_JW^2
    =\mathbb E_U e^{\lambda(U^2/N-1)/2}
    \le A_\lambda:=e^{-\lambda/2}(1-\lambda)^{-1/2}.
 \label{eq:sk-partition-second-moment}
\end{equation}
The bound follows from the same scalar Gaussian calculation as
\eqref{eq:sk-gaussian-graph-bound}.

Differentiate the finite sum in
\eqref{eq:sk-annealed-normalization}. Each derivative inserts
$\beta\sigma_i\sigma_j/\sqrt N$. Squaring, summing over edges, and
averaging the Gaussians gives exactly
\begin{align}
 \mathbb E_J\|\nabla_JW\|_2^2
 &=\frac\lambda2\mathbb E_U
       \left[(U^2/N-1)e^{\lambda(U^2/N-1)/2}\right]
       \notag\\
 &\le B_\lambda
     :=1+\frac\lambda2e^{-\lambda/2}(1-\lambda)^{-3/2}.
 \label{eq:sk-partition-gradient}
\end{align}
To justify the bound, expand the exponential. In each even moment,
independent Rademacher moments are bounded termwise by independent
Gaussian moments, with nonnegative multinomial coefficients. Thus
$\mathbb E(U/\sqrt N)^{2j}\le(2j-1)!!$, and
\[
 \mathbb E\left[(U^2/N)e^{\lambda U^2/(2N)}\right]
       \le(1-\lambda)^{-3/2}.
\]
Dropping the negative term proves the upper bound in
\eqref{eq:sk-partition-gradient}. This does not differentiate an
inequality; the gradient identity was computed directly.

Since $\mathbb EW=1$, Cauchy--Schwarz yields
\[
 1\le\frac12+
         \sqrt{\mathbb EW^2\Pr(W\ge1/2)},\qquad
 \Pr(W\ge1/2)\ge\frac1{4A_\lambda}.
\]
Markov's inequality applied to \eqref{eq:sk-partition-gradient} shows
that the set
\begin{equation}
 E=\{W\ge1/2,\ \|\nabla W\|_2\le\sqrt{8A_\lambda B_\lambda}\}
 \label{eq:sk-good-gradient-set}
\end{equation}
has probability at least $p_\lambda=1/(8A_\lambda)$. On this set,
$\|\nabla\log W\|_2\le L_\lambda:=2\sqrt{8A_\lambda B_\lambda}$.
The function $\log W$ is convex: its Hessian is the covariance matrix
of the coefficient vectors of the affine functions in the log-sum of
exponentials. For every $x$ and $y\in E$, the supporting plane gives
\begin{equation}
 \log W(x)\ge\log W(y)+\nabla\log W(y)\cdot(x-y)
             \ge-\log2-L_\lambda\|x-y\|_2.
 \label{eq:sk-convex-lower-bound}
\end{equation}
Take the distance from $x$ to $E$ and apply
\eqref{eq:sk-set-distance}. After adjusting constants for bounded $u$,
\begin{equation}
 \Pr(\log W<-u)\le C_\beta e^{-c_\beta u^2},\qquad u\ge0.
 \label{eq:sk-logW-tail}
\end{equation}
Only the gradient on the positive-probability set $E$ was bounded;
no dimension-free global gradient bound is asserted.

It remains to pass from $W$ to $D$. Equations
\eqref{eq:sk-spin-graph} and \eqref{eq:sk-annealed-normalization},
together with $\log\cosh x\le x^2/2$, imply
\begin{equation}
 \log D\ge\log W-\frac\lambda{2N}
                       \left(\sum_{e=1}^M J_e^2-M\right).
 \label{eq:sk-logD-comparison}
\end{equation}
For $Q=\sum_eJ_e^2-M$ and $0<t<1/2$, elementary Gaussian integration
and the series for $-\log(1-2t)$ give
\begin{equation}
 \log\mathbb E e^{tQ}
   =M[-t-\tfrac12\log(1-2t)]
   \le\frac{Mt^2}{1-2t}.
 \label{eq:sk-chi-square-mgf}
\end{equation}
For $x>0$, put $a=\sqrt{x/M}$ and $t=a/(1+2a)$ in Chernoff's bound.
The resulting exponent at $2\sqrt{Mx}+2x$ is at most $-x$, so
\begin{equation}
 \Pr(Q>2\sqrt{Mx}+2x)\le e^{-x}.
 \label{eq:sk-chi-square-tail}
\end{equation}
For $0<u\le4\lambda N$, take $x=u^2/(32\lambda^2)$. Since
$M\le N^2/2$, the threshold in \eqref{eq:sk-chi-square-tail} is at
most $Nu/(2\lambda)$. The event $\{\log D<-u\}$ is contained in
\[
 \{\log W<-u/2\}\ \cup\ \{Q>Nu/\lambda\}.
\]
Equations~\eqref{eq:sk-logW-tail} and \eqref{eq:sk-chi-square-tail}
therefore bound its probability by $C_\beta e^{-c_\beta u^2}$ in this
range, after changing constants. Set $u=s\log N$. For every fixed
$s>0$, the range restriction holds for sufficiently large $N$ and
proves \eqref{eq:sk-denominator-tail}.
\end{proof}

\subsection{Finite precision and the polynomial bit algorithm}
\label{sec:sk-bits}

We complete the numerical claim used in
\eqref{eq:sk-arithmetic-budget}. The ideal estimates above are applied
to the underlying continuous Gaussian disorder. Rounding is a
deterministic perturbation on a high-probability input box, not a
replacement disorder distribution.

\paragraph{Parameters and the input box.}
Given a certified approximation procedure for a fixed $\beta<1$,
refine it until rationals $\beta<\bar\beta<\beta'<1$ are known.
The search has a fixed parameter-dependent cost. The integer $k$ in
\eqref{eq:sk-cutoffs} can then be found by repeated rational
multiplication. Because $\bar\rho<1$, this takes $O_\beta(\log N)$
steps. The powers defining $m_s$ have integer exponents, and $p_s$
is rational, so all truncations and family sizes are computable.

Put $L_N=\lceil\log_2(N+2)\rceil$ and use a box
\begin{equation}
 \mathcal B_N=\{\max_{e\in E_N}|J_e|\le C_0 L_N\}.
 \label{eq:sk-input-box}
\end{equation}
A scalar Gaussian tail and a union bound give
\begin{equation}
 \Pr(\mathcal B_N^c)
       \le 2M e^{-C_0^2L_N^2/2}
       \le C e^{-c(\log(N+2))^2}.
 \label{eq:sk-box-tail}
\end{equation}
For every fixed requested exponent, this is smaller than the required
failure budget for sufficiently large $N$. The finite input consists
of rounded entries on this box, with an overflow flag outside it.
One can equivalently test rounded entries in a slightly enlarged box;
the width change affects only the constants in
\eqref{eq:sk-box-tail}.

\paragraph{Sensitivity of the output polynomials.}
For fixed color lists, $\widehat T_A$ is a polynomial in the edge
weights. Each retained graph has degree at most $k$, and its color
multiplier lies between zero and $p_s^{-1}\le e^s$.
There are at most $(k+1)(1+M)^k$ edge subsets of size at most $k$.
Thus a computable common upper bound for the absolute coefficient
sum is
\begin{equation}
 \mathfrak A_N=(k+1)(1+M)^k3^{s_{\max}},
 \qquad
 \mathfrak A_N=\exp[O_{\beta,\ell,T}(\log^2(N+2))].
 \label{eq:sk-coefficient-mass}
\end{equation}
We used $e^s\le3^s$. On the cube $|t_e|\le1$, the sum of the absolute
first derivatives of any output polynomial is at most
$k\mathfrak A_N$. Hence, for two edge arrays in that cube,
\begin{equation}
 |\widehat T_A(t)-\widehat T_A(\widetilde t)|
       \le k\mathfrak A_N\max_e|t_e-\widetilde t_e|.
 \label{eq:sk-polynomial-sensitivity}
\end{equation}
Choose an integer $Q>c+\ell+4$ and choose a dyadic precision $b$ so
that
\begin{equation}
 k\mathfrak A_N2^{-b}\le N^{-Q}.
 \label{eq:sk-bit-choice}
\end{equation}
This choice is explicit, uses only integer or rational comparisons,
and has $b=O_{\beta,\ell,a,c}(\log^2(N+2))$.
Obtain dyadic $\widetilde t_e\in[-1,1]$ with
$|\widetilde t_e-t_e|\le2^{-b}$ and evaluate
\eqref{eq:sk-estimator} exactly on those dyadic weights.
Then each output changes by at most $N^{-Q}$, so
\begin{equation}
 \sum_{A\in\mathcal A_\ell}|\widetilde T_A-\widehat T_A|^2
       +R_\ell|\widetilde T_\varnothing-\widehat T_\varnothing|^2
       \le2R_\ell N^{-2Q}
       \le\frac{N^{-2c-2}}{1024}
 \label{eq:sk-rounding-vector-bound}
\end{equation}
for sufficiently large $N$. This is
\eqref{eq:sk-arithmetic-budget}.

The map $\tanh$ is globally $1$-Lipschitz. On
\eqref{eq:sk-input-box}, approximating $J_e$, $\beta$, and $N^{-1/2}$
to $b+O(\log(N+2))$ fractional bits suffices to obtain the stated
edge accuracy. Square roots can be computed by bisection, and
exponentials by range reduction and their convergent series.
Then $\tanh x=(e^{2x}-1)/(e^{2x}+1)$ is evaluated to a certified
absolute error. Only an approximate dyadic value is needed: the
algorithm never has to decide an exact nearest-dyadic tie. These
operations have polynomial bit cost at the chosen precision.

\paragraph{Exact rational evaluation.}
After constructing the dyadic edge weights, use exact rational
arithmetic for the path, core, and coloring calculations.
An intermediate path or core term has $O(k+\ell)$ edge factors;
its dyadic denominator divides $2^{O(b(k+\ell))}$.
The core factorials involve only the fixed parameters $B,\ell$.
Cycle orientation factors contribute at most $2^{O(k+\ell)}$.
The color averages involve $m_sp_s=m_ss!/s^s$, and
\begin{equation}
 \sum_{s=1}^{s_{\max}}
       \bigl(\log(m_s+1)+s\log(s+1)\bigr)
       =O_{\beta,\ell,T}(\log^2(N+2)\log\log(N+3)).
 \label{eq:sk-color-denominator-bits}
\end{equation}
A common denominator for these factors therefore has
$O_{\beta,\ell,a,c}(\log^3(N+2))$ bits.
Absolute sums of represented intermediate path and core terms are
at most $N^{O_{B,\ell}(k+1)}$; their numerators have the same
$O(\log^3(N+2))$ bit bound after reduction. This argument allows
large signed cancellations and does not assume that intermediate
values have polynomial magnitude. Exact integer and rational
operations at this bit length take polynomial time.

The color-family potentials have the rational bit bounds proved in
Section~\ref{sec:sk-color-proof}. For each $s$, construction costs
$m_s^2N^2\operatorname{poly}(s)2^s$, and the colorful evaluation has
cost \eqref{eq:sk-colorful-cost}. Since $s=O(\log N)$, $m_s=N^{O(1)}$,
and there are at most $C_\ell N^\ell$ boundaries, the total bit cost
is polynomial in $N$.

\paragraph{Ratios, logarithms, and the underlying target.}
On \eqref{eq:sk-success-event}, the denominator exceeds the safe
threshold by \eqref{eq:sk-relative-denominator}. Exact rational
division and clipping give the output vector in
\eqref{eq:sk-stable-ratios}. To compute
\eqref{eq:sk-log-output}, note that $\log\cosh x$ is $1$-Lipschitz.
The input and function-evaluation errors in its $M$ summands can
therefore be made to have total at most $N^{-c}/12$ by adding
$O(\log N)$ accuracy bits. Approximate $N\log2$ to the same tolerance,
and approximate $\log\widetilde T_\varnothing$ to $N^{-c}/12$.
The latter is computable by range reduction and a logarithm series
from a positive rational with polynomial bit length; the lower bound
$\widetilde T_\varnothing\ge1/(2N)$ is available whenever it is used.
These choices give the additional $N^{-c}/4$ error budget in the main
text. Outside the safe threshold, return the stated fallback instead
of taking a logarithm.

The output has polynomial length. On the input box,
\begin{equation}
 |\log Z_N|
       \le N\log2+\frac\beta{\sqrt N}\sum_{e\in E_N}|J_e|
       =O_\beta(N^{3/2}\log(N+2)).
 \label{eq:sk-output-size}
\end{equation}
One may report $\widehat L$ directly. Exponentiating it to the requested
relative accuracy also has polynomial bit cost, using either an
ordinary binary representation or a mantissa and exponent.

Finally, \eqref{eq:sk-total-mean-square} concerns the ideal estimator
at the true unrounded edge weights. Equation
\eqref{eq:sk-rounding-vector-bound} bounds its deterministic change
on the input box. Combining the two only on that box, and adding
\eqref{eq:sk-box-tail} to the failure probability, proves the theorem
for the original continuous-disorder model. No Gaussian orthogonality
claim is made for the rounded weights. This completes the proof of
Theorem~\ref{thm:sk-main} in bit complexity.

\end{document}